\documentclass[graybox, nosecnum]{svmult}

\usepackage{mathptmx}       
\usepackage{helvet}         
\usepackage{courier}        
\usepackage{type1cm}        
\usepackage{makeidx}         
\usepackage{graphicx}        
\usepackage{multicol}        
\usepackage{graphicx,latexsym,amsmath,amssymb}
\usepackage[bottom]{footmisc}
\usepackage{hyperref}        
\usepackage{soul}            
\hypersetup{colorlinks=true,urlcolor=blue}
\usepackage[square,numbers]{natbib}
\makeindex             

\usepackage[export]{adjustbox}

\def\mbul{{M}_{\rm BH}}

\newcommand {\kms}{\ensuremath{{\rm km\,s}^{-1}}}

\newcommand{\sbul}{a_{\bullet}}
\newcommand{\met}{\mbox{g}}

\def\pderiv(#1/#2){\frac{\partial#1}{\partial#2}}
                    
 \newcommand{\derd}{{\rm d}}

\newcommand{\Ms}{{\rm ~M}_\odot}

\newcommand{\gw}{{\rm GW}}
\newcommand{\gc}{{\rm GC}}

\newcommand{\ibh}{{\rm IMBH}}

\newcommand{\imri}{{\rm IMRI}}

\newcommand{\bh}{{\rm BH}}

\def\mnras{Mon. Not. R. Astron. Soc. }
\def\ssr{Space Sci. Rev.}
\def\apj{Astrophys. J.}
\def\apjl{Astrophys. J. Lett.}

\def\aap{Astron. Astrophys.}

\def\araa{Annu. Rev. Astron. Astrophys.}

\def\baas{Bull. Am. Astron. Soc.}
\def\prd{Phys. Rev. D}
\def\nar{New Astron. Rev.}

\begin{document}

\title*{The gravitational capture of compact objects by massive black holes}
\titlerunning{Gravitational capture of COs by SMBHs}

\author{Pau Amaro Seoane 
\thanks{
PAS acknowledges support from the Ram{\'o}n y Cajal Programme of the Ministry
of Economy, Industry and Competitiveness of Spain, as well as the COST Action
GWverse CA16104. This work was supported by the National Key R\&D Program of
China (2016YFA0400702) and the National Science Foundation of China (11721303).
He is indebted with Marta Masini for her support during the lockdown, without
which this work would not have been possible.
}
}
\institute{Pau Amaro Seoane \at 
Universitat Polit{\`e}cnica de Val{\`e}ncia, IGIC, Spain\\
DESY Zeuthen, Germany\\
Kavli Institute for Astronomy and Astrophysics, Beijing, China\\
Institute of Applied Mathematics, Academy of Mathematics and Systems Science, CAS, Beijing, China\\
Zentrum f{\"u}r Astronomie und Astrophysik, TU Berlin, Berlin, Germany\\
\email{amaro@upv.es}
}
\maketitle

\abstract
{
The gravitational capture of a stellar-mass compact object (CO) by a
supermassive black hole is a unique probe of gravity in the strong field
regime.  Because of the large mass ratio, we call these sources extreme-mass
ratio inspirals (EMRIs).  In a similar manner, COs can be captured by
intermediate-mass black holes in globular clusters or dwarf galaxies. The mass
ratio in this case is lower, and hence we refer to the system as an
intermediate-mass ratio inspiral (IMRI).  Also, sub-stellar objects such as a
brown dwarf, with masses much lighter than our Sun, can inspiral into
supermassive black holes such as Sgr A* at our Galactic centre. In this case,
the mass ratio is extremely large and, hence, we call this system ab
extremely-large mass ratio inspirals (XMRIs).  All of these sources of
gravitational waves will provide us with a collection of snapshots of spacetime
around a supermassive black hole that will allow us to do a direct mapping of
warped spacetime around the supermassive black hole, a live cartography of
gravity in this extreme gravity regime.  E/I/XMRIs will be detected by the
future space-borne observatories like LISA.  There has not been any other probe
conceived, planned or even thought of ever that can do the science that we can
do with these inspirals. We will discuss them from a viewpoint of relativistic
astrophysics.
}

\section{Keywords} 
gravitational captures, supermassive black holes, stellar dynamics, LISA, tests of general relativity

\section{Introduction - Why is this important?}

\begin{figure}
          {\includegraphics[width=1.0\textwidth,center]{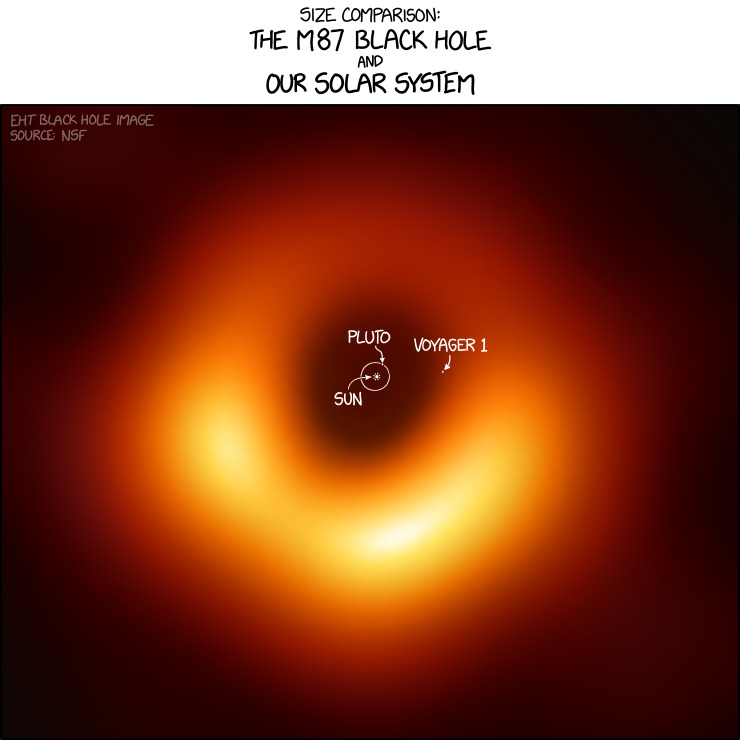}}
\caption
   {
The shadow of the dark object found in M87. From
\href{https://xkcd.com/2135/}{https://xkcd.com/2135/} and with the XKCD
team authorization, using the original figure from the EHT team web page,
\href{https://eventhorizontelescope.org/}{https://eventhorizontelescope.org/}
   }
\label{fig.m87_black_hole_size_comparison}
\end{figure}

For many years we have known that at the centre of most nearby bright galaxies
a very massive, compact and dark object must be lurking
\citep[e.g.,][]{2004cbhg.symp....1K,2010RvMP...82.3121G,2013ARA&A..51..511K}.
Recently, the Event Horizon Telescope finally delivered the result that we were
expecting; the ``shadow'' of one of these dark objects at the centre of the
galaxy M87, with a mass of $6.5\times10^7\,M_{\odot}$
\citep{2019ApJ...875L...1E}.  In
Fig.~(\ref{fig.m87_black_hole_size_comparison}) we show their result with the
position of Voyager1 and Pluto if our Sun was located at the centre of the
image. This is exceptionally \textit{to scale}, Note thoug, as the author of
the comic told us, that the blurring makes the dark shadow look smaller than it
is. But the diameter of the brightest part of the ring should be $\sim 640$ AU.

In 2020, the Nobel prize went to Roger Penrose, Reinhard Genzel and Andrea
Ghez. The two latter have devoted a significant part of their careers to study
an interesting phenomenon happening at our own Galactic Centre: A cluster of
young stars revolving around a point of small size, a radio source, on
observable timescales. After analysing the orbits, Genzel's and Ghez' groups
came to the conclusion that these stars, which are called the S- or SO-stars,
are orbiting a point mass of about $4\times 10^6\,M_{\odot}$ enclosed in a
volume as small as $1/3$ the distance between the Earth and the Sun. We call
this ``point'' mass Sgr A* (see for a review \citealt{2010RvMP...82.3121G}, and
references therein). The properties of these stars are interesting. For
instance, S4714 moves at about 8\% the speed of light, and S62 is as close as
16 AU from SgrA* \citep{2020ApJ...899...50P}. The team of Genzel presented this
year the detection of the Schwarzschild precession in the orbit of S2 after its
second periapsis passage.  According to their measurements, the predictions of
general relativity (GR) and the observations are in agreement by 17\%
\cite{2020A&A...636L...5G}.

Therefore, even if we still cannot yet completly exclude other exotic
possibilities that nature might have come up with, it looks like SgrA* and the
dark object located at the centre of M87 are supermassive black holes, and this
is what we will assume in this chapter. In that case, the Event Horizon
Telescope has measured the event horizon to be of about 0.0013 pc
\citep{2019ApJ...875L...1E}. We will however have a stronger evidence when we
detect the gravitational waves (GWs) from a system involving a supermassive
black hole (SMBH), in particular an E/I/XMRI.

In this chapter we focus on this particular source of GWs. In the literature,
the most studied inspiral has so far been the gravitational capture of a
stellar-mass compact object (i.e. a stellar-mass black hole, a neutron star or
a white dwarf) by a SMBH is. Because in their orbits and in a matter of a few
years, these sources significantly shrink their semi-major axes, precess and if
the SMBH is Kerr (and that is the most likely situation in nature, see e.g.
\cite{SesanaEtAl2014}), the plane of the orbit changes, so that they map
spacetime around SMBHs. The number of cycles that they describe around the SMBH
is inversely proportional to the mass ratio, so that an EMRI will provide us
with a collection of about $\sim 10^5$ snapshots of spacetime around a SMBH.
This translates literally into a cartography of warped spacetime in the strong
regime of gravity. In this sense, these sources of GWs are unique.

The ideal instrument to look for this kind of inspirals is the ESA/NASA mission
The Laser Interferometer Space Antenna \citep[LISA,
see][]{Amaro-SeoaneEtAl2017}, because the total mass is too large for
ground-based detectors (with the exception of some particular cases of IMRIs,
as we will discuss).  LISA consists of three spacecraft arranged in an
equilateral triangle of $2.5\times 10^{\,6}\,\textrm{km}$ armlength flying on a
heliocentric orbit, with a strain sensitivity of less than 1 part in $10^{20}$
in frequencies of about a millihertz.

Most of this chapter will be devoted to addressing the question of how do you
form these sources. We will see that not all phase-space is at our disposition
to create one of these inspirals, and that the conditions change a lot
depending on whether it is an EMRI, an IMRI or an XMRI.

\subsection{Fundamental science}

EMRIs, IMRIs and, within some limitations, XMRIs are probes of fundamental
physics with GWs, mostly due to the fact that these inspirals spend a large
number of cycles very close at the verge of the the last stable orbit, where
precessional are the strongest~\citep[see e.g.][]{2017PhRvD..95j3012B,2019BAAS...51c..42B}.

An EMRI has three fundamental frequencies associated to it
\citep{Amaro-SeoaneEtAl07,Amaro-SeoaneGairPoundHughesSopuerta2015} which are
encoded in the gravitational wave emitted by the system. The associated
timescales depend on the Kerr geometry and will evolve due to gravitational
backreaction, which can evolve in very different ways depending on the type of
gravity one considers, or whether there are extra dimensions or fields. In this
sense, the detection and extraction of this information can be used as a tool
to explore these new ideas, or might turn into something that we had not
previously considered or even thought of.

Because we want to extract information from the modelling of a source that will
last months to years, the waveforms ab definitio must be very precise and
therefore subject to important (accumulated) deviations if what we are
measuring does not correspond to the general relativity picture. In some sense,
this might be regarded as a double-edged sword.  On the one hand we can extract
very accurately gravitational multipole moments describing the SMBH geometry,
as well as other parameters that provide us with information about
modifications of GR, such as coupling constants and extra dimension length
scales. On the other hand, if such modifications are there and are important,
we might lose the source altogether. Therefore, we can in principle use EMRIs
as they generate GWs~\citep{2019RPPh...82a6904B} as tools to explore the nature and geometry of SMBHs, i.e. the no-hair
conjecture\citep[see
e.g.][]{Amaro-SeoaneEtAl07,Amaro-SeoaneGairPoundHughesSopuerta2015},
complementary in a sense with the detection of quasinormal models excited in
the ringdown of a SMBH binary merger~\citep{Berti:2018vdi}, and rule out exotic
objects such as e.g. boson
stars~\citep{Amaro-SeoaneBarrancoBernalRezzolla10,Hannuksela:2018izj}.
This is particularly true for multi-bandwidth detections, i.e. IMRIs, which can
be detected in the early inspiraling phase by LISA and later by ground-based
detectors~\citep{Amaro-Seoane2018,2020arXiv200612137D}, because the combined 
detection would help to break up different degeneracies in the parameter extraction.

EMRI GWs are subject to be affected by additional mechanisms that might alter
them on their way towards us, such as extra polarizations, gravitational parity
violation, breaking of Lorentz invariance, all of them related to high-energy
effects~\citep[see][]{2020GReGr..52...81B}. There are also other potential
effect that could importantly modify the waves, such as the aberrational- and
beaming effects, and their
combination~\citep{Torres-OrjuelaEtAl2020,Torres-OrjuelaEtAl2019}, which emerge
when the source is moving relative to us.

In a more speculative manner, we could also use E/IMRIs to explore the
possibility that primordial black holes do exist~\citep{2020arXiv200212778C},
assuming that we have confidence in our astrophysical modelling of what the
mass of the CO is, which might not necessarily be true. This connects to the
idea that dark matter could be constituted of such primordial black
holes~\citep[see the review of][]{2020ARNPS..7050520C}

For astrophysics, these sources of GWs are important because we can try to
reverse-engineer the information that they will provide us with to probe
regions of the galaxy which are unaccessible to light because of obscuration or
distance. 

\section{Extreme-mass ratio inspirals}

\subsection{A long story short}

The problem of how a CO could become an EMRI in a galactic nucleus is, as of
writing these lines, a problem which goes back in time three decades to the
best of our knowledge
\citep{1997MNRAS.284..318S,HilsBender1995,Haehnelt1994,Schutz1989}. The first
reference uses the fundamentals of the theory of loss-cone in the context of
relaxation to derive the rates and characteristics. It is natural that this
first attempt happened at that time and place, because the study of stellar
disruptions was one of the main interests of the second author of the paper.
However, as compared to a tidal disruption, the slow, progressive inspiral
of a CO is a more challenging problem, as we will see.

We will adopt the Milky Way as a typical, reference galaxy host to the kind
of sources that we are going to address in this chapter. The mass of
the central SMBH will have that of SgrA*, $4\times 10^6\,M_{\odot}$,
or of the order of it. The fiducial CO we will consider is a stellar-mass 
black hole of mass $m_{\rm BH}=10\,M_{\odot}$ for historical reasons: This
was the default assumed mass before the first discoveries of LIGO/Virgo,
since we cannot explain how to form stellar-mass black holes with larger
masses via (reasonable) stellar evolution.

In this chapter we are going to focus in detail on the most well-understood
mechanism (in the sense that the number of free parameters is very small),
which is relaxation, more specifically two-body relaxation. Other mechanisms
have been proposed, in particular the so-called ``resonant relaxation''
\citep{RT96}, which seemed very important a few years ago. Another interesting
scenario has been proposed, which involves the tidal separation of a binary in
which one of the stars is a CO. This possibility is an interesting one, but we
are missing a fundamental piece of information, which is the disitribution of
initial semi-major axes and eccentricities, as well as the fraction of such
binaries harbouring a CO. Nonetheless, since the mechanism is simple and robust
enough, we will also address it in some detail.

To summarise in a few paragraphs what has been the work of decades, the study
of EMRI event rate was addressed in the framework of two-body relaxation, since
the problem shared similarities to the tidal disruption of a star. The event
rate for a typical galaxy (i.e. a Milky-Way-like galaxy, as stated before) is
very low, of the order of $10^{-5}-10^{-6}~\textrm{yr}^{-1}$ \citep[see
e.g.][and references therein]{2018LRR....21....4A}. Because of the dynamical
properties that the COs have when they form a potential EMRI source with the
central SMBH, i.e. have very large eccentricities and semi-major axis of about
$0.1-1~\textrm{pc}$, at apocentre they risk being scattered off the EMRI orbit
via the accumulation of gravitational tugs from other stars (we will elaborate
this later). If we perturb the apocentre, the pericentre is also perturbed and
the fate of the potential EMRI is twofold: It can either increase more and more
the initial eccentricity to the point that it forms almost a straight line
which leads it to a directly cross the event horizon of the SMBH or it can
simply be reabsorbed by the stellar system. In the first case, it would produce
one burst of gravitational radiation and then be lost in terms of GWs
\citep{2013MNRAS.429..589B,2007MNRAS.378..129H}.  It is important to note here
that such orbits have unfortunately been dubbed in the (astrophysical)
literature as ``plunges'' or ``direct plunges'', which leads to confusion and
should be avoided, since any EMRI will eventually cross the event horizon of
the SMBH.  This is referred to in other texts as the plunge of the orbit. We
will only use the term ``plunge'' in the sense that the CO crosses the event
horizon of the SMBH.  In the second, it would never emit (detectable) GWs. A
better term would be a ``1-burst orbit'', because the system emits one intense
burst of GWs and then it is lost. We note there are scenarios in which we can
have repeated bursts coming parabolic orbits, not bound ones, and the rate is
of about one burst per observation year \citep[see references in][]{Amaro-SeoaneLRR}. We are not
referring to these here.

The fact is that the assumptions under the classification of orbits which led
to consider 1-burst orbits have been highly simplified in the literature. In
reality, both for Schwarzschild and Kerr SMBHs, 1-burst orbits are extremely
difficult to achieve. The work of \citep{Amaro-SeoaneSopuertaFreitag2013}
illustrates this with an example. The authors take a SMBH with no spin,
Schwarzschild, and analyse orbits for a stellar-mass black hole with such an eccentricity that would
lead to a 1-burst scenarios. 

The authors then calculate the number of periapsis passages before it plunges
through the event horizon. From the initial semi-lactus rectum, eccentricity
and inclination ${p}$, ${e}$, ${\iota}$, they calculate the constants of motion
${E}$, ${L}_z$, $C$ (energy, z-component of the angular momentum and Carter's
constant), the average flux of these ``constants'', i.e. the average time
evolution, $\dot{E}$, $\dot{L}_z$, and $\dot{C}$. This allows them to calculate
the time to to go from apoapsis and back (radial period), and thus the change
in ${E}$, ${L}_z$, $C$ and so the new constants of motion, $p_{\rm new}$,
$e_{\rm new}$, ${\iota}_{\rm new}$. The authors find that the family of 
separatrices in the Kerr case deviate from the Schwarzschild case depending
on (1) the spin of the SMBH, (2) the inclination of the orbit and (3) whether
the orbit is pro- or retrograde.

\begin{figure}
          {\includegraphics[width=0.8\textwidth,center]{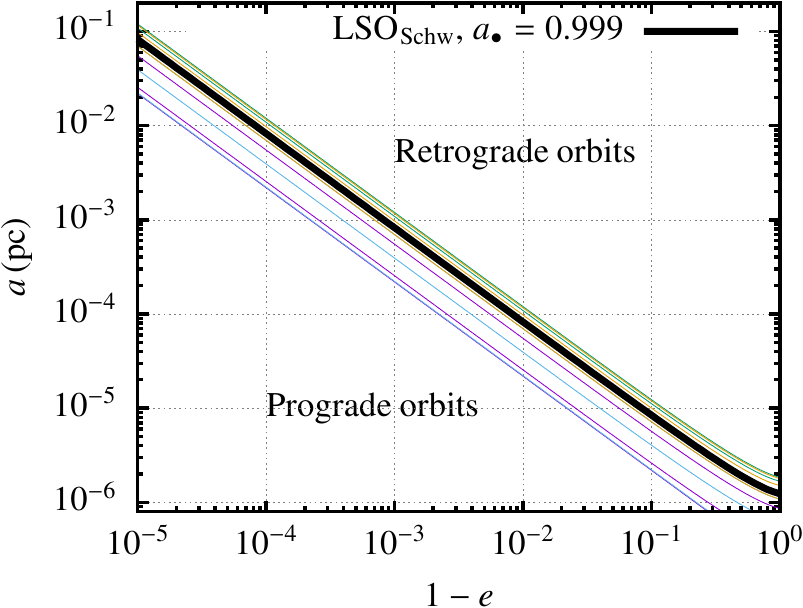}}
\caption
   {
Different separatrices for a Kerr SMBH of $a_{\bullet}=0.999$. The separatrices which
correspond to retrograde orbits are shown above the Schwarzschild LSO (black, thick
line) and prograde below it.
   }
\label{fig.LSO}
\end{figure}

For prograde orbits, the last stable orbit (LSO) is closer to the event horizon
and, hence, the possibility that the CO is on a 1-burst orbit becomes more
rare. Retrograde orbits, on the contrary, have LSOs farther away than in the
Schwarzschild case. Therefore, it is easier to find 1-burst situations.
Nevertheless, the situation is not symmetric. The farthest possible away
separatrix for a retrograde orbit is much closer to the Schwarzschild
separatrix than the equivalent for a prograde orbit.  In Fig.~(\ref{fig.LSO}),
which is using the same method as in \citep{Amaro-SeoaneSopuertaFreitag2013},
we can see this. The green separatrices for retrograde orbits are closer the
the Schwarzschild case than the blue ones, which correspond to prograde orbits.

The work of \citep{Amaro-SeoaneSopuertaFreitag2013} shows that the number of
cycles in the band of the detector before plunging through the event horizon is
as large as $10^5$ for some prograde cases, and at least of a few hundreds,
depending on the spin and inclination. Retrograde cases, however, can lead more
easily to 1-burst orbits. However, due to the asymmetry, these are much more
rare.

It must be stressed out that, \textit{even in the Schwarzschild case, it is
difficult to have 1-burst orbits}. This is so because the energy radiated away
at periapsis is proportional to $q$, so that even for orbits with very small
periapsis distances the EMRI in general cannot radiate away the required amount
of energy for the CO to plunge.

There are exceptions to this situations which, however, are fine-tuned, such as
the zoom-whirl orbits \citep{GlampedakisEtAl2002b}.  In this case, even if we
only have one radial oscillation, there is an arbitrary number of (different)
oscillations along the azimuthal axis, which translate into a very strong
precession of the orbit.  

When a Keplerian orbit extends beyond the potential well in radius, the orbits
are non-closed anymore \citep[the Newtonian precession, see
e.g.][]{Amaro-SeoaneLRR}.  Such non-closed orbits have a periapsis advance as a
consequence. There are two periods in these orbits, the orbital period, is the
required time to go from apo- to periapsis and another one is the time to
revolve $2\pi$ in the orbit. These two periods are identical in the Keplerian
case of closed orbits. In the relativistic case these two periods always
differ, and the more relativistic, the larger the difference, to the point of
being arbitrarily different. This leads to a precession of the orbit with an
arbitrary number of cycles in a single radial period, if the initial conditions
are matched. This means that one needs a high degree of fine tuning in the
Hamiltonian system which describes the initial dynamical parameters.  In such a
case, the amount of energy radiated is considerable and it could lead to the CO
to indeed plunge through the event horizon after one single periapsis passage.  

In the Kerr case the physical picture is even more complex, depending on
whether the orbit is pro- or retrograde. If it is prograde, since the orbit
gets closer to the LSO, it will induce a stronger precession of the orbit than
in the Schwarzschild case. If it is retrograde, the number of cycles will be
less than in the Schwarzchild case.

Another possibility would be that the SMBH is Kerr and the orbit of the CO is
perfectly parallel and aligned with the axis of the spin of the SMBH. In that
case, from the perspective of the CO, the SMBH would be de facto a
Schwarzschild SMBH and this pecular situation could have a 1-burst orbit. 

These are nonetheless very peculiar configurations and tuning of the initial
conditions of the orbits, so that the probability for these scenarios to happen
frequently in the formation of EMRIs is low.

The believe that most sources would be lost to 1-burst orbits led a number of
researchers to look at alternative scenarios, in particular to resonant
relaxation.  The idea was that by getting closer and closer to the SMBH, the
number density of stars would decrease and, hence, the danger of turning a
successful EMRI orbit into a 1-burst orbit would decrease. But, at the same time,
because of the drop of density, the timescale associated to relaxation would
increase more and more, so that the event rate would drastically drop. This is
why the concept of (scalar) resonant relaxation as presented by
\cite{1996NewA....1..149R} was envisaged as a possible way to enhance the rate.

This motivation led to interesting discoveries in the field of theoretical
stellar dynamics (see in particular the work of
\citep{2006ApJ...645.1152H,2009ApJ...698..641E,2014CQGra..31x4003B,2016MNRAS.458.4143S,2018ApJ...860L..23B}).
However, even in the range of radii in which scalar resonant relaxation was
thought to be a promising driver of EMRIs, it is now well settled that in the
end two-body relaxation is the main mechanism..  Therefore, in this work we
will focus on two-body relaxation and loss-cone theory. For more details about
resonant and scalar relaxation, and alternative scenarios, we refer the reader
to the review \cite{2018LRR....21....4A}.

\subsection{Stellar tidal disruptions}
\label{Sec.Disr}

The theory on which we address the problem of formation and evolution of EMRIs
is two-body relaxation \citep[see e.g.][]{BinneyTremaine08,Spitzer87} in the
context of stellar tidal disruptions (for a recent review of the
theory and derivation of the rates, and references therein \cite{StoneEtAl2020},
and for some classical references, see \cite{Rees88,MT99,SU99,WM04}).

When an extended star (meaning a star which is not a CO), such as our Sun,
approaches ``too closely'' a SMBH, it will suffer a difference of gravitational
forces on points diametrically separated on its structure due to the fact that
it is not a point particle. Hence, depending on how close the star, typically
on a hyperbolic orbit, gets to the SMBH, it will get disrupted or not, since we
have to compare the work exerted over it by the tidal force with its own
binding energy.  The radius within which this happens is what we call the tidal
radius $r_{\rm t}$, and can be derived to be \citep[see e.g. Sec.~1.3
of][but note the small typo, the factor 2 should be outside of the brackets]{Amaro-SeoaneLRR2012}

\begin{equation}
r_{\rm t}=\Bigg[\frac{(5-n)}{3} \frac{M_{\rm BH}}{m}\Bigg]^{1/3} 2r,
\label{eq.r_tid_bind}
\end{equation}

\noindent 
with $r$ the radius of the star, $m$ its mass, $M_{\rm BH}$ the mass of the SMBH and $n$ the polytropic index.
For our Sun, considering a $n=3$ polytrope, and $M_{\rm BH}=10^6\,M_{\odot}$,

\begin{equation}
r_{\rm t} \simeq 1.83\times 10^{-6}~\textrm{pc} \simeq 0.38~\textrm{AU}.
\label{eq.r_tib_bind_b}
\end{equation}

\begin{figure}
          {\includegraphics[width=0.9\textwidth,center]{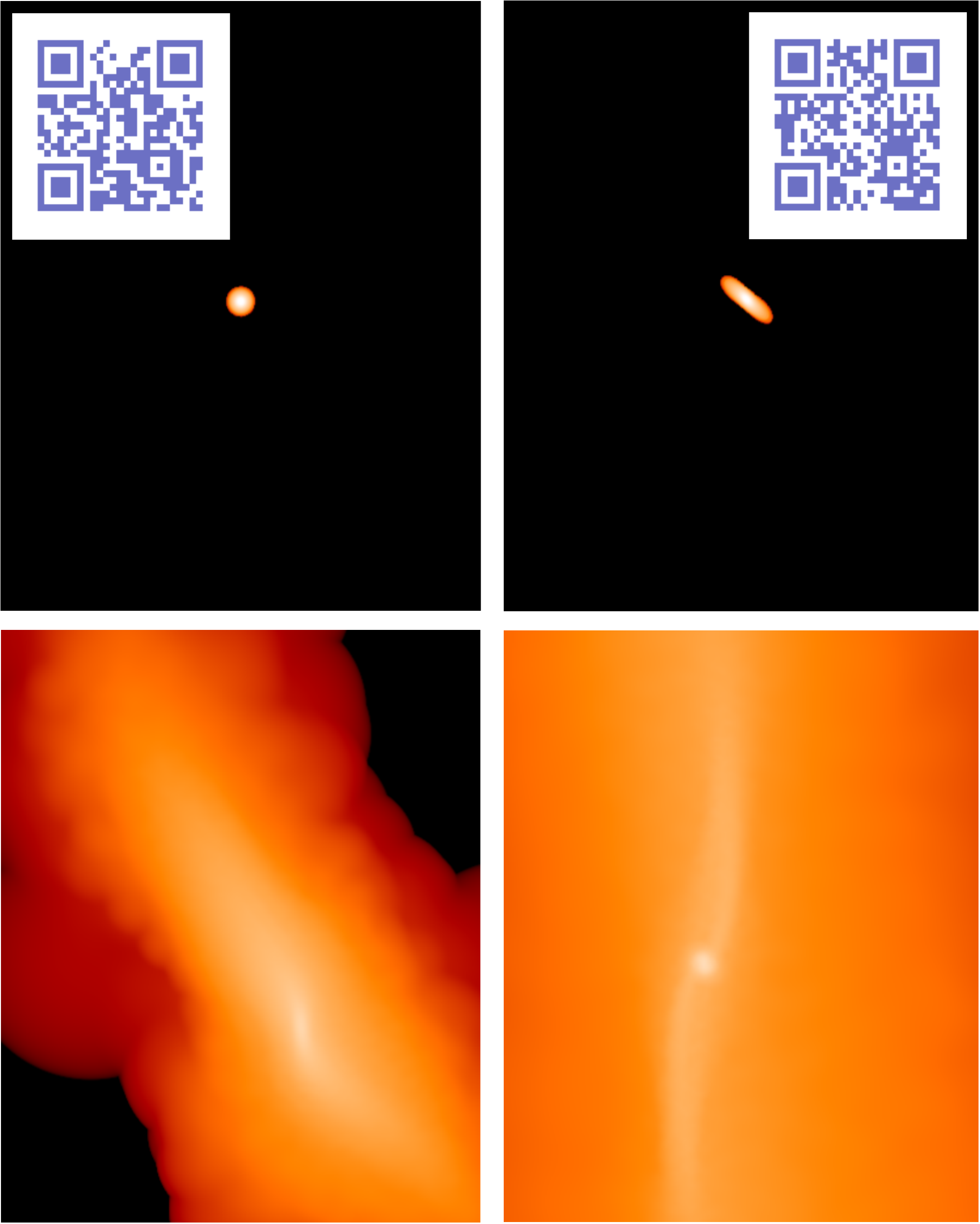}}
\caption
   {
Tidal disruption of a sun-like star around a SMBH of mass $M_{\rm BH}=10^6\,M_{\odot}$, on the frame of the star.
The QRC link to the URLs \url{https://youtu.be/Ryc44v4Eb7I} and \url{https://youtu.be/uZqXBD8R9Dw}.
This is Fig.~(2) of \citep{Amaro-SeoaneLRR}, distributed under the terms of the Creative Commons Attribution 4.0 International License
(\url{http://creativecommons.org/licenses/by/4.0/}).
   }
\label{fig.BUUUM}
\end{figure}

We show in Figure~\ref{fig.BUUUM} a smoothed-particle hydrodynamics simulation
of the tidal disruption of a star of mass $m=1\,M_{\odot}$, which corresponds
to Fig.~2 of \citep{Amaro-SeoaneLRR2012}. We can see that the star adopts a
spheroidal shape after it has passed through periapsis on its orbit around a
SMBH of mass $10^6\,M_{\odot}$. The top, left panel corresponds to the initial
time, and the rest of the panels to later moments.

\begin{figure}
          {\includegraphics[width=0.8\textwidth,center]{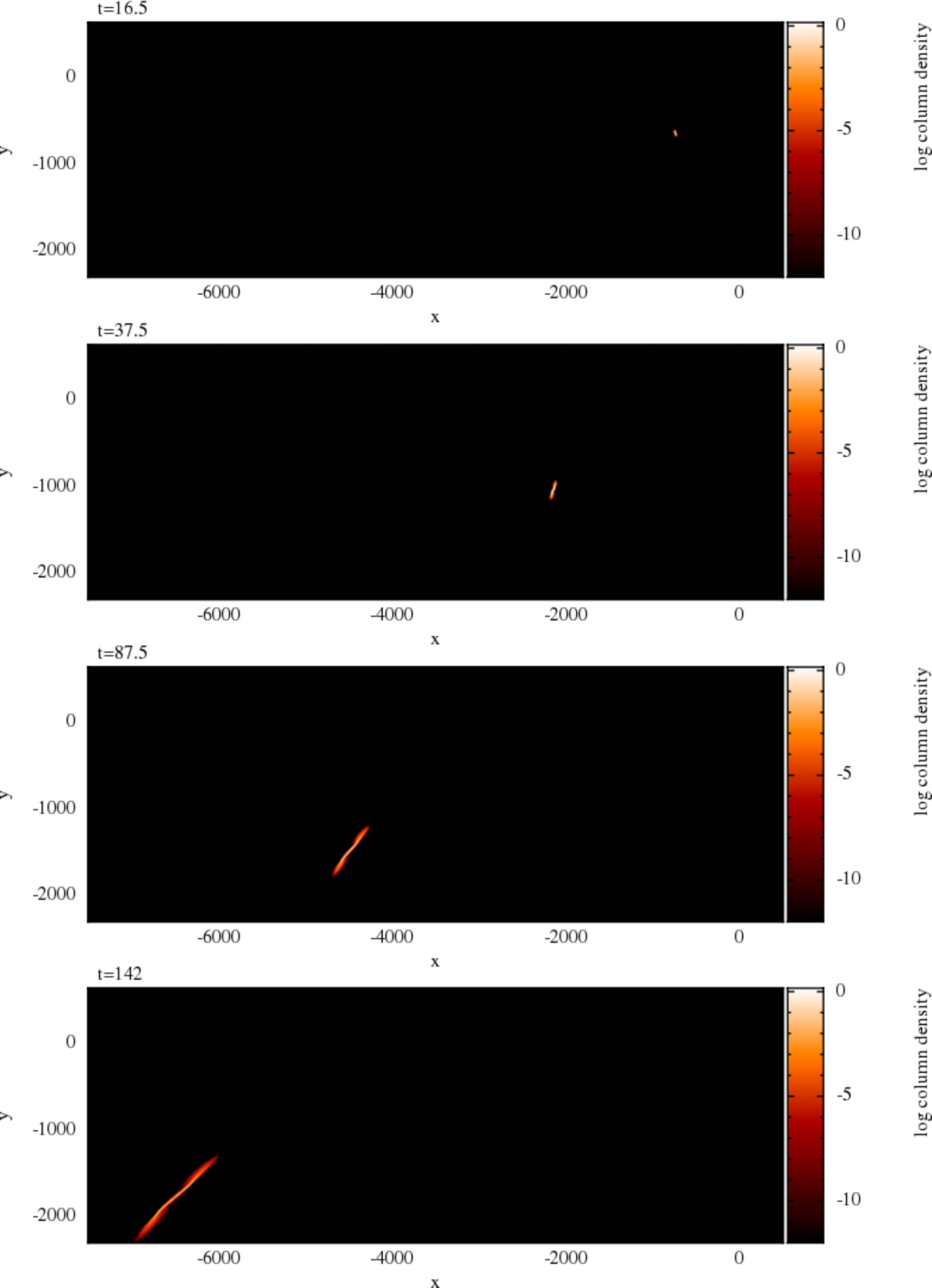}}
\caption
   {
Same as Fig.~(\ref{fig.BUUUM}) but in the general frame. The SMBH is located at the
origin of coordinates. We have selected four
snapshots right after the periapsis passage. The length is expressed in solar
radii. We can see how the star is torn apart and extends to much longer radii
than the initial radius of the star, $m=1\,M_{\odot}$.
   }
\label{fig.BUUUM_external}
\end{figure}

As we can see in Figs.~(\ref{fig.BUUUM}) and (\ref{fig.BUUUM_external}), the Sun
will be able to only describe a close passage around the SMBH to then be tidally
torn apart (if the trajectory crosses the tidal radius, and the disruption degree
depends on how deep the passage is). 

Contrary to extended stars, COs can revolve around the SMBH many times, as we
have already mentioned. This is so because the tidal radius of a neutron star
(NS) is located within the event horizon of the SMBH, so that we will never see
the NS being disrupted. An interesting situation, however, are systems composed
of an IMBH and a white dwarf (WD), because the WD will be tidally disrupted
before plunging through the event horizon \citep[see
e.g.][]{2008NewAR..51..884M,2008MNRAS.391..718S,2020arXiv200512528R,2020SSRv..216...39M}.

\begin{figure}
          {\includegraphics[width=0.8\textwidth,center]{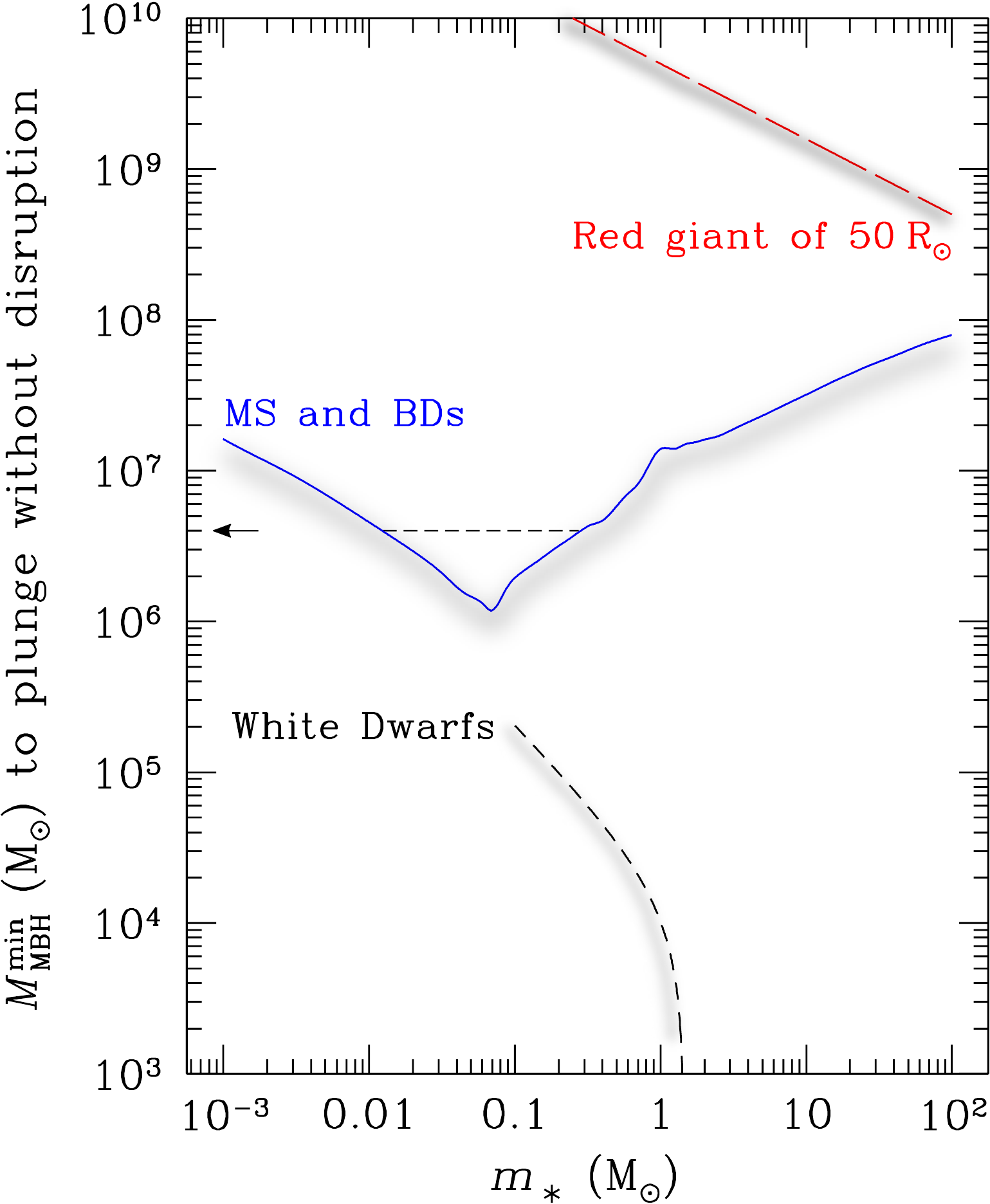}}
\caption
   {
Adapted figure Fig.~1 from \citep{Amaro-Seoane2019}. Minimum mass $M_{\rm BH}$ as a function of the mass of a star, or
sub-stellar object, for the latter to cross the event horizon without suffering significant
tidal stresses. The arrow points the mass of the SMBH at the centre of our Galaxy,
and the dashed lines the interval of masses for stars and brown dwarfs (BD) which
can directly plunge without disruption.
   }
\label{fig.MinMassPlunge_labelled2}
\end{figure}

In Fig.~(\ref{fig.MinMassPlunge_labelled2}), we show the relation between the
mass of a SMBH and the mass of a star, or sub-stellar object, for it to cross
the event horizon without being tidally disrupted. For instance, we can see
that a red giant with a mass of $50\,M_{\odot}$ can plunge directly the event
horizon of a SMBH of mass $\gtrsim 4\times 10^{8}\,M_{\odot}$ without being
disrupted. For this figure we have taken into account the mass-radius relations
from the modelling of stars of \citep{SSMM92,MMSSC94,CDSBMMM99,CB00}. These
relations reproduce almost identically the more recent data of
\cite{ChabrierEtAl2009} for brown-dwarfs.

The problem of an EMRI is conceptually very similar to that of a tidal
disruption. However, instead of one periapsis passage, the source will pass
through periapsis tens to hundreds of thousand of times. EMRIs form initially
with very high eccentricities, as we will see, so that the apocentre can extend
well into the bulk of the stellar system. A perturbation at apocentre can lead
to (i) the EMRI scattering off the orbit, so that we lose the source of GWs or
to (ii) increase even more its eccentricity.

\subsection{Relaxation theory}

In Newtonian physics we cannot analytically solve systems with more than two
bodies (but for very special configurations, in which there is an additional
body which has a much smaller mass than the two other bodies, which is the
restricted three-body problem). However, when we try to understand the
formation of EMRIs, IMRIs or XMRIs, we need to look at dense stellar systems
which might have up to $10^8\,M_{\odot}\,\textrm{pc}^{-3}$. These densities are
orders of magnitude above what we find around our solar system, which typically
has of the order $10^{-2}\,M_{\odot}\,\textrm{pc}^{-3}$. 

Since we cannot solve this problem, we try to address it by borrowing ideas
from other fields in physics, such as plasma physics, because it also shares
the property of inverse square laws. However, plasmas are often nearly uniform,
rest and large spatial extent, which is not the case of a dense stellar system.
Another possibility is thermodynamics because, in some sense, we can think of a
dense stellar system as a gaseous medium. Nonetheless, thermodynamics excludes
a description of self-gravitating systems. We cannot have an asymptotic
thermodynamic limit because the gravitational forces are long-range and we
cannot ignore the effect of a star which is ``far away''. This translates into
the fact that the a thermodynamical equilibrium is ruled out by construction.
In spite of this, using thermodynamics in the context of stellar dynamics has
proven to be an important asset. Nevertheless, surprises arise in this context,
such as the concept of negative heat capacity \citep[``(...) stars act like
donkeys slowing down when pulled forwards and speeding up when held
back.''][]{Lynden-BellKalnajs1972}.

We introduce now the fundamentals of relaxation theory, just as a way to
understand in an intuitive way how to ponder the phenomena that lead, in the
Newtonian limit, to centrophilic orbits; i.e. those which will approach the
SMBH enough so as to stand a chance of forming an EMRI.  For a more detailed
description, we remit the reader to
\citep{Henon73,Saslaw85,Spitzer87,BinneyTremaine08}.

In relaxation, the potential of the dense stellar system is described by
approximating it in the addition of two separated components. One is given by a
dominating smooth contribution, which we will call $\Phi_\mathrm{s}$, plus the
contribution of individual stars, $\delta\Phi$, which are responsible for the
phase-space distribution function to follow the collisionless Boltzmann
equation. 

A way to describe the evolution of these dense stellar systems is the
Fokker-Planck (or Kolmogorov forward-) equation, which is a partial
differential equation that allows one to study the time evolution of the
velocity probability density function of a particle (a star) due to of drag
forces and random forces. When two stars interact with each other, the
associated volume typically has dimensions which are small compared to the
macroscopic lengths of the system, such as the radius of the galactic nucleus.
We can then approximate the evolution by assuming that the entire cumulative
effect of all encounters on our star are identical to an ideal situation. We
assume that the star is embedded in a homogeneous system with the same local
distribution function everywhere \citep[this is usually refered to as the
``local approximation'', see e.g. Sec. 8.5 of][for a detailed
description]{2018LRR....21....4A}.

As time flows, the accumulated variations of $\delta\Phi$ alter $E$ and $J$ and
lead to a modification of the distribution function. We then treat the role of
the individual changes of $\delta\Phi$ as the sum of uncorrelated hyperbolic
encounters between two stars with a small deviation angle. Hence, if we
consider a star ``1'' in a stellar system which is perfectly homogeneous of
stars of the type ``2'', with the same masses and velocities, after a time
$\delta t$ the initial trajectory will suffer a deflection $\theta$ of

\begin{align}
  \left\langle \theta \right\rangle_{\delta t} &= 0, \nonumber \\
  \left\langle\theta^2\right\rangle_{\delta t} &\simeq 
  8\pi N \ln\left(\frac{b_\mathrm{max}}{b_0} \right) 
  \frac{G^2\left(m_1+m_2\right)^2}{v_\mathrm{rel}^3} \delta t,
\end{align}

\noindent 
with $m_1$, $m_2$ the masses of two stars, $N$ the stellar number density,
$b_0$ is the impact parameter that leads to a deflection angle of $\pi/2$,
$b_\mathrm{max}$ is a normalisation factor to avoid logarithmic divergence 
(interpreted as the maximum impact parameter, of the order the size of the stellar system), and $v_\mathrm{rel}$ is the
relative speed between star 1 and the background stars 2. We note that we
obtain $\left\langle \theta \right\rangle_{\delta t} = 0$ because this is
a diffusion process, and hence $\langle\theta^2\rangle_{\delta t} \propto \delta t$.

We define the velocity dispersion $\sigma$ of the nucleus or cluster via the statistical
concept of root mean square dispersion. The variance $\sigma^2$ provides us with a measure
of the dispersion of the observations within the statistical population at our disposal
(i.e. the observational data). This means that

\begin{equation}
\sigma^2= \frac{1}{N} \sum_{i=1}^{N} (V_{i}- \mu)^2. \nonumber
\end{equation}

\noindent
Where $V_{i}$ are the individual stellar velocities and $\mu$ the arithmetic mean,

\begin{equation}
\mu_{a} \equiv  \frac{1}{N} \sum_{i=1}^{N}V_{i}. \nonumber
\end{equation}

If we take $M$, $R$ as the total stellar mass and radius of the nucleus,
respectively, and define $m_*$ as the average stellar mass, then the argument
of the Coulomb logarithm (remember our discussion about plasma physics) is
approximately

\begin{equation}
  \frac{b_\mathrm{max}}{b_0} \simeq
  \frac{v_\mathrm{rel}^2 R}{G\left(m_1+m_2\right)}
  \simeq \frac{\sigma^2 R}{G M} \simeq \gamma \frac{M}{m_*}=\gamma \, N.
  \label{eq:log_coul}
\end{equation}

\noindent 
In the last equation $\gamma$ is a dimensionless proportionality constant. Note
that this proportionality applies only to self-gravitating, virialized stellar
systems.  In order to derive the relaxation timescale of the system, we would
have to integrate over all possible values of $b_\mathrm{max}/b_0$; however in
practise this is again approximated by assuming that $\gamma$ is constant. This
global value can be derived using analytical arguments \citep{SH71a,Henon75},
which has been corroborated with numerical simulations.  For systems composed
of stars with a single type of mass, \cite{Henon75} derived $\gamma \simeq 0.10
- 0.17$.  He also proved that this value should be significantly smaller in the
case that we consider a mass spectrum.

When we analyse relaxation via the local approximation we are implicitly
assuming that the dense stellar system is finite and homogeneous. Real systems
however, are quite the contrary, with important density gradients. By adopting
the local approximation we are imposing the fact that $b \ll R$. This allows
us, as explained, to neglect $\Phi_\mathrm{s}$ -and hence approximate the
trajectories as Keplerian orbits- as well as to use the local properties around
star 1 as representative ones. This a priori conceptual grotesque approximation
is however physically tolerable because we are talking about the argument of a
logarithm, $\ln(b_2/b_1)$.  This assumption has been tested in a large number
of works and seems to be acceptable for pragmatic purposes
\citep{GH94a,SA96,Giersz98,PZHMMcM98,TPZ98,Spurzem99}.

Although there is a conceptual problem in the way we have introduced it
\cite[see discussion in][]{2018LRR....21....4A}, we can use
Eq.~(\ref{eq:log_coul}) to introduce a characteristic timescale associated to
relaxation,

\begin{equation}
  \left\langle\theta^2\right\rangle_{\delta t} \simeq
  \left(\frac{\pi}{2}\right)^2 \frac{\delta t}{{T}_\mathrm{rlx}}.
\end{equation}

\noindent 
The full expression of this timescale is given in e.g. \cite{Chandra42}, and see \cite{Larson70a}. In
this work we will
not derive it. We only introduced the basic notions so that the reader can 
develop a physical intuition for the process. The expression is

\begin{equation}
T_{\rm rlx}= \frac {9}{16 \sqrt {\pi}} \frac {\sigma ^3 }{G^2m \rho \ln(\gamma \, N)},
\label{eq.relax_t}
\end{equation}

\noindent 
which can be rewritten as

\begin{align}
T_{\rm rlx} & \sim {2 \times 10^{10}}\left({\ln \Lambda}\right)^{-1} {\rm yrs}  
               \left( \frac{\sigma}{100 {\rm km/s}} \right)^3 
               \left( \frac{m}{1 M_\odot}  \right)^{-1}
               \left( \frac{r}{1 {\rm pc}}  \right)^\gamma.
\label{eq.relax_t2}
\end{align}

\noindent 
In this expression we have adopted a power-law distribution in the stellar
density, $\rho(r) \sim r^{-\gamma}$, see
\citep{2018A&A...609A..28B,2018A&A...609A..27S,2018A&A...609A..26G,2019MNRAS.484.3279P}.

Within the influence radius of a SMBH $r_{\rm infl}$ (which can be loosely defined as
the radius within which the potential is dominated by the SMBH), we can
approximate as explained before $b_{\rm max}=r_{\rm infl}$, and relaxation leads to a
steady-state distribution of orbital energies on, of course, a timescale
$\approx T_{\rm rlx}$.

So far we have considered that all stars have the same mass in the nucleus.
This is obviously not correct. Stellar-mass black holes, in particular, have a
higher mass than our Sun, which we adopt to be in this work $m_{\rm
bh}=10\,M_{\odot}$. If we consider that a fraction of all stars are
stellar-mass black holes, typically as small as $\approx 10^{-3}$ for a
standard initial mass fraction\citep[see e.g.][]{2001MNRAS.322..231K}, then a
limiting form of relaxation leads to an interesting phenomenon. Stars, or COs,
with masses larger than the average stellar mass, assumed to be of $1\,M_{\odot}$
will segregate in phase-space in a timescale shorter than the associated $T_{\rm rlx}$
by a factor given approximately by the ratio $Q$ of the average stellar mass divided by
the mass of the CO. I.e. if $m_{\rm bh}=10\,M_{\odot}$ and $m=1\,M_{\odot}$, this
timescale is 1/10 of $T_{\rm rlx}$. This limiting form of relaxation was described
by Chandrasekar \citep[see e.g.][]{BinneyTremaine08} and is called dynamical friction,
with the associated timescale

\begin{equation}
T_{\rm DF} = \frac{v_{\rm c}^3}{4 \pi G^2 \rho m_{\rm bh} \ln \Lambda} \left[ {\rm erf(X)} - \frac{2X}{\sqrt{\pi}} e^{-X^2} \right]^{-1}
\sim \frac{\langle m \rangle}{m_{\rm bh}}\,T_{\rm rlx} := \frac{T_{\rm rlx}}{Q}, 
\end{equation}

\noindent 
where $v_{\rm c}$ is the (local, obviously) circular velocity, $\rho$ is the mass density of
the stars others than COs, the mass ratio is $Q=m_{\rm bh}/m$ and $X=v/(\sqrt{2} \sigma)$.  
What this means is that we \textit{do} expect to have stellar-mass black holes
very close to the SMBH in a timescale shorter than the $T_{\rm rlx}$, which is
what we need to create the orbit which will ultimately lead to an EMRI.

This leads us to the question whether nuclei in the bandwidth of LISA will have
$T_{\rm rlx}$ shorter than a Hubble time because otherwise the nuclei will not
be relaxed, and hence we cannot expect stellar-mass black holes close to the
SMBH. If nuclei which are target of LISA, i.e. with masses approximately
ranging between $10^4-10^7\,M_{\odot}$ are not \textit{relaxed}, then we cannot
expect EMRIs to be detected.

If these nuclei follow the so-called mass-sigma relationship (which is not
clear, since they are at the low-mass end of the correlation, see e.g.
\cite{KormendyHo2013}), $M_{\rm BH} \propto \sigma^4$, then we can express the
influence radius assuming that the nucleus is isothermal (i.e. $\rho \sim
r^{-2}$, $\sigma(r) = \rm{constant}$, for $r \gtrsim \rm{few} \times 10^{-1}
r_{\rm infl}$) as $r_{\rm infl} \sim G M_{\rm BH}/\sigma^2 \propto M_{\rm BH}^{1/2}$. On the
other hand, the average stellar density at $r_{\rm infl}$ can be estimated roughly to be
$\rho \sim 2M_{\rm BH}/(r_{\rm infl}^3)$ (since we have the same mass in stars as the mass of the SMBH).  The velocity dispersion within $r_{\rm infl}$ is
$\sigma^2(r) \sim G M_{\rm BH} /r$ (as determined via the Jeans equation, see
\cite{BinneyTremaine08}), so that we finally have $T_{\rm rlx} \sim \sigma^3/\rho$,
which means that

\begin{equation}
T_{\rm rlx} = \frac{17}{100} \left( m \Lambda \right)^{-1} \left( \frac{M_{\rm BH}}{G}\right)^{1/2} r_{\rm infl}^{3/2}
\end{equation}

\noindent 
Adopting $\ln \Lambda \sim \ln (M_{\rm BH}/m_{\rm bh}) \sim 11.5$ (see Eq.~13 of \citep{2018LRR....21....4A}),
and normalising to typical values, we find that

\begin{equation}
T_{\rm rlx} \cong 1.1\times 10^9\,\textrm{yrs} \left( \frac{M_{\rm BH}}{4\times 10^{\,6}\,M_{\odot}} \right)^{1/2} 
                                                \left( \frac{\langle m \rangle}{0.4\,M_{\odot}} \right)^{-1} 
                                                \left( \frac{r_{\rm infl}}{1\,\textrm{pc}}\right)^{3/2},  
\end{equation}

\noindent 
where we have used the average stellar mass at the Galactic Centre. This can be
roughly derived by assuming that $\langle m \rangle = M_{\rm tot}/N_{\rm tot}$,
with $M_{\rm tot}$ the total stellar mass and $N_{\rm tot}$ the total number of
stars. In our approximation, we only consider main sequence stars and
stellar-mass black holes, which represent a fraction $10^{-3}$ of the total
number. Using a typical IMF, the average stellar mass is of $\sim
0.4\,M_{\odot}$, which is in agreement with the Milky-Way models of
\cite{FAK06a}.

Using the fact that stellar distribution follows a power-law expression of the radius,
as explained before, we can furthermore derive that \citep[see][]{Amaro-Seoane2019}

\begin{equation}
r_{\rm infl} \simeq 2.5\, \textrm{pc} \left(\frac{M_{\rm BH}}{4\times 10^6\,M_{\odot}}\right)^{3/5}.
\end{equation}

\noindent
Then, we are finally left with

\begin{equation}
T_{\rm rlx}^{\rm tot} \cong 4.35\times 10^9\,\textrm{yrs} \left( \frac{M_{\rm BH}}{4\times 10^{\,6}\,M_{\odot}} \right)^{11/10}
                                                \left( \frac{\langle m \rangle}{0.4\,M_{\odot}} \right)^{-1}.
\label{eq.Trlx}
\end{equation}

\begin{figure}
          {\includegraphics[width=0.8\textwidth,center]{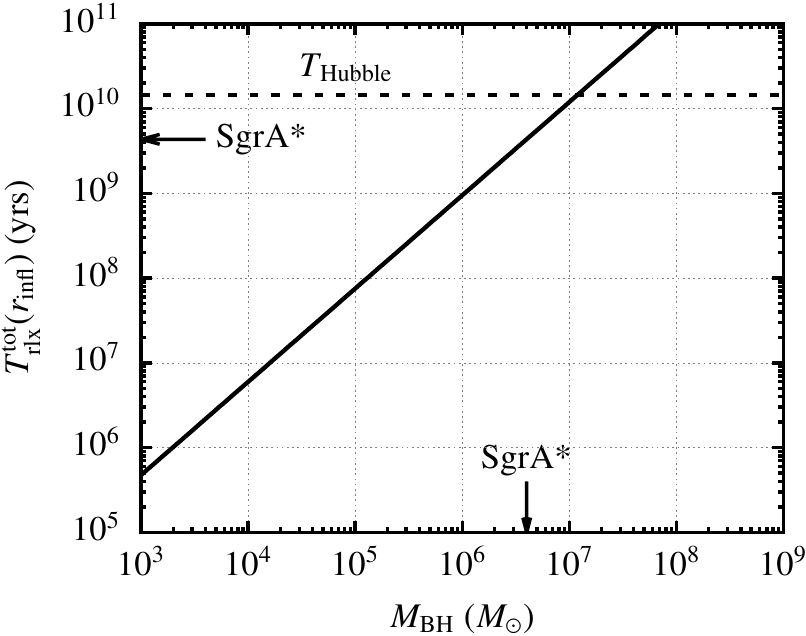}}
\caption
   {
Relaxation time as a function of the SMBH mass $M_{\rm BH}$. The horizontal,
dashed line marks the Hubble time limit, the arrows the position of the SMBH
in our Galaxy and in the relation we have taken into account the influence
radius as a function of $M_{\rm BH}$. Nuclei harboring SMBHs with masses
$M_{\rm BH} \gtrsim 10^7 M_\odot$ have a relaxation time longer than the
age of the Universe and have not had time to converge into a steady-state
and develop segregated stellar cusps. Their stellar distribution keep memory
from their formation and (if any) of the last strong perturbation process.
   }
\label{fig.Trlx}
\end{figure}

\noindent 
In this expression we have the relaxation time as a function only of the mass
of the SMBH, while taking into account the scaling of the influence radius with
this mass. This allows us to have an idea of what nuclei will be relaxed.  We
depict this in Fig~(\ref{fig.Trlx}) for a range of masses, and note that in the
low-end of masses the results should be taken with skepticism, since SMBHs with
light masses will wander off the centre. By wandering off, the massive black
hole (in this case an IMBH) could potentially explore regions of phase-space
with different relaxation times and, since the wandering timescale is much
shorter than the relaxation timescale, the system does not have time to
re-adjust. Hence any capture is led by a dynamical process, and not a
relaxational one. 

It is remarkably interesting to note that the relaxation time will exceed a
Hubble time for nuclei harbouring SMBHs with masses slightly above
$10^7\,M_{\odot}$ (the exact number should not be envisaged as a realistic one
in view of the -few- assumptions that we have made in the derivation).
Surprisingly, this type of masses fit perfectly well in the kind of nuclei that
LISA will probe. If this had not been the case, LISA would never be able to
detect an EMRI. To the best of my knowledge, nobody thought of this potential
risk before designing the sensitivity curve of the instrument. I would
appreciate it if somebody could provide me with more details, in case that
I am wrong (which is a possibility, obviously).

\subsection{The loss-cone}
\label{sec.losscone}

After having conceptually introduced what is the relaxation of a dense stellar
system, we now look at the definition of the region of phase-space denoted as
the ``loss-cone''. Not all stars in a galactic nucleus or a globular cluster
will have such an energy and angular momentum that will bring them very close
to the SMBH; after all, from the point of view of the stellar system, the SMBH
is something microscopic, with a size (i.e. a Schwarzschild radius) of $R_{\rm
Schw} \simeq 3.8 \times 10^{-7}\,\textrm{pc}$ if its mass is that of SgrA*. In
comparison, its influence radius, as we have seen, is about $r_{\rm infl}
\simeq 2.5\,\textrm{pc}$, i.e. seven orders of magnitude bigger in length.

Only a very peculiar subset of all stars with a given semi-major axis will have
an eccentricity high enough to bring them close at periapsis to interact with
the SMBH. Since historically this was studied in the context of stellar
disruptions, interest this subset is dubbed the ``loss-cone'', because (1) the
orbit around the SMBH can be defined in terms of an angle such that will lead
the star to get close enough to the SMBH and (2) if at periapsis it crosses the
tidal radius, the star can be tidally disrupted, as we saw before, and is hence
lost for the system.

Previously we introduced the relaxation time. It is useful to define another
interesting timescale, the dynamical timescale or the ``crossing time''. This
is the time required for a star to cross the host dense stellar system. For a
cluster, $T_{\rm dyn}= {r_{\rm infl}}/{\sigma_{\rm infl}}$, where $\sigma_{\rm
infl}$ is the velocity dispersion at the influence radius. If we impose Virial
equilibrium, then $T_{\rm dyn} \approx \sqrt {r_{\rm infl}^3/(GM_{\rm infl})}
\simeq (G\rho)^{-1/2}$, with $M_{\rm infl}$ the total stellar mass comprised
within $r_{\rm infl}$ (which is in order of magnitude equivalent to the mass of
the SMBH).

Following our previous discussion of (cheating with) thermodynamics, we note
that contrary to a gaseous system, in a stellar system the thermodynamical
equilibrium timescale will be much longer than the crossing one, $T_{\rm rlx}
\ll T_{\rm dyn}$.  If we consider a perfectly homogeneous stellar system, we
will reach a stationary state in the limit $t\rightarrow \infty$. How quickly a
Virial equilibrium is reached depends on the timescale in which a perturbation
in the stellar system is washed out.

\begin{figure}
          {\includegraphics[width=1\textwidth,center]{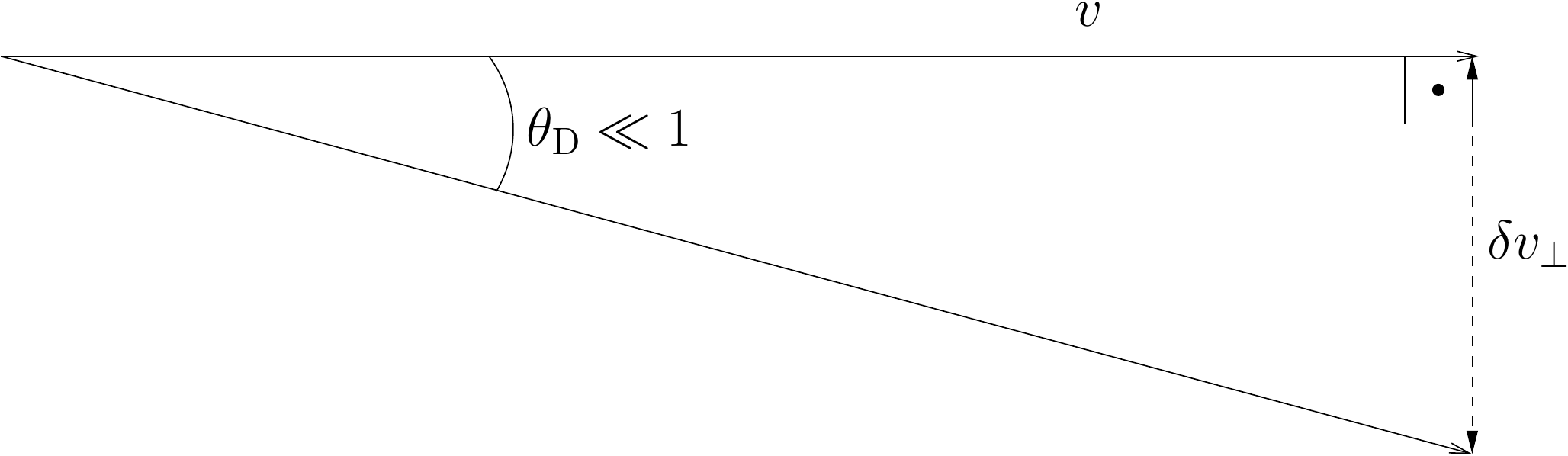}}
\caption
   {
The change in the perpendicular component of the velocity of the star is
assumed to be small after every single encounter with another star. This leads
to the definition of a diffusiong angle $\theta_{\rm D}$ which, ab
definiti$\bar{\rm{o}}$ (see text) must be small. When this component is of the
same order than the perpendicular velocity component itself, a relaxation time
has passed. 
   }
\label{fig.diffusion_angle_xfig.pdf}
\end{figure}

We have introduced the relaxation time as the time that we need to wait for the
system to ``mix up'', i.e. more formally, for the perpendicular velocity
component of a star to be of the same order than the perpendicular velocity
component itself, $\triangle v_{\perp}^2 /v_{\perp}^2 \simeq 1$, as illustrated
in Fig~(\ref{fig.diffusion_angle_xfig.pdf}). I.e. the relaxation time is the
required time to induce a change in the perpendicular velocity component of the
same order as the perpendicular velocity component $v_{\perp}$ itself,
$\triangle v_{\perp}^2 \simeq  v_{\perp}^2$, after a number of dynamical times
(e.g. crossing times) $n$, which allows us to define the associated timescale
for mixing up the system: $\triangle v_{\perp}^2 / v_{\perp}^2 = T/T_{\rm
rlx}$, as in e.g. \cite{BinneyTremaine08}.  This means that we have that
$\triangle  v_{\perp}^2 / v_{\perp}^2 = {n \, \delta v_{\perp}^2}/ {
v_{\perp}^2}=1$, so that $T_{\rm rlx}=n \, T_{\rm dyn} = \left(
{v_{\perp}^2}/{\delta v_{\perp}^2} \right) \, T_{\rm dyn}$.

The mean deviation of a orbit in a dynamical timescale $T_{\rm dyn}$ (which is
typically defined as the time required for a star to cross the system) can be
estimated via a ``diffusion angle'' $\theta_{\rm D}^2:=T_{\rm dyn}/T_{\rm
rlx}$, and we furthermore assume that this angle is very small, following the
line of thought of e.g. \cite{FR76,BW77,LS77,AS01}. Therefore,
$\sin{\theta_{\rm D}} \simeq { \delta v_{\perp}}/{v} \simeq \theta_{\rm D}$,
and so $\theta_{\rm D} \simeq \sqrt{{T_{\rm dyn}}/{T_{\rm rlx}}}$. This angle
is very useful to understand the evolution of the system; it will help us to
quantify how efficiently (or not) the loss-cone can bring stars close to the
SMBH.

Stars which \textit{are} in the loss-cone are lost in a dynamical time, i.e. a
crossing time.  We can define the stars that belong to the loss-cone by
evaluating the angular momentum of the star, the influence radius and the
velocity dispersion of the system. 

\begin{figure}
          {\includegraphics[width=1\textwidth,center]{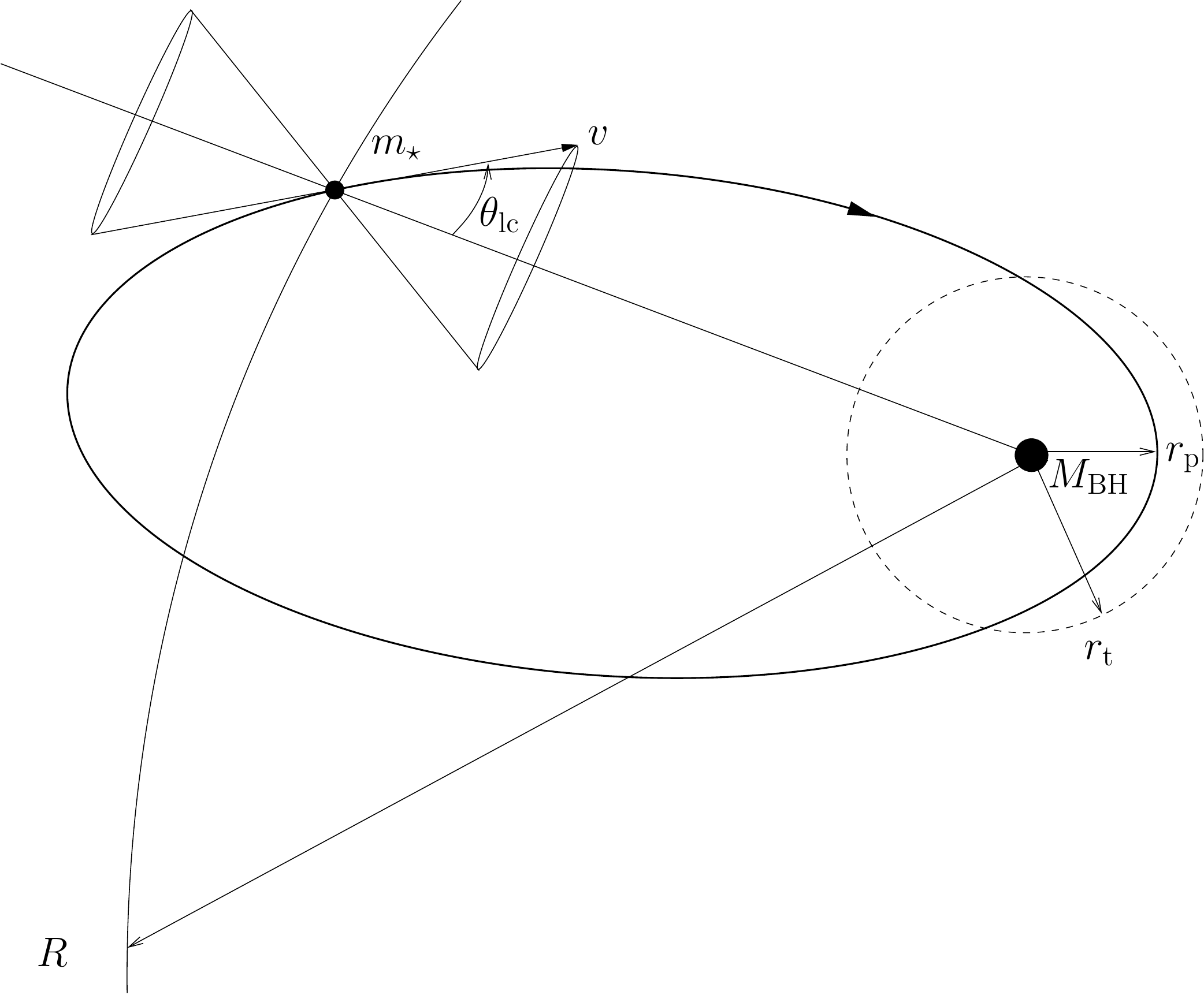}}
\caption
   {
Definition of the loss-cone angle for a star of mass $m_{\star}$ approaching a SMBH of mass $M_{\rm BH}$.
   }
\label{fig.loss_cone_star_xfig}
\end{figure}

We can illustrate this angle in an approximate way as in
Fig.~(\ref{fig.loss_cone_star_xfig}).  If the star of mass $m_{\star}$ has a
velocity vector such that the periapsis distance $r_{\rm p}$ falls within the
tidal radius $r_{\rm t}$, it will suffer important tidal stresses. A total
disruption happens when the so-called ``penetration factor'' $\beta:= r_{\rm
t}/r_{\rm p}> 1.85$, see \cite{GuillochonRamirez-Ruiz2013}. This will happen
when $r_{\rm p}(E,\,L) \leq r_{\rm t}$ and $\theta \leq \theta_{\rm lc}$.  One
can show that the loss-cone angle is $\theta _{\rm lc} = \sqrt { {2r_{\rm
t}}/{(3r)}}$, with $r_{\rm t}$ the tidal radius and $r$ the distance from the
MBH to the star, if $r \le r_{\rm infl}$, and $\theta_{\rm lc} \simeq \sqrt {
2r_{\rm t} r_{\rm infl} /(3r)}$ for $r \ge r_{\rm infl}$ \citep[for details
about this derivation, see e.g.][]{2018LRR....21....4A}.

With these considerations we can define a critical radius which will determine
the future of stars in the loss-cone. If we define the ratio $\xi:=
\theta_{lc}/ \theta_{D}$, when $\xi =1$, $\theta_{lc}=\theta_{D}$, which can be
converted into a distance $r_{\rm crit}$. Stars inside this radius are removed
on a $T_{\rm dyn}$ because $\xi >1$, $\theta_{lc}>\theta_{D}$. 

\begin{figure}
          {\includegraphics[width=1\textwidth,center]{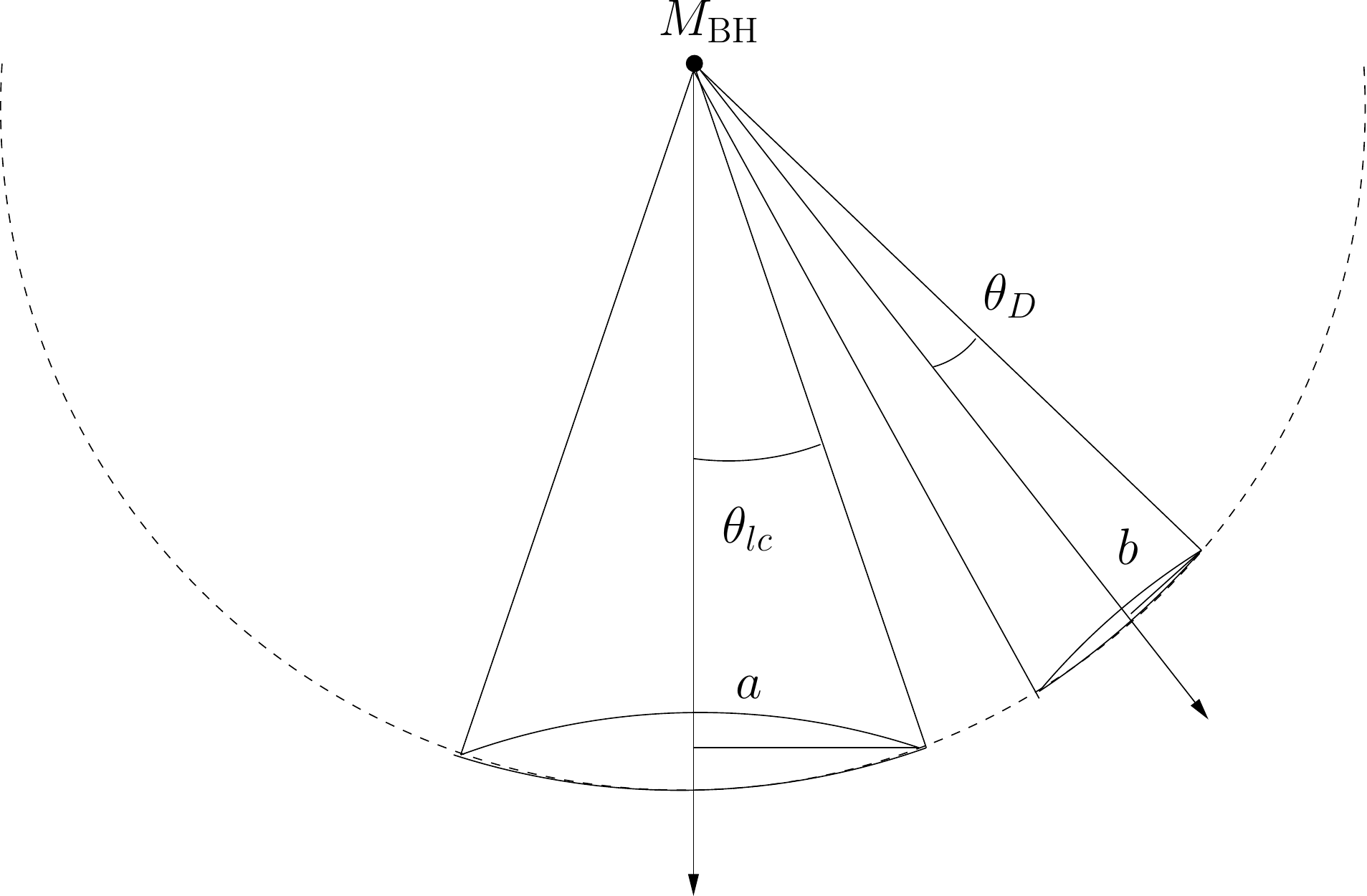}}
\caption
   {
Definition of the critial radius via the loss-cone- and diffusion angles.
   }
\label{fig.replenishment_xfig}
\end{figure}

We can therefore make an educated guess for the loss-cone to be replenished.
The conditions is that $\xi ^2 \times T_{\rm dyn}$, where the square appears
because, as illustrated in Fig.~(\ref{fig.replenishment_xfig}), the surface which
corresponds to the radius $b$ must cover the empty surface of radius $a$.

When a star is outside of the critical radius, even if it is in the loss-cone,
it will be diffused out of the it before it can reach the central SMBH. This is
so because $\xi <1$, $\theta_{lc}<\theta_{D}$, and by definition, $\theta_{D}$
corresponds to the variation of $\theta$ in a $T_{\rm dyn}$. Inside of the
critical radius, however, stars will fall on to the SMBH without being
perturbed. We have that $\xi = {a}/{b}$ because  $a=v \sin \theta_{lc} \approx v \theta_{lc}$, and $b=v
\sin \theta_{D} \approx v \theta_{D}$.

\begin{figure}
          {\includegraphics[width=1\textwidth,center]{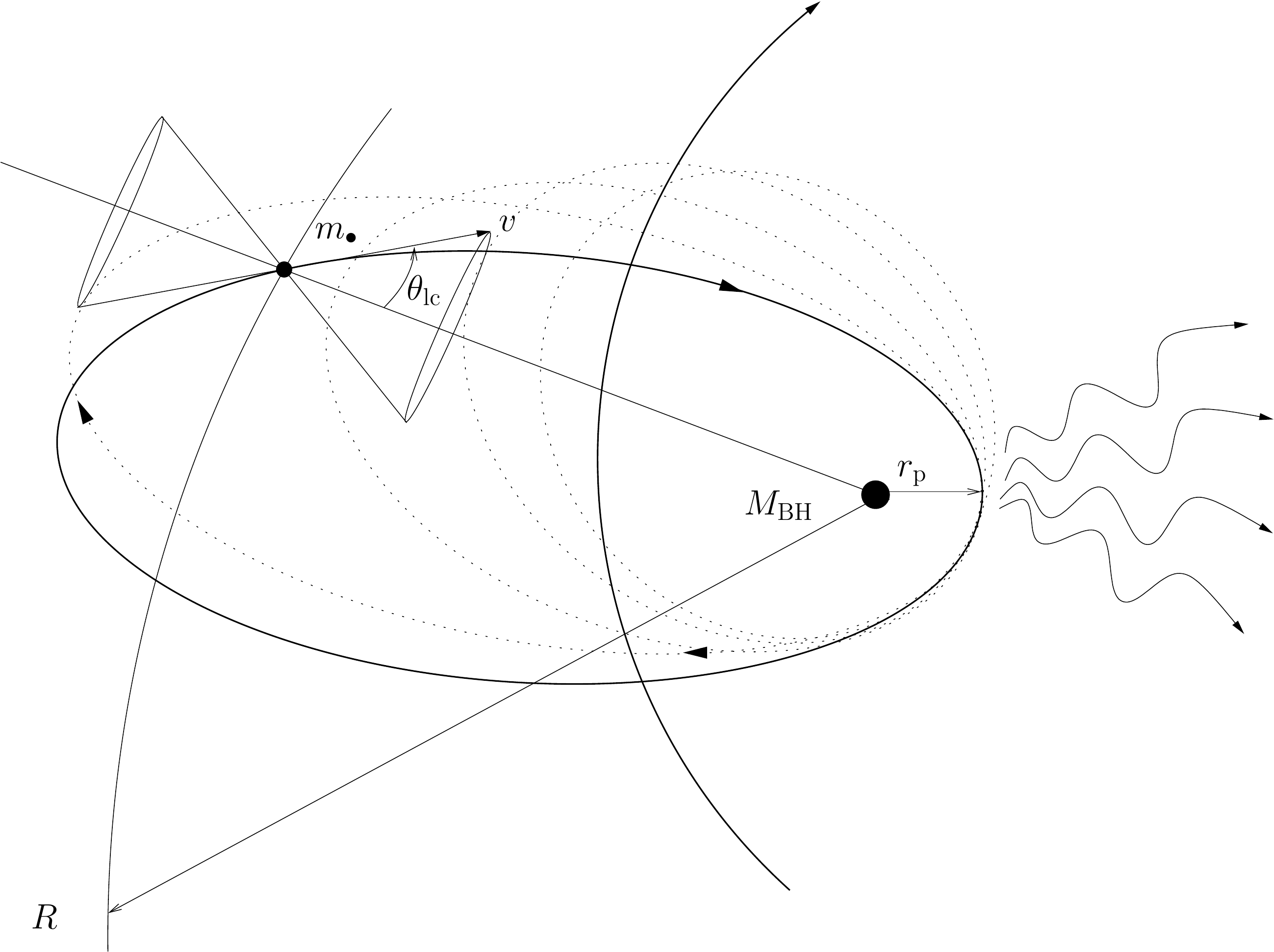}}
\caption
   {
Same as Fig.~(\ref{fig.loss_cone_star_xfig}) but for a CO orbiting around a SMBH.
The orbit precesses, losses energy and thus shrinks over time and the plane of the
orbit basculates due to frame-dragging if the SMBH is spinning, which is not shown
in the illustration because it is a two-dimensional projection. At periapsis, there
is a strong burst of gravitational radiation and the semi-major axis of the orbit
shrinks a bit (the shrinkage is not to scale, as well as the precession). This 
figure illustrates the difficulty of defining analytically the relativistic 
equivalent of the loss-cone treatment for a tidal disruption.
   }
\label{fig.loss_cone_star_rel_xfig}
\end{figure}

Whlist it would be ideal to be able to define a relativistic loss-cone, i.e. a
``GW-cone'', in practise this turns out to be challenging, if not impossible.
In Fig.~(\ref{fig.loss_cone_star_rel_xfig}) we depict the complications of this
situation. The orbits are not closed and shrinking over time and, if the
central SMBH is Kerr, there will be a change in the inclination of the orbit
when the CO approaches the pericentre. 

\subsection{Formation of EMRIs via relaxation}

The evolution of a CO on its way to becoming a source of gravitational waves
is determined by two very different types of physics: stellar dynamics, for
which we do not have to bother with general relativistic effects, but with
two-body relaxation, and gravitational radiation. Since we obviously cannot
solve the $10^6-10^8$ problem in Newtonian gravity and not even the two-body
problem in general relativity (even less at these mass ratios), we need to
work with timescales to derive information about the process. We can define
a threshold that separates the evolution via dynamics from general relativity
by equating the two associated timescales, times a factor $C$ of order 1, 

\begin{equation}
T_{\rm rlx,\,peri } = C\,T_{\rm GW}(a,\,e)
\label{eq.TrlxTGW}
\end{equation}

\begin{figure}
          {\includegraphics[width=1.0\textwidth,center]{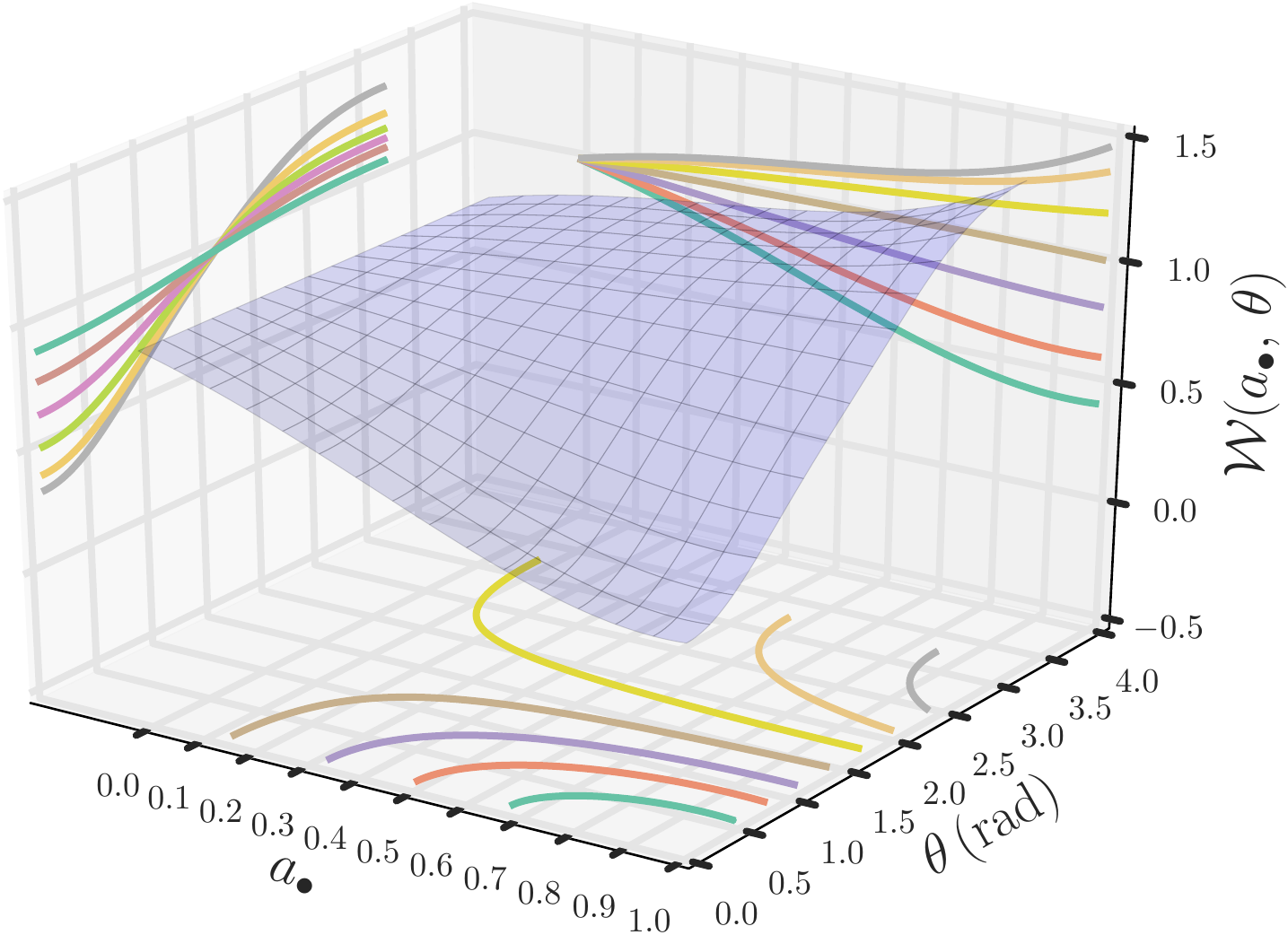}}
\caption
   {
Function $\cal{W}(\theta,\,{\rm a}_{\bullet})$ which gives us the multiplying
factor to identify the location of the LSO for a Kerr SMBH relative to the
Schwarzschild case for prograde and retrograde orbits.
   }
\label{fig.W}
\end{figure}

\noindent
In this equation, $T_{\rm rlx,\,peri}$ is the relaxation time at pericentre,
i.e., $T_{\rm rlx,\,peri }:=T_{\rm rlx}(a)\times (1-e)$ \citep[][]{Amaro-SeoaneLRR}, 
and $T_{\rm GW}$ the time derived in the approximation of Keplerian ellipses of \cite{Peters64}.
Since the initial eccentricities of EMRIs are typically very large, as we will explain later,
the function $F(e)$ of $T_{\rm GW}(a,\,e)$ can be estimated to be $F(e) = 425/(768\,\sqrt{2})$.
We also have to equate

\begin{equation}
\frac{8\,G{M}_{\rm BH}}{c^2} = a\,(1-e),
\label{eq.PeriLSO}
\end{equation}

\noindent
with $c$ the speed of light and we note that the position of the last-stable orbit (LSO) depends on (i) the
spin of the central SMBH $a_{\bullet}$, (ii)the inclination of the orbit
$\theta$, and (iii) whether the orbit is pro- or retrograde, as described in
the work of \cite{Amaro-SeoaneSopuertaFreitag2013}. This is of course only true
for Kerr SMBHs. The Schwarzschild case serves as a reference point, which has
the LSO at a distance of $4\,R_{\rm S}$, with $R_{\rm S}$ the Schwarzschild
radius.  The work of \cite{Amaro-SeoaneSopuertaFreitag2013} provides us with
the function $\cal{W}(\theta,\,{\rm a}_{\bullet})$ which captures this
information and accordingly modifies the position of the LSO. This position is
crucial because it determines the upper integration limit when calculating the
event rate, as we will discuss. In Fig.~(\ref{fig.W}) we give the full shape of
the function.

The GW timescale is given by

\begin{equation}
T^{}_{\rm GW}(a,\,e) \sim \sqrt{2}\,\frac{24}{85}\frac{c^5}{G^3}
                        \frac{a^4\,\left(1-e\right)^{7/2}}{m^{}_{\,\rm CO}\,M_{\rm BH}^2},
\label{eq.Tgwae}
\end{equation}

\noindent 
where $m^{}_{\,\rm CO}$ is the mass of the CO in consideration. It could be a
neutron star, a white dwarf, a stellar-mass black hole or even a brown-dwarf,
which is not a compact object, but in the case of masses in the range of SgrA*,
as we have discussed, a powerful source of gravitational radiation. This is
important, because the timescale is given by two bodies only, the SMBH and the
CO. Moreover, we note that stars with different masses will segregate in
different fashions. I.e. the density profile will follow a different power-law
but, still, relaxation will be dominated by stellar-mass black holes, as
discussed in \cite{Amaro-Seoane2019}. This implies that, while relaxation is
the driving mechanism dominating the stellar dynamical evolution, and is
provided by the stellar-mass black holes at the radii of relevance, the CO must
not necessarily follow the same density distribution as that of the
stellar-mass black holes. This is why we will use two different power indices,
$\gamma$ for the distribution of stellar-mass black holes, and $\beta$ for the
distribution of the other species, the COs. Similarly, we will use two
separated masses, one for the stellar-mass black holes, $m_{\rm bh}$ and
another for the COs, $m_{\,\rm CO}$. If we are interested in EMRIs, then
obviously $m_{\rm bh}=m_{\,\rm CO}$ and $\gamma = \beta$.

As shown in \cite{Amaro-Seoane2019}, we can define the threshold between dynamics
and general relativity taking into account the change of relaxation as a function
of the radius with

\begin{equation}
\left( 1-e \right)^{5/2}  = \frac{4.26}{(3-\gamma)(1+\gamma)^{3/2}}\frac{85}{24}
                           \frac{1}{\sqrt{2}\,c^5}\frac{G^{\,5/2}}{\ln(\Lambda)}
                           \frac{M_{\rm bh}^{7/2}}{N_{\rm BH}}
                           \frac{m_{\,\rm CO}}{m_{\rm bh}}
                           R_0^{3-\gamma}a^{\gamma-11/2}.
\label{eq.1_e_a}
\end{equation}

\noindent 
In this equation $R_0$ is the radius within which relaxation, as discussed before,
is dominated by stellar-mass black holes. Accordingly, $N_{\rm BH}$ is how many of them
are enclosed within $R_0$. We choose $R_0 \equiv r_{\rm infl}$.

This allows us to solve for the critical semi-major axis, which is given by

\begin{align}
a_{\rm crit} &=  r_{\rm infl} \, \left[\frac{C}{4.26}\frac{6144}{85}(3-\gamma)(1+\gamma)^{3/2} \right]^{\frac{1}{\gamma-3}} \nonumber \\
               &                   \times  \left[
                                           {\cal W}(\theta,\,{\rm a}_{\bullet})^{5/2}\,N_{\rm BH} \ln(\Lambda)
                                           \left(\frac{M_{\rm BH}}{m_{\,\rm CO}}\right)
                                           \left(\frac{M_{\rm BH}}{m_{\rm bh}}\right)^{-2}
                                            \right]^{\frac{1}{\gamma-3}},
\label{eq.acrit}
\end{align}

\begin{figure}
          {\includegraphics[width=0.8\textwidth,center]{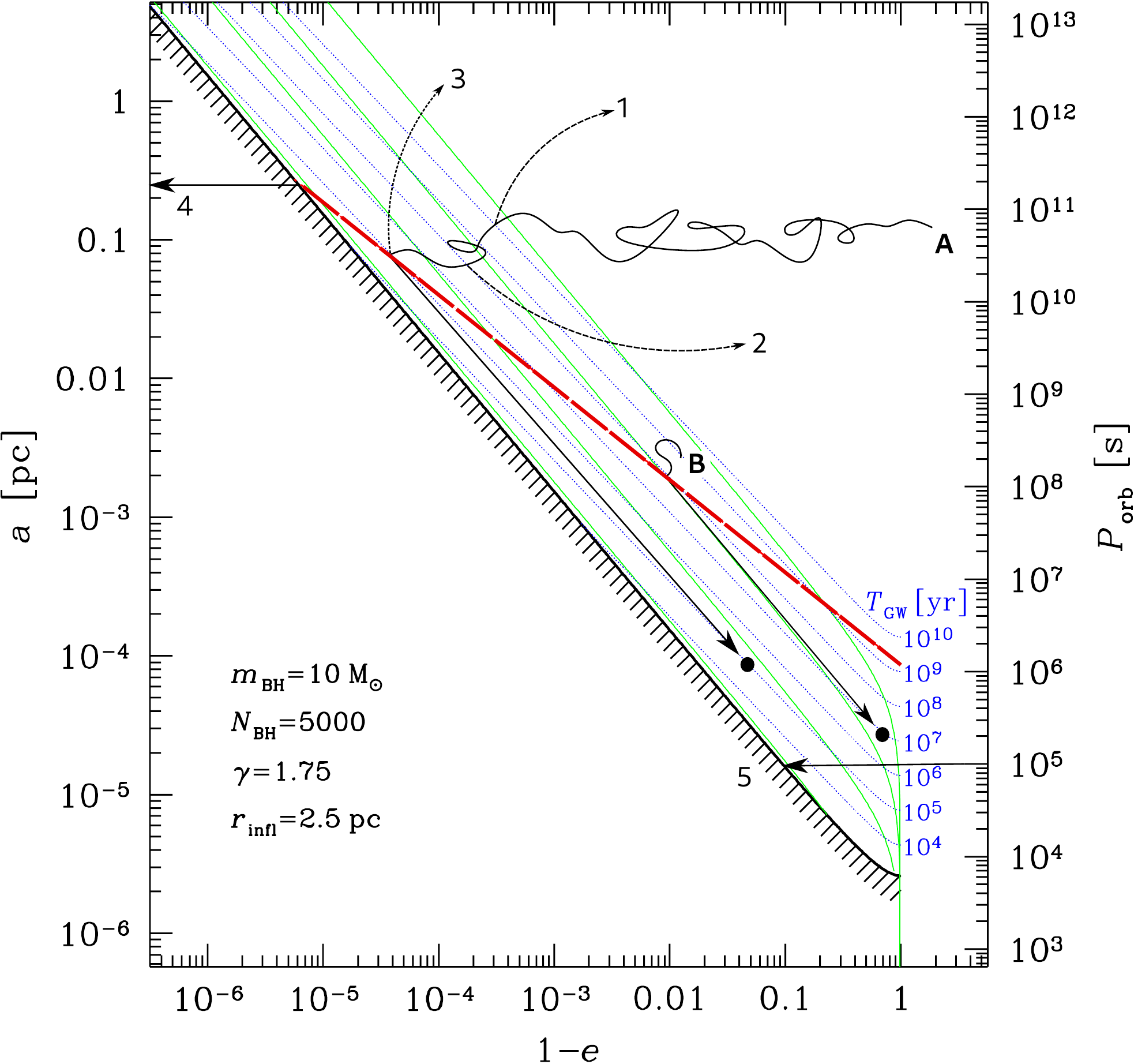}}
\caption
   {
Evolution in phase-space (semi-major axis and orbital period as a function of
the eccentricity) of a potential source of GWs, represented with a black
circle. The mass of the SMBH is of $M_{\rm BH}=4\times10^{4}\,M_{\odot}$. 
The green lines show the correlation between $a$ and $e$ as given in
\cite{Peters64}, and the blue, dotted lines are the isochrones for a given
$T_{\rm GW}$, as described in Eq.~(\ref{eq.Tgwae}). 
\textit{The standard relaxation scenario is marked with the letter ``A'': }
We have marked a few
important points. The first and the second points show the dynamical parameters
for which the CO would merge in less than $10^{10}$ and $10^{8}$ respectively,
if it only evolved due to the emission of gravitational radiation, which is not
the case, because the source is on the right of the oblique, dashed red line, where
dynamics dominate the evolution. Point three represents the crossing of this very
line, from which the evolution is dominated uniquely by GWs. Point four is the
conjunction of the LSO (for a Schwarzschild SMBH) and the threshold line, which
defines in the Y-axis the critical semi-major axis $a_{\rm crit}$, and finally
point five is a rough location of where the LISA bandwidth stars (based on the orbital
period, which will change depending on the dynamical properties of the CO).
\textit{The tidal separation scenario, which will be discussed later, is marked with
the letter ``B'': } We can see that, when entering the LISA band, these sources will
have signifincatly lower eccentricities.
The values which we have chosen
for this plot are shown in the left, bottom corner.
   }
\label{fig.EMRI}
\end{figure}
With the threshold between dynamics and general relativity and $a_{\rm crit}$,
we can now plot the evolution of a potential source of GWs in phase-space. We
represent this in Fig.~(\ref{fig.EMRI}). The source describes a
random-walk-like evolution in energy and angular momentum due to the
interaction with other stars in the stellar system, since we are on the
right-hand of the red line, with a larger scatter in eccentricity than in
semi-major axis until it crosses the threshold given by Eq.~(\ref{eq.1_e_a}).
From that moment onwards, on the left-hand of the red line, the driving
mechanism is gravitational radiation and the evolution in phase-space follows
very closely one of the green lines, which gives the relation between the
semi-major axis and the eccentricity in the two-body problem as approximated by
\cite{Peters64}. We note however, that for strong-field and fast-motion orbits,
radiation reaction will enhance the eccentricity for very small values of the
semi-major axis (i.e. if the semi-lactus rectum $p$ is close to its minimum
value $6+2e$), which can be used as an indicator of the imminent plunge of the
orbit \cite{CKP94}. We give an example of this in Fig.~(\ref{fig.ecc_EMRI_Spin0p2Ecc0p5}),
and see the discussion in the text of that section. 

EMRIs with $a\ll a_{\rm crit}$ will have a much lower event rate because in the
power-law solution for the stellar distribution, we have that the numerical
density of stars $n\propto r^{-\gamma}$, so that the total number of stars $N$
per unit $\log (a)$ scales as ${\rm d}N/{\rm d}(\log a)\propto a^{(3-\gamma)}$.
Moreover, as we move to deeper and deeper radii,  the value of $\gamma$ is
lowered \cite{LS77,ASEtAl04}.  This is so because the loss-cone is much more
quickly depleted and in order to re-populate it we need to wait for several
relaxation times. Also, as we get closer to the SMBH, the value of the
relaxation time increases, as we can see from Eq.~(\ref{eq.relax_t}) and
Eq.~(\ref{eq.relax_t2}). Indeed, unless SMBHs in the Universe are
Schwarzschild, which is very unlikely \cite{SesanaEtAl2014}, EMRIs will
originate at large semi-major axes with very high eccentricities \cite[see
discussion in][]{Amaro-SeoaneSopuertaFreitag2013}.

\subsection{Formation of EMRIs via tidal separation of binaries}

An interesting idea about how to produce an EMRI has its origins in the work of
\cite{Hills88}, who estimated that, in the same way a star can be tidally
disrupted, as we have seen in Sec.~(\ref{Sec.Disr}), a binary could be
separated, ``ionised'', via the same process: The gravitational forces acting
on to one of the two companions would be different than on the other one, and
depending on the distance to the SMBH, this difference of forces could overcome
the binding energy of the star. He predicted that this would lead to the creation
of the so-called ``hyper-velocity stars'', stars with a velocity of $> 10^3\,\kms$.
The discovery of these stars in the work of \cite{BrownEtAl09} led to literally
an avalanche of theoretical and observational work.
Among these works, Miller and collaborators in \cite{MillerEtAl05} presented an
interesting idea. If one of the two stars happened to be a CO, more
specifically a stellar-mass black hole. Because the separation happens very
close to the SMBH, the stellar-mass black hole could eventually become an EMRI.
These EMRIs, contrary to those produced by relaxation, would have a much lower
eccentricity when entering the LISA band, as we can see in
Fig.~(\ref{fig.EMRI}), case ``B''. Because the initial semi-major axis is
smaller than in the relaxation case ``A'', the eccentricity at bandwidth
entrance is much lower.

\begin{figure}
          {\includegraphics[width=0.5\textwidth,center]{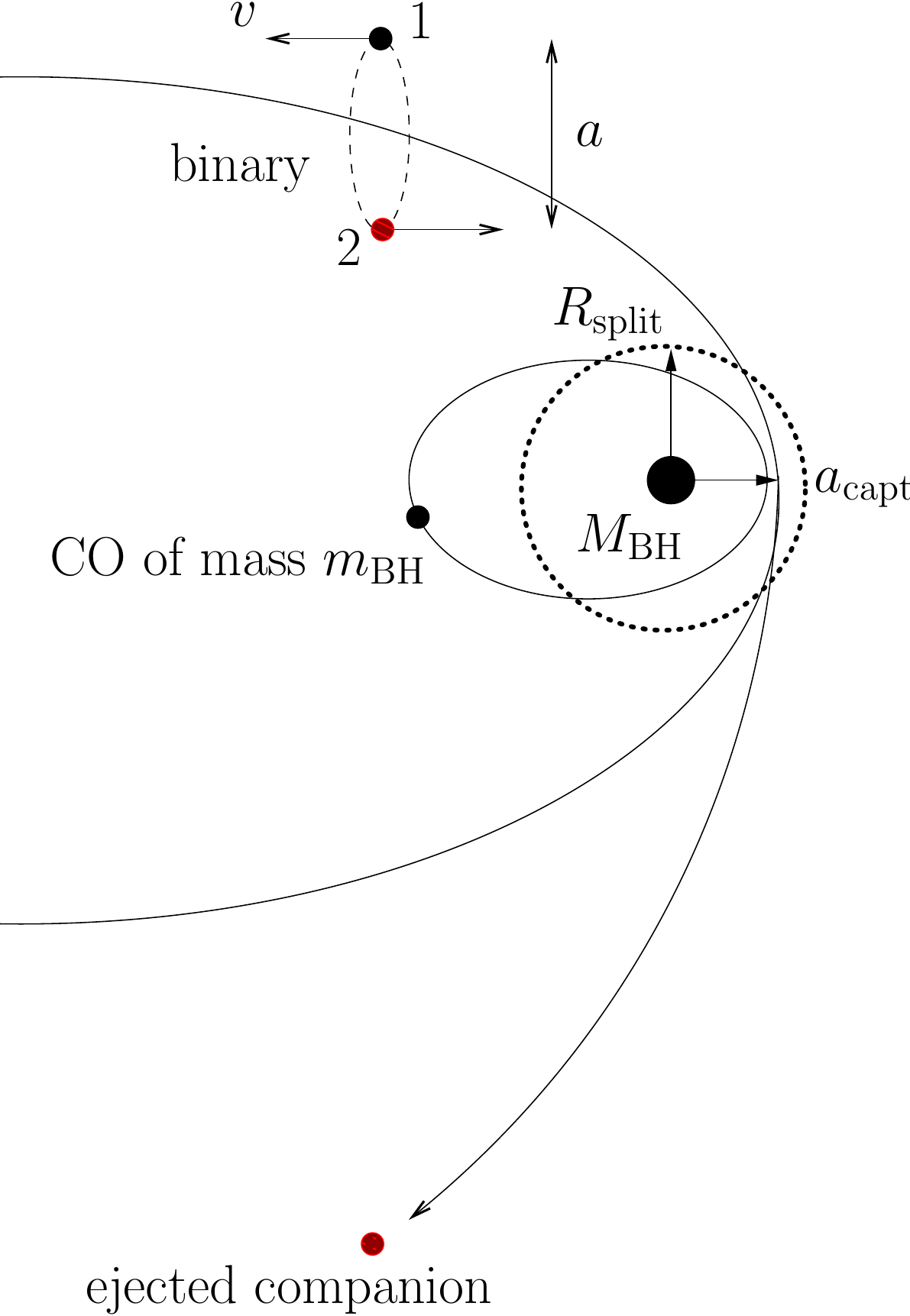}}
\caption
   {
Tidal separation of a binary of two stars of semi-major axis $a$. We consider star ``1'' to be a compact
object of mass $m_{\rm BH}$ which is gravitationally bound to the SMBH of mass $M_{\rm BH}$
because the orbit crosses the splitting radius $R_{\rm split}$. This radius is equivalent
to the tidal disruption radius in the tidal disruption problem. After the separation, star 2
is ejected and star 1 forms a new binary around the SMBH with a new semi-major axis $a_{\rm capt}$,
which typically is much smaller than $a_{\rm crit}$ as defined in Eq.~(\ref{eq.acrit}). This
leads to much smaller eccentricities in the LISA band.
   }
\label{fig.Tidal_separation_binary_xfig.pdf}
\end{figure}

As in the tidal disruption problem, we can derive the splitting radius, which obviously is very similar to the former one,

\begin{equation}
R_{\rm split} \sim a \left(\frac{{M}_{\rm BH}}{m_{\rm bin}}\right)^{1/3},
\label{eq.Rsplit}
\end{equation}

\noindent 
with e.g. $m_{\rm bin}=m_{\star} + m_{\rm CO}$, if one of the companions is a star
and the other one a CO. To get an idea of the ejection velocity that the
companion will receive, we note that the orbital velocity is $v_{\rm orb} \sim
\sqrt{{G\,m_{\rm bin}}/{a}}$, i.e. 

\begin{equation}
v_{\rm orb} \simeq 312\,{\rm km\,s}^{-1}\left( \frac{m_{\rm bin}}{11\,M_{\odot}} \right)^{1/2} 
                    \left(\frac{a}{0.1{\rm AU}} \right)^{-1/2}, 
\end{equation}

\noindent 
assuming that $m_{\star}=1\,M_{\odot}$ and $m_{\rm CO}=10\,M_{\odot}$. Note that
this is just an example, because in a nucleus it is more likely that the mass
of the extended star has a similar mass to that of the CO.  Assuming a
parabolic encounter, which is the most probable scenario, the velocity of the
centre-of-mass $V_{\rm CoM}$ can be estimated to be

\begin{equation}
V_{\rm CoM} \gtrsim \sqrt{\frac{G\,{M}_{\rm BH}}{R_{\rm split}}}\sim\,v_{\rm orb}
\left(\frac{{M}_{\rm BH}}{m_{\rm bin}}\right)^{1/3} \simeq 1.4\times 10^{4}\,{\rm km\,s}^{-1} \gg v_{\rm orb},
\label{eq.}
\end{equation}

\noindent 
for a $M_{\rm BH}=10^6\,M_{\odot}$. Following the derivations of \cite{2018LRR....21....4A}, we have that the ejection velocity
of the companion is $v_{\rm eject} \gtrsim \left({{M}_{\rm}}/{m_{\rm bin}}\right)^{1/6}$ and the new semi-major axis of the binary
$a_{\rm capt} \approx a \left({{M}_{\rm BH}}/{m_{\rm bin}} \right)^{2/3} \simeq 10^4\,a \simeq 5\times 10^{-4}\,\textrm{pc}$.
One can approximate the separation radius $R_{\rm split}$ to the periapsis distance, and with that information we can derive the
initial eccentricity when the CO is captured by the SMBH is of $e_{\rm capt} = 1-\left({{M}_{\rm BH}}/{m_{\rm bin}} \right)^{-1/3} \simeq 0.97$
for the values that we have adopted here. 

These values for the initial semi-major axis and eccentricity are responsible
for the source to enter the LISA band at much lower eccentricities, as we can
see in Fig.~(\ref{fig.EMRI}), case B (which correspond to a slighter more
massive SMBH, of $4\times 10^{4}\,M_{\odot}$). This signature in the
eccentricity as compared to a relaxation EMRI is what will allow us to extract
information about the mechanism that produced the source in the first place.
This is so because if we set all dynamical parameters of two EMRIs identical
but for the eccentricity, we can calculate the mismatch $M$ between them, as
done in \cite{2018LRR....21....4A}, to find 

\begin{equation}
  {M} := 1 -
           \frac{\left<h_{\rm rlx}\left|h_{\rm bin}\right.\right>}
                     {\sqrt{\left<h_{\rm rlx}\left|h_{\rm rlx}\right.\right>
                     \left<h_{\rm bin}\left|h_{\rm bin}\right.\right>}} = 99.9971\%,
\label{eq.mismatch}
\end{equation}

\noindent 
with $h_{\rm rlx}$ and $h_{\rm bin}$ the waveforms of an EMRI produced via
relaxation and binary separation respectively, using the kludge approximation
of \cite{GG06}. The natural scalar product product $<|>$ is introduced by
treating the waveforms as vectors in a Hilbert space \cite{Helstrom68}, and we
refer to this work as well as to \cite{Thorne87,Finn92} for further details.
For this kind of sources, as a rule of thumb, a mismatches $M<0.1$ will make
detection impossible, and a mismatch of $M<10^{-3}$ will make parameter
extraction challenging \citep[see
e.g.][]{ChuaEtAl2017,ChuaEtAl2020,LindblomEtAl2008,CutlerVallisneri2007}.

\subsection{Geodesic motion and relativistic precession}

In this section, which profits from parts of
\citep{Amaro-SeoanePretoSopuerta2010}, we give a summary of the concepts that
we will need for the evolution of an EMRI around a Schwarzschild SMBH, the
geodesic motion and relativistic precession. The Kerr case is much more complex
and we will simply give an illustrative summary. We then give examples that
solve the evolution of an EMRI (or an IMRI, XMRI) in phase-space by resorting
to approximate, semy-analitical and numerical techniques.

\subsubsection{Geodesic motion around a Schwarzschild black hole}

The only spherically symmetric solution of Einstein's equations for vacuum is the
Schwarzschild solution (Birkhoff's theorem)\footnote{Using normal units we have
\[f = 1 - \frac{2G\mbul}{c^{2}r}\]
where $G$ is Newton's constant and $c$ denotes the speed of light.}
\begin{equation}
ds^{2} = -f\,dt^{2} +\frac{dr^{2}}{f} + r^{2}d\Omega^{2}\,, \qquad
f = 1 - \frac{2\mbul}{r}\,, \qquad
d\Omega^{2} = d\theta^{2} + \sin^{2}\theta d\varphi^{2}\,.
\label{schmetric}
\end{equation}
This solution predicts the existence of a horizon at $r^{}_{g}= 2\mbul$ 
($=2G\mbul/c^{2}$ in {\em normal} unit conventions) and hence a {\em black hole}
geometry.  We are interested in the motion of a test mass around such black
hole (whose geometry is described by Eq.~(\ref{schmetric})).   As in Newtonian
gravity, the motion takes place in a plane that we can take to be the 
equatorial plane $\theta = \pi/2$.  There are two constants of motion, the
energy ${\cal E}$ and the angular momentum ${\cal J}$, associated with the time and azimuthal
Killing symmetries\footnote{We use the letters $({\cal E},{\cal J})$ to distinguish them
from the Newtonian definitions, $(E,J)$.  Later we will see what are the relations between them.}.  
Thanks to this the motion is completely separable and hence
integrable.  Let us see how this works.  The energy and angular momentum, $({\cal E},{\cal J})$, 
in terms of the Schwarzschild
coordinates of Eq.~(\ref{schmetric}), can be written as:
\begin{equation}
{\cal E} = f\, \frac{dt}{d\tau}\,,\qquad
{\cal J} = r^{2}\,\frac{d\varphi}{d\tau}\,, \label{EJsch}
\end{equation}
where $\tau$ denotes proper time (the time measured by the clocks of an observer
moving with the test mass) and $t$ is the coordinate time of Eq.~(\ref{schmetric}),
the time measured by the clocks of distance observers, at infinity).   Notice that
the constants of motion $({\cal E},{\cal J})$ are also {\em specific} constants
of motion (per unit mass of the test particle), like in the Newtonian situation of
Eqs.~(\ref{hamiltonian}) and~(\ref{newtonianJ}).

\noindent In the language of General Relativity, we have the four velocity
\begin{equation}
u^\mu = \frac{dx^\mu}{d\tau}\,, \qquad
(u^\mu) = \left(\frac{dt}{d\tau},\frac{dr}{d\tau},\frac{d\theta}{d\tau},
\frac{d\varphi}{d\tau}\right)\,,
\end{equation}
which is a normalized vector with respect to the metric (because it has been defined in
terms of proper time)
\begin{equation}
\met_{\mu\nu}u^{\mu}u^{\nu} = -1 \,, \label{unormalization}
\end{equation}
where $\met_{\mu\nu}$ are the metric components, which in the case of Schwarzschild can be
read off from Eq.~(\ref{schmetric}). Then, Eq.~(\ref{unormalization}) is equivalent to 
the following expression:
\begin{equation}
-1 = -f \left(\frac{dt}{d\tau}\right)^{2} + \frac{1}{f}\left(\frac{dr}{d\tau}\right)^{2}
+ r^{2}\left[\left(\frac{d\theta}{d\tau}\right)^{2} +\sin^{2}\theta\left(\frac{d\varphi}{d\tau}
\right)^{2} \right]\,.
\end{equation}
If we substitute in this equation $dt/d\tau$ and $d\varphi/d\tau$ from Eq.~(\ref{EJsch})
and use the choice $\theta=\pi/2$, which implies $d\theta/d\tau = 0$, we obtain the following
equation for the radial motion:
\begin{equation}
\left(\frac{dr}{d\tau}\right)^{2} = {\cal E}^2 - \left( 1 + \frac{{\cal J}^2}{r^2}\right)
\left(1-\frac{2\mbul}{r}\right)\,.
\end{equation}
To summarize, the equations of motion, in terms of proper time are:
\begin{align}
\frac{dt}{d\tau} & =  \frac{{\cal E}}{1-\frac{2\mbul}{r}}\,, \label{dtdtau} \\
\left(\frac{dr}{d\tau}\right)^{2} & =  {\cal E}^2 - \left( 1 + \frac{{\cal J}^2}{r^2}\right)
\left(1-\frac{2\mbul}{r}\right)\,, \label{drdtau}\\
\frac{d\varphi}{d\tau} & =  \frac{{\cal J}}{r^2} \label{dvarphidtau} \,.
\end{align}
Eq.~(\ref{dtdtau}) just gives the relation between coordinate time $t$ and proper time $\tau$.  
We can use it, to rewrite the equations of motion, the geodesics, in terms of coordinate time.

For the purposes of this work, it is very convenient to introduce in an appropriate way a
three-velocity.  This can be done in a {\em natural} way by factoring out the time-component
of the four velocity:
\begin{equation}
u^{\mu} = u^{t}v^{\mu} = u^{t}\,\left(1,v^{i}\right)\,,\qquad
v^{i} \equiv \frac{u^{i}}{u^{t}}\,.
\end{equation}
Here, it is important to notice that this three-velocity, or better spatial velocity, is 
now defined in terms of the coordinate time $t$ as follows
\begin{equation}
v^{i} = \frac{dx^{i}}{dt}\,.
\end{equation}
In this way, and by virtue of the normalization property of $u^{\mu}$, we can interpret $u^{t}$
as the relativistic gamma factor associated with the observers that measure the coordinate time
$t$:
\begin{equation}
\met_{\mu\nu}u^{\mu}u^{\nu} = -1\qquad\Longrightarrow\qquad 
u^{t} = \frac{1}{\sqrt{-\met_{tt}-2\met_{ti}v^{i}-\met_{ij}v^{i}v^{j}}}\equiv\Gamma\,.
\label{gammafactor}
\end{equation}
In special relativity, the gamma factor is just $\Gamma = 1/\sqrt{1-v^{2}}$.  We can check that
this is what we get from Eq.~(\ref{gammafactor}) for the Schwarzschild metric of Eq.~(\ref{schmetric})
when we take the limit of no gravity, i.e. $G\rightarrow 0$ (or, in our units, the limit $\mbul\rightarrow
0$).

Going back to the study of Schwarzschild geodesics, let us now focus on bound
orbits, since they are the ones we are interested in.  By definition the radial
coordinate for these orbits must lie inside a finite interval, $[r_{p},r_{a}]$,
where $r_{p}$ is the pericenter radial coordinate and $r_{a}$ is the apocenter
one.  This means that for these particular values of $r$, $dr/d\tau$ must
vanish.  If we look at Eq.~(\ref{drdtau}) we realize this can only happen if
${\cal E}^{2}\leq 1$.  Assuming ${\cal E}>0$ (negative values corresponds to
the time-reverse orbits), this means $0\leq{\cal E}\leq 1$.  More in detail,
Eq.~(\ref{drdtau}) can be rewritten as follows:

\begin{equation}
\left(\frac{dr}{d\tau}\right)^{2} =  \frac{1}{r^{3}}\left(1-{\cal E}^2\right)\left(r_{a}-r\right)
\left(r-r_{p}\right)\left(r-r_{o}\right)\,,
\end{equation}

\noindent 
only valid for bound geodesics.  Here, $r_{o}$ is a third root that satisfies: $r_{a}>r_{p}>r_{o}$,
and hence it will not be reached during the motion.  By comparing this with Eq.~(\ref{drdtau}) we 
can obtain $(r_{p},r_{a},r_{o})$,

\begin{align}
r_{p}+r_{a}+r_{o} & =  \frac{2\mbul}{1-{\cal E}^{2}}\,, \\
r_{p}r_{a} + r_{o}(r_{p}+r_{a}) & =  \frac{{\cal J}}{1-{\cal E}^{2}}\,,\\
r_{p}r_{a}r_{o} & =  \frac{2\mbul {\cal J}^{2}}{1-{\cal E}^{2}} \,.
\end{align}

\noindent 
In order to solve them, we can introduce the eccentricity and dimensionless semilatus rectum orbital parameters, $(e,p)$,
in the usual way\footnote{The general rule to recover proper units is to make the substitution: $\mbul$
$\longrightarrow$ $G\mbul/c^{2}$.}

\begin{equation}
r_{p} = \frac{p\,\mbul}{1+e}\,, \qquad
r_{a} = \frac{p\,\mbul}{1-e}\qquad \Longleftrightarrow \qquad
p = \frac{2r_{p}r_{a}}{\mbul(r_{p}+r_{a})}\,,\qquad
e = \frac{r_{a}-r_{p}}{r_{a}+r_{p}}\,.
\end{equation}

\noindent 
Then, we can first find the expressions of $({\cal E},{\cal J})$ and $r_{o}$ in terms only of
$(e,\,p)$ and the black hole mass:

\begin{align}
r_{o}        & =   \frac{2\,p\,\mbul}{p-4}\,, \\[1mm]
{\cal E}^{2} & =  \frac{(p-2+2\,e)(p-2-2\,e)}{p(p-3-e^{2})}\,, \\[1mm]
{\cal J}^{2} & =  \frac{p^{2}\,\mbul^{2}}{p-3-e^{2}}\,.
\end{align}

\noindent 
An interesting relation comes out by imposing that $r_{o}<r_{p}$ as we have assumed at the beginning:

\begin{equation}
p-6-2\,e>0\,, \label{separatrix}
\end{equation}

\noindent 
which is a separatrix between stable and unstable bound orbits.

In practice, in order to numerically integrate the equations of motion for bound orbits
we have to take into account that $r$ is not a good coordinate to use due to the 
existence of turning points.  In order to avoid the numerical problems derived from
this, it is very convenient to introduce the following alternative angular variable:

\begin{equation}
r = \frac{p\mbul}{1+e\cos\psi}\,,
\end{equation}

\noindent 
The orbital motion, in terms of the variables $(t,\psi,\varphi)$ is described by
the following set of ODEs

\begin{equation}
\frac{d\psi}{dt} = \frac{(1+e\cos\psi)^{2}}{p^{2}\mbul} 
\frac{(p-2-2e\cos\psi)\sqrt{p-6-2e\cos\psi}}{\sqrt{(p-2)^{2}-4e^{2}}}\,,
\label{dpsidt}
\end{equation}

\begin{equation}
\frac{d\varphi}{dt} = \frac{(1+e\cos\psi)^{2}}{p^{3/2}\mbul}
\frac{(p-2-2e\cos\psi)}{\sqrt{(p-2)^{2}-4e^{2}}}\,.
\label{dvarphidt}
\end{equation}

\subsubsection{Relativistic precession}

In Keplerian motion, the time that a particle take to go from $r_{p}$ to $r_{a}$ and
back to $r_{p}$ (the radial period) is exactly the same as the time it takes to go
$2\pi$ around the central object, that is, to cover a $\varphi$-period.   In General
Relativity, this is not the case, and in the case of a non-spinning black hole these
two periods do not coincide.  The consequences of this is that the orbit does not
close itself and there is precession of the pericenter (it is not located in the
same position with respect to Cartesian coordinates associated with $(r,\varphi)$,
i.e. $(x,y)=(r\cos\varphi,r=\sin\varphi)$).  

Using the equations of motion given in Eqs.~(\ref{dpsidt}) and~(\ref{dvarphidt}), 
the amount of angle $\varphi$ covered during one single radial period is given
by
\begin{equation}
\Delta\varphi = 2 \int^{t_{a}}_{t_{p}} \frac{d\varphi}{dt}\,dt \,,
\end{equation}
where $t_{p}$ and $t_{p}$ indicate the coordinate time corresponding to the apocenter
and pericenter locations.  We can rewrite this integral in terms of $\psi$ to get
\begin{equation}
\Delta\varphi = 2 \int^{\pi}_{0} \frac{d\varphi}{d\psi}\,d\psi = 2 p^{1/2}\int^{\pi}_{0}
\frac{d\psi}{\sqrt{p-6-2e\cos\psi}} \,,
\end{equation}
and using that $\cos\psi = -\cos(\pi-\psi)= -1 + 2\sin^{2}(\pi/2-\psi/2)$ and defining 
$x \equiv(\pi - \psi)/2$ we get
\begin{equation}
\Delta\varphi = \frac{4p^{1/2}}{\sqrt{p-6+2e}}\int^{\pi/2}_{0}\frac{dx}{\sqrt{1-\frac{4e}{p-6+2e}\sin^{2}x}}
= \frac{4p^{1/2}}{\sqrt{p-6+2e}}{K}\left(\sqrt{\frac{4e}{p-6+2e}}\right)\,,
\end{equation}
where ${K}(k)=\int_0^{\pi/2} d\alpha \, (1-k\,\sin^2 \alpha)^{-1/2}$ is the complete elliptic integral
of the first kind.  It turns out that this integral diverges as $p$ approaches $6+2e$, that is,
when we approach the separatrix of Eq.~(\ref{separatrix}).

\subsection{The Kerr case}

Now that we have addressed the Schwarzschild case, which is our reference
point, we will describe the characteristics that distinguish a Kerr SMBH from
it. This is a simple summary focused on the main ideas which are important for
the evolution on phase-space that we address later. For a rigurous, detailed
review, we refer the reader to the reviews of \citep{Visser2007,Teukolsky2015}. 

The most interesting feature is that, whilst in the Schwarzschild case the
geometry is spherically-symmetric, in Kerr it is axisymmetric respect to the
spin axis. This translates into the fact that orbits outside the equatorial
plane are not planar. There is no such (Keplerian- or Schwarzschild-like)
concept in Kerr. As a matter of fact, the concept of ``orbit'' is not exactly
straightforward in the relativistic case if it is not a bound one, and there is
no simple way to compare it with the Keplerian concept. In any case, the
inclination of the orbit $\iota$ with respect to the spin axis is fundamental
in Kerr and decides the dynamics of the system.

Non-equatorial orbits (which are a special case) precess with a given $\iota$
at a frequency $f^{}_{\theta}$, with $\theta$ the polar Boyer-Lindquist
coordinate~\citep{BoyerLindquist1967,MisnerThorneWheeler1973}.  We can define
the inclination for Kerr geodesic orbits in two different ways.  One
possibility is via Carter's constant $Q$,

\begin{equation}
\cos \iota = \frac{L_{\rm z}}{\sqrt{L_{\rm z}^2 + Q}},
\end{equation}

\noindent
and another with the minimum value of $\theta$,

\begin{equation}
\iota = \textrm{sign}\,(L_{\rm z}) \left[ \frac{\pi}{2} - \theta_{\rm min}\right],
\end{equation}

\noindent 
where $\textrm{sign}\,(L_{\rm z})$ allows us to distinguish between prograde (positive)
and retrograde orbits (negative).

\begin{figure}
          {\includegraphics[width=0.8\textwidth,center]{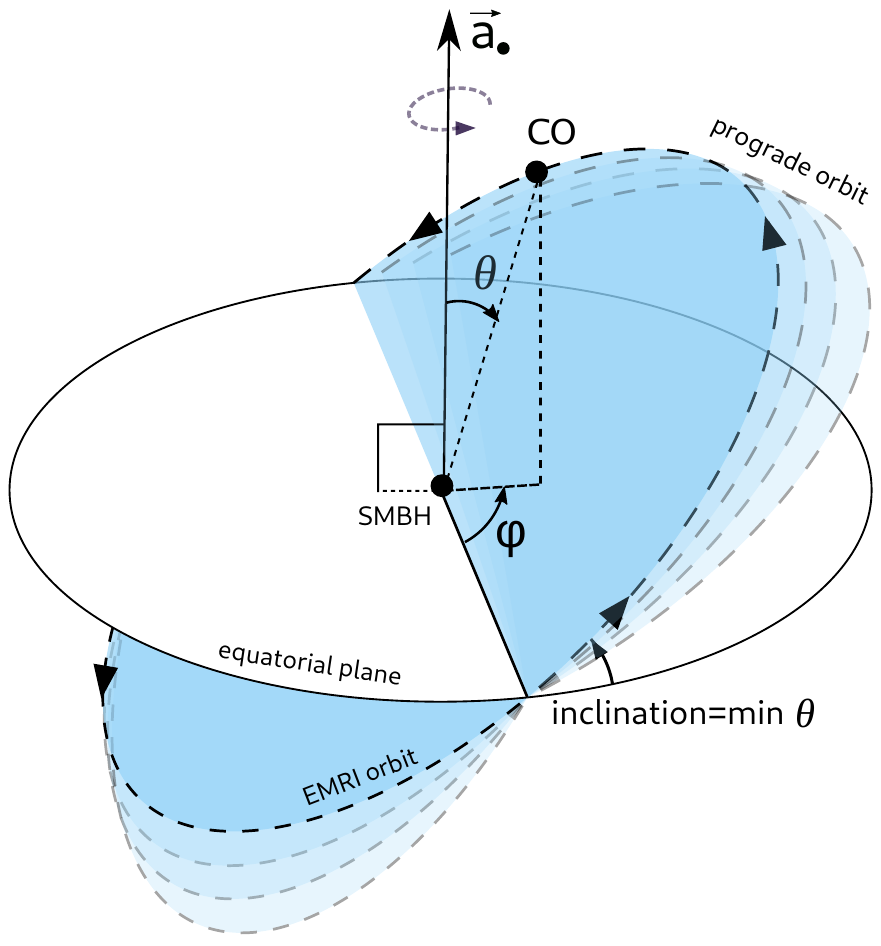}}
\caption
   {
The minimum polar angle $\theta$ from the spin axis is defined as the inclination $\iota$ of the
EMRI. We illustrate this with a prograde orbit of a CO revolving around a SMBH
of spin $a_{\bullet}$ on a prograde orbit. While $\theta$ changes significantly,
the inclination does not, as we can see in Fig.~(\ref{fig.iota_Spin0p9Ecc0p01}).
   }
\label{fig.inclination}
\end{figure}

In Fig.~(\ref{fig.inclination}) we depict the orbit of a prograde EMRI
revolving around a SMBH of spin $a_{\bullet}$. The minimum value of the polar
angle $\theta$, determines $\iota$.  This figure is to be regarded as an
illustration of the process to show the effect of radiation reaction. The
orbital plane however precesses on a much faster timescale, so that the orbit
does not look like depicted for a static observer. In general it fills the
volume of a torus. For a more realistic illustration, see e.g. Fig.~(4) of
\citep{BarackPound2019}. The inclination $\iota$ must vary as the EMRI
approaches more and more the LSO, it is a constant of motion for geodesics, not
to be misinterpreted with the instantaneous orbital colatitude $\theta$ of
Boyer-Lindquist coordinates.

As in the Schwarzschild, the periapsis precesses, so that we need to introduce
two additional frequencies. One is related to the radial motion and the time to
go from periapsis, $r^{}_{\rm peri}$, to apoapsis, $r^{}_{\rm apo}$, and back.
The second one is linked to the azimuthal motion around the spin axis and the
required time to describe $2\pi$ around it, i.e. the time for the azimuthal
angle $\varphi$ to increase $2\pi$ radians. We call these two frequencies
$f^{}_{r}$ and $f^{}_{\varphi}$, respectively. These three fundamental
frequencies and their harmonics are responsible to make so rich in information
an EMRI, because they allow the CO, the EMRI, to probe the geometry around the
SMBH with gravitational radiation, which encodes this cartography of warped
spacetime.

The emitted gravitational waves backreact on to the CO itself and lead to a
change of its orbital parameters which can be estimated by the energy and
angular momentum carried away from the radiation, namely the semi-lactus rectum
-or alternatively the semi-major axis-, the eccentricity and the inclination of
the orbit, $(p,\,e,\,\iota)$.

The geodesic motion around a Kerr SMBH has three constants of motion, the
energy per unit mass $E$ (which we normalise to $m_{\rm CO}$), $L^{}_{\,z}$,
and the Carter constant per unit mass square. This last is linked to an extra
symmetry of the Kerr geometry, in the same way that some axisymmetric Newtonian
potentials display~\citep[see e.g.][]{BinneyTremaine08,Amaro-SeoaneLRR}.  These
three constants are modified due to the emission of gravitational radiation.
The work of \citep{Schmidt2002} shows that there is a connection between the
constants $(E,L^{}_{\,z},C)$ and the set of orbital parameters $(p,e,\iota)$;
more precisely, there is a (bijective, an injective-surjective) mapping between
both sets, which is useful to analyse the evolution of the EMRI without having
to explicitely solve the evolution of the orbit. This mapping is however
complex and we refer the reader to e.g. the implementation of
\citep{SopuertaYunes2011} for a reference.

Again, as a reference, we illustrate this mapping for a simpler case, a Schwarzschild SMBH. 
It is simpler because $\iota$ and $C$ are not requied. In this case,

\begin{align}
\frac{E^{2}}{c^{2}} = & \frac{(p-2-2e)(p-2+2e)}{p\,(p-3-e^2)}\,,\\
L^{2}_{\,z} = & \frac{G^{2}M^2_{\rm BH}\,p^2}{c^{2}(p-3-e^2)}\,.
\end{align}

Using the symmetries of the geometry of a Kerr SMBH we can separate the equations for
geodesic orbital motion so that the trajectory of a massive body, described in terms
of Boyer-Lindquist coordinates $\{t,r,\theta,\varphi\}$, can be written as follows

\begin{align}
\rho^{2}\, \frac{d{t}}{d\tau}                        & = \frac{1}{\Delta}\left( \Sigma^{2}\frac{E}{c} - 2 a^{}_{\bullet} 
                                                         r^{}_{\bullet} \frac{{L}^{}_{\,z}}{c} r  \right)  \label{tdot-GR} \\
\rho^{4}\, \left(\frac{d {r}}{d\tau} \right)^{2}      & = \left[ \left( r^{2} + a^{2}_{\bullet} \right) \frac{E}{c} - a^{}_{\bullet} 
                                                          \frac{L^{}_{\,z}}{c} \right]^{2} -\left(\frac{Q}{c^{2}}+r^{2}\right)\Delta
                                                          \equiv  R(r) \label{rdot-GR} \\
\rho^{4}\, \left(\frac{d{\theta}}{d \tau} \right)^{2} & = \frac{C}{c^{2}} - \frac{{L}^{2}_{\,z}}{c^{2}} \cot^{2}{\theta} - 
                                                          a^{2}_{\bullet} \left(1 - \frac{{E}^{2}}{c^{2}} \right)
                                                          \cos^{2}{\theta} \label{thetadot-GR}\\
\rho^{2}\, \frac{d \varphi}{d{\tau}}                  & = \frac{1}{\Delta}\left[ 2a^{}_{\bullet}r^{}_{\bullet} \frac{E}{c} r + 
                                                          \frac{{L}^{}_{\,z}}{c} \frac{ \Delta -a^{2}_{\bullet}{\sin^{2}{\theta}}}
                                                          {\sin^{2}\theta}  \right] \label{phidot-GR}.
\end{align}

\noindent
Where we have defined the gravitational radius $r^{}_{\bullet} \equiv
{GM^{}_{\bullet}}/{c^{2}}$, $Q \equiv C +  \left(L^{}_{\,z} - a^{}_{\bullet} E
\right)^{2}$, $\rho^{2} \equiv r^{2} + a^{2}_{\bullet} \cos^{2}{\theta}$ and
$\Delta \equiv r^{2}-2r^{}_{\bullet}r+a^{2}_{\bullet} = r^{2} f +
a^{2}_{\bullet}$, with $f \equiv 1 - {2\, r^{}_{\bullet}}/{r}$, and $\Sigma^{2}
\equiv (r^2 + a^2_{\bullet})^{2} - a^{2}_{\bullet}\Delta\,\sin^{2}\theta\,$.
In this set of equations, Eq.~(\ref{tdot-GR}) gives us the link between the
change of coordinate time $t$, the time of observers at infinity, and the
proper time $\tau$, while Eqs.~(\ref{rdot-GR}), (\ref{thetadot-GR}) and
(\ref{phidot-GR}) give us the link for the spatial trajectory. This set can be
combined to derive the spatial trajectory in terms of coordinate time $t$,
$(r(t),\theta(t),\varphi(t))$.

\subsection{Evolution in phase-space}

We can analyse the the evolution of the radiation emitted by an EMRI thanks to
the approximation of Keplerian ellipses of \cite{PM63}. In this approximation,
the orbital parameters change slowly because of GWs, which are emitted at every
integer multiple of the orbital frequency, $\omega_{\,n}=n\,\sqrt{G\,M_{\rm
BH}/a^3}$.  At a given distance $D$, the strain amplitude in the n-th harmonic
is 

\begin{align}
    h_n &= g(n,e) \frac{G^2\,M_{\rm BH} m_{\rm BH}}{D\,a\,c^4} \\
        \nonumber &\simeq  1.6\times 10^{-22} g(n,e)
    \left(\frac{D}{1\,\mathrm{Gpc}}\right)^{-1}
    \left(\frac{a}{10^{-2}\,\mathrm{pc}}\right)^{-1} \nonumber \\
    & \left(\frac{M_{\rm BH}}{4\times 10^4\,M_{\odot}}\right)
    \left(\frac{m_{\rm BH}}{10\,M_{\odot}}\right).
\label{eq.hn}
\end{align}

\noindent
In this expression $g(n,\,e)$ is a function of the harmonic number $n$ and the
eccentricity $e$ \citep[see][]{PM63}. We consider the RMS amplitude averaged
over the two GW polarizations and all directions.Whilst there are more accurate
descriptions of the very few last orbits, as e.g.
\cite{PPSLR01,GlampedakisEtAl2002,BarackCutler2004,GG06}, the scheme of
\cite{PM63} yields a qualitatively correct estimation of the frequency cutoff
at the innermost stable circular orbit.

\begin{figure}
          {\includegraphics[width=0.8\textwidth,center]{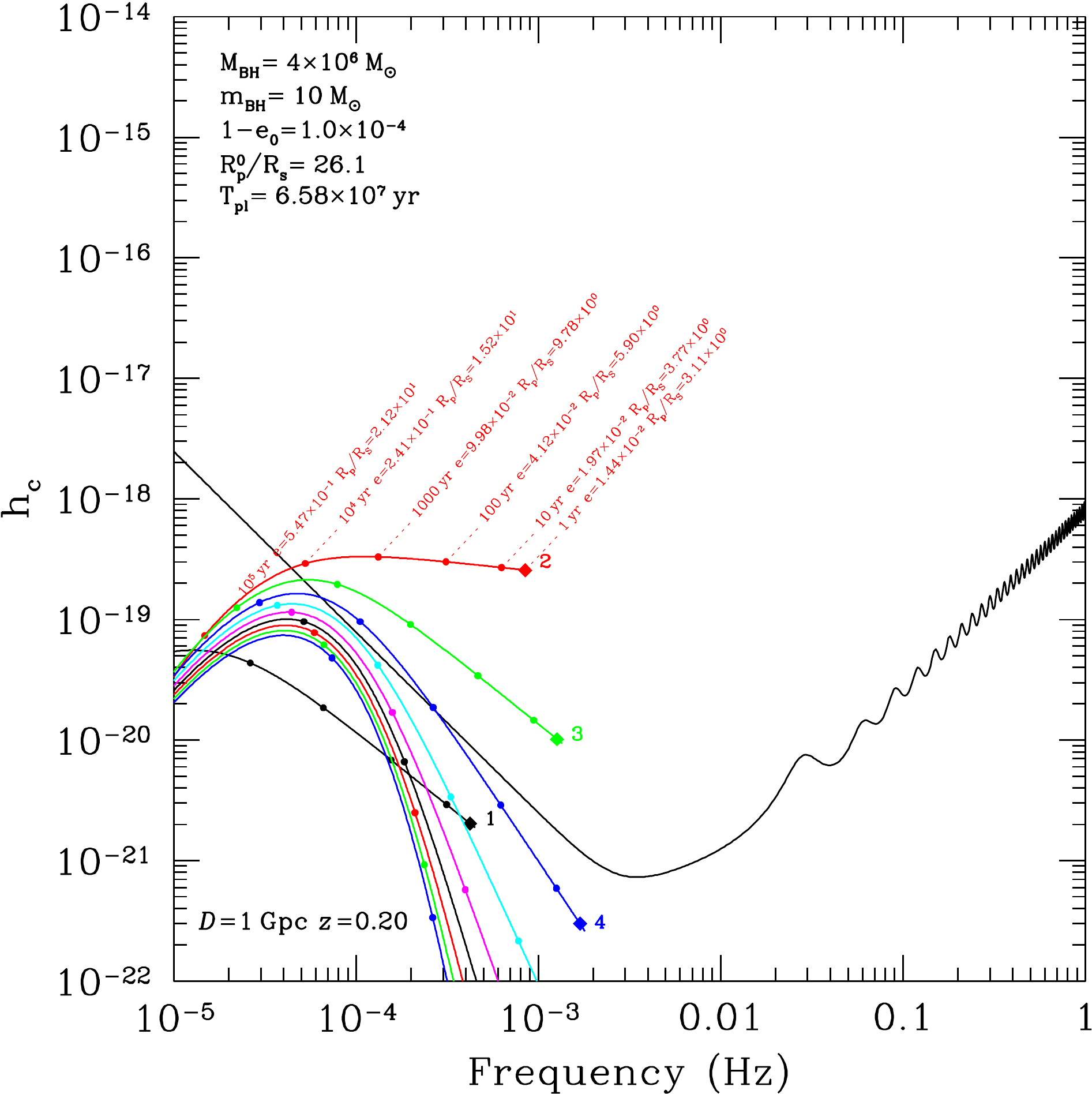}}
\caption
   {
Cascade of the first 10 harmonics in the approximation of \cite{Peters64}. The
Y-axis shows the characteristic, dimensionless amplitude (how much the length of the
arms of the detector changes divided by the length) as a function of the frequency
in Hz. The CO has been set on an orbit whose dynamical parameters are summarised
on the top, left corner, with $e_0$ the initial eccentricity, $R^0_{\rm p}$ the
initial periapsis distance, and $T_{\rm pl}$ corresponds to the associated timescale
$T_{\rm GW}$ as described in Eq.~(\ref{eq.Tgwae}).    
We have adopted values corresponding to a relaxation
EMRI, as in Fig.~(\ref{fig.EMRI}). The upper, red curve corresponds to the
second harmonic, which is dominant in this case because, although the CO
initially has a large eccentricity, it has circularised. Different points on
this curve display information about the system as it evolves with time. The
diamond corresponds to one year before the plunge on to the SMBH.
}
\label{fig.EMRI_harmonics}
\end{figure}

In Fig.~(\ref{fig.EMRI_harmonics}) we show the first ten harmonics in the plane
of characteristic amplitude as a function of the frequency. We can see that
at detector ``entrance'' the source is already quite circular, even if the initial
eccentricity was set to what we expect from a relaxation EMRI. This is why the
second harmonic dominates over the rest of them. Even if initially the periapsis
distance was of $R^0_{p}=26.1\,R_{\rm S}$, $10^4$~yrs before the plunge (i.e.
crossing the event horizon of the SMBH), the EMRI has an eccentricity of $e=0.547$
and a periapsis distance of $R_{p}=21.2\,R_{\rm S}$. This depicts quite clearly
the slow shrinkage and circularization of EMRIs. Ten years before the plunge,
we have that $e=1.97\times 10^{-2}$, with $R_{p}=3.77\,R_{\rm S}$, and nine
years later, $e=1.44\times 10^{-2}$, with $R_{p}=3.11\,R_{\rm S}$. This means
that the CO, a stellar-mass black hole in this case, is at the verge of falling
on to the SMBH but for a small epsilon, and this distance becomes smaller by
extremely amounts over a timescale of 9 years. This translates into a very 
large number of strong bursts from a coherent source at the verge of the abyss,
being transported to us in an almost unperturbed way and carrying information
about spacetime from a region inaccessible to the photon. Moreover, since the
CO is changing the plane of the orbit (for a Kerr SMBH), shrinking the semi-major
axis and precessing, we are de facto gathering a cartography of warped spacetime.

We now give a few examples for the dynamical parameters of relaxation EMRIs as
well as their polarizations. We follow the ``numerical kludge scheme'', which
conceptually computes the trajectory of the CO using Boyer-Lindquist
coordinates \citep{BoyerLindquist1967}, to then identify them with flat-space
spherical polar ones. This allows us to then derive the waveform from the
multipole moments in the context of linearised gravitational perturbation
theory in flat spacetime.  This waveform is computed with a slow-motion
quadrupole formula, a quadrupole/octupole formula, and the weak-field approach
for fast-motion of \cite{Press1977}. This scheme has evolved over the years and
has been significantly improved, see
\cite{BarackCutler2004,BabakEtAl2007,ChuaGair2015,ChuaEtAl2017}. For the
particular calculations we have done, we have used the ``EMRI Kludge Suite'' of
\cite{ChuaGair2015,ChuaEtAl2017}.

\begin{figure}
          {\includegraphics[width=0.7\textwidth,center]{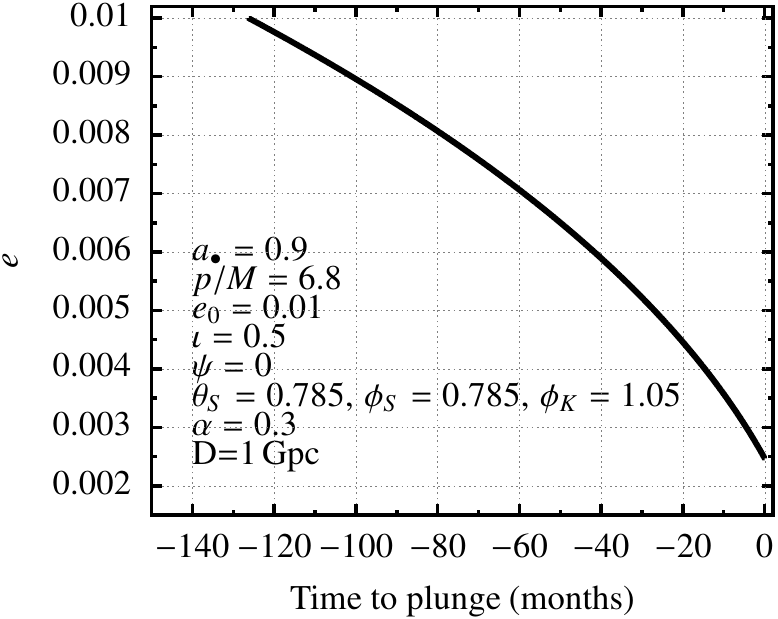}}
\caption
   {
Eccentricity evolution as a function of time to plunge in months. The labels
correspond (from the top to the bottom, left to right) to the initial values of the semi-latus rectum $p/M$,
the spin of the SMBH $a_{\bullet}$, the inclination $\iota$, the true anomaly $\psi$,
the source polar angle $\theta_S$, the azimuthal angle $\phi_S$, the SMBH spin azimuthal angle
$\phi_K$ (all of them in ecliptic coordinates), the azimuthal orientation $\alpha$,
as defined in Eq~(18) of \cite{BarackCutler2004}, and the distance to the source D.
Note that we do not see an increase in the eccentricity as in Fig.~(\ref{fig.ecc_EMRI_Spin0p2Ecc0p5}) 
because the separatrix for $a_{\bullet}=0.9$ located at a different minimum value of 
the semi-lactus rectum $p$.
   }
\label{fig.ecc_Spin0p9Ecc0p01}
\end{figure}

\begin{figure}
          {\includegraphics[width=0.7\textwidth,center]{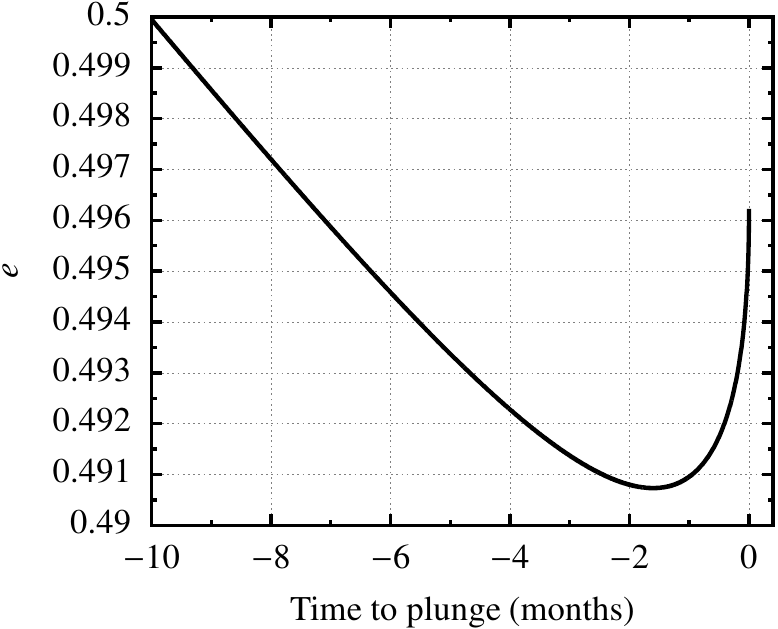}}
\caption
   {
Evolution of the eccentricity of an EMRI displaying the increase in the last orbits for
a SMBH of mass $M_{\rm BH}=4\times10^6\,M_{\odot}$, of spin $a_{\bullet}=0.2$, and 
a CO which initially had $e_0=0.5$. The rest of the parameters are identical to those
of Fig.~(\ref{fig.ecc_Spin0p9Ecc0p01}).
   }
\label{fig.ecc_EMRI_Spin0p2Ecc0p5}
\end{figure}

Fig.~(\ref{fig.ecc_Spin0p9Ecc0p01}) corresponds to the evolution of the
eccentricity for an EMRI with characteristics similars to those of
Fig.~(\ref{fig.EMRI_harmonics}), at detector entrance, when the eccentricity is
of about $e\simeq 10^{-2}$. In Fig.~(\ref{fig.ecc_EMRI_Spin0p2Ecc0p5}) we show
the same but for a system of a much lower eccentricity, so that the separatrix
is located very close to the minimum value of the semi-latus rectum of a 
Schwarzschild black hole, $p=6+2e$.

\begin{figure}
          {\includegraphics[width=0.7\textwidth,center]{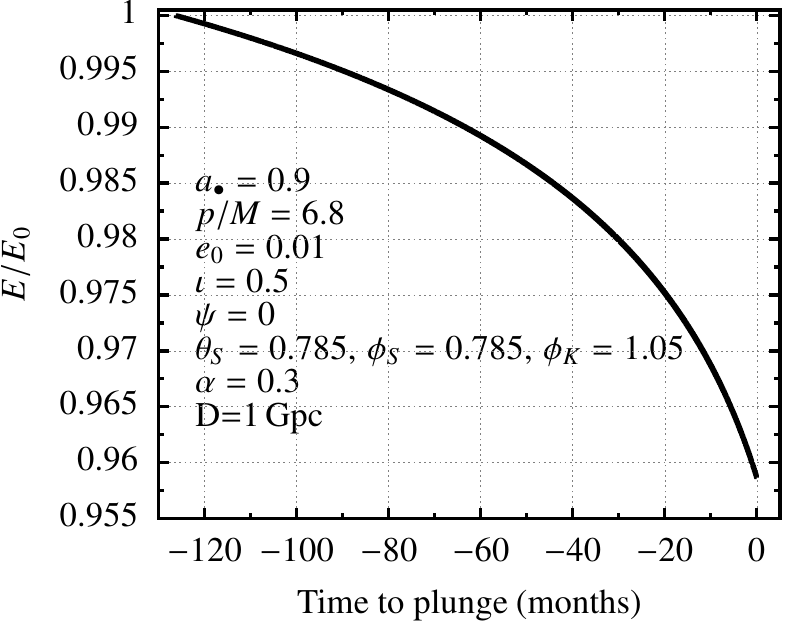}}
\caption
   {
Evolution of the energy of the EMRI normalised to the initial energy
at the beginning of the integration. This system corresponds to the one
described in Fig.~(\ref{fig.ecc_Spin0p9Ecc0p01}).
   }
\label{fig.E}
\end{figure}

In Fig.~(\ref{fig.E}) we depict the evolution of the energy $E$ of the same
system as a function of time. We normalise to the initial value of the energy
when the system still has $126.23$ months ($\sim 10.5$ yrs) to go before the
plunge.

\begin{figure}
          {\includegraphics[width=0.7\textwidth,center]{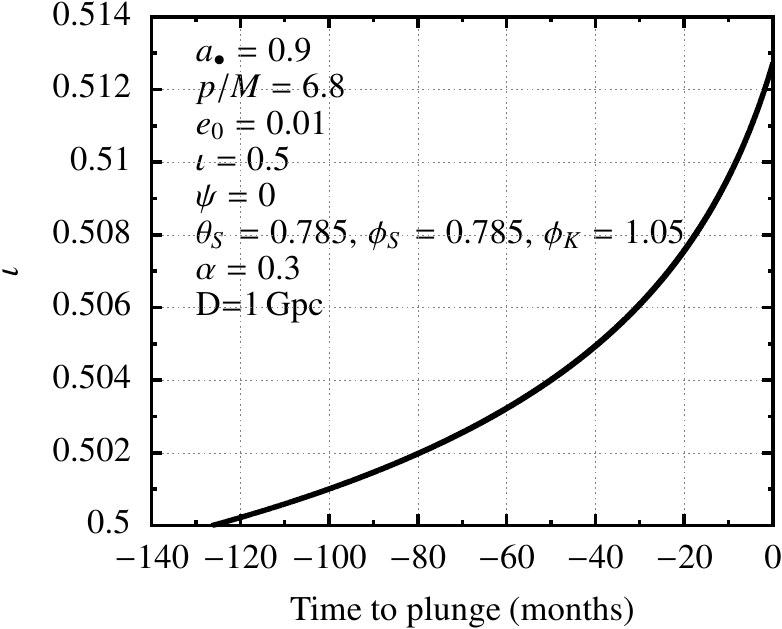}}
\caption
   {
Same as Fig.~(\ref{fig.ecc_Spin0p9Ecc0p01}) but for the evolution of the inclination.
   }
\label{fig.iota_Spin0p9Ecc0p01}
\end{figure}

In Fig.~(\ref{fig.iota_Spin0p9Ecc0p01}) we show the evolution of the inclination for
the same system. At this point, it is important to note that when we say inclination
we are not refering to the polar angle, but to the maximum value reached by the polar
angle with respect to the equatorial plane of the orbit if the SMBH was Schwarzschild
(or if the direction of the spin was perfectly aligned with the z-component of the angular
momentum of the orbital plane, $L_{\rm z}$). 

Because the minimum value of $\theta$, i.e.  the inclination $\iota$, see
Fig.~(\ref{fig.inclination}), changes very slowly \citep[see e.g. Fig.~2
of][]{GlampedakisEtAl02}, it has been assumed to be fixed by the kludge models
in the past, as e.g. in the Teukolsky approach of Fig.~2 of \cite{Hughes2001}.
In that figure the dashed line is the location of the innermost stable
spherical orbit. We can see that the secondary explores the maximum and the
corresponding minimum value during every period.

\begin{figure}
          {\includegraphics[width=1.4\textwidth,center]{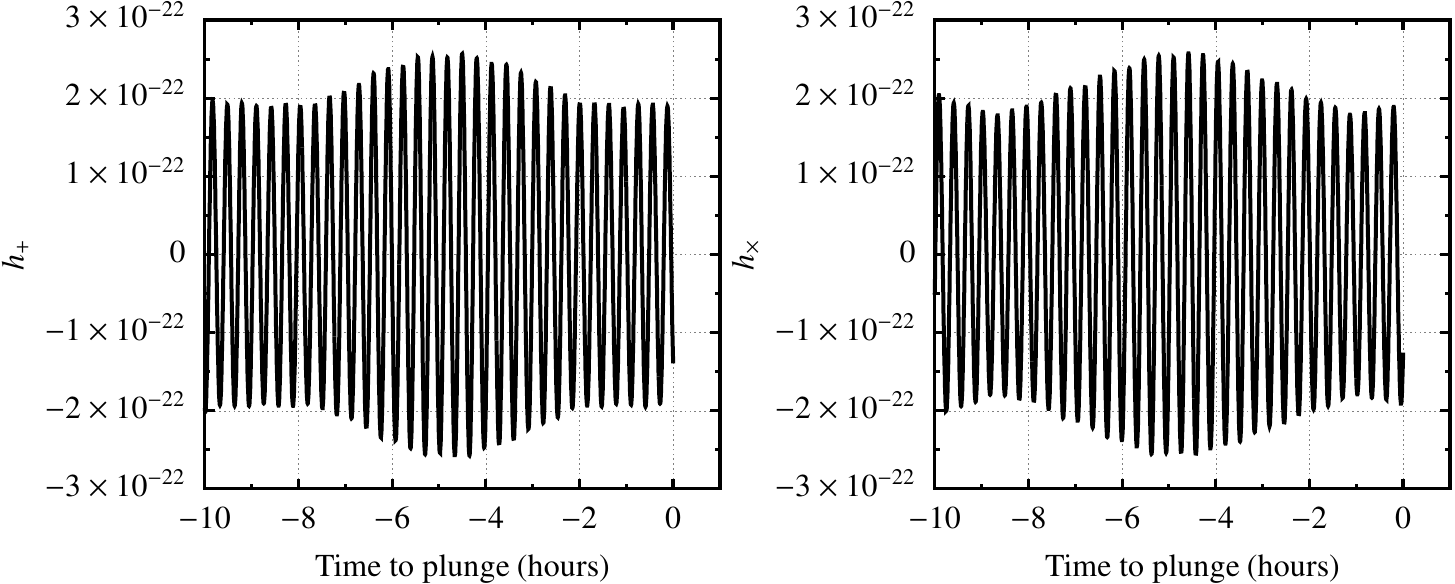}}
\caption
   {
The two waveform polarizations $h_{+}$ (left panel) and $h_{\times}$ (right
panel) as a function of time to merger. The system correspond to that of
Fig.~(\ref{fig.ecc_Spin0p9Ecc0p01}). 
   }
\label{fig.waveforms}
\end{figure}

In Fig.~(\ref{fig.waveforms}) we show the evolution of the polarizations only
for the few last hours because otherwise the figure is too crowded. The system
corresponds to the same one as in the rest of the previous figures.

\subsection{Accumulated phase shift}

We have seen that there are different mechanisms to form EMRIs. In this chapter
we are focusing only on relaxation EMRIs and those formed via tidal separation
of binaries. These, but also othe scenarios, leave a fingerprint in the
dynamical parameters of the EMRIs when it enters the LISA band that allows us
to reverse-engineer the properties of the host environment, which is
interesting from a point of view of astrophysics.  The same applies to IMRIs,
as we will discuss later.

In particular, we can look at the impact on how the phase shifts due to any
residual eccentricity that the source might have. As an example, we have seen
that relaxation EMRIs will typically have higher eccentricities as those
produced via the tidal separation of a binary. This eccentricity will induce a
difference in the phase evolution of the signal as compared to a circular
source. 

We can estimate the accumulated phase shift to lowest post-Newtonian order and
to first order in $e^2$ (via the derivation of \citep{KrolakEtAl1995} of
the phase correction),

\begin{equation}
\Delta \Psi_{e}(f) = \Psi_{\,\rm last} - \Psi_{\,\rm i} \cong - \Psi_{\,\rm i} =
                     \frac{7065}{187136}\,e_i^2\left(\pi\,f\,M_{\rm z} \right)^{-5/3}.
\label{eq.Psi}
\end{equation}

\noindent
Here $e_i$ represents the eccentricity at the frequency of the dominant harmonic
when it enters the detector, we have introduced the quantity 

\begin{equation}
M_{\rm z}:= \frac{G(1+z)}{c^3} \frac{\left( M_{\rm BH}\,m_{\rm CO}\right)^{3/5}}{(M_{\rm BH}+m_{\rm CO})^{1/5}},
\end{equation}

\noindent 
and $f$ is the frequency for the $n=2$ harmonic. We have furthermore assumed that

\begin{equation}
\Delta \Psi_{e}(f) = \Psi_{\,\rm last} - \Psi_{\,\rm i} \simeq -\Psi_{\,\rm i}, 
\end{equation}

\noindent 
with $\Psi_{\,\rm last}$ and $\Psi_{\,\rm i}$ the final and initial phase. As shown
in section B.2 of \cite{CutlerHarms2006}, $\Psi_{e}(f)$ has a pronounced
fall-off with increasing frequency. Hence, to derive the accumulated phase shift in terms of $f$ and the remaining
time to merger, we now recall from \citep{Kepler1619} that the semi-major axis
of the binary is

\begin{equation}
a^3 = \frac{G\left(M_{\rm BH}+m_{\rm CO} \right)}{\left(\pi\,f\right)^2},
\label{eq.K}
\end{equation}

\noindent 
so that we can derive the accumulated phase shift in terms of $f$ and the remaining
time to merger, because this last quantity is given by \citep{Peters64}

\begin{equation}
T_{\rm mrg}  \cong \frac{5}{256} \frac{c^5}{G^3M_{\rm BH} \, m_{\rm CO}
\left(M_{\rm BH}+m_{\rm CO} \right)}
             \left[\frac{G(M_{\rm BH}+m_{\rm CO})}{(\pi\,f)^2} \right]^{4/3},
\label{eq.TmrgLowEcc}
\end{equation}

\noindent 
and we will elaborate on this in the section about the rates. From the same reference,
relation 5.12, assuming $1/(1-e^2) \simeq 1$ and Eq.~(\ref{eq.K}), we have that
$e^2\,f^{19/9} \cong \textrm{constant}$, which means that $a \propto f^{-2/3}$.
Hence, from Eq.~(\ref{eq.TmrgLowEcc}) and  Eq.~(\ref{eq.K}),

\begin{equation}
\pi f \cong \left( \frac{5}{256} \right)^{3/8} M_{\rm z}^{-5/8} T_{\rm mrg}^{-3/8},
\label{eq.pif}
\end{equation}

\noindent 
so that the accumulated phase shift is

\begin{align}
\Delta \Psi_{e}(f) & = \left(\frac{5}{256}\right)^{-17/12}\frac{7065}{187136}
                    \left(\pi f_i \right)^{19/9}e_i^2 M_{\rm z}^{25/36} T_{\rm mrg}^{17/12} \nonumber \\
                   & \cong 10 \left(\pi f_i \right)^{19/9}e_i^2 M_{\rm z}^{25/36} T_{\rm mrg}^{17/12},
\end{align}

\noindent 
and is detectable if $\gtrsim \pi$.

\subsection{Event rate of relaxation EMRIs}

In this section, and as mentioned in the introduction, we will focus on relaxation as the mechanism for producing EMRIs.
To derive the event rate, we have to solve the following integral,

\begin{equation}
\dot{\Gamma}_{\,\rm CO} \simeq \int^{a_{\rm crit}}_{a_{\rm min}} \frac{dn_{\,\rm CO}(a)}
                     {T^{}_{\rm rlx}(a)\,\ln{\left(\theta_{\rm lc}^{-2}\right)}}\,.
\label{eq.EventRate}
\end{equation}

\noindent
where the loss-cone angle can be estimated to be $\theta_{\rm lc} \simeq (J^{}_{\rm max}/J^{}_{\rm lc})^{-1}$,
as can be derived from the previous discussion, and see \cite{2018LRR....21....4A} for more details. 
Following \citep[][]{AL01} $\theta^{2}_{\rm lc} \simeq \sqrt{{8\,R_{\rm S}}/{a}}$. 
The numerator is given by \cite[see][]{Amaro-Seoane2019} 

\begin{equation}
dn^{}_{\,\rm CO}(a) = f_{\rm sub}^{\,\rm CO}\,(3-\beta)\frac{N^{\,\rm CO}_{\rm infl\,\textrm{MS}}}{r^{}_{\rm infl}}
                   \left(\frac{a}{r_{\rm infl}}\right)^{2-\beta}da \,.
\end{equation}

\noindent 
with $f_{\rm sub}^{\,\rm CO}$ the fraction of the type of CO in consideration in the stellar system
(e.g. for stellar-mass black holes $\sim 10^{-3}$) and $N^{\rm infl}_{\, \rm CO\,\textrm{MS}}$
the total number of objects (main-sequence stars and COs, or brown-dwarfs) within $r_{\rm infl}$.

The lower limit in the integral can be estimated by calculating the radius
within which we expect to have at least one CO of the type we are considering.
Since $N^{\,\rm CO}_{\rm infl\,\textrm{MS}}={M_{\rm BH}}/{\bar{m}_{\ast}}$,
with $\bar{m}_{\ast}$ the average stellar mass, we have that $a_{\rm min}
\simeq 1.65\times 10^{-5}\,\textrm{pc}\,f_{\,\rm
CO,\,\textrm{sub}}^{\beta}\left[r_{\rm infl}/(1\,\textrm{pc})\right]$, with
$f_{\,\rm CO,\,\textrm{sub}}$ the the fraction of CO taken into consideration.
With these values, we can solve analytically Eq.~(\ref{eq.EventRate}), which
yields

\begin{equation}
\dot{\Gamma}_{\,\rm CO} =  \frac{3-\beta}{2\,\lambda}\frac{N^{\,\rm CO}_{\rm infl\,\textrm{MS}}}
                                                {T_0\,r_{\rm infl}^{\lambda}} f_{\rm sub}^{\,\rm CO} 
                                  \left\{
                                       a_{\rm crit}^{\lambda}\left[\ln({a_{\rm crit}}/{8\,R_{\rm S}}) -\frac{1}{\lambda}\right] -
                                       a_{\rm min}^{\lambda}\left[\ln({a_{\rm min}}/{8\,R_{\rm S}})   -\frac{1}{\lambda}\right]
                                  \right\},
\label{eq.FinalGamma}
\end{equation}

\noindent 
where we have introduced \citep{Amaro-Seoane2019}

\begin{equation}
T^{}_{0} \simeq \frac{4.26}{(3-\gamma)(1+\gamma)^{3/2}}
             \frac{\sqrt{r_{\rm infl}^3(G M^{}_{\rm BH})^{-1}}}{\ln (\Lambda)\,N^{}_{\rm bh}}
             \left(\frac{M^{}_{\rm BH}}{m^{}_{\rm bh}}\right)^2 \,.
\label{eq.t0full}
\end{equation}

These equations are not easy to interpret in terms of rates, so that we will
give now investigate Eq.~(\ref{eq.FinalGamma}) for the standard EMRI scenario,
i.e. we choose the CO to be a stellar-mass black hole with a mass of
$10\,M_{\odot}$. As for the power indeces, we address two possibilities. One is
a classical solution, the so-called Bahcall-Wolf result \cite[BW,][]{BW77}.  We
remark, however, that their results for two-mass components (the stellar-mass
black holes and the main-sequence stars, all with the same type of mass,
$1\,M_{\odot}$) are heuristically derived from their earlier work of
\cite{BW76}. In their paper they derive $\gamma=7/4$ and $\beta \rightarrow
3/2$.  This result of is based only on the mass ratio of the two populations
and assumes that stellar-mass black holes have a fraction as high as 50\% of
all stars. A physically realistic solution of the problem must require that
stellar-mass black holes have a fraction at most of $10^{-3}$ of all stars, as
derived from a standard IMF. When using this occupation fraction,
\cite{AlexanderHopman09} and \cite{PretoAmaroSeoane10,Amaro-SeoanePreto11}
found that diffusion is more efficient and $\gamma = 2$ and $\beta=3/4$. We
shall call this solution the strong-mass segregation result (SM). For
legibility reasons, we introduce the quantities
$\hat{\Lambda}:=\ln(\Lambda)/13$, $\hat{N}_{\rm infl}:={N}_{\rm
infl}/12\times10^3$, $\hat{r}_{\rm infl}:={r}_{\rm infl}/(1\textrm{pc})$ and
$\hat{m}_{\rm bh}:={m}_{\rm bh}/(10\,M_{\odot})$. 

With these definitions, we derive that for the (mathematically correct but
physically unrealistic) Bahcall \& Wolf solution,

\begin{align}
\dot{\Gamma}_{\rm BW,\,bh} & \sim 2.63\times 10^{-6}\,\textrm{yrs}^{-1}\hat{N}_{\rm infl}\,\hat{\Lambda}\,\hat{r}_{\rm infl}^{-5/2}\,\hat{m}_{\rm bh}^2 \times \nonumber\\
                           & \Bigg\{
                                    5\times 10^{-2} \hat{r}_{\rm infl}\hat{N}_{\rm infl}^{-4/5}\hat{\Lambda}^{-4/5}\hat{m}_{\rm bh}^{4/5}\,{\cal W}(\theta,\,{\rm a}_{\bullet})^{-2}\times \nonumber\\
                           &         \left[\ln\left(16318\, \hat{r}_{\rm infl}\, \hat{N}_{\rm infl}^{-4/5} \hat{\Lambda}^{-4/5}\hat{m}_{\rm bh}^{4/5}\,{\cal W}(\theta,\,{\rm a}_{\bullet})^{-2}\right) - 1 \right] \times \nonumber \\
                           &  2\times 10^{-3}\hat{r}_{\rm infl}\times \left[\ln\left(618\, \hat{r}_{\rm infl}\right) - 1 \right]
                             \Bigg\}.
\label{eq.GammaBWbh}
\end{align}

\noindent 
Setting $\hat{m}_{\rm bh}=1$ and all of the other quantities with a hat to unity as
well, $\dot{\Gamma}_{\rm BW,\,bh}\sim 10^{-6}\,\textrm{yr}^{-1}$, which is the usual
solution \citep[see][]{Amaro-SeoaneLRR2012}. As for the strong mass segregation result, which \textit{does}
represent correctly the segregation in a
galactic nucleus,

\begin{align}
\dot{\Gamma}_{\rm SM,\,bh} & \sim 1.92\times 10^{-6}\,\textrm{yrs}^{-1}\hat{N}_{\rm infl}\,\hat{\Lambda}\,\hat{r}_{\rm infl}^{-2}\,\hat{m}_{\rm bh}^2 \times \nonumber\\
                           & \Bigg\{
                                    1.6\times 10^{-1} \hat{r}_{\rm infl}^{1/2}\hat{N}_{\rm infl}^{-1/2}\hat{\Lambda}^{-1/2}\hat{m}_{\rm bh}^{1/2}\,{\cal W}(\theta,\,{\rm a}_{\bullet})^{-5/4}\times \nonumber\\
                           &         \left[\ln\left(9138\, \hat{r}_{\rm infl}\, \hat{N}_{\rm infl}^{-1} \hat{\Lambda}^{-1}\hat{m}_{\rm bh}\,{\cal W}(\theta,\,{\rm a}_{\bullet})^{-5/2}\right) - 2 \right] - \nonumber \\
                           &         4\times 10^{-2}\hat{r}_{\rm infl}^{1/2}\times \left[\ln\left(618\, \hat{r}_{\rm infl}\right) - 2 \right]
                             \Bigg\},
\label{eq.GammaSMsbh}
\end{align}

\noindent 
which is then $\dot{\Gamma}_{\rm SM,\,bh}\sim 2\times
10^{-6}\,\textrm{yr}^{-1}$.  Note that for both results we have adopted a
Schwarzschild solution, which means that we have set ${\cal W}(\theta,\,{\rm
a}_{\bullet})=1$. The rates for a Kerr SMBH will have a multiplying factor that
depends on the inclination of the orbit and spin of the SMBH.  A detailed
analysis of (i) the relativistic evolution in phase-space and (ii) the rates
depending on the mass species and distribution is in preparation and will be
submitted elsewhere (Amaro Seoane \& Sopuerta in preparation).

It is important to note that these results are subject to be revisited, since
recently \cite{ZwickEtAl2019} derived an improved timescale $T_{\rm GW}$ which
differs from the results of \cite{Peters64} because it includes the effects of
the first-order post-Newtonian perturbation and additionally provides a simple
fit to account for the Newtonian self-consistent evolution of the eccentricity.
These improvements can be captured via relatively trivial modifications to the
usual timescale, which must be multiplied by two factors,

\begin{align}
R(e_{\rm 0}) & = 8^{1-\sqrt{1-e_{\rm 0}}} \nonumber \\
Q_{\rm f} (p_{0}) & = \rm exp \left(\frac{2.5\,R_\textrm{S}}{p_0} \right),
\end{align}

\noindent 
where $e_{\rm 0}$ is the initial eccentricity and $p_{\rm 0}=a_{\rm 0}(1-e_{\rm
0})$ the periapsis. The final corrected expression for the GW-induced
decay of two orbiting bodies $M_{\rm BH}$ and $m_{\rm CO}$, is

\begin{equation}
T_{\rm GW}=  \frac{24 \; \sqrt{2} }{85}\;  \frac{c^4 a^5}{G^3} \frac{(1-e)^{7/2}}{m_{\rm CO}\, M_{\rm BH}}\;  8^{1-\sqrt{1-e_{\rm 0}}} \rm \; exp \left( \frac{2.5\, R_{\rm S}}{p_{\rm 0}}\right),
\end{equation} 

\noindent 
as derived in \cite{ZwickEtAl2019}. Since we are modifying $T_{\rm GW}$, this
will be propagated into the rates, as we can see from Eq.~(\ref{eq.TrlxTGW})
For high eccentricity orbits and spin, the correction factors are (Zwick et al to be submitted 2020)

\begin{align}
 R(e_{\rm 0})       & = 8^{\sqrt{1-e_0}} \nonumber \\
 Q_{\rm h}Q_{\rm s} & \to \exp \left( \frac{2.8 R_{\rm S}}{p_0} + 
                          s_1\left( \frac{0.3 R_{\rm S}}{p_0}\right) + 
                          \lvert s_1 \rvert ^{3/2} \left( \frac{1.1 R_{\rm S}}{p_0} \right)^{5/2} \right) 
\end{align}

And accordingly, the corrected timescale is

\begin{align}
T_{\rm GW} & =  \frac{24 \; \sqrt{2} }{85}\;  \frac{c^4 a^5}{G^3} \frac{(1-e)^{7/2}}{m_{\rm BD} M_{\rm BH}}\;  8^{1-\sqrt{1-e_{\rm 0}}}\nonumber \\
           &    \times \exp \left[ \frac{2.8 R_{\rm S}}{p_0} + s_1\left( \frac{0.3 R_{\rm S}}{p_0}\right) + \lvert s_1 \rvert ^{3/2} \left( \frac{1.1 R_{\rm S}}{p_0} \right)^{5/2} \right],
\end{align} 

\noindent 
where $s_1$ is a spin parameter defined as $s_1 := s\,\cos \theta$, with $s$
the magnitude of the spin and $\theta$ the angle between the SMBH spin vector
and the angular momentum vector of the orbit.  The detailed derivation of this
result will be presented soon as Zwick et al. 2020 and the impact on the event
rates as V{\'a}zquez et al 2020. We see that when we take into account both
corrections, namely the first-order post-Newtonian perturbation with a
self-consistent evolution recipe for the evolution of the eccentricity and the
spin effects on that timescale (additionally to the location of the LSO, via
the ${\cal W}(\theta,\,{\rm a}_{\bullet})$ function), the results do not differ
significantly from the location of $a_{\rm crit}$ in Eq.~(\ref{eq.acrit}).

\begin{figure}
          {\includegraphics[width=1.4\textwidth,center]{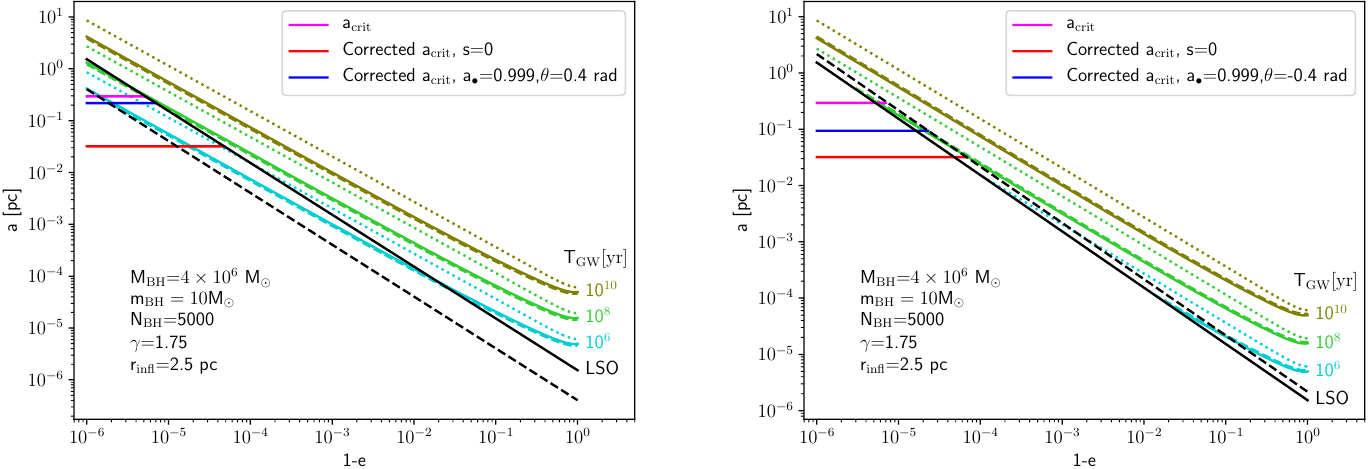}}
\caption
   {
Derivation of $a_{\rm crit}$ as in Eq.~(\ref{eq.acrit}) and in Fig.~(\ref{fig.EMRI}) with
the modified $T_{\rm GW}$, which takes into account both  the effects of
the first-order post-Newtonian perturbation with a simple
fit to account for the Newtonian self-consistent evolution of the eccentricity, and the
effect of the spin as well. Note that the correction for the spin is a different one as
the location of the LSO via the ${\cal W}(\theta,\,{\rm a}_{\bullet})$ function presented
in Fig.~(\ref{fig.W}), altough we do use this function as well (for details, see the upcoming
work of Zwick et al 2020 and V{\'a}zquez et al 2020). The left panel corresponds to prograde orbits and the right panel
to retrograde ones. 
   }
\label{fig.NewAcrit}
\end{figure}

\section{Intermediate-mass ratio inspirals}

\subsection{Intermediate-mass black holes}

For an IMRI to exist, we first have to introduce what intermediate-mass black
holes are, what motivates this search and how many of these have been detected.
On the one hand, for a long time we know that stellar-mass black holes must be
present thanks to electromagnetic observations. Ground-based gravitational wave
observatories have opened up a new window and have corroborated this.  

On the other hand, the understanding of galactic nuclei (the inner-most cores
of galaxies) has advanced rapidly during the past decade, not least due to
major advances in high angular resolution instrumentation at a variety of
wavelengths. As we have previously mentioned, the overwhelming evidence is that
supermassive black holes (with masses between a million and ten thousand
million Suns), occupy the centers of most galaxies for which such observations
can be made.  

Moreover, an intimate link exists between the central supermassive black hole
and its host galaxy \citep{KormendyHo2013}, as exemplified by the discovery of
correlations between the mass of the supermassive black hole and global
properties of the surrounding stellar system, e.g. the velocity dispersion
$\sigma$ of the spheroid of the galaxy.  Despite much progress in recent
decades, many fundamental questions about these relations remain open.  

These two flavors of black holes have masses that differ by up to nine orders
of magnitude.  The same way humans grow from babies to teenagers and, later, to
adults, black holes must also exist in the intermediate regime.  Such
``teenager'', intermediate-mass black holes (IMBHs) must have masses typically
ranging between $\sim 10^{2} - 10^{4}\,M_\odot$, and in fact we have detected
high X-ray luminosities not coincident with the nucleus of the host galaxy,
which translate into these masses under the assumption that they are black
holes. Theoretically, we know that IMBHs form and are located at the center of
dense stellar systems such as globular clusters, young clusters, or the cores
of dwarf galaxies, and indeed this is what the correlations we mentioned before
predict \citep[see][]{GRH02,GerssenEtAl02}, and the reviews of
\citep{Mezcua2017,LuetzgendorfEtAl2013}.

IMBHs are key to understand how supermassive black holes gained their titanic
masses from stellar-mass black holes, but they are elusive and we do not have
any conclusive evidence of their existence in X-rays or radio, although they
should accrete matter. For decades, in spite of the effort with electromagnetic
radiation, we lacked an evidence of their existence. Recently this has changed,
thanks to the observation of GW190521, a gravitational wave signal resulting
from the merger of two progenitor black holes of $85\,M_{\odot}$ and
$66\,M_{\odot}$, resulting into an IMBH of mass $142\,M_{\odot}$
\citep{AbbottEtAlIMBH2020}, which means that a total of $9\,M_{\odot}$ were
completely transformed into energy in the form of gravitational waves
\citep{AbbottEtAlIMBHb2020}.

These massive black holes must be lurking in star clusters, as we explained
before. These clusters are complicated and interesting, not least because they
allow us to probe the event horizon. They are, however, key to understand IMBHs
but difficult to simulate due to the many orders of magnitude we have to
overcome.

If we want to observe IMBHs with light in globular clusters, we need
ultra-precise astronomy, because the influence radius of an IMBH of mass
$10^4\,M_{\odot}$, $r_h \sim 5''$, assuming a central velocity dispersion of
$\sigma = 20\,{\rm km\,s}^{-1}$ and a distance of $\sim 5$ kpc. Within that
radius we only have a few stars. At such a distance and assuming an
observational timescale of 10 yrs, with adaptive optics we could in the best of
the cases have a couple of measurements of velocities. The sensitivity limits
correspond to a K-band magnitude of $\sim$ 15, like the kind of stars that we
observe in our own Galactic Centre, at 8 kpc and of type B- MS. To derive the
influence of the mass of the IMBH on the stellar population around it, as it
has been done in our Galactic Centre with SgrA*, one needs instruments such as
the VSI or GRAVITY \citep{GillessenEtAl06,EisenhauerEtAl08}.
Until the recent discovery of the LIGO/Virgo team, IMBHs were simply a logical
conjecture which has now been corroborated. 

\subsection{Wandering of IMBHs}

It is relatively simple to estimate in an analytical way how far away an IMBH
can wander off the centre due to Browninan motion. For simplicity, we assume
that the stellar cluster is described by a so-called eta model
\citep{Dehnen93,Tremaine94} with enclosed stellar mass

\def\mcl{{M}_{\rm tot}}
\def\m{{M}}

\begin{equation}
\m_\ast(r) = \mcl \left(\frac{r/R_{\rm cut}}{1+r/R_{\rm cut}}\right)^{\eta},
\end{equation}

\noindent 
where $M_{\rm tot}$ is the total mass in stars and $R_{\rm cut}$ the cut-off radius.
For $r\ll R_{\rm cut}$, we have the usual power-law distribution $\rho \propto
r^{-\gamma}$ with $\gamma=3-\eta$. The massive black hole will wander within a
given radius $R_{\rm w}$, where it can capture COs by perturbing stellar orbits
(i.e. dynamically) for light enough IMBHs. We simplify the problem by assuming
a constant stellar density, so that the gravitational potential in which the
IMBH is located can be described as a harmonic oscillator, with angular
frequency $\omega = \sqrt{G\m_\ast(R_{\rm w})/R_{\rm w}^3}$.  Given an
equilibrium point, the maximum speed $V_{\rm osc}$ achieved by the oscillator
for an amplitude (the maximum displacement from the equilibrium position)
$R_{\rm osc}$ is $V_{\rm osc}^2 = \omega^2 R_{\rm osc}^2$. This value, commonly
referred to as the root mean square amplitude of the oscillations in velocity,
is linked to its equivalent in space,

\begin{equation}
 V_{\rm osc}^2 = \omega^2 R_{\rm osc}^2 \approx \omega R_{\rm w}^2 =
\frac{G\m_\ast(R_{\rm w})}{R_{\rm w}}.
\label{eq:Vrms}
\end{equation}

\noindent 
Assuming equipartition of kinetic energy between the IMBH and the stellar 
component, we have that \citep[see][for the specific case $\eta=1.5$]{DHM03}

\begin{equation}
M_{\rm BH} V_{\rm osc}^2 \simeq m_\ast \sigma(R_{\rm cut})^2,
\label{eq:Vrms2}
\end{equation}

\noindent 
where $\sigma$ is the stellar velocity dispersion at $r=R_{\rm cut}$. We note that
$\sigma^2 \simeq 0.1 G\,M_{\rm tot}/R_{\rm cut}$.
Since we can assume that typically $R_{\rm w} \ll R_{\rm cut}$ and, hence, $\m_\ast(R_{\rm
  w}) \simeq \mcl \left(R_{\rm w}/R_{\rm cut}\right)^\eta$ and combining
Eq.~(\ref{eq:Vrms}), (\ref{eq:Vrms2}) and the expression for $\sigma^2$, we obtain

\begin{equation}
R_{\rm w} \propto R_{\rm cut}\left(\frac{m_\ast}{M_{\rm BH}}\right)^{1/(2-\gamma)}.
\end{equation}

\begin{figure}
          {\includegraphics[width=1.2\textwidth,center]{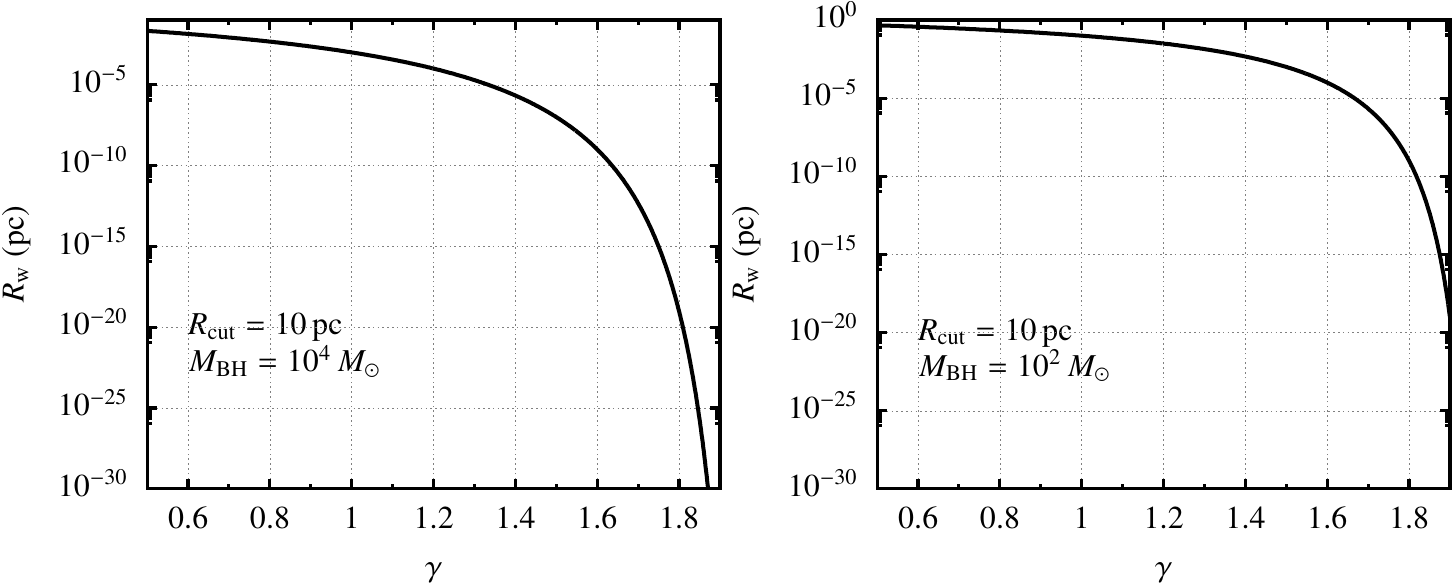}}
\caption
   {
Wandering radius $R_{\rm w}$ of an IMBH due to Brownian motion in a dense stellar cluster
as a function of the power-law exponent $\gamma$.
We use a so-called ``eta'' model with cutoff radius $R_{\rm cut}$ and the
mass of the IMBH is shown as $M_{\rm BH}$.
   }
\label{fig.Rw}
\end{figure}

In Fig.~(\ref{fig.Rw}) we show the wandering radius of an IMBH with two
different masses for different values of the power-law index $\gamma$. As we
can see, for shallow values the IMBH can reach significant values. As we have
seen in Eq.~(\ref{eq.Trlx} and in Fig.~(\ref{fig.Trlx}), the system sould be
relaxed, but this implicitly assumes that the mass-$\sigma$ relationship holds
at the low-end of masses, which has not been yet confirmed observationally.
This means that we expect a cusp to build around the IMBH, so that for a
single-mass population, $\gamma=1.75$, and in the case of a mass spectrum, it
will be typically of the order of $\gamma \gtrsim 1.75$ for the heavier
components, and $\gamma \simeq 1.5$ for the lighter ones. 

This quick estimate agrees well with the much more detailed work of
\cite{LinTremaine1980}.  The authors identify three main mechanisms for the
wandering of the IMBH: (1) Brownian motion, (2) the effect of the segregated
stellar cusp and (3) three-body interactions between stars and the IMBH. The
authors however find that Brownian motion is the most important one. Later,
\cite{GuelketinEtAl2004} ran scattering, numerical experiments of of single
stars on to a binary formed by the IMBH and another star. Their results are
that only IMBHs with masses above $300\,M_{\odot}$ have small wandering radii.
Below this limit, the IMBHs wander off the centre of the cluster. Hence, if
Brownian motion was the only phenomenon to take into account, we could assume
that the IMBH is fixed at the centre, since for any realistic value of $\gamma$
the displacement is negligible, and so the treatment of loss-cone could be
applied but in view of the results of \cite{GuelketinEtAl2004}, this can only
be done for rather massive IMBHs.

\subsection{Numerical simulations of IMRIs}

Because of the challenges related to the analytical approach in the case of an
IMBH, in particular the wandering of the massive black hole, we need to resort
to numerical simulations to at least win some intuition about the physics at
role.

For this aim, and as discussed in detail in the relativistic context in
\citep{Amaro-SeoaneLRR}, the most accurate integration one can do are the
so-called direct-summation $N-$body algorithms. These integrators have been
used for decades in astronomy and have been put to test a number of times with
observations and analytical techniques. It all boils down at integrating
Newton's equations of motion between all stars in the stellar system at every
timestep, with a regularisation algorithm for binaries. Since we are
integrating directly, all phenomena purely related to dynamics (i.e. a star is
a point) emerge naturally \citep[see e.g.][and the latter for the concept of
regularisation]{Aarseth99,Aarseth03,AarsethZare74}. Nowadays many of these
Newtonian integrators have adapted a post-Newtonian correction, first
implemented in \citep{KupiEtAl06}. We note however, that the post-Newtonian
expansion assumes the bodies to be completely isolated and without additional
perturbations. This has been neglected in all astrophysical integrations with
relativistic corrections of this type, but it has been shown that it can induce
important errors. Indeed, it has been shown that the so-called cross-terms must
be taken into account \citep{Will2014}. These are terms that represent a
coupling between the SMBH and the stellar potential. In the case of having a
hierarchical triple system, the author shows that the effects of such terms can
actually be enhanced to amplitudes of Newtonian order.

But, in general, and even for a binary in vacuum, the post-Newtonian approach
is not yet fully understood and as a matter of fact, it has received criticism.
In particular, the work of \citep{EhlersEtAl1976} pointed that the derivations
either contained inconsistencies or are incomplete. These points have been
addressed almost completely, already in the 1990's. There remains however one
important open question, about the nature of the sequence itself, because we do
not know yet whether it converges, diverges or is asymptotic. However, on the
calculation side, we can avoid divergences by carefully constructing the
hierarchy of the equations \citep[see e.g.][]{AsadaFutamase1997}

Nonetheless, it is \textit{for some reason} a very efficient tool to address
the evolution of isolated binaries.  I refer the reader to the elegant work
``On the unreasonable effectiveness of the post-Newtonian approximation in
gravitational physics'' presented in \citep{Will2011}. The summary could be the
following sentence from the article: ``The reasons for this effectiveness are
largely unknown.''.

We note that an interesting alternative is implementing a geodesic solver to
the Newtonian algorithm in the regime of interest; i.e. when the CO is detached
from the rest of the stellar system, at distances close to periapsis from the
SMBH (or IMBH, in this case). The details are given in Sec.~(8.8.2) of
\citep{Amaro-SeoaneLRR}.

In the context of stellar dynamics and IMRIs, the first numerical work that
addressed the formation and evolution of an IMRI in a cluster is the work of
\cite{KonstantinidisEtAl2013}. They observe the formation of an IMRI with
masses $M_{\rm BH}=500\,M_{\odot}$ and $m_{\rm CO}=26\,M_{\odot}$. After some
50  Myrs, the IMRI merges and the IMBH receives a relativistic recoil
\citep{CampanelliEtAl2006,BakerEtAl2006,GonzalezEtAl2007}. The result is that,
due to the low escape velocity of the cluster, the IMBH leaves the host stellar
system. They noticed in their simulations that the IMBH forms a binary with a
(relatively massive) stellar-mass black hole for about 90\% of the time of the
simulation. 

The interesting fact is that the stellar-mass black hole that forms the IMRI is
the result of an abrupt exchange of companion with the IMBH. In view of the
initial semi-major axis and eccentricity, $a \sim 10^{-5}$ pc and $e=0.999$,
this capture seems to follow the ``gravitational brake'' capture of the
parabolic orbits: A sudden loss of energy via gravitational radiation can lead
to the spontaneous formation of a relativistic binary. This was first
presented in the work of \cite{QuinlanShapiro1989}, while the energy and
angular momentum changes in the case of a hyperbolic orbit presented in
\cite{Hansen1972}. This scenario has been recently explored numerically by
\cite{KocsisEtAl2006,MandelEtAl2008,OlearyEtAl09,LeeEtAl2010,HongLee2015}.

The work of \cite{KonstantinidisEtAl2013} has been confirmed by
\cite{LeighEtAl2014}, who finds very similar results but using a different
numerical scheme. Also, the work of \cite{HasterEtAl2016} is basically a
reproduction of \cite{KonstantinidisEtAl2013} but with a different integrator,
which leads the authors to similar results as well. Later, the detailed
analysis of \cite{MacLeodEtAl2016} explore lighter IMBHs and also find that the
IMBH spends about 90\% of the integration time on a binary with a CO, also with
a distribution of semi-major peaking at values of $\lesssim 10^{-5}$ pc.

\subsection{Event rate of IMRIs}

As we have discussed previously, we rely on numerical simulations to address the
formation and evolution of IMRIs. Therefore, the derivation of the event rate must
accordingly rely on a large sample of numerical simulations that allows us to
at least have an educated guess about the free parameters that riddle the problem.

\begin{figure}
          {\includegraphics[width=1.3\textwidth,center]{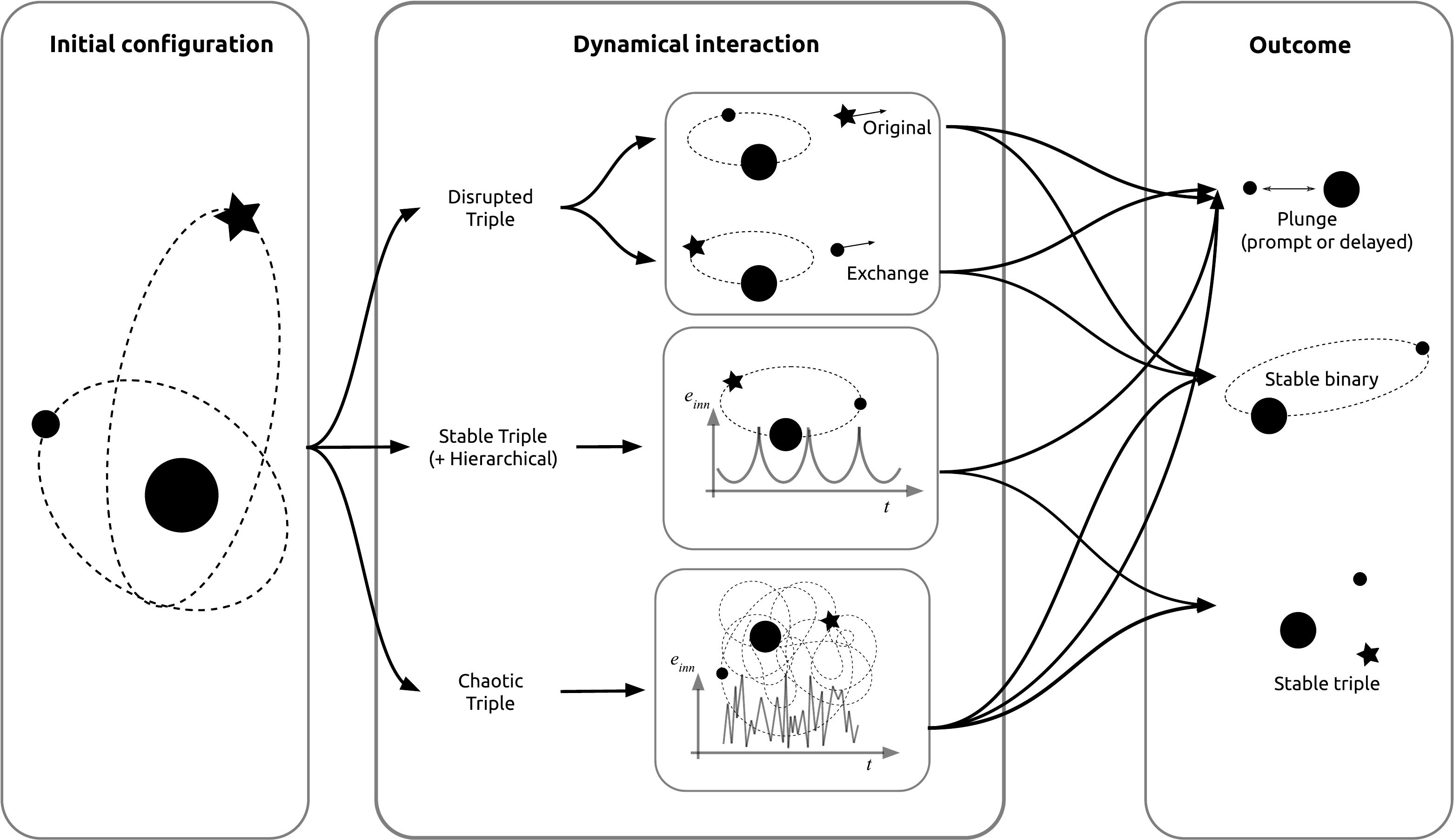}}
\caption
   {
Sketch of the set of $\sim 3\times10^4$ numerical simulations of
\citep{ArcaSeddaEtAl2020}.  The three boxes display the different stages in the
simulation. The first one corresponds to the initial conditions, which are
based on previous works, in particular on \citep{KonstantinidisEtAl2013}, which
has been confirmed by independent groups. An IMBH (the larger black circle)
forms a binary with a stellar-mass black hole (smaller black circle) and a
perturber, which is another stellar-mass black hole, but in principle could be
another type of star with the proviso that the mass is large enough to have
been segregated to small distances from the IMBH (black star). This three-body
problem is embedded in an external stellar potential representing the
background stellar component, and the evolution can be approximately divided in
three different possibilities, as we see in the ``dynamical interaction'' box.
The last box gives us the result of the interaction depending on the details of
the initial configuration.
   }
\label{fig.Sketch_IMRI}
\end{figure}

In Fig.~(\ref{fig.Sketch_IMRI}) we show a cartoon of the set of $3\times10^4$
simulations of \citep{ArcaSeddaEtAl2020}, which sweep the parameter space based
on the initial conditions motivated by the previous work of
\cite{KonstantinidisEtAl2013,LeighEtAl2014,MacLeodEtAl2016,HasterEtAl2016}.
We elaborate more on this in the next section. 

Thanks to the outcome of the numerical study we can make a guess on the event
rate that we expect for different detectors. I.e. depending on the horizon
distance $z_{\rm hor}$ of the detector we consider for a given type of mass, the rates will
be different. 

The work of \citep{ArcaSeddaEtAl2020} derive the rates by estimating the total
number of IMRIs within a given cosmological volume of radius $z_{\rm hor}$. We
reproduce here their calculation as a way to exhibit clearly the number of free
factors that are at play in the calculation. The number of IMRIs,
$\Gamma_{\imri}$, can be obtained via the following integral

\begin{equation}
\Gamma_{\imri} =  \Omega_s \int_{M_{1}}^{M_{2}} \int_{0}^{z_{\rm hor}} \frac{\derd\,n_{\imri}}{\derd M_\ibh\derd\,z} 
                   \frac{\derd V_c}{\derd\,z}  \frac{\derd\,z}{1+z} \derd M_\ibh,
\end{equation}

\noindent 
where $\derd V_c/\derd\,z$ is the comoving cosmological volume element,
$(1+z)^{-1}$ takes into account the time dilation and $\derd\,n_{\imri}/\derd
M_\ibh$ expresses how many IMRIs we have per unit of mass. This, as pointed by
the authors, can be rewritten as

\begin{equation}
\frac{\derd\,n_{\imri}}{\derd M_\ibh} = \xi_\bh f_{\gw} p_\ibh n_{\rm rep}
        \frac{\derd\,n}{\derd M_g\derd\,z}\frac{\derd\,n_\gc}{\derd M_\gc}\frac{\derd M_\gc}{\derd M_\ibh}.
\end{equation}

\noindent 
In this last expression $\derd\,n/(\derd M_g \derd\,z)$ is the number of galaxies per $z$, $\xi_\bh$
is (a priori) a free parameter that gives us the the probability for an IMBH to be in a binary with a 
stellar-mass black hole, $\derd\,n/\derd M_\gc$ represents the number of clusters per cluster mass in a
certain host galaxy, $\derd M_\gc/\derd M_\ibh$ links the mass of the GC to that of the IMBHs, $n_{\rm rep}$
is another unknown factor that expresses how often an IMBH can for a binary with a CO, $f_{\gw}$ is the fraction of 
successful IMRIs (i.e. those which merger within a Hubble time, an information drawn from their simulations).
Finally, $p_\ibh$ is the probability that an IMBH is indeed lurking in the considered GC. This is very likely
the most unknown parameter. The authors rely on the set of Monte-Carlo simulations of \citep{GierszEtAl2015} and decide
to set it to $p_\ibh = 0.2$, i.e. 20\% of all GC are supposed to be hosting an IMBH. 
The assume a power-law for $\derd\,n/\derd M_\gc$ of the following fashion,

\begin{equation}
\frac{\derd\,n}{\derd M_\gc} = k M_\gc^{-s},
\end{equation}

\noindent 
with a slope $s = 2.2$, for consistency reasons with the initial mass function of young and old star clusters in galaxies, as
derived in the work of \citep[e.g.][]{Gieles2009}, with the normalisation factor $k$ defined as

\begin{equation*}
k = \frac{\delta M_g (2-s)}{\left(M_{\gc 2}^{2-s} - M_{\gc 1}^{2-s}\right)}.
\end{equation*}

\noindent 
We need to assume some reference values for this normalisation. The authors
adopt Galactic values, so that $M_g = 6\times 10^{10}\Ms$ and $M_{\gc 1,2} =
(5\times 10^3 - 8\times 10^6) \Ms$, and derive $M_\ibh \simeq (30 - 4.6\times
10^4)\Ms$, as well as $M_\gc^{-s} = aM_\ibh^{-bs}$, which again, contain two
parameters. As for $\derd\,n/(\derd M_g\derd\,z)$, they resort to the results of
\cite{ConseliceEtAl2016}.

Hence, the number of IMRIs contained in a given observable volume is

\begin{equation}
N_{\imri} = k a^{1-s} b p_\ibh\, n_{\rm rep}\, \xi_\bh \int_{M_{1}}^{M_{2}} 
            \int_0^{z_{hor}} f_{\gw} \, M_\ibh^{\,(1-s)\,b-1}  
            \derd M_\ibh \frac{\phi(z)}{1+z} \frac{\derd V_c}{\derd\,z}\derd\,z .
\label{eq:nimri1}
\end{equation}

\noindent 
with

\begin{equation}
\phi(z) = -\frac{\phi_* 10^{(\alpha_* + 1)(M_2-M_*)}}{\alpha_* + 1},
\end{equation}

\noindent 
a parametric expression of galaxies number density, where
$\phi_*,~\alpha_*,~M_*$ a function of $z$ \citep[see Table 1
in][]{ConseliceEtAl2016}, and $M_2 = 12$.  The estimation limits $z_{\rm hor}$ to
either $z=2$ (which corresponds to the peak of GC formation), or $z=6$
(formation of the first stars).  The probability that the IMBH forms a tight
binary with a CO is given by $\xi_\bh$, and it is assumed that $\xi_\bh
\rightarrow 1$, based on initial mass function arguments but, more importantly,
on the results of the different numerical simulations performed by different
groups.

The last piece of information required is the timescale for the IMRIs to form,
which can be expressed as the accumulative sum of the timescales for cluster
formation, IMBH formation, IMRIs formation and coalescence. Since we are, to
put it mildly, at debate about the process that leads to IMBH formation,
\citep{ArcaSeddaEtAl2020} adopt a weighted timescale for it of 2 Gyr based on
the numerical experiments of \citep{GierszEtAl2015,ArcaSeddaEtAl2019}. They take the
mass-segregation timescale for the IMRI formation, which is sensible, of $\sim
0.1-1$ Gyr, and they derive from their simulations that $t_\gw = 0.6-1.5$ Gyr.

As \citep{ArcaSeddaEtAl2020} point out, an alternative way of deriving the merger rate
is via the cosmological GC star formation rate $\rho_{\rm SFR}(z)$. This can be
used to obtain the total number of GCs as a function of $z$,

\begin{equation}
N(z_{\rm max}) =  \int_0^{z_{\rm max}} \frac{\rho_{\rm SFR}(z)}{<M_\gc>}\frac{\derd V_c}{\derd\,z}\frac{\derd\,z}{1+z}.
\end{equation} 

\noindent 
If one adopts the power-law GCs mass function, the normalisation in this case is

\begin{align}
k  =&\frac{(1-s)}{M_{\gc, 1}^{1-s} - M_{\gc ,2}^{1-s}},
\end{align}

\noindent 
so that the total number of IMRIs within a given volume is

\begin{equation}
N_{\imri} = k a^{1-s} b p_\ibh n_{\rm rep}\, \int_{M_{1}}^{M_{2}} \int_0^{z_{hor}} M_\ibh^{\,(1-s)\,b-1} 
            f_{\gw}\, \rho_{\rm SFR}(z)\frac{\derd V_c}{\derd\,z} \frac{\derd\,z}{1+z} \derd M_\ibh.
\label{eq:nimri2}
\end{equation}

\noindent
Therefore, so as to obtain the merger rate, we convert Eqs.~(\ref{eq:nimri1})
and (\ref{eq:nimri2}) at a given $z$ with the relation $\Gamma_\imri = N_\imri
/ T$. Therefore, we can convert the rates to specific detectors, and we have
that

\begin{itemize}
    \setlength{\itemsep}{1pt}
    \setlength{\parskip}{0pt}
    \setlength{\parsep}{0pt}

 \item \textbf{LIGO/Virgo} should detect up to $1-5$ IMRIs per yr with $M_\ibh \simeq 100\Ms$
 \item \textbf{LISA} is in the position of observing up to $\sim 6-60$ events per yr
 \item \textbf{ET} could detect $100-800$ IMRIs per yr with masses $M_\imri < 2000 \Ms$
 \item \textbf{Decihertz detectors} could detect up to $3800$ events per yr

\end{itemize}

\noindent 
Obviously, in view of the detection of \citep{AbbottEtAlIMBH2020} of the system
GW190521 by LIGO/Virgo with a mass $142\,M_{\odot}$ suits well these predictions
but we are talking about single-point statistics. We need to wait to have more 
detections to derive any conclusion.

\subsection{Multibandwidth IMRIs}
\label{sec.multiband}

Intermediate-mass ratio inspirals are particularly interesting because, as
discussed in \citep{Amaro-Seoane2018}, they can not only be detected by future
space-borne gravitational wave observatories. Depending on their masses and, in
general, dynamical parameters, they could already be present in ground-based
detectors, such as LIGO/Virgo. Moreover, also depending on their dynamical
characteristics, a percentage of them will be detectable by both space-borne-
and ground-based facilities.

\begin{figure}
          {\includegraphics[width=1.3\textwidth,center]{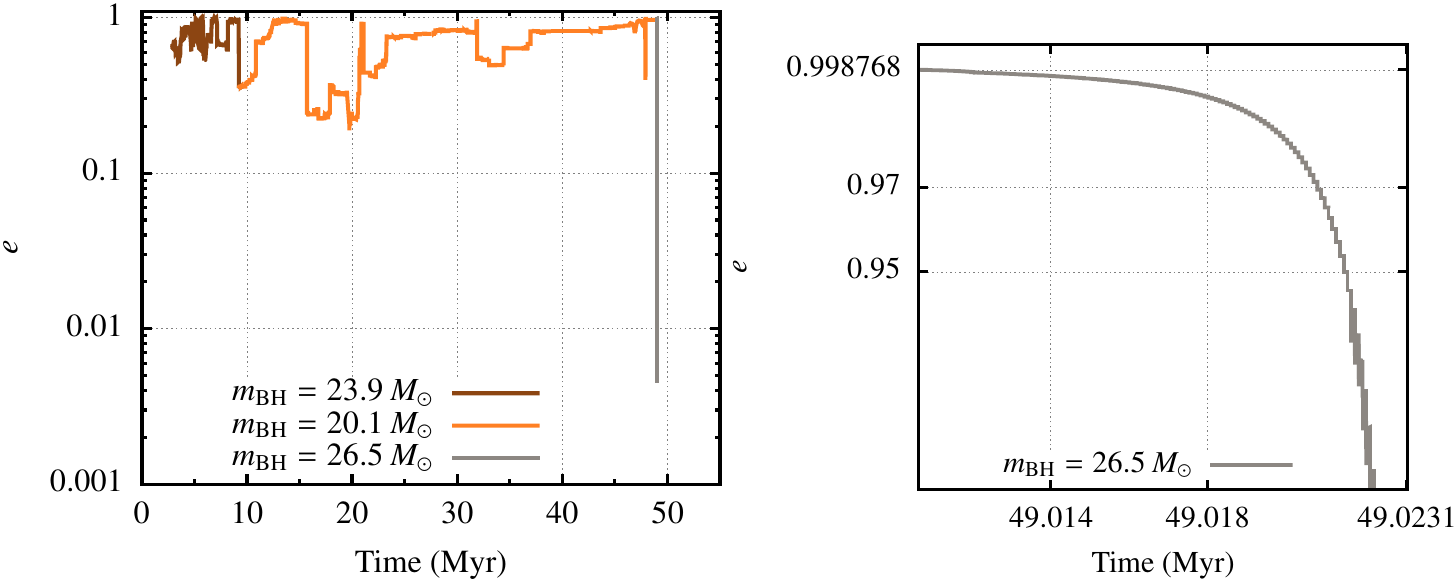}}
\caption
   {
Evolution of the eccentricity for the three different stellar-mass black holes
which form a binary with the IMBH (data from \citep{KonstantinidisEtAl2013}),
as a function of time. The masses of the three different COs is shown in the
left panel, which is the complete evolution of the three systems. On the right
panel we have a zoom in of the last binary, which lasts $1.25\times 10^4\,\textrm{yr}$
from the formation to the merger. We can see the extreme eccentricity that it achieves
and how quickly it circularises as an effect of the radiation of energy.
   }
\label{fig.e_IMRI_zoom}
\end{figure}

\begin{figure}
          {\includegraphics[width=1.3\textwidth,center]{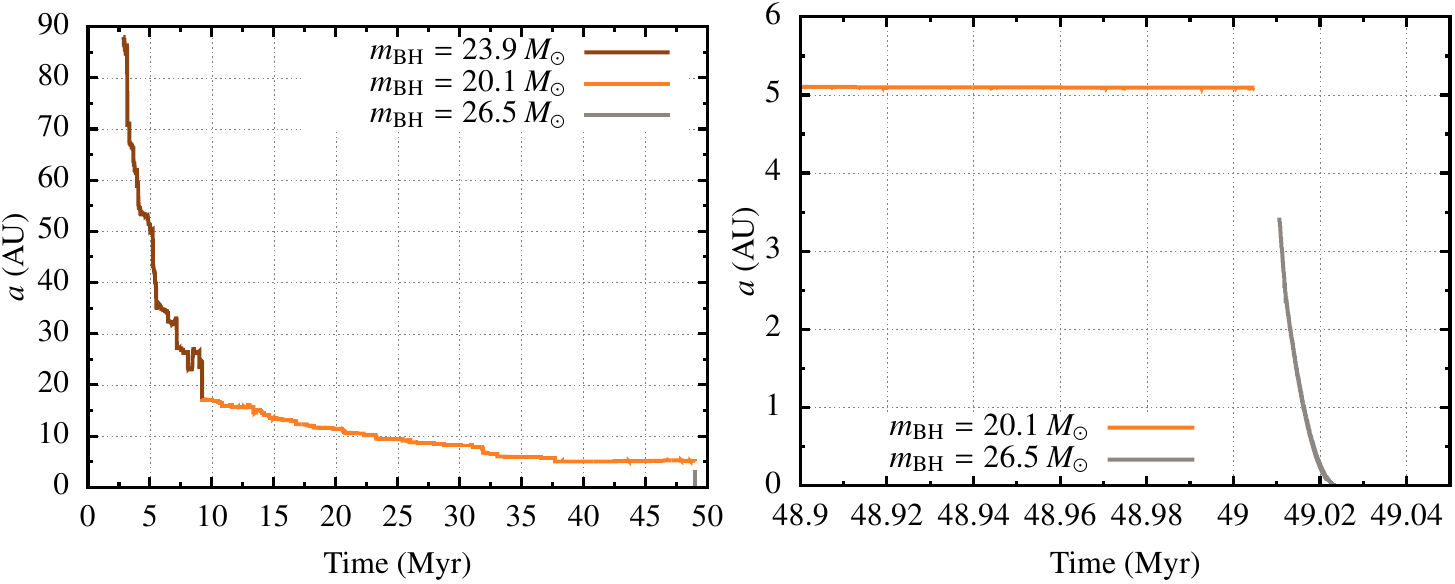}}
\caption
   {
Same as Fig.~(\ref{fig.e_IMRI_zoom}) but for the semi-major axes. On the right
panel we only show the evolution of the two last COs in the order of the dynamical
interactions with the IMBH.
   }
\label{fig.a_IMRI_zoom}
\end{figure}

As we have already mentioned in the previous section, we rely on numerical
simulations to address this kind of source. We stress out that the findings of
the first numerical simulation by \citep{KonstantinidisEtAl2013} have been
confirmed by at least another three different groups. This is important,
because IMRIs seem to form at very high eccentricities and very small
semi-major axes. To illustrate this, in Fig.~(\ref{fig.e_IMRI_zoom}) and
Fig.~(\ref{fig.a_IMRI_zoom}) we show the eccentricities and semi-major axes of
one of the simulations of \citep{KonstantinidisEtAl2013}, which led to the
formation of an IMRI. This particular example is a representative one of what
has been found by other groups. The IMBH forms a binary with a compact object,
a stellar-mass black hole of mass $\sim 20\,M_{\odot}$ for most of the time of
the simulation. The binary is ionised and the CO replaced with another one,
which is also of the same type and mass, while the semi-major axes become
smaller and smaller. Eventually, after an abrupt interaction, a final binary
forms with an initial eccentricity of $e\sim 0.999$ and the IMRI forms and
merges in a fraction of the total simulation time ($\sim 2.5\%$ of the total).
In these figures we can see how quickly both the semi-major axis and eccentricity
decay. 

\begin{figure}
          {\includegraphics[width=1.4\textwidth,center]{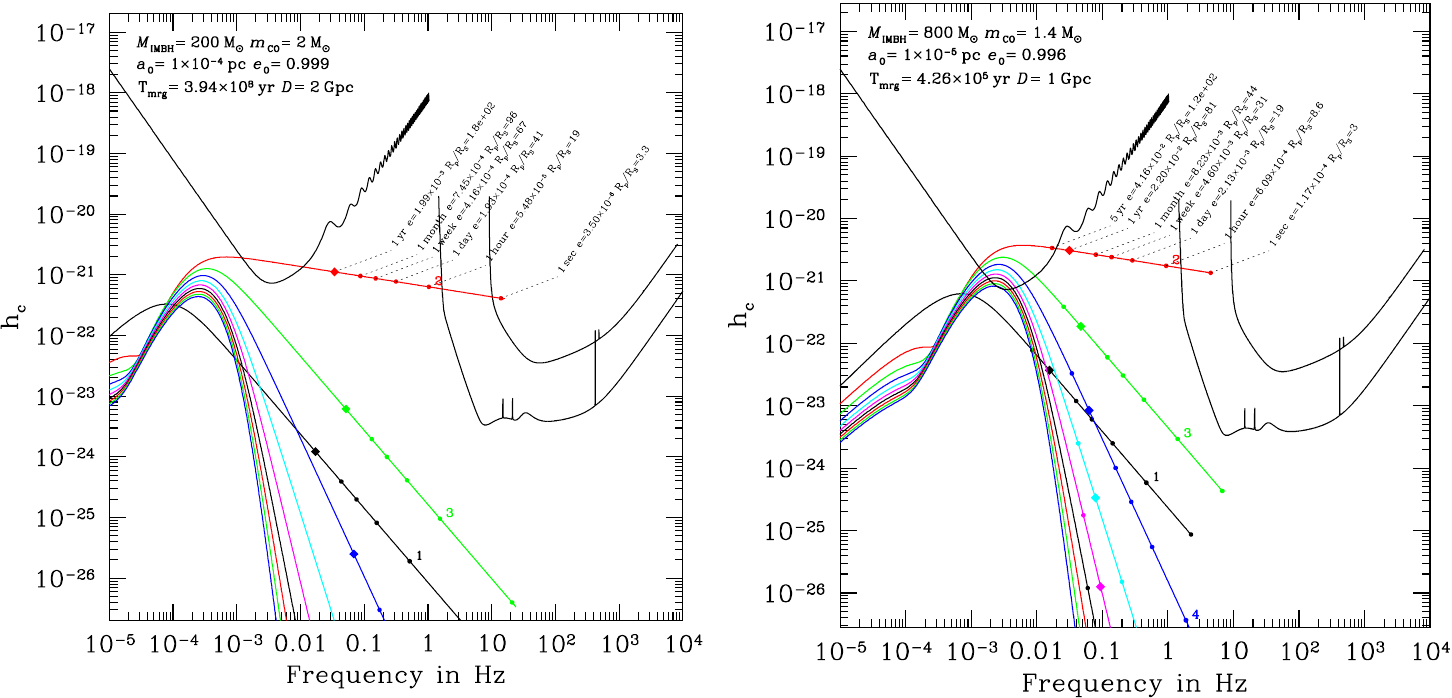}}
\caption
   {
Evolution of the first harmonics of IMRIs with typical values for the
semi-major axis and eccentricity. As in Fig.~(\ref{fig.EMRI_harmonics}), we
show different moments in the evolution, as well as the periapsis distances,
distance to the source, and the merger time.  Additionally, we also display the
Einstein Telescope \citep{2012CQGra..29l4013S,HildEtAl2011} sensitivity curve
on the right, at higher frequencies, and above it, the LIGO sensitivity curve.
   }
\label{fig.multiband_IMRI}
\end{figure}

Based on these results, and in view of the similarity with other groups, we can
take these initial values as representative and ask the question how these
systems would be observed from the space and from the ground, as in the work of
\citep{Amaro-Seoane2018}.  In Fig.~(\ref{fig.multiband_IMRI}) we follow the
approach of Eq.~(\ref{eq.hn}). We can see that the dominant harmonic $n=2$,
crosses a range of frequencies in a timescale of a few years. This means that
if LISA and ground-based observatories are operational at the same time, a
LISA detection would translate into a forewarn to decihertz observatories (if any)
and the groud-based LIGO/Virgo (or ET).

This is interesting because the combination of these detections would allow us
to impose enhanced constraints on the system's parameters. Since LISA can
observe the inspiral, we can derive parameters such as the chirp mass easily.
Ground-based detectors are in the position of observing the merger and
ringdown, and hence one can derive additional parameters such as the final mass
and spin. This compound detection has also the advantage of splitting various
degeneracies and therefore to achieve better measurements of the parameters, as
compared to individual detections.

In the work of \citep{Amaro-Seoane2018} it is shown that IMBHs with light and
medium-size can be observed by space-borne and ground-based observatories with
eccentricity values ranging between $e \in [0.99,\,0.9995]$. The most eccentric
ones can only be seen from the ground \citep[see discussion
in][]{ChenAmaro-Seoane2017}. This is an implication of the characteristic
frequency changing as the pericenter distance decreases because eccentricity
increases. The peak frequency is therefore shifted \citep[see Eq.~37
of][]{Wen2003}. Heavier IMBHs have their frequency peak receding in frequency
as compared to medium-size ones, and therefore the harmonics are embedded in the
LISA range. 

We can estimate the signal-to-noise ratio (SNR or $\rho$) as follows. Assuming
that the $h_{\rm c}$ of an IMRI is, at a given frequency $f$ \citep{FinnThorne2000}

\begin{equation}
h_{\rm c} = \sqrt{(2\dot E/\dot f)}/(\pi D),
\label{eq.SNRhc}
\end{equation}

\noindent
where $\dot E$ is the power emitted, and $\dot f$ the time derivative of the frequency,
the sky- and orientation-averaged $\rho$ of a monochromatic source (assuming the the ansatz of ideal
signal processing) is \citep{FinnThorne2000}

\begin{equation}
\left(\rho\right)^2 = \frac{4}{\pi D^2} \int \frac{\dot{E}}{\dot{f} \, S_h^{SA}(f)} \frac{{\rm d}f}{f^2}
\end{equation}

\noindent
where $S_h^{SA}(f) \approx 5 S_h(f)$ is the sky- and orientation average noise
spectral density. For a source with multiple frequency components, the total
value of $\rho^2$ is obtained by summing the former equation for the different
components.

For high-eccentricity sources we use the expression \citep[see][for an explanation]{Amaro-Seoane2018}

\begin{equation}
\left(\rho \right)^2_n = \int^{f_n(\rm t_{fin})}_{f_n(\rm t_{ini})}
\left(\frac{h_{\rm c,\,n}(f_n)}{h_{\rm det}(f_n)} \right)^2
{\frac{1}{f_n}\,d\left(\ln(f_n) \right)},
\end{equation}

\noindent
where $f_n(t)$ is the (redshifted) frequency of the n harmonic at time $t$
($f_n=n \times f_{\rm orb}$, with $f_{\rm orb}$ the orbital frequency), $h_{\rm
c,\,n}(f_n)$ the characteristic amplitude of the $n$ harmonic when the
frequency associated to that component is $f_n$, and $h_{\rm det}$ is the
square root of the sensitivity curve of the detectors.

With this in mind, we can estimate typical SNRs for different detectors and
sources. Typically, the values are from tens of LISA, LIGO to thousand in ET,
at $D=500\,\textrm{Mpc}$ \citep{Amaro-Seoane2018}.

We therefore have that IMRIs are (a) sources that might already be present in
current ground-based observatories and (b) multiband probes. The fact that no
IMRI has been yet detected (and we are excluding here the ``mysterious source''
\citep{AbbottEtAlGWTC-2_2020}, because it has $q \simeq 0.1$) is very likely
due to the reasons given in the next section.

However, there is another particular characteristic of IMRIs that makes them
interesting, at least from the point of view of modelling. This is the fact
that they are ``clean sources''. What this means is that, contrary to EMRIs,
which might have their inspiral affected by gas effects (or even dark-matter,
if one invokes such a possibility), IMRIs can and must be considered as a
relativistic two-body problem in vacuum from the moment in which it is formed
and the driving mechanism is the emission of GWs (i.e. once it has dettached
from the stellar dynamics regime, see Fig.~(\ref{fig.EMRI}) and related
discussion).

This is so because (a) globular clusters have a negligible amount of gas and
(b) the timescale for merger, once the IMRI has formed, is very small as
compared to other typical timescales associated to perturbations. Indeed,
\citep{Amaro-Seoane2018} shows this by analysing the evolution of a completely
isolated IMRI and another one embedded in a dense stellar system with typical
densities of $\sim 10^5-10^6\,M_{\odot}\textrm{pc}^{-3}$. When comparing the
evolution of the isolated case with the multi-body one, there is no difference.

Finally, we note that it is possible to use IMRIs to extract properties of the
host environment by following the method of accumulated phase shift. Indeed,
\citep{Amaro-Seoane2018} finds that, imposing a threshold $\textrm{SNR}=5$, for
LISA and the ET $\Delta \Psi_f(e) \gtrsim 10^{4}$ radians, and some systems,
typically the light ones, in LIGO have $\Delta \Psi_f(e) \gtrsim 10$ radians.
This means that we can extract information from the host cluster from IMRI
detections.

\subsection{Modelling IMRIs}

Detecting IMRI systems in GW-data will require us to push data analysis
techniques to beyond the state-of-the-art.  In particular, we need to include
higher signal harmonics and to combine data from multiple detectors coherently.
All the gravitational-wave events found so far are from binaries in which the
two components have comparable masses, in any case very different from the
asymmetric systems we are considering here.  

This asymmetry has two implications. First, the gravitational wave signal is
much more complicated, mostly due to the higher modes and harmonics becoming
more important.  Secondly, the higher asymmetry implies that the signals have a
lower amplitude (compared to a more symmetric system which is equally far
away). Thus, the analysis must be more sensitive by including more physical
effects, and by combining the data from multiple detectors more optimally.  

All of the searches thus far have only looked for the dominant harmonic of the
signal with additional physical restrictions (especially assuming
no-precession) and moreover, they have combined the data from different
detectors without considering phase coherence. It is thus not a surprise that
the highly asymmetric systems have not yet been found. This would require
dropping these simplifying assumptions. We would need to include higher signal
harmonics and precession, and also combine data from the different detectors
coherently.

However, even assuming that the signal detection problem have been understood
and addressed, we will still not be able to find IMRIs.  This requires critical
advances in general relativity and numerical analysis. To explain this, we need
to briefly discuss the existing methods and their range of validity. 

Comparable mass-ratios, as exemplified by the current LIGO/Virgo detections,
are modeled by post-Newtonian methods \citep[see][for a review]{Blanchet2013}
for a review, while the extreme-mass ratio systems relevant for LISA require
the self-force calculations in general relativity \citep[see][for a
review]{BarackPound2019,PoissonEtal2011}.  

The intermediate-mass ratio systems fall in the middle and represent an
unexplored part of parameter space.  The most reliable calculations of the
gravitational signal are by solving the Einstein equations numerically.  For
comparable mass-ratio systems, it is now routine to solve the equations with
astrophysical initial conditions and to calculate the full gravitational wave
signal.  

Current methods do not work when the mass ratio $q = m_{\rm CO}/M_{\rm BH}$ is
small; it is computationally far too expensive to simulate even a fraction of
an orbit for $q \sim 10^{-3}$.  The most extensive publicly available
catalogues of gravitational waveforms \cite{BBHCatalog,AbbottEtAlGWTC-2_2020}
currently only go until $q = 0.1$, and $q = 0.01$ represents the smallest ever
mass ratio simulated \citep[][unfortunately covering too few
orbits]{Lousto2011}.  The most extreme mass ratio found by the LIGO/Virgo
collaboration is GW190814, the ``mystery object''
\citep{AbbottEtAlGWTC-2_2020}, and it has $q \simeq 0.1$.

The presence of the much smaller black hole introduces a small length scale in
the problem which, in the standard approach, needs to be resolved.  This
affects then the time steps in the integration of the Einstein equations and
the computational cost increases correspondingly. Going to $q \sim 10^{-3}$ or
$10^{-4}$ is well beyond the capabilities of current methods.  At the other
end, very small values of $q \lesssim 10^{-5}$ can be studied using the
self-force calculations, analogous to the self-force due to electromagnetic
radiation in classical electrodynamics.  Here $q$ is a small parameter linearly
perturbing a background solution \citep{BarackPound2019,Poisson2011}. While several
difficulties remain, it is reasonable to expect that current methods can
address the problem.  These methods cannot however be extended for $q \sim
10^{-3}$ as this would necessitate going well beyond the linear perturbation
approximation.

It is worth pointing out here that there are attempts to interpolate between
the two regimes, in particular by the ``effective-one-body'' approach
\citep[see e.g.][]{BuonannoDamour1999,BoheEtAl2017}.  In this approach, as the
name sugegsts, the idea is to replace the true 2-body system by an effective
description in terms of a 1-body system with a modified Hamiltonian.  The
results of this approach have proven to be effective in the comparable mass and
extreme mass ratio regimes, however it is not clear that this will work in
the intermediate-mass ratio regime.

\section{Extremely-large mass ratio inspirals}

We have see in Fig.~(\ref{fig.MinMassPlunge_labelled2}) the range of masses for
a stellar object to plunge directly through the event horizon of a SMBH without
suffering significant tidal stresses, as a function of the mass of the object.
We note that there is a range of masses which correspond to sub-stellar
objects, in particular to brown-dwarfs (BDs) which will inspiral and cross the
event horizon of SgA* in our Galaxy without being disrupted.

The typical mass of a BD translates into a mass ratio, for SgrA* of $q\sim 10^{-8}$.
This is particularly interesting because if one of these BDs was to form a relativistic
binary with the SMBH in our Galaxy, because of the following points,

\begin{enumerate}

\item The number of cycles that it would revolve until plunging into the event
horizon is roughly inversely proportional to $q$ \citep[see
e.g.][]{Maggiore022018}.  This means that a BD would describe of the order of
$\sim 10^8$ cycles in the band of the detector (in this case LISA). Because of
this value, these sources are called ``extremely-large mass ratio inspirals''
(XMRIs).

\item Due to the previous point, XMRIs would have a much longer life in the
band of the detector as compared to an EMRI.

\item Because of the distance to our Galactic Centre, of $\sim
8\,\textrm{kpc}$, the signal-to-noise ratio (SNR) of an XMRI would be extremely
high, achieving that of binaries of SMBHs or even exceeding it.

\item Since backreaction is proportional to the mass ratio, the modelling of
XMRIs would be much easier than that of EMRIs and, in any case, trivial
as compared to IMRIs. The orbit approaches more and more a pure geodesic and
is, hence, trivial to calculate.

\end{enumerate}

These points were realised by \citep{Amaro-Seoane2019}. In this work, the author 
addresses the possibility that one of these sources exists in our Galactic Centre,
and estimates the associated SNR. 

The mass of a BD ranges approximately between $0.012\,M_{\odot}$ to
$0.076\,M_{\odot}$. We note that this is a lower limit, since they can also
have masses in the range $0.07-0.15\,M_{\odot}$ through the BD formation
process, see \cite{KroupaEtAl2013}, so that the results here are to be
regarded as conservative, since we assume masses lower than $0.07\,M_{\odot}$.  

By following a similar procedure as the one presented starting from
Eq.~(\ref{eq.SNRhc}), and in the same approach, \citep{Amaro-Seoane2019}
calculates the SNR for conservative values of XMRI systems and finds that LISA
would be able to detect them with SNRs ranging from 10 millions of years before
the final plunge, to values as high as SNR $\gtrsim 10^4$ thousands of years
before it, assuming only one year of observation. We can see this in
Fig.~(\ref{SNR_SgrA_BD0p06Msun_1yr-N1000.pdf}).

\begin{figure}
          {\includegraphics[width=0.9\textwidth,center]{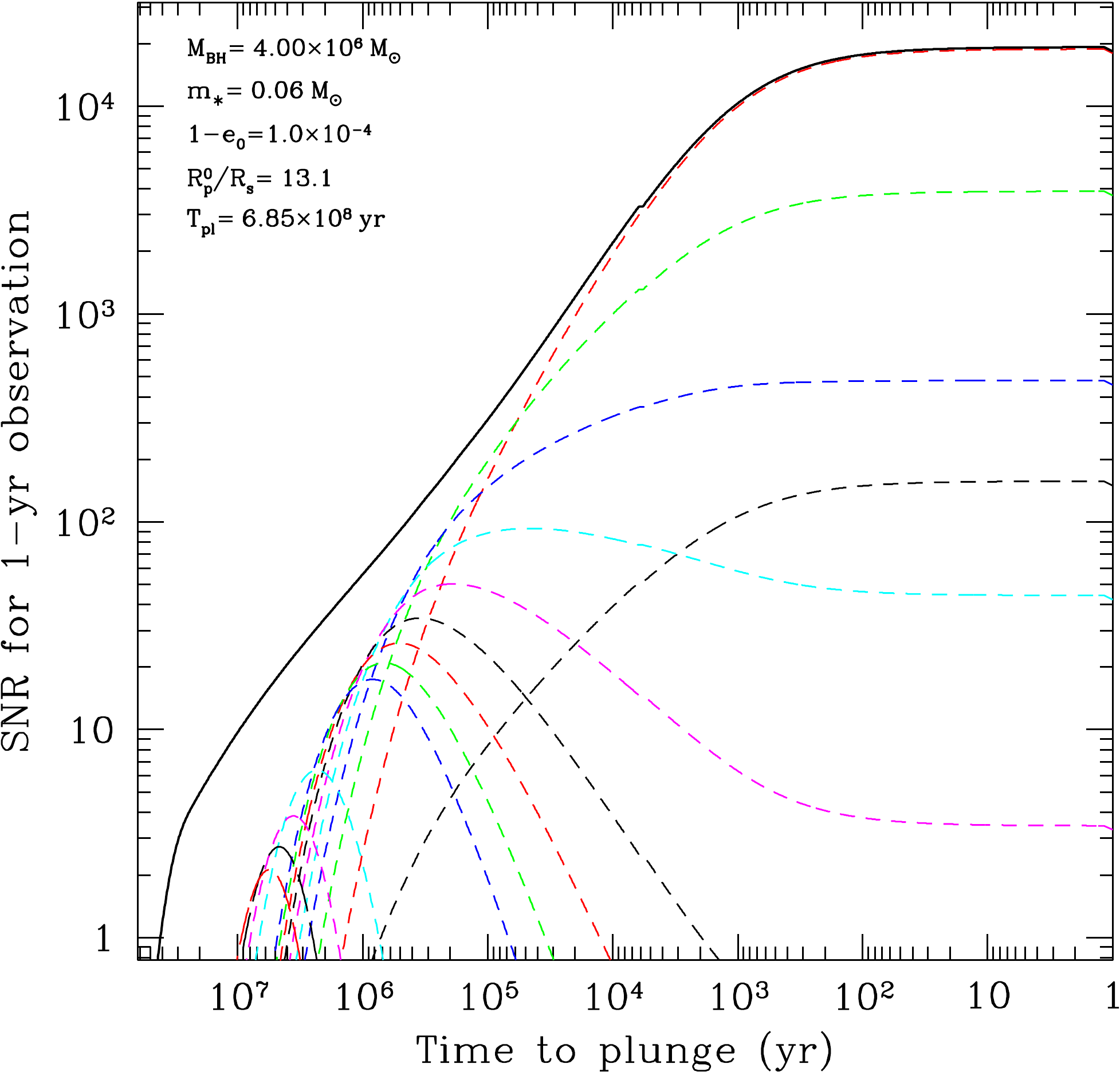}}
\caption
   {
SNR of an XMRI with typical parameters in our Galactic Centre. About $10^7$ yrs before the
BD plunges through the event horizon, and assuming that we can only follow the XMRI for one
year, it can be observed with a SNR of 10. Different lines correspond to different harmonics,
while the black, upper line is the sum of them. One million of years before the plunge, it achieves
SNR $\simeq 60$. From a thousand years before the plunge, it is SNR $\gtrsim 10^4$.
   }
\label{fig.SNR_SgrA_BD0p06Msun_1yr-N1000}
\end{figure}

\begin{figure}
          {\includegraphics[width=1.0\textwidth,center]{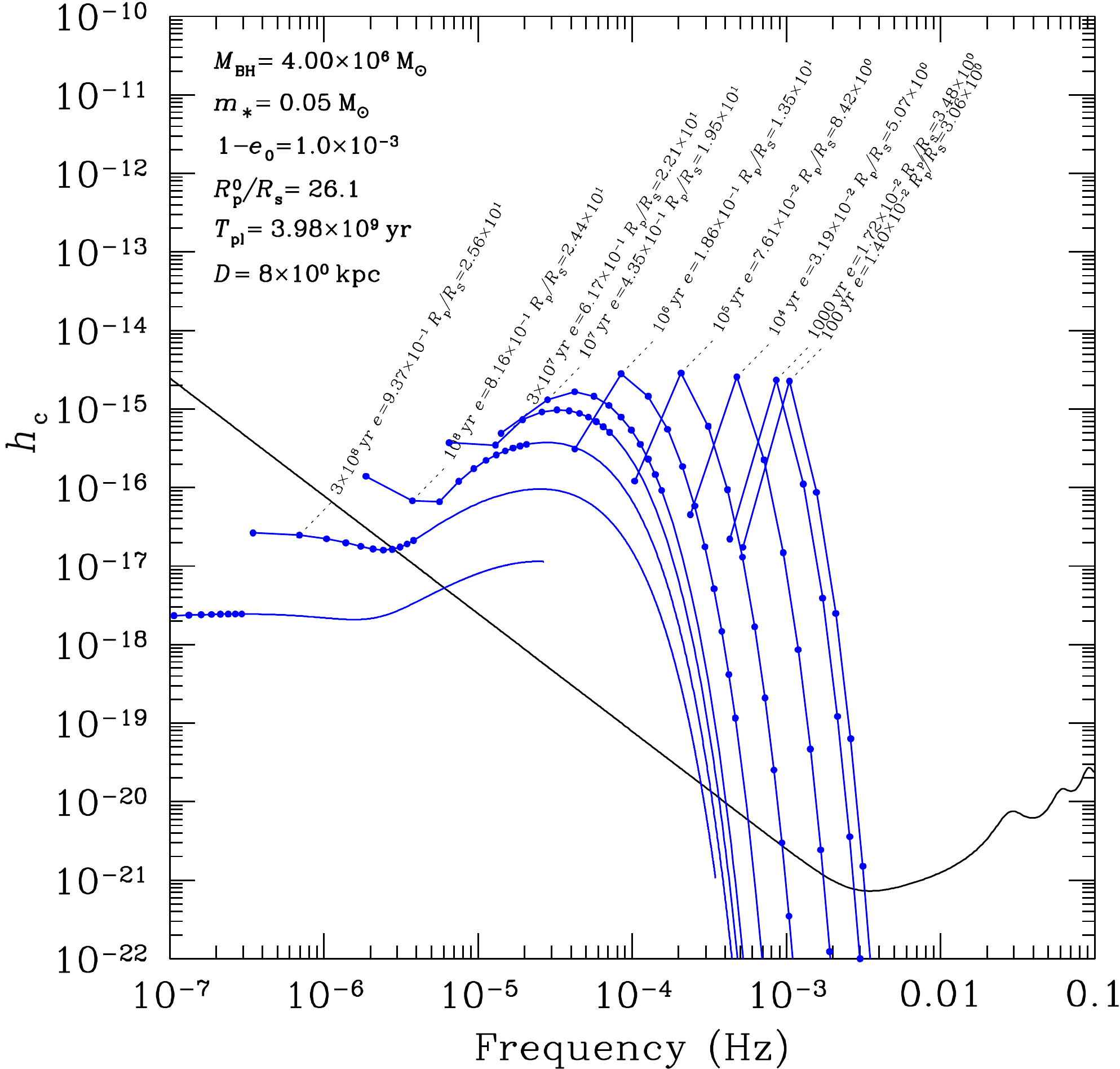}}
\caption
   {
Evolution in phase-space of an XMRI following the same approach and convention
as in Fig.~(\ref{fig.EMRI}). However, instead of showing the different harmonics as
in that figure, we display the isochrones. This means that a curve depicts the position
of the different contribution of the different harmonics at the same time. At the time
of formation, the XMRI time to plunge is of $T_{\rm pl} \sim 4\times 10^9$ yrs.
   }
\label{fig.isochr}
\end{figure}

In Fig.~(\ref{fig.isochr}) we show the evolution in phase-space of a typical
XMRI. We can see that the XMRI can spend millions of years in the band of the
detector. This is the reason (and the proximity in distance) of the large SNR
of Fig.~(\ref{fig.SNR_SgrA_BD0p06Msun_1yr-N1000}).

The event rate for an XMRI to plunge into the SMBH of our Galaxy is not
drastically higher than what can be expected for an EMRI, given in
Eq.~(\ref{eq.GammaBWbh}) and Eq.~(\ref{eq.GammaSMSbh}).  The prcedure to derive
the event rate is very similar to that of stellar-mass black holes, i.e. EMRIs,
with a crucial subtlety related to the dynamics of BDs, which we summarise here
for the conveniece of the reader, and refer to details to the work of
\citep{Amaro-Seoane2019}. 

BDs have their own initial mass function, which is not well-constrained, but it
can be approximated by a single power law \citep[Eq.~4.55 of][]{KroupaEtAl2013},
and this is consistent with the observational data of \citep{WeggEtAl2017}.
Moreover, we can assume that relaxation is dominated, driven, by stellar-mass black holes.
This implicitely assumes that relaxation can be added up individually from these two
mass species, the BDs and the stellar-mass black holes. Indeed, close to the central
region, energy equipartition is found only among the largest masses, and this equipartition
is spread then to velocity at low masses first \citep[][]{BianchiniEtAl2016}. This means
that for a given distribution $f(m,\,v)$ with $v$ the velocity and $m$ the mass, the
moment of the change of velocities is

\begin{equation}
<dv^2> = \int dv^2 f(m,\,v)\, dm\, dv.\nonumber
\end{equation}

\noindent
Because we can neglect equipartition among the low-mass object, we can then express this as

\begin{equation}
<dv^2> = \sum_m n(m) \left(\int dv^2 f(v) dv\right), \nonumber
\end{equation}

\noindent
with $n(m)$ the density of stars of mass $m$. This is crucial, because it implies that
(1) BDs must be close to the centre and (2) we can ignore the contribution of relaxation
provided by them.

The event rate, as derived in \citep{Amaro-Seoane2019}, and following the same nomenclature starting
from Eq.~(\ref{eq.FinalGamma}), is 

\begin{equation}
\dot{\Gamma}_{\rm X-MRI} =  \frac{3-\beta}{2\,\lambda}\frac{N^{\rm BD}_{0\,\textrm{MS}}}
                                                {T_0\,R_{\rm h}^{\lambda}} f_{\rm sub}^{\rm BD} \times \nonumber 
                                  \left\{
                                       a_{\rm crit}^{\lambda}\left[\ln(\Lambda_{\rm crit}) -\frac{1}{\lambda}\right] -
                                       a_{\rm min}^{\lambda}\left[\ln(\Lambda_{\rm min}) -\frac{1}{\lambda}\right]
                                  \right\},
\label{eq.FinalGammaBD}
\end{equation}

\noindent
where we have introduced $\lambda =  9/2-\beta-\gamma$, $\Lambda_{\rm crit} = a_{\rm crit}/(8\,R_{\rm S})$ and
$\Lambda_{\rm min} = a_{\rm min}/(8\,R_{\rm S})$.

\noindent 
As with the EMRI rates, and following the same notation, we give now two examples, the BW- and SM-solution,

\begin{align}
\dot{\Gamma}_{\rm BW} & \sim 1.8\times 10^{-4}\,\textrm{yrs}^{-1}\hat{N}_{0}\,\hat{\Lambda}\,\hat{r}_{\rm infl}^{-11/4} \times \nonumber\\
                           & \Bigg\{
                                    1.34\times 10^{-4} \hat{r}_{\rm infl}^{5/4}\hat{N}_{0}^{-1}\hat{\Lambda}^{-1}\hat{m}_{\rm BD}\,{\cal W}(\iota,\,{\rm a}_{\bullet})\times \nonumber\\
                           &         \left[\ln\left(262\, \hat{r}_{\rm infl}\, \hat{N}_{0}^{-4/5} \hat{\Lambda}^{-4/5}\hat{m}_{\rm BD}^{4/5}\,{\cal W}(\iota,\,{\rm a}_{\bullet})^{-2}\right) - \frac{4}{5} \right] - \nonumber \\
                           &         6.86\times 10^{-25/4}\hat{r}_{\rm infl}^{5/4}\times \left[\ln\left(15.22\, \hat{r}_{\rm infl}\right) - \frac{4}{5} \right]
                             \Bigg\}.
\end{align}

\noindent 
Where we have introduced $\tilde{m}_{\rm BD}:={m_{\rm BD}}/({0.05\,M_{\odot}})$.
In the case of SM, we have that

\begin{align}
\dot{\Gamma}_{\rm SM} & \sim 2.3\times 10^{-3}\,\textrm{yrs}^{-1} \hat{r}_{\rm infl}^{-5/2}\hat{N}_{0}\hat{\Lambda} \times \nonumber\\
                           & \Bigg\{
                                    1.4\times 10^{-4} \hat{r}_{\rm infl}\, \hat{N}_{0}^{-1} \hat{\Lambda}^{-1} \hat{m}_{\rm BD}\,{\cal W}(\iota,\,{\rm a}_{\bullet})^{-5/2}\times \nonumber\\
                           &         \left[\ln\left(46\,\hat{r}_{\rm infl}\, \hat{N}_{0}^{-1}\hat{\Lambda}^{-1}\hat{m}_{\rm BD}\,{\cal W}(\iota,\,{\rm a}_{\bullet})^{-5/2}\right) - 1  \right] - \nonumber \\
                           &         4.67\times 10^{-7} \hat{r}_{\rm infl} \times \left[\ln\left(15.24\, \hat{r}_{\rm infl}\right) - 1\right]
                             \Bigg\}.
\end{align}

\begin{figure}
          {\includegraphics[width=1.3\textwidth,center]{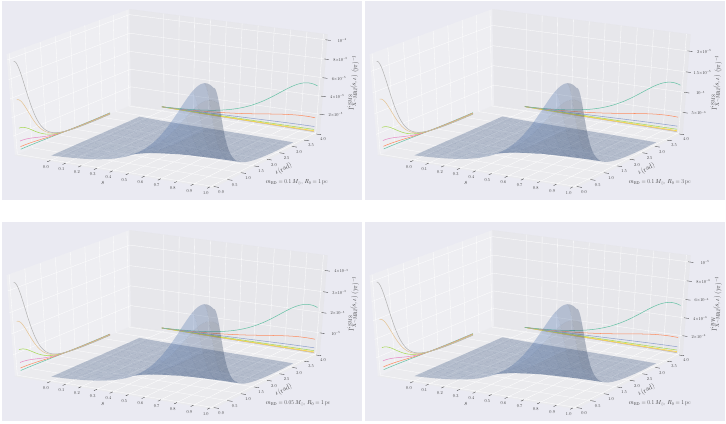}}
\caption
   {
Event rate for XMRIs at the Galactic Centre in the BW- and SM solution. 
   }
\label{fig.Gamma_Various_Cases}
\end{figure}

We reproduce here the results of \citep{Amaro-Seoane2019}, in
Fig.~(\ref{fig.Gamma_Various_Cases}). We can see that the event rate is not
much higher as for the EMRI case, see Eq.~(\ref{eq.FinalGamma}) and the
examples.

The important, and interesting point about XMRIs is that, contrary to more
massive EMRIs, XMRIs spend a long time on band, as we have seen. They start
with a SNR $>10$ in LISA $T\sim 10^6\,{\rm yr^{-1}}$ before the final plunge, and
the event rate $\dot{\Gamma}\cong 10^{-5}\,{\rm yr}^{-1}$.  Therefore, the
number of sources in band at \textit{any given moment} (with the proviso that the nucleus
is relaxed) is the multiplication of the event rate times their lifetime,
i.e. $\sim \textrm{~a~few~}10$. 

From the continuity equation of the events we can derive the relative
occupation fractions of the (line) density $g=dN/da$, with $N$ the number of
sources and $a$ their semi-major axes.  Taking into account the eccentricity of
the sources when integrating $N$, \citep{Amaro-Seoane2019} finds 

\begin{itemize}
    \setlength{\itemsep}{1pt}
    \setlength{\parskip}{0pt}
    \setlength{\parsep}{0pt}

 \item[$\bullet$]  $\gtrsim 15$ X-MRIs at low frequencies with high
eccentricities and associated SNRs of $\simeq ~{\rm a~few~} 100$.
 \item[$\bullet$]  $\gtrsim 5$, at higher frequencies, i.e. at very high SNRs (from ${\rm a~few~} 100$ up
to $2\times 10^4$), in circular-, or almost circular orbits.

\end{itemize}

\noindent 
Is is important to stress that this derivation has been done assuming a
steady-state distribution; it is a statistical representation of the system.
This means that, \textit{at any given moment of the Milky Way} -under the
assumption that our Galactic Centre is in the range of relaxed nuclei, see
Fig.~(\ref{fig.Trlx})-, \textit{the numbers of sources that we quote above
hold}. If this requirement is fulfilled, this means that millions of years ago
and millions of years in the future, those are the numbers of XMRIs that we can
expect to be populating the Galactic Centre.

We stress that these are conservative values. These numbers can be multiplied
by a factor of a few depending on the eccentricity of the sources when they
form. Also, this approach is purely analytical and based on pure power-laws.
These should decrease as one approaches the innermost radii, which translates
into a more efficient diffusion of stellar-mass black holes. Hence, we are
artificially enhancing the relaxation time and therefore decreasing the event
rate.

While the SNRs are very high, to the point of being able to bury the SNR of
binaries of SMBHs in some cases, the distance plays a crucial role. This means
that, whilst we can expect to detect them in our own Galactic Centre, other
galaxies are out of question but for, maybe, satellite galaxies of M82 with
SMBHs in the range of LISA.

Because of their extreme mass ratios, these systems are closer to a geodesic
than EMRIs, which makes it easier to model them. However, because XMRIs evolve
so slowly, in the lifetime of LISA they can be envisaged as monochromatic
sources. This, in turn poses the question of how many of them and in what
distribution of orbital parameters would we need to reproduce the exploration
of timespace that an EMRI can do. An EMRI probes a vast range of values in the
semi-major axis, eccentricity and change of plane of the orbit in as short as a
few months or years. Each burst can be considered like a picture of warped
spacetime, and hence it delivers us with thousands of pictures to allow us to
do a cartography of that spacetime around the SMBH. An XMRI, on the contrary,
is ``fixed'' in phase-space and the question arises of how many of them, and in
what distribution, would we need to derive that mapping of spacetime

There is an additional problem related to XMRIs. Since we can expect the Milky
Way to be a proxy galaxy of the type that LISA will probe, it is very likely
that an important fraction of all of them will have a few XMRIs distributed
around their SMBHs. As we have noted before, this is a steady-state
representation of relaxed nuclei. In other words, we can expect that all
\textit{relaxed} nuclei in the range of masses that LISA will observe will have
a few tens of XMRIs distributed around the SMBH moving at a fraction of the
speed of light.

Any infalling EMRI will therefore inevitably encounter them on its orbit
torwards the final plunge into the SMBH. Because the XMRIs are moving at such
high velocities, their effective masses are similar to those of the EMRI, which
means that an interaction EMRI-XMRI can significantly alter the orbit of the
EMRI or, even worse, scatter it completely off, cancelling the merger of the
system.

\section{A Relativistic Fokker-Planck Algorithm}

As we have seen in the previous sections, the problem of forming an EMRI (or
IMRI/XMRI) involves both astrophysics (in particular stellar dynamics) and
general relativity. This is challenging because this translates into addressing
the two-body problem in general relativity, which is unsolved at these mass
ratios, as we have briefly described before, and the $10^6$--$10^8$ body
problem in Newtonian gravity.  The host galactic nucleus (in the case of an
EMRI or XMRI) or globular cluster (for an IMRI) has that order of stars.  While
we care about relativistic effects between COs and the central SMBH/IMBH,
relativity does not play a significant role in general among the stars that
form part of the stellar system in which the inspiral takes place.

In this section, which gives the primary idea of
\citep{Amaro-SeoanePretoSopuerta2010}, we describe what ought to be done to
address this problem with a two-dimensional Fokker-Planck integrator, which has
shown to be very robust even compared to direct $N-$body simulations \citep[see
e.g.][]{PretoAmaroSeoane10,Amaro-SeoanePreto11}, with the advantage that (i)
one has a control over the physics as it is implemented into the integrator
scheme and (ii) the Fokker-Planck approach is orders of magnitude faster than
any brute force, direct-summation integrator.

Before we describe the main idea, we need to introduce a few physical concepts,
and the connection between the Newtonian and relativistic regimes.

\subsection{Newtonian motion around a Newtonian potential}
We start from the Hamiltonian of Eq.~(\ref{hamiltonian}).  In this case, the 
equation for radial motion can be written as:
\begin{equation}
\left(\frac{dr}{dt}\right)^{2} = -2E + \frac{2G\mbul}{r} - \frac{J^{2}}{r^{2}}
= \frac{2E}{r^{2}}(r_{a}-r)(r-r_{p})\,.
\end{equation}
From here, and using the definitions of the previous section of the pericenter and
apocenter radial coordinates, we find
\begin{align}
E & =  \frac{G\mbul}{r_{p}+r_{a}} = \frac{G(1-e^{2})}{2p}\,, \\
J^{2} & = \frac{2G\mbul r_{p}r_{a}}{r_{p}+r_{a}} = G\mbul^{2}p\,.
\end{align}

\subsection{Relations between the Relativistic and Newtonian parameters for a Schwarzschild SMBH}
\label{relationNewtonEinstein}

In order to connect the Newtonian and relativistic calculations into a single
Fokker-Planck integrator, it is important first to have a clear way of
connecting the Newtonian and relativistic parameters describing the orbital
motion.
Here, we want to relate the Newtonian constants of motion $(E,J)$ of Eqs.~(\ref{hamiltonian}) and~(\ref{newtonianJ})
with the relativistic constants of motion $({\cal E},{\cal J})$ of Eqs.~(\ref{EJsch}).  Let us start first
with the angular momentum. For that, we need to compare the Newtonian definition of 
the angular momentum in terms of the spherical coordinates of motion

\begin{equation}
J = r^{2}\dot\varphi\,,
\end{equation}

\noindent 
with the second equation in~(\ref{EJsch}).  We realize that, by identifying the radial and azimuthal coordinates
(there is a justification for this), we can just identify the angular momentum parameters, i.e.

\begin{equation}
{\cal J} \equiv J\,.
\end{equation}

\noindent 
The case of the energy is more comples, in part because of the intrinsic four-dimensional character
of general relativity.  Moreover, the properties of the Einsteinian and Newtonian gravitational fields
are different and complicate the comparison since, at finite distance, the energy involves the gravitational
potential.  Hence, the ideal way of comparing things is by going to spatial infinity, which means analyzing
the energy at $r\rightarrow\infty$.  At infinity, the relativistic specific energy satisfies 

\begin{align}
\Gamma_{\infty} & =  u^{t}_{\infty} =  {\cal E}\,, \label{gammaEinfinity} \\
\Gamma_{\infty} & =  \frac{1}{\sqrt{1-v^{2}_{\infty}}} \label{gammaVinfinity} \,,
\end{align}

\noindent 
where we have used Eqs.~(\ref{EJsch})
and~(\ref{gammafactor}) in the case of the Schwarzschild metric. Here
$v^{2}_{\infty}$ is the spatial velocity at infinity, given by

\begin{equation}
v^{2}_{\infty} \equiv \met^{\infty}_{ij}\,v^{i}_{\infty}v^{j}_{\infty} = \left(\frac{dr}{d\tau}\right)^{}_{\infty}
+ \left[ r^{2}\left(\frac{d\varphi}{d\tau}\right)^{2}\right]^{}_{\infty}    \,.
\end{equation}

\noindent 
Then, combining Eqs.~(\ref{gammaEinfinity}) and~(\ref{gammaVinfinity}) we obtain the well-known relation
in special relativity:

\begin{equation}
v^{2}_{\infty} =  \frac{ \Gamma^{2}_{\infty}-1}{\Gamma^{2}_{\infty}} = \frac{ {\cal E}^{2}-1}{{\cal E}^{2}}\,.
\end{equation}

\noindent 
In the Newtonian case, from Eq.~(\ref{hamiltonian}), we have

\begin{equation}
v^{2}_{\infty} = - 2\,E\,.
\end{equation}

\noindent 
Hence, by comparing we can relate both energies in the following way

\begin{equation}
E = \frac{1- {\cal E}^{2}}{2\,{\cal E}^{2}}\,,\qquad
{\cal E} = \frac{1}{\sqrt{1+2\,E}}\,. \label{energyrelation}
\end{equation}

\noindent 
Since we are analyzing the equations at infinity, this is only
valid for unbound orbits, which satisfy $E<0$.   For bound orbits we have to make a different
analysis. A possibility is to identify the $(e,p)$ parameters as they have the same meaning both
in Newtonian and Einsteinian gravity, as they determine the turning points of the motion,
i.e. $r_{p}$ and $r_{a}$.  However, this produces ${\cal J}\neq J$, which in principle is not
a problem by itself.  By doing this identification one can get the following relations

\begin{align}
\frac{{\cal J}^{2} - J^{2}}{{\cal J}^{2}} & =  \frac{4+e}{p} \,, \\
\frac{E - \frac{1-{\cal E}^{2}}{{\cal E}^{2}}}{\frac{1-{\cal E}^{2}}{{\cal E}^{2}}} 
& =   \frac{4(1-e^{2})}{p(p-4)}\,,
\end{align}

\noindent 
where we are comparing the Newtonian energy $E$ with $(1-{\cal E}^{2})/{\cal E}^{2}$ 
[as in Eq.~(\ref{energyrelation})].  We have normalized with respect to the relativistic
values as they are supposed to be closer to reality.   Notice that for orbits with big
$p$, the identification can be reasonable, as both expressions go to zero as $p$ goes
to infinity.

\subsection{Relations between the Relativistic and Newtonian parameters for a Kerr SMBH}

In this case the situation is a bit more complicated, as we have an additional
constant of motion.  In a spherically symmetric gravitational potential the
constants of motion are just two, $(E,\,J)$, but in the case of a spinning SMBH
(Kerr geometry) there are three: $(E,{\bf J}_{z},C)$, where ${\bf J}_{z}$ is
the angular momentum component in the spin direction and $C$, the
Carter constant, related to the total angular momentum. 

In the Newtonian limit ${\bf J}^{2}_{z}+C \approx {\bf J}^{2}$.  For some
triaxial Newtonian potentials there is an analogous to the Carter constant.  An
important point is that in the spherically symmetric case $(E,{\bf J}_{z})$ is
equivalent to $(e,r_{p})$, where $r_{p}$ denotes the pericenter radial
coordinate.  With spin, $(E,{\bf J}_{z},C)$ is equivalent to
$(e,\,r_{p},\iota)$, where $\iota$ denotes the inclination of the orbit with
respect to the equatorial plane (the plane orthogonal to the spin).  This means
that the inclination plays a different role in the spherical symmetric case
with respect to the axisymmetric one. 

\subsection{A possible scheme}

The first step is to obtain the steady state distribution, defined by the
density profile $\rho(r)$ and the phase space distribution function (DF)
$f(E)$, of stars around the SMBH.  This is given by the solution of the
Fokker-Planck equation (or by N-body simulations). We have solutions for
clusters which are spherically symmetric and which have an isotropic
distribution of velocities.  Here, we assume that the cluster is spherical and
the SMBH spin does not change this appreciably. It is possible that the
distribution of inclinations for stars with lowest $J$ will deviate from
sphericity due to close interactions with the hole, but not the cluster as a
whole.

We now define ${\bf J} =(J_x,J_y,J_z)$. We note that we use specific energy $E$
and  angular momentum ${\bf J}$.  An  $E>0$ corresponds to bound orbits, so for
an elliptical orbit in the field of the SMBH, the Hamiltonian of the system
becomes

\begin{equation}
H=-\frac{1}{2}v^2 + \frac{G\mbul}{r} = E>0\,, \label{hamiltonian}
\end{equation} 

\noindent 
and the magnitude of total angular momentum is 

\begin{equation}
J= \frac{G\mbul}{\sqrt{2E}}\,. \label{newtonianJ}
\end{equation}

Since the stars originate from a spherical cluster, all components of ${\bf J}$
are allowed to diffuse independently under two-body relaxation and, as a
result, orbital inclinations diffuse with ${\bf J}$. We define $J_z \parallel
\sbul$, where $\sbul$ is the spin axis of the SMBH. 

Stars interact relativistically with the Kerr SMBH, this leads to deterministic
changes in ($E, {\bf J}_z, C$); we need $(E,{\bf J})$ to evolve them in the
cluster.  This needs to establish a correspondence between Newtonian and
relativistic constants of motion (either for spinning or non-spinning black
holes).  With this at hand, we can then try to distinguish between changes in
different components of the angular momentum, $\Delta {\bf J}$.

Then, we want to determine the region of phase space $(E,{\bf J})$ from which
EMRIs originate. For such calculation we need to include not only the effect
of scattering on the EMRI candidate due to the other stars, but also the
deterministic loss $(\Delta E_{GW}, \Delta {\bf J}_{GW})$ that results from the
emission of GWs at close passages to the SMBH. 

We need to then estimate the changes in the constants of motion due to
gravitational wave emission, $(\Delta E^{GW}, \Delta {\bf J}^{GW}_{z},\Delta
C^{GW})$. This means that we do not need $\Delta C^{GW}$ for the Fokker-Planck
simulations.  Nevertheless, in order to work with the orbital parameters
$(p,\,e,\,\iota)$ we have to calculate $\Delta C^{GW}$ to estimate the changes
in the orbital parameters.  

The idea is to study the phase space evolution of many individual EMRI
candidates in a Monte Carlo fashion. One starts by sampling the stellar
population according to the DF given by FP calculation. So, each EMRI candidate
has an initial set of phase space coordinates such as  $(E_0,\,{\bf J}_o)$. In
principle, we should be interested in orbits which are bound to the SMBH, but
one is also interested in the case when they are unbound to the hole, but bound
to the cluster. 

The FP solution $\rho(r)$ and $f(E)$ completely determines the diffusion
coefficients for scattering with the cluster stars---assuming that two-body
relaxation is the only dynamical mechanism.  The EMRI candidate will undergo a
random walk in phase space plus a determination loss of energy and angular
momentum due to radiation reaction. The angular momentum jump $\delta
J_P(E,\,J)$, per orbital period $P(E)$, is given by

\begin{align}
\delta J_P(E,J) & =  \langle \Delta J \rangle_t P(E) + \xi_J \left(\langle (\Delta J)^2 \rangle_t P(E) \right)^{1/2} -
                      \Delta J_{GW}(E,J) =  \nonumber   \\
                & =    \frac{J_c^2(E)}{2 ~ J ~ T_{\rm rlx}(E)} P(E) + \xi_J \left(\frac{J_c^2(E)}{T_{\rm rlx}(E)} P(E) \right)^{1/2} -  
                          \Delta J_{GW}(E,J),
\label{eq1}
\end{align}

\noindent 
where $\xi_J=\pm 1$ and the quantities inside $\langle ~. ~ \rangle$ are the
first and second diffusion coefficients per unit time. The second equality is
obtained from simple scaling arguments. The orbital energy jump, per orbital
period, is given by

\begin{equation}
\delta E_P(E,\,J) = \langle \Delta E \rangle_t P(E) + \xi_E \left(\langle (\Delta E)^2 \rangle_t P(E) \right)^{1/2}
                                 + \Delta E_{GW}(E,\,J),
\label{eq2}
\end{equation}

\noindent 
In this equation $D_{EE}$ and $D_E$ are the diffusion coefficients,
introduced when defining the time-dependent, orbit-averaged,
isotropic, Fokker-Planck equation in energy space. For each component
this is \citep{Spitzer87,ChernoffWeinberg1990}

\begin{equation}
p(E) \pderiv(f_i/t) =  - \pderiv(F_{E,i}/E),\,F_{E,i}  =  -D_{EE,i} \pderiv(f_i/E) - D_E f_i,
\end{equation}

\noindent 
with

\begin{align}
D_{EE,i} & =  4\pi^2 G^2 m_*^2 \mu_i^2 \ln \Lambda \left(\frac{\mu_j}{\mu_i} \right)^2 \sum_j^{N_c} \left[ q(E) \int_{-\infty}^E dE' f_j(E')
+ \int_E^{+\infty} dE' q(E') f_j(E') \right], \nonumber \\
D_{E,i} & =  - 4\pi^2 G^2 m_*^2 \mu_i^2 \ln \Lambda \sum_j^{N_c} \left( \frac{\mu_j}{\mu_i} \right) \int_E^{+\infty} dE' p(E') f_j(E').                                     
\end{align}

\noindent 
The indices $i,j$ run from $1$ to the number of mass components considered, $N_c$, and we define $\mu_i=m_i/m_{*}$,
with $m_*=1/N$. One can obtain gravitational potential $\Phi$ from Poisson's equation $\nabla^2 \Phi(r) = 4 \pi G \rho(r)$,
which needs to be updated as $\rho(r)$ evolves over time. This can be achieved following the technique of operator-splitting  by \cite{Cohn80} and
\cite{ChernoffWeinberg1990}. One successively updates the distribution function $f(E)$ through the diffusion equation
and the gravitation potential $\Phi(r)$ via the Poisson equation. In each diffusion step $\Phi(r)$ is kept constant, whilst
$f(E)$ and the diffusion coefficients are updated. Correspondingly, for every Poisson step, $\Phi(r)$ is updated and $f(E)$
is kept constant as a function of the adiabatic invariant.

One then needs to take into account the deterministic losses $(\Delta E_{GW},
\Delta {\bf J}_{GW})$ due to GW emission that need to be added to the right
hand sides of the equations. These could be obtained from the numerical
calculations given some $(E,{\bf J})$ and some interval of time $\Delta t$. In
the simplest case, $\Delta t=P(E)$, i.e.  one orbital period; but, in general,
we need to have $P(E) \ll \Delta t \ll T_{rlx}(E)$. 

This can be achieved by reading the values from the Fokker-Planck calculations
to derive the new values $(E,{\bf J})$, after computing  $(\Delta E_{GW},
\Delta {\bf J}_{GW})_{\Delta t}$ over some $N$ orbital periods.  

Let us consider $N^{orb}_{p}$ the number of radial periods, i.e. the time to
go from $r_{p}$ to $r_{a}$ and back to $r_{p}$.  This has to be distinguished
from the other two periods present in generic orbits around a spinning SMBH.

Given $N^{orb}_{p}$ (or alternatively $\Delta t$), the relativistic Fokker-Planck 
integrator computes $(\Delta E^{GW},\Delta {\bf J}^{GW}_{z}, \Delta C^{GW})$ and/or
the new $(p,e,\iota)$. Again, besides the radial
period, $T_{r}$, we have the polar and azimuthal periods.  The first one, the
polar $T_{\theta}$, is related with the fact that when we have a spinning black
hole the motion, in general, is not confined in a plane, as mentioned previously.  

Nevertheless, one can define an instantaneous plane of the orbit, which
oscillates in a symmetric way around the equatorial plane. In that case
$T_{\theta}$ is the period of such oscillations.  The third period, the
azimuthal one $T_{\varphi}$, corresponds to the time that the azimuthal angle
$\varphi$ takes to run $2\pi$.  For generic orbits the three periods are
different.  For a non-spinning black hole we just have $T_{r}$ and
$T_{\varphi}$ which are different because of the relativistic precission.  In
Newtonian dynamics these two are identical.

The algorithm could then be summarised with the following steps:

\begin{itemize}

\item[1.] Obtain the initial steady state distribution of stars around the
SMBH, described by ($\rho(r)$,\,$f(E)$) via a Newtonian, two-dimensional
Fokker-Planck integrator. This allows us to have $(E,{\bf J})$ for each star in
the distribution.  I.e. $\{(E_{i},{\bf J}_i)\}^{}_{i=1,...,N_{\ast}}$, where
$N_{\ast}$ is the number of stars. Assume $J_z \parallel {\bf \sbul}$. The
orbital inclinations are already encoded in the angular momentum vector.

\item[2.] Establish a correspondence between $(E,{\bf J})_{N}$ and
$(E,J_{z},C)_{E}$ (where the subscripts $N$ and $E$ stand for Newtonian and
Einsteinian respectively). We note that $N^{orb}_{p}$ is not the same for all
stars.  A solution to this is to make $\Delta t$ equal to a fixed fraction of
the (local) relaxation time. For a cusp with $f(E) \propto E^p$, $T_{\rm rlx}
\propto E^{-p}$ \citep{Peebles1972}, where we expect $p$ to range between
$\sim 0$ and $\sim 1/2$.  The Newtonian orbital periods are $P(E) \propto
E^{-3/2}$, so that the steps will change significantly between different
particles and also during the inspiral.

\item[3.] Given the relativistic parameters, and given $N^{orb}_{p}$, we need
to compute the changes due to gravitational wave emission on them,
$(\Delta E^{GW},\Delta J^{GW}_{z},\Delta C^{GW})$. The relativistic
Fokker-Planck scheme then delivers the new $(E,J_{z},C)_{E}$.

\item[4.] Using the new $(\Delta E^{GW},\Delta J^{GW}_{z},\Delta C^{GW})$ map
back to obtain the new $(E,{\bf J})_{N}$.  In this way, we compute $\{(\Delta
E^{GW}_{i}, \Delta {\bf J}^{GW}_i)\}^{}_{i=1,...,N_{\ast}}$

\item[5.] Use the results of Point (4) in the equations for $(\delta E_P(E,J),\delta J_P(E,J))$.

\item[6.] Iterate from point (2). Here we should pass the information of when one of the EMRIs has 
plunged.  This is decided by the relativistic integrator. Obviously, we can derive the
corresponding orbital parameters, so as to have a statistical distribution.

\end{itemize}

\subsection{\color{red}Citations}


\begin{thebibliography}{183}
\ifx \bisbn   \undefined \def \bisbn  #1{ISBN #1}\fi
\ifx \binits  \undefined \def \binits#1{#1} \fi
\ifx \bauthor  \undefined \def \bauthor#1{#1} \fi
\ifx \bjtitle  \undefined \def \bjtitle#1{\textrm{#1}}\fi
\ifx \batitle  \undefined \def \batitle#1{#1} \fi
\ifx \bctitle  \undefined \def \bctitle#1{#1} \fi
\ifx \bvolume  \undefined \def \bvolume#1{\textbf{#1}}\fi
\ifx \byear  \undefined \def \byear#1{#1} \fi
\ifx \bissue  \undefined \def \bissue#1{#1} \fi
\ifx \bfpage  \undefined \def \bfpage#1{#1} \fi
\ifx \blpage  \undefined \def \blpage #1{#1} \fi
\ifx \burl  \undefined \def \burl#1{#1} \fi
\ifx \doiurl  \undefined \def \doiurl#1{#1} \fi
\ifx \betal  \undefined \def \betal{et al.} \fi
\ifx \binstitute  \undefined \def \binstitute#1{#1} \fi
\ifx \beditor  \undefined \def \beditor#1{#1} \fi
\ifx \bpublisher  \undefined \def \bpublisher#1{#1} \fi
\ifx \bbtitle  \undefined \def \bbtitle#1{\textit{#1}} \fi
\ifx \bedition  \undefined \def \bedition#1{#1} \fi
\ifx \bseriesno  \undefined \def \bseriesno#1{#1} \fi
\ifx \blocation  \undefined \def \blocation#1{#1} \fi
\ifx \bsertitle  \undefined \def \bsertitle#1{#1} \fi
\ifx \bsnm \undefined \def \bsnm#1{#1} \fi
\ifx \bsuffix \undefined \def \bsuffix#1{#1} \fi
\ifx \bparticle \undefined \def \bparticle#1{#1} \fi
\ifx \barticle \undefined \def \barticle#1{#1} \fi
\ifx \botherref \undefined \def \botherref #1{#1} \fi
\ifx \url \undefined \def \url#1{#1} \fi
\ifx \bchapter \undefined \def \bchapter#1{#1} \fi
\ifx \bbook \undefined \def \bbook#1{#1} \fi
\ifx \bcomment \undefined \def \bcomment#1{#1} \fi
\ifx \oauthor \undefined \def \oauthor#1{#1} \fi
\ifx \citeauthoryear \undefined \def \citeauthoryear#1{#1} \fi
\ifx \texttildelow  \undefined \def \texttildelow{\symbol{126}} \fi
\def \endbibitem {}
\ifx \bconflocation  \undefined \def \bconflocation#1{#1} \fi

\bibitem[\protect\citeauthoryear{{Aarseth}}{1999}]{Aarseth99}
\begin{barticle}
\bauthor{\binits{S.J.} \bsnm{{Aarseth}}},
\batitle{{From NBODY1 to NBODY6: The Growth of an Industry}}.
\bjtitle{The Publications of the Astronomical Society of the Pacific}
\bvolume{111},
\bfpage{1333}--\blpage{1346}
(\byear{1999})
\end{barticle}
\endbibitem

\bibitem[\protect\citeauthoryear{{Aarseth}}{2003}]{Aarseth03}
\begin{bbook}
\bauthor{\binits{S.J.} \bsnm{{Aarseth}}},
\bbtitle{{Gravitational N-Body Simulations}}
(\bpublisher{ISBN 0521432723.~Cambridge, UK: Cambridge University Press,
  November 2003.}, \blocation{???}, \byear{2003})
\end{bbook}
\endbibitem

\bibitem[\protect\citeauthoryear{{Aarseth} and {Zare}}{1974}]{AarsethZare74}
\begin{barticle}
\bauthor{\binits{S.J.} \bsnm{{Aarseth}}},
\bauthor{\binits{K.} \bsnm{{Zare}}},
\batitle{{A regularization of the three-body problem}}.
\bjtitle{Celestial Mechanics}
\bvolume{10},
\bfpage{185}--\blpage{205}
(\byear{1974}).
doi:\doiurl{10.1007/BF01227619}
\end{barticle}
\endbibitem

\bibitem[\protect\citeauthoryear{Alexander and
  Hopman}{2009}]{AlexanderHopman09}
\begin{barticle}
\bauthor{\binits{T.} \bsnm{Alexander}},
\bauthor{\binits{C.} \bsnm{Hopman}},
\batitle{{Strong mass segregation around a massive black hole}}.
\bjtitle{ApJ}
\bvolume{697},
\bfpage{1861}--\blpage{1869}
(\byear{2009}).
doi:\doiurl{10.1088/0004-637X/697/2/1861}
\end{barticle}
\endbibitem

\bibitem[\protect\citeauthoryear{{Alexander} and {Livio}}{2001}]{AL01}
\begin{barticle}
\bauthor{\binits{T.} \bsnm{{Alexander}}},
\bauthor{\binits{M.} \bsnm{{Livio}}},
\batitle{{Tidal Scattering of Stars on Supermassive Black Holes in Galactic
  Centers}}.
\bjtitle{ApJ Lett.}
\bvolume{560},
\bfpage{143}--\blpage{146}
(\byear{2001})
\end{barticle}
\endbibitem

\bibitem[\protect\citeauthoryear{{Amaro-Seoane}}{2018a}]{Amaro-Seoane2018}
\begin{barticle}
\bauthor{\binits{P.} \bsnm{{Amaro-Seoane}}},
\batitle{{Detecting intermediate-mass ratio inspirals from the ground and
  space}}.
\bjtitle{Phys. Rev. D.}
\bvolume{98}(\bissue{6}),
\bfpage{063018}
(\byear{2018}a).
doi:\doiurl{10.1103/PhysRevD.98.063018}
\end{barticle}
\endbibitem

\bibitem[\protect\citeauthoryear{{Amaro-Seoane}}{2018b}]{Amaro-SeoaneLRR2012}
\begin{barticle}
\bauthor{\binits{P.} \bsnm{{Amaro-Seoane}}},
\batitle{{Relativistic dynamics and extreme mass ratio inspirals}}.
\bjtitle{Living Reviews in Relativity}
\bvolume{21},
\bfpage{4}
(\byear{2018}b).
doi:\doiurl{10.1007/s41114-018-0013-8}
\end{barticle}
\endbibitem

\bibitem[\protect\citeauthoryear{{Amaro-Seoane}}{2018c}]{Amaro-SeoaneLRR}
\begin{barticle}
\bauthor{\binits{P.} \bsnm{{Amaro-Seoane}}},
\batitle{{Relativistic dynamics and extreme mass ratio inspirals}}.
\bjtitle{Living Reviews in Relativity}
\bvolume{21},
\bfpage{4}
(\byear{2018}c).
doi:\doiurl{10.1007/s41114-018-0013-8}
\end{barticle}
\endbibitem

\bibitem[\protect\citeauthoryear{{Amaro-Seoane} and
  {Preto}}{2011}]{Amaro-SeoanePreto11}
\begin{barticle}
\bauthor{\binits{P.} \bsnm{{Amaro-Seoane}}},
\bauthor{\binits{M.} \bsnm{{Preto}}},
\batitle{{The impact of realistic models of mass segregation on the event rate
  of\ extreme-mass ratio inspirals and cusp re-growth}}.
\bjtitle{Classical and Quantum Gravity}
\bvolume{28}(\bissue{9}),
\bfpage{094017}
(\byear{2011}).
doi:\doiurl{10.1088/0264-9381/28/9/094017}
\end{barticle}
\endbibitem

\bibitem[\protect\citeauthoryear{{Amaro-Seoane} and {Spurzem}}{2001}]{AS01}
\begin{barticle}
\bauthor{\binits{P.} \bsnm{{Amaro-Seoane}}},
\bauthor{\binits{R.} \bsnm{{Spurzem}}},
\batitle{{The loss-cone problem in dense nuclei}}.
\bjtitle{MNRAS}
\bvolume{327},
\bfpage{995}--\blpage{1003}
(\byear{2001})
\end{barticle}
\endbibitem

\bibitem[\protect\citeauthoryear{{Amaro-Seoane} et~al.}{2004}]{ASEtAl04}
\begin{botherref}
\oauthor{\binits{P.} \bsnm{{Amaro-Seoane}}},
\oauthor{\binits{M.} \bsnm{{Freitag}}},
\oauthor{\binits{R.} \bsnm{{Spurzem}}},
{Accretion of stars on to a massive black hole: A realistic diffusion model and
  numerical studies}.
MNRAS
(2004)
\end{botherref}
\endbibitem

\bibitem[\protect\citeauthoryear{{Amaro-Seoane}
  et~al.}{2010}]{Amaro-SeoanePretoSopuerta2010}
\begin{botherref}
\oauthor{\binits{P.} \bsnm{{Amaro-Seoane}}},
\oauthor{\binits{M.} \bsnm{{Preto}}},
\oauthor{\binits{C.} \bsnm{{Sopuerta}}},
{Notes on a relativistic Fokker-Planck integrator for EMRIs}.
Classical and Quantum Gravity
(2010)
\end{botherref}
\endbibitem

\bibitem[\protect\citeauthoryear{{Amaro-Seoane}
  et~al.}{2013}]{Amaro-SeoaneSopuertaFreitag2013}
\begin{barticle}
\bauthor{\binits{P.} \bsnm{{Amaro-Seoane}}},
\bauthor{\binits{C.F.} \bsnm{{Sopuerta}}},
\bauthor{\binits{M.D.} \bsnm{{Freitag}}},
\batitle{{The role of the supermassive black hole spin in the estimation of the
  EMRI event rate}}.
\bjtitle{MNRAS}
\bvolume{429},
\bfpage{3155}--\blpage{3165}
(\byear{2013}).
doi:\doiurl{10.1093/mnras/sts572}
\end{barticle}
\endbibitem

\bibitem[\protect\citeauthoryear{{Amaro-Seoane}
  et~al.}{2007}]{Amaro-SeoaneEtAl07}
\begin{barticle}
\bauthor{\binits{P.} \bsnm{{Amaro-Seoane}}},
\bauthor{\binits{J.R.} \bsnm{{Gair}}},
\bauthor{\binits{M.} \bsnm{{Freitag}}},
\bauthor{\binits{M.C.} \bsnm{{Miller}}},
\bauthor{\binits{I.} \bsnm{{Mandel}}},
\bauthor{\binits{C.J.} \bsnm{{Cutler}}},
\bauthor{\binits{S.} \bsnm{{Babak}}},
\batitle{{Intermediate and extreme mass-ratio inspirals. Astrophysics,\ science
  applications and detection using LISA}}.
\bjtitle{Classical and Quantum Gravity}
\bvolume{24},
\bfpage{113}
(\byear{2007}).
doi:\doiurl{10.1088/0264-9381/24/17/R01}
\end{barticle}
\endbibitem

\bibitem[\protect\citeauthoryear{{Amaro-Seoane}
  et~al.}{2010}]{Amaro-SeoaneBarrancoBernalRezzolla10}
\begin{barticle}
\bauthor{\binits{P.} \bsnm{{Amaro-Seoane}}},
\bauthor{\binits{J.} \bsnm{{Barranco}}},
\bauthor{\binits{A.} \bsnm{{Bernal}}},
\bauthor{\binits{L.} \bsnm{{Rezzolla}}},
\batitle{{Constraining scalar fields with stellar kinematics and collisional
  dark matter}}.
\bjtitle{Journal of Cosmology and Astroparticle Physics}
\bvolume{11},
\bfpage{2}
(\byear{2010}).
doi:\doiurl{10.1088/1475-7516/2010/11/002}
\end{barticle}
\endbibitem

\bibitem[\protect\citeauthoryear{{Amaro-Seoane}
  et~al.}{2015}]{Amaro-SeoaneGairPoundHughesSopuerta2015}
\begin{barticle}
\bauthor{\binits{P.} \bsnm{{Amaro-Seoane}}},
\bauthor{\binits{J.R.} \bsnm{{Gair}}},
\bauthor{\binits{A.} \bsnm{{Pound}}},
\bauthor{\binits{S.A.} \bsnm{{Hughes}}},
\bauthor{\binits{C.F.} \bsnm{{Sopuerta}}},
\batitle{{Research Update on Extreme-Mass-Ratio Inspirals}}.
\bjtitle{Journal of Physics Conference Series}
\bvolume{610}(\bissue{1}),
\bfpage{012002}
(\byear{2015}).
doi:\doiurl{10.1088/1742-6596/610/1/012002}
\end{barticle}
\endbibitem

\bibitem[\protect\citeauthoryear{{Amaro-Seoane}
  et~al.}{2017}]{Amaro-SeoaneEtAl2017}
\begin{botherref}
\oauthor{\binits{P.} \bsnm{{Amaro-Seoane}}},
\oauthor{\binits{H.} \bsnm{{Audley}}},
\oauthor{\binits{S.} \bsnm{{Babak}}},
\oauthor{\binits{J.} \bsnm{{Baker}}},
\oauthor{\binits{E.} \bsnm{{Barausse}}},
\oauthor{\binits{P.} \bsnm{{Bender}}},
\oauthor{\binits{E.} \bsnm{{Berti}}},
\oauthor{\binits{P.} \bsnm{{Binetruy}}},
\oauthor{\binits{M.} \bsnm{{Born}}},
\oauthor{\binits{D.} \bsnm{{Bortoluzzi}}},
\oauthor{\binits{J.} \bsnm{{Camp}}},
\oauthor{\binits{C.} \bsnm{{Caprini}}},
\oauthor{\binits{V.} \bsnm{{Cardoso}}},
\oauthor{\binits{M.} \bsnm{{Colpi}}},
\oauthor{\binits{J.} \bsnm{{Conklin}}},
\oauthor{\binits{N.} \bsnm{{Cornish}}},
\oauthor{\binits{C.} \bsnm{{Cutler}}},
\oauthor{\binits{K.} \bsnm{{Danzmann}}},
\oauthor{\binits{R.} \bsnm{{Dolesi}}},
\oauthor{\binits{L.} \bsnm{{Ferraioli}}},
\oauthor{\binits{V.} \bsnm{{Ferroni}}},
\oauthor{\binits{E.} \bsnm{{Fitzsimons}}},
\oauthor{\binits{J.} \bsnm{{Gair}}},
\oauthor{\binits{L.} \bsnm{{Gesa Bote}}},
\oauthor{\binits{D.} \bsnm{{Giardini}}},
\oauthor{\binits{F.} \bsnm{{Gibert}}},
\oauthor{\binits{C.} \bsnm{{Grimani}}},
\oauthor{\binits{H.} \bsnm{{Halloin}}},
\oauthor{\binits{G.} \bsnm{{Heinzel}}},
\oauthor{\binits{T.} \bsnm{{Hertog}}},
\oauthor{\binits{M.} \bsnm{{Hewitson}}},
\oauthor{\binits{K.} \bsnm{{Holley-Bockelmann}}},
\oauthor{\binits{D.} \bsnm{{Hollington}}},
\oauthor{\binits{M.} \bsnm{{Hueller}}},
\oauthor{\binits{H.} \bsnm{{Inchauspe}}},
\oauthor{\binits{P.} \bsnm{{Jetzer}}},
\oauthor{\binits{N.} \bsnm{{Karnesis}}},
\oauthor{\binits{C.} \bsnm{{Killow}}},
\oauthor{\binits{A.} \bsnm{{Klein}}},
\oauthor{\binits{B.} \bsnm{{Klipstein}}},
\oauthor{\binits{N.} \bsnm{{Korsakova}}},
\oauthor{\binits{S.L.} \bsnm{{Larson}}},
\oauthor{\binits{J.} \bsnm{{Livas}}},
\oauthor{\binits{I.} \bsnm{{Lloro}}},
\oauthor{\binits{N.} \bsnm{{Man}}},
\oauthor{\binits{D.} \bsnm{{Mance}}},
\oauthor{\binits{J.} \bsnm{{Martino}}},
\oauthor{\binits{I.} \bsnm{{Mateos}}},
\oauthor{\binits{K.} \bsnm{{McKenzie}}},
\oauthor{\binits{S.T.} \bsnm{{McWilliams}}},
\oauthor{\binits{C.} \bsnm{{Miller}}},
\oauthor{\binits{G.} \bsnm{{Mueller}}},
\oauthor{\binits{G.} \bsnm{{Nardini}}},
\oauthor{\binits{G.} \bsnm{{Nelemans}}},
\oauthor{\binits{M.} \bsnm{{Nofrarias}}},
\oauthor{\binits{A.} \bsnm{{Petiteau}}},
\oauthor{\binits{P.} \bsnm{{Pivato}}},
\oauthor{\binits{E.} \bsnm{{Plagnol}}},
\oauthor{\binits{E.} \bsnm{{Porter}}},
\oauthor{\binits{J.} \bsnm{{Reiche}}},
\oauthor{\binits{D.} \bsnm{{Robertson}}},
\oauthor{\binits{N.} \bsnm{{Robertson}}},
\oauthor{\binits{E.} \bsnm{{Rossi}}},
\oauthor{\binits{G.} \bsnm{{Russano}}},
\oauthor{\binits{B.} \bsnm{{Schutz}}},
\oauthor{\binits{A.} \bsnm{{Sesana}}},
\oauthor{\binits{D.} \bsnm{{Shoemaker}}},
\oauthor{\binits{J.} \bsnm{{Slutsky}}},
\oauthor{\binits{C.F.} \bsnm{{Sopuerta}}},
\oauthor{\binits{T.} \bsnm{{Sumner}}},
\oauthor{\binits{N.} \bsnm{{Tamanini}}},
\oauthor{\binits{I.} \bsnm{{Thorpe}}},
\oauthor{\binits{M.} \bsnm{{Troebs}}},
\oauthor{\binits{M.} \bsnm{{Vallisneri}}},
\oauthor{\binits{A.} \bsnm{{Vecchio}}},
\oauthor{\binits{D.} \bsnm{{Vetrugno}}},
\oauthor{\binits{S.} \bsnm{{Vitale}}},
\oauthor{\binits{M.} \bsnm{{Volonteri}}},
\oauthor{\binits{G.} \bsnm{{Wanner}}},
\oauthor{\binits{H.} \bsnm{{Ward}}},
\oauthor{\binits{P.} \bsnm{{Wass}}},
\oauthor{\binits{W.} \bsnm{{Weber}}},
\oauthor{\binits{J.} \bsnm{{Ziemer}}},
\oauthor{\binits{P.} \bsnm{{Zweifel}}},
{Laser Interferometer Space Antenna}.
ArXiv e-prints
(2017)
\end{botherref}
\endbibitem

\bibitem[\protect\citeauthoryear{{Amaro-Seoane}}{2018}]{2018LRR....21....4A}
\begin{barticle}
\bauthor{\binits{P.} \bsnm{{Amaro-Seoane}}},
\batitle{{Relativistic dynamics and extreme mass ratio inspirals}}.
\bjtitle{Living Reviews in Relativity}
\bvolume{21}(\bissue{1}),
\bfpage{4}
(\byear{2018}).
doi:\doiurl{10.1007/s41114-018-0013-8}
\end{barticle}
\endbibitem

\bibitem[\protect\citeauthoryear{{Amaro-Seoane}}{2019}]{Amaro-Seoane2019}
\begin{barticle}
\bauthor{\binits{P.} \bsnm{{Amaro-Seoane}}},
\batitle{{Extremely large mass-ratio inspirals}}.
\bjtitle{Phys.Rev.D.}
\bvolume{99}(\bissue{12}),
\bfpage{123025}
(\byear{2019}).
doi:\doiurl{10.1103/PhysRevD.99.123025}
\end{barticle}
\endbibitem

\bibitem[\protect\citeauthoryear{{Arca-Sedda} et~al.}{2020}]{ArcaSeddaEtAl2020}
\begin{botherref}
\oauthor{\binits{M.} \bsnm{{Arca-Sedda}}},
\oauthor{\binits{P.} \bsnm{{Amaro-Seoane}}},
\oauthor{\binits{X.} \bsnm{{Chen}}},
{Detecting intermediate-mass black holes in Milky Way globular clusters and the
  Local Volume with LISA and other gravitational wave detectors}.
arXiv e-prints,
2007--13746
(2020)
\end{botherref}
\endbibitem

\bibitem[\protect\citeauthoryear{{Arca Sedda} et~al.}{2019}]{ArcaSeddaEtAl2019}
\begin{botherref}
\oauthor{\binits{M.} \bsnm{{Arca Sedda}}},
\oauthor{\binits{A.} \bsnm{{Askar}}},
\oauthor{\binits{M.} \bsnm{{Giersz}}},
{MOCCA-SURVEY Database I. Intermediate mass black holes in Milky Way globular
  clusters and their connection to supermassive black holes}.
arXiv e-prints,
1905--00902
(2019)
\end{botherref}
\endbibitem

\bibitem[\protect\citeauthoryear{{Asada} and
  {Futamase}}{1997}]{AsadaFutamase1997}
\begin{barticle}
\bauthor{\binits{H.} \bsnm{{Asada}}},
\bauthor{\binits{T.} \bsnm{{Futamase}}},
\batitle{{Chapter 2. Post-Newtonian Approximation ---Its Foundation and
  Applications---}}.
\bjtitle{Progress of Theoretical Physics Supplement}
\bvolume{128},
\bfpage{123}--\blpage{181}
(\byear{1997}).
doi:\doiurl{10.1143/PTPS.128.123}
\end{barticle}
\endbibitem

\bibitem[\protect\citeauthoryear{{Babak} et~al.}{2007}]{BabakEtAl2007}
\begin{barticle}
\bauthor{\binits{S.} \bsnm{{Babak}}},
\bauthor{\binits{H.} \bsnm{{Fang}}},
\bauthor{\binits{J.R.} \bsnm{{Gair}}},
\bauthor{\binits{K.} \bsnm{{Glampedakis}}},
\bauthor{\binits{S.A.} \bsnm{{Hughes}}},
\batitle{{``Kludge'' gravitational waveforms for a test-body orbiting a Kerr
  black hole}}.
\bjtitle{Ph.Rv.D}
\bvolume{75}(\bissue{2}),
\bfpage{024005}
(\byear{2007}).
doi:\doiurl{10.1103/PhysRevD.75.024005}
\end{barticle}
\endbibitem

\bibitem[\protect\citeauthoryear{{Babak} et~al.}{2017}]{2017PhRvD..95j3012B}
\begin{barticle}
\bauthor{\binits{S.} \bsnm{{Babak}}},
\bauthor{\binits{J.} \bsnm{{Gair}}},
\bauthor{\binits{A.} \bsnm{{Sesana}}},
\bauthor{\binits{E.} \bsnm{{Barausse}}},
\bauthor{\binits{C.F.} \bsnm{{Sopuerta}}},
\bauthor{\binits{C.P.L.} \bsnm{{Berry}}},
\bauthor{\binits{E.} \bsnm{{Berti}}},
\bauthor{\binits{P.} \bsnm{{Amaro-Seoane}}},
\bauthor{\binits{A.} \bsnm{{Petiteau}}},
\bauthor{\binits{A.} \bsnm{{Klein}}},
\batitle{{Science with the space-based interferometer LISA. V. Extreme
  mass-ratio inspirals}}.
\bjtitle{\prd}
\bvolume{95}(\bissue{10}),
\bfpage{103012}
(\byear{2017}).
doi:\doiurl{10.1103/PhysRevD.95.103012}
\end{barticle}
\endbibitem

\bibitem[\protect\citeauthoryear{{Bahcall} and {Wolf}}{1976}]{BW76}
\begin{barticle}
\bauthor{\binits{J.N.} \bsnm{{Bahcall}}},
\bauthor{\binits{R.A.} \bsnm{{Wolf}}},
\batitle{Star distribution around a massive black hole in a globular cluster}.
\bjtitle{ApJ}
\bvolume{209},
\bfpage{214}--\blpage{232}
(\byear{1976})
\end{barticle}
\endbibitem

\bibitem[\protect\citeauthoryear{{Bahcall} and {Wolf}}{1977}]{BW77}
\begin{barticle}
\bauthor{\binits{J.N.} \bsnm{{Bahcall}}},
\bauthor{\binits{R.A.} \bsnm{{Wolf}}},
\batitle{The star distribution around a massive black hole in a globular
  cluster. {II} unequal star masses}.
\bjtitle{ApJ}
\bvolume{216},
\bfpage{883}--\blpage{907}
(\byear{1977})
\end{barticle}
\endbibitem

\bibitem[\protect\citeauthoryear{{Baker} et~al.}{2006}]{BakerEtAl2006}
\begin{barticle}
\bauthor{\binits{J.G.} \bsnm{{Baker}}},
\bauthor{\binits{J.} \bsnm{{Centrella}}},
\bauthor{\binits{D.-I.} \bsnm{{Choi}}},
\bauthor{\binits{M.} \bsnm{{Koppitz}}},
\bauthor{\binits{J.R.} \bsnm{{van Meter}}},
\bauthor{\binits{M.C.} \bsnm{{Miller}}},
\batitle{{Getting a Kick Out of Numerical Relativity}}.
\bjtitle{ApJ Lett.}
\bvolume{653},
\bfpage{93}--\blpage{96}
(\byear{2006}).
doi:\doiurl{10.1086/510448}
\end{barticle}
\endbibitem

\bibitem[\protect\citeauthoryear{{Bar-Or} and
  {Alexander}}{2014}]{2014CQGra..31x4003B}
\begin{barticle}
\bauthor{\binits{B.} \bsnm{{Bar-Or}}},
\bauthor{\binits{T.} \bsnm{{Alexander}}},
\batitle{{The statistical mechanics of relativistic orbits around a massive
  black hole}}.
\bjtitle{Classical and Quantum Gravity}
\bvolume{31}(\bissue{24}),
\bfpage{244003}
(\byear{2014}).
doi:\doiurl{10.1088/0264-9381/31/24/244003}
\end{barticle}
\endbibitem

\bibitem[\protect\citeauthoryear{{Bar-Or} and
  {Fouvry}}{2018}]{2018ApJ...860L..23B}
\begin{barticle}
\bauthor{\binits{B.} \bsnm{{Bar-Or}}},
\bauthor{\binits{J.-B.} \bsnm{{Fouvry}}},
\batitle{{Scalar Resonant Relaxation of Stars around a Massive Black Hole}}.
\bjtitle{\apjl}
\bvolume{860}(\bissue{2}),
\bfpage{23}
(\byear{2018}).
doi:\doiurl{10.3847/2041-8213/aac88e}
\end{barticle}
\endbibitem

\bibitem[\protect\citeauthoryear{{Barack} and
  {Cutler}}{2004}]{BarackCutler2004}
\begin{barticle}
\bauthor{\binits{L.} \bsnm{{Barack}}},
\bauthor{\binits{C.} \bsnm{{Cutler}}},
\batitle{{LISA capture sources: Approximate waveforms, signal-to-noise ratios,
  and parameter estimation accuracy}}.
\bjtitle{Phys. Rev. D}
\bvolume{69}(\bissue{8}),
\bfpage{082005}
(\byear{2004}).
doi:\doiurl{10.1103/PhysRevD.69.082005}
\end{barticle}
\endbibitem

\bibitem[\protect\citeauthoryear{{Barack} and
  {Pound}}{2019a}]{2019RPPh...82a6904B}
\begin{barticle}
\bauthor{\binits{L.} \bsnm{{Barack}}},
\bauthor{\binits{A.} \bsnm{{Pound}}},
\batitle{{Self-force and radiation reaction in general relativity}}.
\bjtitle{Reports on Progress in Physics}
\bvolume{82}(\bissue{1}),
\bfpage{016904}
(\byear{2019}a).
doi:\doiurl{10.1088/1361-6633/aae552}
\end{barticle}
\endbibitem

\bibitem[\protect\citeauthoryear{{Barack} and {Pound}}{2019b}]{BarackPound2019}
\begin{barticle}
\bauthor{\binits{L.} \bsnm{{Barack}}},
\bauthor{\binits{A.} \bsnm{{Pound}}},
\batitle{{Self-force and radiation reaction in general relativity}}.
\bjtitle{Reports on Progress in Physics}
\bvolume{82},
\bfpage{016904}
(\byear{2019}b).
doi:\doiurl{10.1088/1361-6633/aae552}
\end{barticle}
\endbibitem

\bibitem[\protect\citeauthoryear{{Barausse} et~al.}{2020}]{2020GReGr..52...81B}
\begin{barticle}
\bauthor{\binits{E.} \bsnm{{Barausse}}},
\bauthor{\binits{E.} \bsnm{{Berti}}},
\bauthor{\binits{T.} \bsnm{{Hertog}}},
\bauthor{\binits{S.A.} \bsnm{{Hughes}}},
\bauthor{\binits{P.} \bsnm{{Jetzer}}},
\bauthor{\binits{P.} \bsnm{{Pani}}},
\bauthor{\binits{T.P.} \bsnm{{Sotiriou}}},
\bauthor{\binits{N.} \bsnm{{Tamanini}}},
\bauthor{\binits{H.} \bsnm{{Witek}}},
\bauthor{\binits{K.} \bsnm{{Yagi}}},
\bauthor{\binits{N.} \bsnm{{Yunes}}},
\bauthor{\binits{T.} \bsnm{{Abdelsalhin}}},
\bauthor{\binits{A.} \bsnm{{Achucarro}}},
\bauthor{\binits{K.} \bsnm{{van Aelst}}},
\bauthor{\binits{N.} \bsnm{{Afshordi}}},
\bauthor{\binits{S.} \bsnm{{Akcay}}},
\bauthor{\binits{L.} \bsnm{{Annulli}}},
\bauthor{\binits{K.G.} \bsnm{{Arun}}},
\bauthor{\binits{I.} \bsnm{{Ayuso}}},
\bauthor{\binits{V.} \bsnm{{Baibhav}}},
\bauthor{\binits{T.} \bsnm{{Baker}}},
\bauthor{\binits{H.} \bsnm{{Bantilan}}},
\bauthor{\binits{T.} \bsnm{{Barreiro}}},
\bauthor{\binits{C.} \bsnm{{Barrera-Hinojosa}}},
\bauthor{\binits{N.} \bsnm{{Bartolo}}},
\bauthor{\binits{D.} \bsnm{{Baumann}}},
\bauthor{\binits{E.} \bsnm{{Belgacem}}},
\bauthor{\binits{E.} \bsnm{{Bellini}}},
\bauthor{\binits{N.} \bsnm{{Bellomo}}},
\bauthor{\binits{I.} \bsnm{{Ben-Dayan}}},
\bauthor{\binits{I.} \bsnm{{Bena}}},
\bauthor{\binits{R.} \bsnm{{Benkel}}},
\bauthor{\binits{E.} \bsnm{{Bergshoefs}}},
\bauthor{\binits{L.} \bsnm{{Bernard}}},
\bauthor{\binits{S.} \bsnm{{Bernuzzi}}},
\bauthor{\binits{D.} \bsnm{{Bertacca}}},
\bauthor{\binits{M.} \bsnm{{Besancon}}},
\bauthor{\binits{F.} \bsnm{{Beutler}}},
\bauthor{\binits{F.} \bsnm{{Beyer}}},
\bauthor{\binits{S.} \bsnm{{Bhagwat}}},
\bauthor{\binits{J.} \bsnm{{Bicak}}},
\bauthor{\binits{S.} \bsnm{{Biondini}}},
\bauthor{\binits{S.} \bsnm{{Bize}}},
\bauthor{\binits{D.} \bsnm{{Blas}}},
\bauthor{\binits{C.} \bsnm{{Boehmer}}},
\bauthor{\binits{K.} \bsnm{{Boller}}},
\bauthor{\binits{B.} \bsnm{{Bonga}}},
\bauthor{\binits{C.} \bsnm{{Bonvin}}},
\bauthor{\binits{P.} \bsnm{{Bosso}}},
\bauthor{\binits{G.} \bsnm{{Bozzola}}},
\bauthor{\binits{P.} \bsnm{{Brax}}},
\bauthor{\binits{M.} \bsnm{{Breitbach}}},
\bauthor{\binits{R.} \bsnm{{Brito}}},
\bauthor{\binits{M.} \bsnm{{Bruni}}},
\bauthor{\binits{B.} \bsnm{{Br{\"u}gmann}}},
\bauthor{\binits{H.} \bsnm{{Bulten}}},
\bauthor{\binits{A.} \bsnm{{Buonanno}}},
\bauthor{\binits{L.M.} \bsnm{{Burko}}},
\bauthor{\binits{C.} \bsnm{{Burrage}}},
\bauthor{\binits{F.} \bsnm{{Cabral}}},
\bauthor{\binits{G.} \bsnm{{Calcagni}}},
\bauthor{\binits{C.} \bsnm{{Caprini}}},
\bauthor{\binits{A.} \bsnm{{C{\'a}rdenas-Avenda{\~n}o}}},
\bauthor{\binits{M.} \bsnm{{Celoria}}},
\bauthor{\binits{K.} \bsnm{{Chatziioannou}}},
\bauthor{\binits{D.} \bsnm{{Chernoff}}},
\bauthor{\binits{K.} \bsnm{{Clough}}},
\bauthor{\binits{A.} \bsnm{{Coates}}},
\bauthor{\binits{D.} \bsnm{{Comelli}}},
\bauthor{\binits{G.} \bsnm{{Comp{\`e}re}}},
\bauthor{\binits{D.} \bsnm{{Croon}}},
\bauthor{\binits{D.} \bsnm{{Cruces}}},
\bauthor{\binits{G.} \bsnm{{Cusin}}},
\bauthor{\binits{C.} \bsnm{{Dalang}}},
\bauthor{\binits{U.} \bsnm{{Danielsson}}},
\bauthor{\binits{S.} \bsnm{{Das}}},
\bauthor{\binits{S.} \bsnm{{Datta}}},
\bauthor{\binits{J.} \bsnm{{de Boer}}},
\bauthor{\binits{V.} \bsnm{{De Luca}}},
\bauthor{\binits{C.} \bsnm{{De Rham}}},
\bauthor{\binits{V.} \bsnm{{Desjacques}}},
\bauthor{\binits{K.} \bsnm{{Destounis}}},
\bauthor{\binits{F.} \bsnm{{Di Filippo}}},
\bauthor{\binits{A.} \bsnm{{Dima}}},
\bauthor{\binits{E.} \bsnm{{Dimastrogiovanni}}},
\bauthor{\binits{S.} \bsnm{{Dolan}}},
\bauthor{\binits{D.} \bsnm{{Doneva}}},
\bauthor{\binits{F.} \bsnm{{Duque}}},
\bauthor{\binits{R.} \bsnm{{Durrer}}},
\bauthor{\binits{W.} \bsnm{{East}}},
\bauthor{\binits{R.} \bsnm{{Easther}}},
\bauthor{\binits{M.} \bsnm{{Elley}}},
\bauthor{\binits{J.R.} \bsnm{{Ellis}}},
\bauthor{\binits{R.} \bsnm{{Emparan}}},
\bauthor{\binits{J.M.} \bsnm{{Ezquiaga}}},
\bauthor{\binits{M.} \bsnm{{Fairbairn}}},
\bauthor{\binits{S.} \bsnm{{Fairhurst}}},
\bauthor{\binits{H.F.} \bsnm{{Farmer}}},
\bauthor{\binits{M.R.} \bsnm{{Fasiello}}},
\bauthor{\binits{V.} \bsnm{{Ferrari}}},
\bauthor{\binits{P.G.} \bsnm{{Ferreira}}},
\bauthor{\binits{G.} \bsnm{{Ficarra}}},
\bauthor{\binits{P.} \bsnm{{Figueras}}},
\bauthor{\binits{S.} \bsnm{{Fisenko}}},
\bauthor{\binits{S.} \bsnm{{Foffa}}},
\bauthor{\binits{N.} \bsnm{{Franchini}}},
\bauthor{\binits{G.} \bsnm{{Franciolini}}},
\bauthor{\binits{K.} \bsnm{{Fransen}}},
\bauthor{\binits{J.} \bsnm{{Frauendiener}}},
\bauthor{\binits{N.} \bsnm{{Frusciante}}},
\bauthor{\binits{R.} \bsnm{{Fujita}}},
\bauthor{\binits{J.} \bsnm{{Gair}}},
\bauthor{\binits{A.} \bsnm{{Ganz}}},
\bauthor{\binits{P.} \bsnm{{Garcia}}},
\bauthor{\binits{J.} \bsnm{{Garcia-Bellido}}},
\bauthor{\binits{J.} \bsnm{{Garriga}}},
\bauthor{\binits{R.} \bsnm{{Geiger}}},
\bauthor{\binits{C.} \bsnm{{Geng}}},
\bauthor{\binits{L.{\'A}.} \bsnm{{Gergely}}},
\bauthor{\binits{C.} \bsnm{{Germani}}},
\bauthor{\binits{D.} \bsnm{{Gerosa}}},
\bauthor{\binits{S.B.} \bsnm{{Giddings}}},
\bauthor{\binits{E.} \bsnm{{Gourgoulhon}}},
\bauthor{\binits{P.} \bsnm{{Grand clement}}},
\bauthor{\binits{L.} \bsnm{{Graziani}}},
\bauthor{\binits{L.} \bsnm{{Gualtieri}}},
\bauthor{\binits{D.} \bsnm{{Haggard}}},
\bauthor{\binits{S.} \bsnm{{Haino}}},
\bauthor{\binits{R.} \bsnm{{Halburd}}},
\bauthor{\binits{W.-B.} \bsnm{{Han}}},
\bauthor{\binits{A.J.} \bsnm{{Hawken}}},
\bauthor{\binits{A.} \bsnm{{Hees}}},
\bauthor{\binits{I.S.} \bsnm{{Heng}}},
\bauthor{\binits{J.} \bsnm{{Hennig}}},
\bauthor{\binits{C.} \bsnm{{Herdeiro}}},
\bauthor{\binits{S.} \bsnm{{Hervik}}},
\bauthor{\binits{J.v.} \bsnm{{Holten}}},
\bauthor{\binits{C.J.D.} \bsnm{{Hoyle}}},
\bauthor{\binits{Y.} \bsnm{{Hu}}},
\bauthor{\binits{M.} \bsnm{{Hull}}},
\bauthor{\binits{T.} \bsnm{{Ikeda}}},
\bauthor{\binits{M.} \bsnm{{Isi}}},
\bauthor{\binits{A.} \bsnm{{Jenkins}}},
\bauthor{\binits{F.} \bsnm{{Juli{\'e}}}},
\bauthor{\binits{E.} \bsnm{{Kajfasz}}},
\bauthor{\binits{C.} \bsnm{{Kalaghatgi}}},
\bauthor{\binits{N.} \bsnm{{Kaloper}}},
\bauthor{\binits{M.} \bsnm{{Kamionkowski}}},
\bauthor{\binits{V.} \bsnm{{Karas}}},
\bauthor{\binits{S.} \bsnm{{Kastha}}},
\bauthor{\binits{Z.} \bsnm{{Keresztes}}},
\bauthor{\binits{L.} \bsnm{{Kidder}}},
\bauthor{\binits{T.} \bsnm{{Kimpson}}},
\bauthor{\binits{A.} \bsnm{{Klein}}},
\bauthor{\binits{S.} \bsnm{{Klioner}}},
\bauthor{\binits{K.} \bsnm{{Kokkotas}}},
\bauthor{\binits{H.} \bsnm{{Kolesova}}},
\bauthor{\binits{S.} \bsnm{{Kolkowitz}}},
\bauthor{\binits{J.} \bsnm{{Kopp}}},
\bauthor{\binits{K.} \bsnm{{Koyama}}},
\bauthor{\binits{N.V.} \bsnm{{Krishnendu}}},
\bauthor{\binits{J.A.V.} \bsnm{{Kroon}}},
\bauthor{\binits{M.} \bsnm{{Kunz}}},
\bauthor{\binits{O.} \bsnm{{Lahav}}},
\bauthor{\binits{A.} \bsnm{{Landragin}}},
\bauthor{\binits{R.N.} \bsnm{{Lang}}},
\bauthor{\binits{C.} \bsnm{{Le Poncin-Lafitte}}},
\bauthor{\binits{J.} \bsnm{{Lemos}}},
\bauthor{\binits{B.} \bsnm{{Li}}},
\bauthor{\binits{S.} \bsnm{{Liberati}}},
\bauthor{\binits{M.} \bsnm{{Liguori}}},
\bauthor{\binits{F.} \bsnm{{Lin}}},
\bauthor{\binits{G.} \bsnm{{Liu}}},
\bauthor{\binits{F.S.N.} \bsnm{{Lobo}}},
\bauthor{\binits{R.} \bsnm{{Loll}}},
\bauthor{\binits{L.} \bsnm{{Lombriser}}},
\bauthor{\binits{G.} \bsnm{{Lovelace}}},
\bauthor{\binits{R.P.} \bsnm{{Macedo}}},
\bauthor{\binits{E.} \bsnm{{Madge}}},
\bauthor{\binits{E.} \bsnm{{Maggio}}},
\bauthor{\binits{M.} \bsnm{{Maggiore}}},
\bauthor{\binits{S.} \bsnm{{Marassi}}},
\bauthor{\binits{P.} \bsnm{{Marcoccia}}},
\bauthor{\binits{C.} \bsnm{{Markakis}}},
\bauthor{\binits{W.} \bsnm{{Martens}}},
\bauthor{\binits{K.} \bsnm{{Martinovic}}},
\bauthor{\binits{C.J.A.P.} \bsnm{{Martins}}},
\bauthor{\binits{A.} \bsnm{{Maselli}}},
\bauthor{\binits{S.} \bsnm{{Mastrogiovanni}}},
\bauthor{\binits{S.} \bsnm{{Matarrese}}},
\bauthor{\binits{A.} \bsnm{{Matas}}},
\bauthor{\binits{N.E.} \bsnm{{Mavromatos}}},
\bauthor{\binits{A.} \bsnm{{Mazumdar}}},
\bauthor{\binits{P.D.} \bsnm{{Meerburg}}},
\bauthor{\binits{E.} \bsnm{{Megias}}},
\bauthor{\binits{J.} \bsnm{{Miller}}},
\bauthor{\binits{J.P.} \bsnm{{Mimoso}}},
\bauthor{\binits{L.} \bsnm{{Mittnacht}}},
\bauthor{\binits{M.M.} \bsnm{{Montero}}},
\bauthor{\binits{B.} \bsnm{{Moore}}},
\bauthor{\binits{P.} \bsnm{{Martin-Moruno}}},
\bauthor{\binits{I.} \bsnm{{Musco}}},
\bauthor{\binits{H.} \bsnm{{Nakano}}},
\bauthor{\binits{S.} \bsnm{{Nampalliwar}}},
\bauthor{\binits{G.} \bsnm{{Nardini}}},
\bauthor{\binits{A.} \bsnm{{Nielsen}}},
\bauthor{\binits{J.} \bsnm{{Nov{\'a}k}}},
\bauthor{\binits{N.J.} \bsnm{{Nunes}}},
\bauthor{\binits{M.} \bsnm{{Okounkova}}},
\bauthor{\binits{R.} \bsnm{{Oliveri}}},
\bauthor{\binits{F.} \bsnm{{Oppizzi}}},
\bauthor{\binits{G.} \bsnm{{Orlando}}},
\bauthor{\binits{N.} \bsnm{{Oshita}}},
\bauthor{\binits{G.} \bsnm{{Pappas}}},
\bauthor{\binits{V.} \bsnm{{Paschalidis}}},
\bauthor{\binits{H.} \bsnm{{Peiris}}},
\bauthor{\binits{M.} \bsnm{{Peloso}}},
\bauthor{\binits{S.} \bsnm{{Perkins}}},
\bauthor{\binits{V.} \bsnm{{Pettorino}}},
\bauthor{\binits{I.} \bsnm{{Pikovski}}},
\bauthor{\binits{L.} \bsnm{{Pilo}}},
\bauthor{\binits{J.} \bsnm{{Podolsky}}},
\bauthor{\binits{A.} \bsnm{{Pontzen}}},
\bauthor{\binits{S.} \bsnm{{Prabhat}}},
\bauthor{\binits{G.} \bsnm{{Pratten}}},
\bauthor{\binits{T.} \bsnm{{Prokopec}}},
\bauthor{\binits{M.} \bsnm{{Prouza}}},
\bauthor{\binits{H.} \bsnm{{Qi}}},
\bauthor{\binits{A.} \bsnm{{Raccanelli}}},
\bauthor{\binits{A.} \bsnm{{Rajantie}}},
\bauthor{\binits{L.} \bsnm{{Randall}}},
\bauthor{\binits{G.} \bsnm{{Raposo}}},
\bauthor{\binits{V.} \bsnm{{Raymond}}},
\bauthor{\binits{S.} \bsnm{{Renaux-Petel}}},
\bauthor{\binits{A.} \bsnm{{Ricciardone}}},
\bauthor{\binits{A.} \bsnm{{Riotto}}},
\bauthor{\binits{T.} \bsnm{{Robson}}},
\bauthor{\binits{D.} \bsnm{{Roest}}},
\bauthor{\binits{R.} \bsnm{{Rollo}}},
\bauthor{\binits{S.} \bsnm{{Rosofsky}}},
\bauthor{\binits{J.J.} \bsnm{{Ruan}}},
\bauthor{\binits{D.} \bsnm{{Rubiera-Garc{\'\i}a}}},
\bauthor{\binits{M.} \bsnm{{Ruiz}}},
\bauthor{\binits{M.} \bsnm{{Rusu}}},
\bauthor{\binits{F.} \bsnm{{Sabatie}}},
\bauthor{\binits{N.} \bsnm{{Sago}}},
\bauthor{\binits{M.} \bsnm{{Sakellariadou}}},
\bauthor{\binits{I.D.} \bsnm{{Saltas}}},
\bauthor{\binits{L.} \bsnm{{Sberna}}},
\bauthor{\binits{B.} \bsnm{{Sathyaprakash}}},
\bauthor{\binits{M.} \bsnm{{Scheel}}},
\bauthor{\binits{P.} \bsnm{{Schmidt}}},
\bauthor{\binits{B.} \bsnm{{Schutz}}},
\bauthor{\binits{P.} \bsnm{{Schwaller}}},
\bauthor{\binits{L.} \bsnm{{Shao}}},
\bauthor{\binits{S.L.} \bsnm{{Shapiro}}},
\bauthor{\binits{D.} \bsnm{{Shoemaker}}},
\bauthor{\binits{A.d.} \bsnm{{Silva}}},
\bauthor{\binits{C.} \bsnm{{Simpson}}},
\bauthor{\binits{C.F.} \bsnm{{Sopuerta}}},
\bauthor{\binits{A.} \bsnm{{Spallicci}}},
\bauthor{\binits{B.A.} \bsnm{{Stefanek}}},
\bauthor{\binits{L.} \bsnm{{Stein}}},
\bauthor{\binits{N.} \bsnm{{Stergioulas}}},
\bauthor{\binits{M.} \bsnm{{Stott}}},
\bauthor{\binits{P.} \bsnm{{Sutton}}},
\bauthor{\binits{R.} \bsnm{{Svarc}}},
\bauthor{\binits{H.} \bsnm{{Tagoshi}}},
\bauthor{\binits{T.} \bsnm{{Tahamtan}}},
\bauthor{\binits{H.} \bsnm{{Takeda}}},
\bauthor{\binits{T.} \bsnm{{Tanaka}}},
\bauthor{\binits{G.} \bsnm{{Tantilian}}},
\bauthor{\binits{G.} \bsnm{{Tasinato}}},
\bauthor{\binits{O.} \bsnm{{Tattersall}}},
\bauthor{\binits{S.} \bsnm{{Teukolsky}}},
\bauthor{\binits{A.L.} \bsnm{{Tiec}}},
\bauthor{\binits{G.} \bsnm{{Theureau}}},
\bauthor{\binits{M.} \bsnm{{Trodden}}},
\bauthor{\binits{A.} \bsnm{{Tolley}}},
\bauthor{\binits{A.} \bsnm{{Toubiana}}},
\bauthor{\binits{D.} \bsnm{{Traykova}}},
\bauthor{\binits{A.} \bsnm{{Tsokaros}}},
\bauthor{\binits{C.} \bsnm{{Unal}}},
\bauthor{\binits{C.S.} \bsnm{{Unnikrishnan}}},
\bauthor{\binits{E.C.} \bsnm{{Vagenas}}},
\bauthor{\binits{P.} \bsnm{{Valageas}}},
\bauthor{\binits{M.} \bsnm{{Vallisneri}}},
\bauthor{\binits{J.} \bsnm{{Van den Brand}}},
\bauthor{\binits{C.} \bsnm{{Van den Broeck}}},
\bauthor{\binits{M.} \bsnm{{van de Meent}}},
\bauthor{\binits{P.} \bsnm{{Vanhove}}},
\bauthor{\binits{V.} \bsnm{{Varma}}},
\bauthor{\binits{J.} \bsnm{{Veitch}}},
\bauthor{\binits{B.} \bsnm{{Vercnocke}}},
\bauthor{\binits{L.} \bsnm{{Verde}}},
\bauthor{\binits{D.} \bsnm{{Vernieri}}},
\bauthor{\binits{F.} \bsnm{{Vernizzi}}},
\bauthor{\binits{R.} \bsnm{{Vicente}}},
\bauthor{\binits{F.} \bsnm{{Vidotto}}},
\bauthor{\binits{M.} \bsnm{{Visser}}},
\bauthor{\binits{Z.} \bsnm{{Vlah}}},
\bauthor{\binits{S.} \bsnm{{Vretinaris}}},
\bauthor{\binits{S.} \bsnm{{V{\"o}lkel}}},
\bauthor{\binits{Q.} \bsnm{{Wang}}},
\bauthor{\binits{Y.-T.} \bsnm{{Wang}}},
\bauthor{\binits{M.C.} \bsnm{{Werner}}},
\bauthor{\binits{J.} \bsnm{{Westernacher}}},
\bauthor{\binits{R.v.d.} \bsnm{{Weygaert}}},
\bauthor{\binits{D.} \bsnm{{Wiltshire}}},
\bauthor{\binits{T.} \bsnm{{Wiseman}}},
\bauthor{\binits{P.} \bsnm{{Wolf}}},
\bauthor{\binits{K.} \bsnm{{Wu}}},
\bauthor{\binits{K.} \bsnm{{Yamada}}},
\bauthor{\binits{H.} \bsnm{{Yang}}},
\bauthor{\binits{L.} \bsnm{{Yi}}},
\bauthor{\binits{X.} \bsnm{{Yue}}},
\bauthor{\binits{D.} \bsnm{{Yvon}}},
\bauthor{\binits{M.} \bsnm{{Zilh{\~a}o}}},
\bauthor{\binits{A.} \bsnm{{Zimmerman}}},
\bauthor{\binits{M.} \bsnm{{Zumalacarregui}}},
\batitle{{Prospects for fundamental physics with LISA}}.
\bjtitle{General Relativity and Gravitation}
\bvolume{52}(\bissue{8}),
\bfpage{81}
(\byear{2020}).
doi:\doiurl{10.1007/s10714-020-02691-1}
\end{barticle}
\endbibitem

\bibitem[\protect\citeauthoryear{{Baumgardt}
  et~al.}{2018}]{2018A&A...609A..28B}
\begin{barticle}
\bauthor{\binits{H.} \bsnm{{Baumgardt}}},
\bauthor{\binits{P.} \bsnm{{Amaro-Seoane}}},
\bauthor{\binits{R.} \bsnm{{Sch{\"o}del}}},
\batitle{{The distribution of stars around the Milky Way's central black hole.
  III. Comparison with simulations}}.
\bjtitle{\aap}
\bvolume{609},
\bfpage{28}
(\byear{2018}).
doi:\doiurl{10.1051/0004-6361/201730462}
\end{barticle}
\endbibitem

\bibitem[\protect\citeauthoryear{{Berry} and
  {Gair}}{2013}]{2013MNRAS.429..589B}
\begin{barticle}
\bauthor{\binits{C.P.L.} \bsnm{{Berry}}},
\bauthor{\binits{J.R.} \bsnm{{Gair}}},
\batitle{{Observing the Galaxy's massive black hole with gravitational wave
  bursts}}.
\bjtitle{\mnras}
\bvolume{429}(\bissue{1}),
\bfpage{589}--\blpage{612}
(\byear{2013}).
doi:\doiurl{10.1093/mnras/sts360}
\end{barticle}
\endbibitem

\bibitem[\protect\citeauthoryear{{Berry} et~al.}{2019}]{2019BAAS...51c..42B}
\begin{barticle}
\bauthor{\binits{C.} \bsnm{{Berry}}},
\bauthor{\binits{S.} \bsnm{{Hughes}}},
\bauthor{\binits{C.} \bsnm{{Sopuerta}}},
\bauthor{\binits{A.} \bsnm{{Chua}}},
\bauthor{\binits{A.} \bsnm{{Heffernan}}},
\bauthor{\binits{K.} \bsnm{{Holley-Bockelmann}}},
\bauthor{\binits{D.} \bsnm{{Mihaylov}}},
\bauthor{\binits{C.} \bsnm{{Miller}}},
\bauthor{\binits{A.} \bsnm{{Sesana}}},
\batitle{{The unique potential of extreme mass-ratio inspirals for
  gravitational-wave astronomy}}.
\bjtitle{\baas}
\bvolume{51}(\bissue{3}),
\bfpage{42}
(\byear{2019})
\end{barticle}
\endbibitem

\bibitem[\protect\citeauthoryear{Berti et~al.}{2018}]{Berti:2018vdi}
\begin{barticle}
\bauthor{\binits{E.} \bsnm{Berti}},
\bauthor{\binits{K.} \bsnm{Yagi}},
\bauthor{\binits{H.} \bsnm{Yang}},
\bauthor{\binits{N.} \bsnm{Yunes}},
\batitle{{Extreme Gravity Tests with Gravitational Waves from Compact Binary
  Coalescences: (II) Ringdown}}.
\bjtitle{Gen. Rel. Grav.}
\bvolume{50}(\bissue{5}),
\bfpage{49}
(\byear{2018}).
doi:\doiurl{10.1007/s10714-018-2372-6}
\end{barticle}
\endbibitem

\bibitem[\protect\citeauthoryear{{Bianchini} et~al.}{2016}]{BianchiniEtAl2016}
\begin{barticle}
\bauthor{\binits{P.} \bsnm{{Bianchini}}},
\bauthor{\binits{G.} \bsnm{{van de Ven}}},
\bauthor{\binits{M.A.} \bsnm{{Norris}}},
\bauthor{\binits{E.} \bsnm{{Schinnerer}}},
\bauthor{\binits{A.L.} \bsnm{{Varri}}},
\batitle{{A novel look at energy equipartition in globular clusters}}.
\bjtitle{MNRAS}
\bvolume{458},
\bfpage{3644}--\blpage{3654}
(\byear{2016}).
doi:\doiurl{10.1093/mnras/stw552}
\end{barticle}
\endbibitem

\bibitem[\protect\citeauthoryear{{Binney} and
  {Tremaine}}{2008}]{BinneyTremaine08}
\begin{bbook}
\bauthor{\binits{J.} \bsnm{{Binney}}},
\bauthor{\binits{S.} \bsnm{{Tremaine}}},
\bbtitle{{Galactic Dynamics: Second Edition}}
(\bpublisher{Princeton University Press}, \blocation{???}, \byear{2008})
\end{bbook}
\endbibitem

\bibitem[\protect\citeauthoryear{Blanchet}{2014}]{Blanchet2013}
\begin{barticle}
\bauthor{\binits{L.} \bsnm{Blanchet}},
\batitle{{Gravitational Radiation from Post-Newtonian Sources and Inspiralling
  Compact Binaries}}.
\bjtitle{Living Rev. Rel.}
\bvolume{17},
\bfpage{2}
(\byear{2014}).
doi:\doiurl{10.12942/lrr-2014-2}
\end{barticle}
\endbibitem

\bibitem[\protect\citeauthoryear{{Boh{\'e}} et~al.}{2017}]{BoheEtAl2017}
\begin{barticle}
\bauthor{\binits{A.} \bsnm{{Boh{\'e}}}},
\bauthor{\binits{L.} \bsnm{{Shao}}},
\bauthor{\binits{A.} \bsnm{{Taracchini}}},
\bauthor{\binits{A.} \bsnm{{Buonanno}}},
\bauthor{\binits{S.} \bsnm{{Babak}}},
\bauthor{\binits{I.W.} \bsnm{{Harry}}},
\bauthor{\binits{I.} \bsnm{{Hinder}}},
\bauthor{\binits{S.} \bsnm{{Ossokine}}},
\bauthor{\binits{M.} \bsnm{{P{\"u}rrer}}},
\bauthor{\binits{V.} \bsnm{{Raymond}}},
\bauthor{\binits{T.} \bsnm{{Chu}}},
\bauthor{\binits{H.} \bsnm{{Fong}}},
\bauthor{\binits{P.} \bsnm{{Kumar}}},
\bauthor{\binits{H.P.} \bsnm{{Pfeiffer}}},
\bauthor{\binits{M.} \bsnm{{Boyle}}},
\bauthor{\binits{D.A.} \bsnm{{Hemberger}}},
\bauthor{\binits{L.E.} \bsnm{{Kidder}}},
\bauthor{\binits{G.} \bsnm{{Lovelace}}},
\bauthor{\binits{M.A.} \bsnm{{Scheel}}},
\bauthor{\binits{B.} \bsnm{{Szil{\'a}gyi}}},
\batitle{{Improved effective-one-body model of spinning, nonprecessing binary
  black holes for the era of gravitational-wave astrophysics with advanced
  detectors}}.
\bjtitle{Ph. Rv. D}
\bvolume{95}(\bissue{4}),
\bfpage{044028}
(\byear{2017}).
doi:\doiurl{10.1103/PhysRevD.95.044028}
\end{barticle}
\endbibitem

\bibitem[\protect\citeauthoryear{{Boyer} and
  {Lindquist}}{1967}]{BoyerLindquist1967}
\begin{barticle}
\bauthor{\binits{R.H.} \bsnm{{Boyer}}},
\bauthor{\binits{R.W.} \bsnm{{Lindquist}}},
\batitle{{Maximal Analytic Extension of the Kerr Metric}}.
\bjtitle{Journal of Mathematical Physics}
\bvolume{8}(\bissue{2}),
\bfpage{265}--\blpage{281}
(\byear{1967}).
doi:\doiurl{10.1063/1.1705193}
\end{barticle}
\endbibitem

\bibitem[\protect\citeauthoryear{{Brown} et~al.}{2009}]{BrownEtAl09}
\begin{barticle}
\bauthor{\binits{W.R.} \bsnm{{Brown}}},
\bauthor{\binits{M.J.} \bsnm{{Geller}}},
\bauthor{\binits{S.J.} \bsnm{{Kenyon}}},
\bauthor{\binits{B.C.} \bsnm{{Bromley}}},
\batitle{{The Anisotropic Spatial Distribution of Hypervelocity Stars}}.
\bjtitle{ApJ Lett.}
\bvolume{690},
\bfpage{69}--\blpage{71}
(\byear{2009}).
doi:\doiurl{10.1088/0004-637X/690/1/L69}
\end{barticle}
\endbibitem

\bibitem[\protect\citeauthoryear{{Buonanno} and
  {Damour}}{1999}]{BuonannoDamour1999}
\begin{barticle}
\bauthor{\binits{A.} \bsnm{{Buonanno}}},
\bauthor{\binits{T.} \bsnm{{Damour}}},
\batitle{{Effective one-body approach to general relativistic two-body
  dynamics}}.
\bjtitle{Ph. Rv. D}
\bvolume{59}(\bissue{8}),
\bfpage{084006}
(\byear{1999}).
doi:\doiurl{10.1103/PhysRevD.59.084006}
\end{barticle}
\endbibitem

\bibitem[\protect\citeauthoryear{{Campanelli}
  et~al.}{2006}]{CampanelliEtAl2006}
\begin{barticle}
\bauthor{\binits{M.} \bsnm{{Campanelli}}},
\bauthor{\binits{C.O.} \bsnm{{Lousto}}},
\bauthor{\binits{P.} \bsnm{{Marronetti}}},
\bauthor{\binits{Y.} \bsnm{{Zlochower}}},
\batitle{{Accurate Evolutions of Orbiting Black-Hole Binaries without
  Excision}}.
\bjtitle{Physical Review Letters}
\bvolume{96}(\bissue{11}),
\bfpage{111101}
(\byear{2006}).
doi:\doiurl{10.1103/PhysRevLett.96.111101}
\end{barticle}
\endbibitem

\bibitem[\protect\citeauthoryear{{Carr} and
  {K{\"u}hnel}}{2020}]{2020ARNPS..7050520C}
\begin{barticle}
\bauthor{\binits{B.} \bsnm{{Carr}}},
\bauthor{\binits{F.} \bsnm{{K{\"u}hnel}}},
\batitle{{Primordial Black Holes as Dark Matter: Recent Developments}}.
\bjtitle{Annual Review of Nuclear and Particle Science}
\bvolume{70}(\bissue{1}),
(\byear{2020}).
doi:\doiurl{10.1146/annurev-nucl-050520-125911}
\end{barticle}
\endbibitem

\bibitem[\protect\citeauthoryear{{Carr} et~al.}{2020}]{2020arXiv200212778C}
\begin{botherref}
\oauthor{\binits{B.} \bsnm{{Carr}}},
\oauthor{\binits{K.} \bsnm{{Kohri}}},
\oauthor{\binits{Y.} \bsnm{{Sendouda}}},
\oauthor{\binits{J.} \bsnm{{Yokoyama}}},
{Constraints on Primordial Black Holes}.
arXiv e-prints,
2002--12778
(2020)
\end{botherref}
\endbibitem

\bibitem[\protect\citeauthoryear{{Chabrier} and {Baraffe}}{2000}]{CB00}
\begin{barticle}
\bauthor{\binits{G.} \bsnm{{Chabrier}}},
\bauthor{\binits{I.} \bsnm{{Baraffe}}},
\batitle{Theory of low-mass stars and substellar objects}.
\bjtitle{ARA\&A}
\bvolume{38},
\bfpage{337}--\blpage{377}
(\byear{2000})
\end{barticle}
\endbibitem

\bibitem[\protect\citeauthoryear{{Chabrier} et~al.}{2009}]{ChabrierEtAl2009}
\begin{bchapter}
\bauthor{\binits{G.} \bsnm{{Chabrier}}},
\bauthor{\binits{I.} \bsnm{{Baraffe}}},
\bauthor{\binits{J.} \bsnm{{Leconte}}},
\bauthor{\binits{J.} \bsnm{{Gallardo}}},
\bauthor{\binits{T.} \bsnm{{Barman}}},
\bctitle{{The mass-radius relationship from solar-type stars to terrestrial
  planets: a review}},
in \bbtitle{15th Cambridge Workshop on Cool Stars, Stellar Systems, and the
  Sun},
ed. by \beditor{\binits{E.} \bsnm{{Stempels}}}
\bsertitle{American Institute of Physics Conference Series},
vol. \bseriesno{1094},
\byear{2009},
pp. \bfpage{102}--\blpage{111}.
doi:\doiurl{10.1063/1.3099078}
\end{bchapter}
\endbibitem

\bibitem[\protect\citeauthoryear{{Chandrasekhar}}{1942}]{Chandra42}
\begin{botherref}
\oauthor{\binits{S.} \bsnm{{Chandrasekhar}}},
{Principles of stellar dynamics}.
Physical Sciences Data
(1942)
\end{botherref}
\endbibitem

\bibitem[\protect\citeauthoryear{{Charbonnel} et~al.}{1999}]{CDSBMMM99}
\begin{barticle}
\bauthor{\binits{C.} \bsnm{{Charbonnel}}},
\bauthor{\binits{W.} \bsnm{{D\"{a}ppen}}},
\bauthor{\binits{D.} \bsnm{{Schaerer}}},
\bauthor{\binits{P.A.} \bsnm{{Bernasconi}}},
\bauthor{\binits{A.} \bsnm{{Maeder}}},
\bauthor{\binits{G.} \bsnm{{Meynet}}},
\bauthor{\binits{N.} \bsnm{{Mowlavi}}},
\batitle{Grids of stellar models. {VIII}. from 0.4 to 1.0
  $\{$M\_$\{$sun$\}$$\}$ at z=0.020 and z=0.001, with the mhd equation of
  state}.
\bjtitle{A\&AS}
\bvolume{135},
\bfpage{405}--\blpage{413}
(\byear{1999})
\end{barticle}
\endbibitem

\bibitem[\protect\citeauthoryear{{Chen} and
  {Amaro-Seoane}}{2017}]{ChenAmaro-Seoane2017}
\begin{barticle}
\bauthor{\binits{X.} \bsnm{{Chen}}},
\bauthor{\binits{P.} \bsnm{{Amaro-Seoane}}},
\batitle{{Revealing the Formation of Stellar-mass Black Hole Binaries: The Need
  for Deci-Hertz Gravitational-wave Observatories}}.
\bjtitle{ApJ Lett.}
\bvolume{842},
\bfpage{2}
(\byear{2017}).
doi:\doiurl{10.3847/2041-8213/aa74ce}
\end{barticle}
\endbibitem

\bibitem[\protect\citeauthoryear{{Chernoff} and
  {Weinberg}}{1990}]{ChernoffWeinberg1990}
\begin{barticle}
\bauthor{\binits{D.F.} \bsnm{{Chernoff}}},
\bauthor{\binits{M.D.} \bsnm{{Weinberg}}},
\batitle{{Evolution of globular clusters in the Galaxy}}.
\bjtitle{ApJ}
\bvolume{351},
\bfpage{121}--\blpage{156}
(\byear{1990}).
doi:\doiurl{10.1086/168451}
\end{barticle}
\endbibitem

\bibitem[\protect\citeauthoryear{{Chua} and {Gair}}{2015}]{ChuaGair2015}
\begin{barticle}
\bauthor{\binits{A.J.K.} \bsnm{{Chua}}},
\bauthor{\binits{J.R.} \bsnm{{Gair}}},
\batitle{{Improved analytic extreme-mass-ratio inspiral model for scoping out
  eLISA data analysis}}.
\bjtitle{Classical and Quantum Gravity}
\bvolume{32}(\bissue{23}),
\bfpage{232002}
(\byear{2015}).
doi:\doiurl{10.1088/0264-9381/32/23/232002}
\end{barticle}
\endbibitem

\bibitem[\protect\citeauthoryear{{Chua} et~al.}{2017}]{ChuaEtAl2017}
\begin{barticle}
\bauthor{\binits{A.J.K.} \bsnm{{Chua}}},
\bauthor{\binits{C.J.} \bsnm{{Moore}}},
\bauthor{\binits{J.R.} \bsnm{{Gair}}},
\batitle{{Augmented kludge waveforms for detecting extreme-mass-ratio
  inspirals}}.
\bjtitle{Phys. Rev. D}
\bvolume{96}(\bissue{4}),
\bfpage{044005}
(\byear{2017}).
doi:\doiurl{10.1103/PhysRevD.96.044005}
\end{barticle}
\endbibitem

\bibitem[\protect\citeauthoryear{{Chua} et~al.}{2020}]{ChuaEtAl2020}
\begin{botherref}
\oauthor{\binits{A.J.K.} \bsnm{{Chua}}},
\oauthor{\binits{M.L.} \bsnm{{Katz}}},
\oauthor{\binits{N.} \bsnm{{Warburton}}},
\oauthor{\binits{S.A.} \bsnm{{Hughes}}},
{Rapid generation of fully relativistic extreme-mass-ratio-inspiral waveform
  templates for LISA data analysis}.
arXiv e-prints,
2008--06071
(2020)
\end{botherref}
\endbibitem

\bibitem[\protect\citeauthoryear{{Cohn}}{1980}]{Cohn80}
\begin{barticle}
\bauthor{\binits{H.} \bsnm{{Cohn}}},
\batitle{Late core collapse in star clusters and the gravothermal instability}.
\bjtitle{ApJ}
\bvolume{242},
\bfpage{765}--\blpage{771}
(\byear{1980})
\end{barticle}
\endbibitem

\bibitem[\protect\citeauthoryear{{Conselice} et~al.}{2016}]{ConseliceEtAl2016}
\begin{barticle}
\bauthor{\binits{C.J.} \bsnm{{Conselice}}},
\bauthor{\binits{A.} \bsnm{{Wilkinson}}},
\bauthor{\binits{K.} \bsnm{{Duncan}}},
\bauthor{\binits{A.} \bsnm{{Mortlock}}},
\batitle{{The Evolution of Galaxy Number Density at z < 8 and Its
  Implications}}.
\bjtitle{ApJ}
\bvolume{830}(\bissue{2}),
\bfpage{83}
(\byear{2016}).
doi:\doiurl{10.3847/0004-637X/830/2/83}
\end{barticle}
\endbibitem

\bibitem[\protect\citeauthoryear{{Cutler} and {Harms}}{2006}]{CutlerHarms2006}
\begin{barticle}
\bauthor{\binits{C.} \bsnm{{Cutler}}},
\bauthor{\binits{J.} \bsnm{{Harms}}},
\batitle{{Big Bang Observer and the neutron-star-binary subtraction problem}}.
\bjtitle{Ph. Rev. D}
\bvolume{73}(\bissue{4}),
\bfpage{042001}
(\byear{2006}).
doi:\doiurl{10.1103/PhysRevD.73.042001}
\end{barticle}
\endbibitem

\bibitem[\protect\citeauthoryear{{Cutler} et~al.}{1994}]{CKP94}
\begin{barticle}
\bauthor{\binits{C.} \bsnm{{Cutler}}},
\bauthor{\binits{D.} \bsnm{{Kennefick}}},
\bauthor{\binits{E.} \bsnm{{Poisson}}},
\batitle{{Gravitational radiation reaction for bound motion around a
  Schwarzschild black hole}}.
\bjtitle{Ph. Rev. D}
\bvolume{50},
\bfpage{3816}--\blpage{3835}
(\byear{1994}).
doi:\doiurl{10.1103/PhysRevD.50.3816}
\end{barticle}
\endbibitem

\bibitem[\protect\citeauthoryear{{Cutler} and
  {Vallisneri}}{2007}]{CutlerVallisneri2007}
\begin{barticle}
\bauthor{\binits{C.} \bsnm{{Cutler}}},
\bauthor{\binits{M.} \bsnm{{Vallisneri}}},
\batitle{{LISA detections of massive black hole inspirals: Parameter extraction
  errors due to inaccurate template waveforms}}.
\bjtitle{Ph. Rv. D.}
\bvolume{76}(\bissue{10}),
\bfpage{104018}
(\byear{2007}).
doi:\doiurl{10.1103/PhysRevD.76.104018}
\end{barticle}
\endbibitem

\bibitem[\protect\citeauthoryear{{Datta} et~al.}{2020}]{2020arXiv200612137D}
\begin{botherref}
\oauthor{\binits{S.} \bsnm{{Datta}}},
\oauthor{\binits{A.} \bsnm{{Gupta}}},
\oauthor{\binits{S.} \bsnm{{Kastha}}},
\oauthor{\binits{K.G.} \bsnm{{Arun}}},
\oauthor{\binits{B.S.} \bsnm{{Sathyaprakash}}},
{Tests of general relativity using multiband observations of intermediate mass
  binary black hole mergers}.
arXiv e-prints,
2006--12137
(2020)
\end{botherref}
\endbibitem

\bibitem[\protect\citeauthoryear{{Dehnen}}{1993}]{Dehnen93}
\begin{barticle}
\bauthor{\binits{W.} \bsnm{{Dehnen}}},
\batitle{{A Family of Potential-Density Pairs for Spherical Galaxies and
  Bulges}}.
\bjtitle{MNRAS}
\bvolume{265},
\bfpage{250}
(\byear{1993})
\end{barticle}
\endbibitem

\bibitem[\protect\citeauthoryear{{Dorband} et~al.}{2003}]{DHM03}
\begin{barticle}
\bauthor{\binits{E.N.} \bsnm{{Dorband}}},
\bauthor{\binits{M.} \bsnm{{Hemsendorf}}},
\bauthor{\binits{D.} \bsnm{{Merritt}}},
\batitle{{Systolic and hyper-systolic algorithms for the gravitational N-body
  problem, with an application to Brownian motion}}.
\bjtitle{Journal of Computational Physics}
\bvolume{185},
\bfpage{484}--\blpage{511}
(\byear{2003})
\end{barticle}
\endbibitem

\bibitem[\protect\citeauthoryear{{Ehlers} et~al.}{1976}]{EhlersEtAl1976}
\begin{barticle}
\bauthor{\binits{J.} \bsnm{{Ehlers}}},
\bauthor{\binits{A.} \bsnm{{Rosenblum}}},
\bauthor{\binits{J.N.} \bsnm{{Goldberg}}},
\bauthor{\binits{P.} \bsnm{{Havas}}},
\batitle{{Comments on gravitational radiation damping and energy loss in binary
  systems.}}
\bjtitle{ApJ Lett.}
\bvolume{208},
\bfpage{77}--\blpage{81}
(\byear{1976}).
doi:\doiurl{10.1086/182236}
\end{barticle}
\endbibitem

\bibitem[\protect\citeauthoryear{{Eilon} et~al.}{2009}]{2009ApJ...698..641E}
\begin{barticle}
\bauthor{\binits{E.} \bsnm{{Eilon}}},
\bauthor{\binits{G.} \bsnm{{Kupi}}},
\bauthor{\binits{T.} \bsnm{{Alexander}}},
\batitle{{The Efficiency of Resonant Relaxation Around a Massive Black Hole}}.
\bjtitle{\apj}
\bvolume{698}(\bissue{1}),
\bfpage{641}--\blpage{647}
(\byear{2009}).
doi:\doiurl{10.1088/0004-637X/698/1/641}
\end{barticle}
\endbibitem

\bibitem[\protect\citeauthoryear{{Eisenhauer} et~al.}{2008}]{EisenhauerEtAl08}
\begin{bchapter}
\bauthor{\binits{F.} \bsnm{{Eisenhauer}}},
\bauthor{\binits{G.} \bsnm{{Perrin}}},
\bauthor{\binits{W.} \bsnm{{Brandner}}},
\bauthor{\binits{C.} \bsnm{{Straubmeier}}},
\bauthor{\binits{A.} \bsnm{{Richichi}}},
\bauthor{\binits{S.} \bsnm{{Gillessen}}},
\bauthor{\binits{J.P.} \bsnm{{Berger}}},
\bauthor{\binits{S.} \bsnm{{Hippler}}},
\bauthor{\binits{A.} \bsnm{{Eckart}}},
\bauthor{\binits{M.} \bsnm{{Sch{\"o}ller}}},
\bauthor{\binits{S.} \bsnm{{Rabien}}},
\bauthor{\binits{F.} \bsnm{{Cassaing}}},
\bauthor{\binits{R.} \bsnm{{Lenzen}}},
\bauthor{\binits{M.} \bsnm{{Thiel}}},
\bauthor{\binits{Y.} \bsnm{{Cl{\'e}net}}},
\bauthor{\binits{J.R.} \bsnm{{Ramos}}},
\bauthor{\binits{S.} \bsnm{{Kellner}}},
\bauthor{\binits{P.} \bsnm{{F{\'e}dou}}},
\bauthor{\binits{H.} \bsnm{{Baumeister}}},
\bauthor{\binits{R.} \bsnm{{Hofmann}}},
\bauthor{\binits{E.} \bsnm{{Gendron}}},
\bauthor{\binits{A.} \bsnm{{Boehm}}},
\bauthor{\binits{H.} \bsnm{{Bartko}}},
\bauthor{\binits{X.} \bsnm{{Haubois}}},
\bauthor{\binits{R.} \bsnm{{Klein}}},
\bauthor{\binits{K.} \bsnm{{Dodds-Eden}}},
\bauthor{\binits{K.} \bsnm{{Houairi}}},
\bauthor{\binits{F.} \bsnm{{Hormuth}}},
\bauthor{\binits{A.} \bsnm{{Gr{\"a}ter}}},
\bauthor{\binits{L.} \bsnm{{Jocou}}},
\bauthor{\binits{V.} \bsnm{{Naranjo}}},
\bauthor{\binits{R.} \bsnm{{Genzel}}},
\bauthor{\binits{P.} \bsnm{{Kervella}}},
\bauthor{\binits{T.} \bsnm{{Henning}}},
\bauthor{\binits{N.} \bsnm{{Hamaus}}},
\bauthor{\binits{S.} \bsnm{{Lacour}}},
\bauthor{\binits{U.} \bsnm{{Neumann}}},
\bauthor{\binits{M.} \bsnm{{Haug}}},
\bauthor{\binits{F.} \bsnm{{Malbet}}},
\bauthor{\binits{W.} \bsnm{{Laun}}},
\bauthor{\binits{J.} \bsnm{{Kolmeder}}},
\bauthor{\binits{T.} \bsnm{{Paumard}}},
\bauthor{\binits{R.-R.} \bsnm{{Rohloff}}},
\bauthor{\binits{O.} \bsnm{{Pfuhl}}},
\bauthor{\binits{K.} \bsnm{{Perraut}}},
\bauthor{\binits{J.} \bsnm{{Ziegleder}}},
\bauthor{\binits{D.} \bsnm{{Rouan}}},
\bauthor{\binits{G.} \bsnm{{Rousset}}},
\bctitle{{GRAVITY: getting to the event horizon of Sgr A*}},
in \bbtitle{Society of Photo-Optical Instrumentation Engineers (SPIE)
  Conference Series}.
\bsertitle{Society of Photo-Optical Instrumentation Engineers (SPIE) Conference
  Series},
vol. \bseriesno{7013},
\byear{2008}.
doi:\doiurl{10.1117/12.788407}
\end{bchapter}
\endbibitem

\bibitem[\protect\citeauthoryear{{Event Horizon Telescope Collaboration}
  et~al.}{2019}]{2019ApJ...875L...1E}
\begin{barticle}
\bauthor{\bsnm{{Event Horizon Telescope Collaboration}}},
\bauthor{\binits{K.} \bsnm{{Akiyama}}},
\bauthor{\binits{A.} \bsnm{{Alberdi}}},
\bauthor{\binits{W.} \bsnm{{Alef}}},
\bauthor{\binits{K.} \bsnm{{Asada}}},
\bauthor{\binits{R.} \bsnm{{Azulay}}},
\bauthor{\binits{A.-K.} \bsnm{{Baczko}}},
\bauthor{\binits{D.} \bsnm{{Ball}}},
\bauthor{\binits{M.} \bsnm{{Balokovi{\'c}}}},
\bauthor{\binits{J.} \bsnm{{Barrett}}},
\bauthor{\binits{D.} \bsnm{{Bintley}}},
\bauthor{\binits{L.} \bsnm{{Blackburn}}},
\bauthor{\binits{W.} \bsnm{{Boland}}},
\bauthor{\binits{K.L.} \bsnm{{Bouman}}},
\bauthor{\binits{G.C.} \bsnm{{Bower}}},
\bauthor{\binits{M.} \bsnm{{Bremer}}},
\bauthor{\binits{C.D.} \bsnm{{Brinkerink}}},
\bauthor{\binits{R.} \bsnm{{Brissenden}}},
\bauthor{\binits{S.} \bsnm{{Britzen}}},
\bauthor{\binits{A.E.} \bsnm{{Broderick}}},
\bauthor{\binits{D.} \bsnm{{Broguiere}}},
\bauthor{\binits{T.} \bsnm{{Bronzwaer}}},
\bauthor{\binits{D.-Y.} \bsnm{{Byun}}},
\bauthor{\binits{J.E.} \bsnm{{Carlstrom}}},
\bauthor{\binits{A.} \bsnm{{Chael}}},
\bauthor{\binits{C.-k.} \bsnm{{Chan}}},
\bauthor{\binits{S.} \bsnm{{Chatterjee}}},
\bauthor{\binits{K.} \bsnm{{Chatterjee}}},
\bauthor{\binits{M.-T.} \bsnm{{Chen}}},
\bauthor{\binits{Y.} \bsnm{{Chen}}},
\bauthor{\binits{I.} \bsnm{{Cho}}},
\bauthor{\binits{P.} \bsnm{{Christian}}},
\bauthor{\binits{J.E.} \bsnm{{Conway}}},
\bauthor{\binits{J.M.} \bsnm{{Cordes}}},
\bauthor{\binits{G.B.} \bsnm{{Crew}}},
\bauthor{\binits{Y.} \bsnm{{Cui}}},
\bauthor{\binits{J.} \bsnm{{Davelaar}}},
\bauthor{\binits{M.} \bsnm{{De Laurentis}}},
\bauthor{\binits{R.} \bsnm{{Deane}}},
\bauthor{\binits{J.} \bsnm{{Dempsey}}},
\bauthor{\binits{G.} \bsnm{{Desvignes}}},
\bauthor{\binits{J.} \bsnm{{Dexter}}},
\bauthor{\binits{S.S.} \bsnm{{Doeleman}}},
\bauthor{\binits{R.P.} \bsnm{{Eatough}}},
\bauthor{\binits{H.} \bsnm{{Falcke}}},
\bauthor{\binits{V.L.} \bsnm{{Fish}}},
\bauthor{\binits{E.} \bsnm{{Fomalont}}},
\bauthor{\binits{R.} \bsnm{{Fraga-Encinas}}},
\bauthor{\binits{W.T.} \bsnm{{Freeman}}},
\bauthor{\binits{P.} \bsnm{{Friberg}}},
\bauthor{\binits{C.M.} \bsnm{{Fromm}}},
\bauthor{\binits{J.L.} \bsnm{{G{\'o}mez}}},
\bauthor{\binits{P.} \bsnm{{Galison}}},
\bauthor{\binits{C.F.} \bsnm{{Gammie}}},
\bauthor{\binits{R.} \bsnm{{Garc{\'\i}a}}},
\bauthor{\binits{O.} \bsnm{{Gentaz}}},
\bauthor{\binits{B.} \bsnm{{Georgiev}}},
\bauthor{\binits{C.} \bsnm{{Goddi}}},
\bauthor{\binits{R.} \bsnm{{Gold}}},
\bauthor{\binits{M.} \bsnm{{Gu}}},
\bauthor{\binits{M.} \bsnm{{Gurwell}}},
\bauthor{\binits{K.} \bsnm{{Hada}}},
\bauthor{\binits{M.H.} \bsnm{{Hecht}}},
\bauthor{\binits{R.} \bsnm{{Hesper}}},
\bauthor{\binits{L.C.} \bsnm{{Ho}}},
\bauthor{\binits{P.} \bsnm{{Ho}}},
\bauthor{\binits{M.} \bsnm{{Honma}}},
\bauthor{\binits{C.-W.L.} \bsnm{{Huang}}},
\bauthor{\binits{L.} \bsnm{{Huang}}},
\bauthor{\binits{D.H.} \bsnm{{Hughes}}},
\bauthor{\binits{S.} \bsnm{{Ikeda}}},
\bauthor{\binits{M.} \bsnm{{Inoue}}},
\bauthor{\binits{S.} \bsnm{{Issaoun}}},
\bauthor{\binits{D.J.} \bsnm{{James}}},
\bauthor{\binits{B.T.} \bsnm{{Jannuzi}}},
\bauthor{\binits{M.} \bsnm{{Janssen}}},
\bauthor{\binits{B.} \bsnm{{Jeter}}},
\bauthor{\binits{W.} \bsnm{{Jiang}}},
\bauthor{\binits{M.D.} \bsnm{{Johnson}}},
\bauthor{\binits{S.} \bsnm{{Jorstad}}},
\bauthor{\binits{T.} \bsnm{{Jung}}},
\bauthor{\binits{M.} \bsnm{{Karami}}},
\bauthor{\binits{R.} \bsnm{{Karuppusamy}}},
\bauthor{\binits{T.} \bsnm{{Kawashima}}},
\bauthor{\binits{G.K.} \bsnm{{Keating}}},
\bauthor{\binits{M.} \bsnm{{Kettenis}}},
\bauthor{\binits{J.-Y.} \bsnm{{Kim}}},
\bauthor{\binits{J.} \bsnm{{Kim}}},
\bauthor{\binits{J.} \bsnm{{Kim}}},
\bauthor{\binits{M.} \bsnm{{Kino}}},
\bauthor{\binits{J.Y.} \bsnm{{Koay}}},
\bauthor{\binits{P.M.} \bsnm{{Koch}}},
\bauthor{\binits{S.} \bsnm{{Koyama}}},
\bauthor{\binits{M.} \bsnm{{Kramer}}},
\bauthor{\binits{C.} \bsnm{{Kramer}}},
\bauthor{\binits{T.P.} \bsnm{{Krichbaum}}},
\bauthor{\binits{C.-Y.} \bsnm{{Kuo}}},
\bauthor{\binits{T.R.} \bsnm{{Lauer}}},
\bauthor{\binits{S.-S.} \bsnm{{Lee}}},
\bauthor{\binits{Y.-R.} \bsnm{{Li}}},
\bauthor{\binits{Z.} \bsnm{{Li}}},
\bauthor{\binits{M.} \bsnm{{Lindqvist}}},
\bauthor{\binits{K.} \bsnm{{Liu}}},
\bauthor{\binits{E.} \bsnm{{Liuzzo}}},
\bauthor{\binits{W.-P.} \bsnm{{Lo}}},
\bauthor{\binits{A.P.} \bsnm{{Lobanov}}},
\bauthor{\binits{L.} \bsnm{{Loinard}}},
\bauthor{\binits{C.} \bsnm{{Lonsdale}}},
\bauthor{\binits{R.-S.} \bsnm{{Lu}}},
\bauthor{\binits{N.R.} \bsnm{{MacDonald}}},
\bauthor{\binits{J.} \bsnm{{Mao}}},
\bauthor{\binits{S.} \bsnm{{Markoff}}},
\bauthor{\binits{D.P.} \bsnm{{Marrone}}},
\bauthor{\binits{A.P.} \bsnm{{Marscher}}},
\bauthor{\binits{I.} \bsnm{{Mart{\'\i}-Vidal}}},
\bauthor{\binits{S.} \bsnm{{Matsushita}}},
\bauthor{\binits{L.D.} \bsnm{{Matthews}}},
\bauthor{\binits{L.} \bsnm{{Medeiros}}},
\bauthor{\binits{K.M.} \bsnm{{Menten}}},
\bauthor{\binits{Y.} \bsnm{{Mizuno}}},
\bauthor{\binits{I.} \bsnm{{Mizuno}}},
\bauthor{\binits{J.M.} \bsnm{{Moran}}},
\bauthor{\binits{K.} \bsnm{{Moriyama}}},
\bauthor{\binits{M.} \bsnm{{Moscibrodzka}}},
\bauthor{\binits{C.} \bsnm{{M{\"u}ller}}},
\bauthor{\binits{H.} \bsnm{{Nagai}}},
\bauthor{\binits{N.M.} \bsnm{{Nagar}}},
\bauthor{\binits{M.} \bsnm{{Nakamura}}},
\bauthor{\binits{R.} \bsnm{{Narayan}}},
\bauthor{\binits{G.} \bsnm{{Narayanan}}},
\bauthor{\binits{I.} \bsnm{{Natarajan}}},
\bauthor{\binits{R.} \bsnm{{Neri}}},
\bauthor{\binits{C.} \bsnm{{Ni}}},
\bauthor{\binits{A.} \bsnm{{Noutsos}}},
\bauthor{\binits{H.} \bsnm{{Okino}}},
\bauthor{\binits{H.} \bsnm{{Olivares}}},
\bauthor{\binits{G.N.} \bsnm{{Ortiz-Le{\'o}n}}},
\bauthor{\binits{T.} \bsnm{{Oyama}}},
\bauthor{\binits{F.} \bsnm{{{\"O}zel}}},
\bauthor{\binits{D.C.M.} \bsnm{{Palumbo}}},
\bauthor{\binits{N.} \bsnm{{Patel}}},
\bauthor{\binits{U.-L.} \bsnm{{Pen}}},
\bauthor{\binits{D.W.} \bsnm{{Pesce}}},
\bauthor{\binits{V.} \bsnm{{Pi{\'e}tu}}},
\bauthor{\binits{R.} \bsnm{{Plambeck}}},
\bauthor{\binits{A.} \bsnm{{PopStefanija}}},
\bauthor{\binits{O.} \bsnm{{Porth}}},
\bauthor{\binits{B.} \bsnm{{Prather}}},
\bauthor{\binits{J.A.} \bsnm{{Preciado-L{\'o}pez}}},
\bauthor{\binits{D.} \bsnm{{Psaltis}}},
\bauthor{\binits{H.-Y.} \bsnm{{Pu}}},
\bauthor{\binits{V.} \bsnm{{Ramakrishnan}}},
\bauthor{\binits{R.} \bsnm{{Rao}}},
\bauthor{\binits{M.G.} \bsnm{{Rawlings}}},
\bauthor{\binits{A.W.} \bsnm{{Raymond}}},
\bauthor{\binits{L.} \bsnm{{Rezzolla}}},
\bauthor{\binits{B.} \bsnm{{Ripperda}}},
\bauthor{\binits{F.} \bsnm{{Roelofs}}},
\bauthor{\binits{A.} \bsnm{{Rogers}}},
\bauthor{\binits{E.} \bsnm{{Ros}}},
\bauthor{\binits{M.} \bsnm{{Rose}}},
\bauthor{\binits{A.} \bsnm{{Roshanineshat}}},
\bauthor{\binits{H.} \bsnm{{Rottmann}}},
\bauthor{\binits{A.L.} \bsnm{{Roy}}},
\bauthor{\binits{C.} \bsnm{{Ruszczyk}}},
\bauthor{\binits{B.R.} \bsnm{{Ryan}}},
\bauthor{\binits{K.L.J.} \bsnm{{Rygl}}},
\bauthor{\binits{S.} \bsnm{{S{\'a}nchez}}},
\bauthor{\binits{D.} \bsnm{{S{\'a}nchez-Arguelles}}},
\bauthor{\binits{M.} \bsnm{{Sasada}}},
\bauthor{\binits{T.} \bsnm{{Savolainen}}},
\bauthor{\binits{F.P.} \bsnm{{Schloerb}}},
\bauthor{\binits{K.-F.} \bsnm{{Schuster}}},
\bauthor{\binits{L.} \bsnm{{Shao}}},
\bauthor{\binits{Z.} \bsnm{{Shen}}},
\bauthor{\binits{D.} \bsnm{{Small}}},
\bauthor{\binits{B.W.} \bsnm{{Sohn}}},
\bauthor{\binits{J.} \bsnm{{SooHoo}}},
\bauthor{\binits{F.} \bsnm{{Tazaki}}},
\bauthor{\binits{P.} \bsnm{{Tiede}}},
\bauthor{\binits{R.P.J.} \bsnm{{Tilanus}}},
\bauthor{\binits{M.} \bsnm{{Titus}}},
\bauthor{\binits{K.} \bsnm{{Toma}}},
\bauthor{\binits{P.} \bsnm{{Torne}}},
\bauthor{\binits{T.} \bsnm{{Trent}}},
\bauthor{\binits{S.} \bsnm{{Trippe}}},
\bauthor{\binits{S.} \bsnm{{Tsuda}}},
\bauthor{\binits{I.} \bsnm{{van Bemmel}}},
\bauthor{\binits{H.J.} \bsnm{{van Langevelde}}},
\bauthor{\binits{D.R.} \bsnm{{van Rossum}}},
\bauthor{\binits{J.} \bsnm{{Wagner}}},
\bauthor{\binits{J.} \bsnm{{Wardle}}},
\bauthor{\binits{J.} \bsnm{{Weintroub}}},
\bauthor{\binits{N.} \bsnm{{Wex}}},
\bauthor{\binits{R.} \bsnm{{Wharton}}},
\bauthor{\binits{M.} \bsnm{{Wielgus}}},
\bauthor{\binits{G.N.} \bsnm{{Wong}}},
\bauthor{\binits{Q.} \bsnm{{Wu}}},
\bauthor{\binits{K.} \bsnm{{Young}}},
\bauthor{\binits{A.} \bsnm{{Young}}},
\bauthor{\binits{Z.} \bsnm{{Younsi}}},
\bauthor{\binits{F.} \bsnm{{Yuan}}},
\bauthor{\binits{Y.-F.} \bsnm{{Yuan}}},
\bauthor{\binits{J.A.} \bsnm{{Zensus}}},
\bauthor{\binits{G.} \bsnm{{Zhao}}},
\bauthor{\binits{S.-S.} \bsnm{{Zhao}}},
\bauthor{\binits{Z.} \bsnm{{Zhu}}},
\bauthor{\binits{J.-C.} \bsnm{{Algaba}}},
\bauthor{\binits{A.} \bsnm{{Allardi}}},
\bauthor{\binits{R.} \bsnm{{Amestica}}},
\bauthor{\binits{J.} \bsnm{{Anczarski}}},
\bauthor{\binits{U.} \bsnm{{Bach}}},
\bauthor{\binits{F.K.} \bsnm{{Baganoff}}},
\bauthor{\binits{C.} \bsnm{{Beaudoin}}},
\bauthor{\binits{B.A.} \bsnm{{Benson}}},
\bauthor{\binits{R.} \bsnm{{Berthold}}},
\bauthor{\binits{J.M.} \bsnm{{Blanchard}}},
\bauthor{\binits{R.} \bsnm{{Blundell}}},
\bauthor{\binits{S.} \bsnm{{Bustamente}}},
\bauthor{\binits{R.} \bsnm{{Cappallo}}},
\bauthor{\binits{E.} \bsnm{{Castillo-Dom{\'\i}nguez}}},
\bauthor{\binits{C.-C.} \bsnm{{Chang}}},
\bauthor{\binits{S.-H.} \bsnm{{Chang}}},
\bauthor{\binits{S.-C.} \bsnm{{Chang}}},
\bauthor{\binits{C.-C.} \bsnm{{Chen}}},
\bauthor{\binits{R.} \bsnm{{Chilson}}},
\bauthor{\binits{T.C.} \bsnm{{Chuter}}},
\bauthor{\binits{R.} \bsnm{{C{\'o}rdova Rosado}}},
\bauthor{\binits{I.M.} \bsnm{{Coulson}}},
\bauthor{\binits{T.M.} \bsnm{{Crawford}}},
\bauthor{\binits{J.} \bsnm{{Crowley}}},
\bauthor{\binits{J.} \bsnm{{David}}},
\bauthor{\binits{M.} \bsnm{{Derome}}},
\bauthor{\binits{M.} \bsnm{{Dexter}}},
\bauthor{\binits{S.} \bsnm{{Dornbusch}}},
\bauthor{\binits{K.A.} \bsnm{{Dudevoir}}},
\bauthor{\binits{S.A.} \bsnm{{Dzib}}},
\bauthor{\binits{A.} \bsnm{{Eckart}}},
\bauthor{\binits{C.} \bsnm{{Eckert}}},
\bauthor{\binits{N.R.} \bsnm{{Erickson}}},
\bauthor{\binits{W.B.} \bsnm{{Everett}}},
\bauthor{\binits{A.} \bsnm{{Faber}}},
\bauthor{\binits{J.R.} \bsnm{{Farah}}},
\bauthor{\binits{V.} \bsnm{{Fath}}},
\bauthor{\binits{T.W.} \bsnm{{Folkers}}},
\bauthor{\binits{D.C.} \bsnm{{Forbes}}},
\bauthor{\binits{R.} \bsnm{{Freund}}},
\bauthor{\binits{A.I.} \bsnm{{G{\'o}mez-Ruiz}}},
\bauthor{\binits{D.M.} \bsnm{{Gale}}},
\bauthor{\binits{F.} \bsnm{{Gao}}},
\bauthor{\binits{G.} \bsnm{{Geertsema}}},
\bauthor{\binits{D.A.} \bsnm{{Graham}}},
\bauthor{\binits{C.H.} \bsnm{{Greer}}},
\bauthor{\binits{R.} \bsnm{{Grosslein}}},
\bauthor{\binits{F.} \bsnm{{Gueth}}},
\bauthor{\binits{D.} \bsnm{{Haggard}}},
\bauthor{\binits{N.W.} \bsnm{{Halverson}}},
\bauthor{\binits{C.-C.} \bsnm{{Han}}},
\bauthor{\binits{K.-C.} \bsnm{{Han}}},
\bauthor{\binits{J.} \bsnm{{Hao}}},
\bauthor{\binits{Y.} \bsnm{{Hasegawa}}},
\bauthor{\binits{J.W.} \bsnm{{Henning}}},
\bauthor{\binits{A.} \bsnm{{Hern{\'a}ndez-G{\'o}mez}}},
\bauthor{\binits{R.} \bsnm{{Herrero-Illana}}},
\bauthor{\binits{S.} \bsnm{{Heyminck}}},
\bauthor{\binits{A.} \bsnm{{Hirota}}},
\bauthor{\binits{J.} \bsnm{{Hoge}}},
\bauthor{\binits{Y.-D.} \bsnm{{Huang}}},
\bauthor{\binits{C.M.V.} \bsnm{{Impellizzeri}}},
\bauthor{\binits{H.} \bsnm{{Jiang}}},
\bauthor{\binits{A.} \bsnm{{Kamble}}},
\bauthor{\binits{R.} \bsnm{{Keisler}}},
\bauthor{\binits{K.} \bsnm{{Kimura}}},
\bauthor{\binits{Y.} \bsnm{{Kono}}},
\bauthor{\binits{D.} \bsnm{{Kubo}}},
\bauthor{\binits{J.} \bsnm{{Kuroda}}},
\bauthor{\binits{R.} \bsnm{{Lacasse}}},
\bauthor{\binits{R.A.} \bsnm{{Laing}}},
\bauthor{\binits{E.M.} \bsnm{{Leitch}}},
\bauthor{\binits{C.-T.} \bsnm{{Li}}},
\bauthor{\binits{L.C.-C.} \bsnm{{Lin}}},
\bauthor{\binits{C.-T.} \bsnm{{Liu}}},
\bauthor{\binits{K.-Y.} \bsnm{{Liu}}},
\bauthor{\binits{L.-M.} \bsnm{{Lu}}},
\bauthor{\binits{R.G.} \bsnm{{Marson}}},
\bauthor{\binits{P.L.} \bsnm{{Martin-Cocher}}},
\bauthor{\binits{K.D.} \bsnm{{Massingill}}},
\bauthor{\binits{C.} \bsnm{{Matulonis}}},
\bauthor{\binits{M.P.} \bsnm{{McColl}}},
\bauthor{\binits{S.R.} \bsnm{{McWhirter}}},
\bauthor{\binits{H.} \bsnm{{Messias}}},
\bauthor{\binits{Z.} \bsnm{{Meyer-Zhao}}},
\bauthor{\binits{D.} \bsnm{{Michalik}}},
\bauthor{\binits{A.} \bsnm{{Monta{\~n}a}}},
\bauthor{\binits{W.} \bsnm{{Montgomerie}}},
\bauthor{\binits{M.} \bsnm{{Mora-Klein}}},
\bauthor{\binits{D.} \bsnm{{Muders}}},
\bauthor{\binits{A.} \bsnm{{Nadolski}}},
\bauthor{\binits{S.} \bsnm{{Navarro}}},
\bauthor{\binits{J.} \bsnm{{Neilsen}}},
\bauthor{\binits{C.H.} \bsnm{{Nguyen}}},
\bauthor{\binits{H.} \bsnm{{Nishioka}}},
\bauthor{\binits{T.} \bsnm{{Norton}}},
\bauthor{\binits{M.A.} \bsnm{{Nowak}}},
\bauthor{\binits{G.} \bsnm{{Nystrom}}},
\bauthor{\binits{H.} \bsnm{{Ogawa}}},
\bauthor{\binits{P.} \bsnm{{Oshiro}}},
\bauthor{\binits{T.} \bsnm{{Oyama}}},
\bauthor{\binits{H.} \bsnm{{Parsons}}},
\bauthor{\binits{S.N.} \bsnm{{Paine}}},
\bauthor{\binits{J.} \bsnm{{Pe{\~n}alver}}},
\bauthor{\binits{N.M.} \bsnm{{Phillips}}},
\bauthor{\binits{M.} \bsnm{{Poirier}}},
\bauthor{\binits{N.} \bsnm{{Pradel}}},
\bauthor{\binits{R.A.} \bsnm{{Primiani}}},
\bauthor{\binits{P.A.} \bsnm{{Raffin}}},
\bauthor{\binits{A.S.} \bsnm{{Rahlin}}},
\bauthor{\binits{G.} \bsnm{{Reiland}}},
\bauthor{\binits{C.} \bsnm{{Risacher}}},
\bauthor{\binits{I.} \bsnm{{Ruiz}}},
\bauthor{\binits{A.F.} \bsnm{{S{\'a}ez-Mada{\'\i}n}}},
\bauthor{\binits{R.} \bsnm{{Sassella}}},
\bauthor{\binits{P.} \bsnm{{Schellart}}},
\bauthor{\binits{P.} \bsnm{{Shaw}}},
\bauthor{\binits{K.M.} \bsnm{{Silva}}},
\bauthor{\binits{H.} \bsnm{{Shiokawa}}},
\bauthor{\binits{D.R.} \bsnm{{Smith}}},
\bauthor{\binits{W.} \bsnm{{Snow}}},
\bauthor{\binits{K.} \bsnm{{Souccar}}},
\bauthor{\binits{D.} \bsnm{{Sousa}}},
\bauthor{\binits{T.K.} \bsnm{{Sridharan}}},
\bauthor{\binits{R.} \bsnm{{Srinivasan}}},
\bauthor{\binits{W.} \bsnm{{Stahm}}},
\bauthor{\binits{A.A.} \bsnm{{Stark}}},
\bauthor{\binits{K.} \bsnm{{Story}}},
\bauthor{\binits{S.T.} \bsnm{{Timmer}}},
\bauthor{\binits{L.} \bsnm{{Vertatschitsch}}},
\bauthor{\binits{C.} \bsnm{{Walther}}},
\bauthor{\binits{T.-S.} \bsnm{{Wei}}},
\bauthor{\binits{N.} \bsnm{{Whitehorn}}},
\bauthor{\binits{A.R.} \bsnm{{Whitney}}},
\bauthor{\binits{D.P.} \bsnm{{Woody}}},
\bauthor{\binits{J.G.A.} \bsnm{{Wouterloot}}},
\bauthor{\binits{M.} \bsnm{{Wright}}},
\bauthor{\binits{P.} \bsnm{{Yamaguchi}}},
\bauthor{\binits{C.-Y.} \bsnm{{Yu}}},
\bauthor{\binits{M.} \bsnm{{Zeballos}}},
\bauthor{\binits{S.} \bsnm{{Zhang}}},
\bauthor{\binits{L.} \bsnm{{Ziurys}}},
\batitle{{First M87 Event Horizon Telescope Results. I. The Shadow of the
  Supermassive Black Hole}}.
\bjtitle{\apjl}
\bvolume{875}(\bissue{1}),
\bfpage{1}
(\byear{2019}).
doi:\doiurl{10.3847/2041-8213/ab0ec7}
\end{barticle}
\endbibitem

\bibitem[\protect\citeauthoryear{{Finn}}{1992}]{Finn92}
\begin{barticle}
\bauthor{\binits{L.S.} \bsnm{{Finn}}},
\batitle{{Detection, measurement, and gravitational radiation}}.
\bjtitle{prd}
\bvolume{46},
\bfpage{5236}--\blpage{5249}
(\byear{1992}).
doi:\doiurl{10.1103/PhysRevD.46.5236}
\end{barticle}
\endbibitem

\bibitem[\protect\citeauthoryear{{Finn} and {Thorne}}{2000}]{FinnThorne2000}
\begin{barticle}
\bauthor{\binits{L.S.} \bsnm{{Finn}}},
\bauthor{\binits{K.S.} \bsnm{{Thorne}}},
\batitle{{Gravitational waves from a compact star in a circular, inspiral
  orbit, in the equatorial plane of a massive, spinning black hole, as observed
  by LISA}}.
\bjtitle{Phys. Rev. D}
\bvolume{62}(\bissue{12}),
\bfpage{124021}
(\byear{2000}).
doi:\doiurl{10.1103/PhysRevD.62.124021}
\end{barticle}
\endbibitem

\bibitem[\protect\citeauthoryear{{Frank} and {Rees}}{1976}]{FR76}
\begin{barticle}
\bauthor{\binits{J.} \bsnm{{Frank}}},
\bauthor{\binits{M.J.} \bsnm{{Rees}}},
\batitle{{Effects of massive central black holes on dense stellar systems}}.
\bjtitle{MNRAS}
\bvolume{176},
\bfpage{633}--\blpage{647}
(\byear{1976}).
doi:\doiurl{10.1093/mnras/176.3.633}
\end{barticle}
\endbibitem

\bibitem[\protect\citeauthoryear{{Freitag} et~al.}{2006}]{FAK06a}
\begin{barticle}
\bauthor{\binits{M.} \bsnm{{Freitag}}},
\bauthor{\binits{P.} \bsnm{{Amaro-Seoane}}},
\bauthor{\binits{V.} \bsnm{{Kalogera}}},
\batitle{{Stellar Remnants in Galactic Nuclei: Mass Segregation}}.
\bjtitle{ApJ}
\bvolume{649},
\bfpage{91}--\blpage{117}
(\byear{2006}).
doi:\doiurl{10.1086/506193}
\end{barticle}
\endbibitem

\bibitem[\protect\citeauthoryear{{Gair} and {Glampedakis}}{2006}]{GG06}
\begin{barticle}
\bauthor{\binits{J.R.} \bsnm{{Gair}}},
\bauthor{\binits{K.} \bsnm{{Glampedakis}}},
\batitle{{Improved approximate inspirals of test bodies into Kerr black
  holes}}.
\bjtitle{Ph. Rev. D}
\bvolume{73}(\bissue{6}),
\bfpage{064037}
(\byear{2006}).
doi:\doiurl{10.1103/PhysRevD.73.064037}
\end{barticle}
\endbibitem

\bibitem[\protect\citeauthoryear{{Gallego-Cano}
  et~al.}{2018}]{2018A&A...609A..26G}
\begin{barticle}
\bauthor{\binits{E.} \bsnm{{Gallego-Cano}}},
\bauthor{\binits{R.} \bsnm{{Sch{\"o}del}}},
\bauthor{\binits{H.} \bsnm{{Dong}}},
\bauthor{\binits{F.} \bsnm{{Nogueras-Lara}}},
\bauthor{\binits{A.T.} \bsnm{{Gallego-Calvente}}},
\bauthor{\binits{P.} \bsnm{{Amaro-Seoane}}},
\bauthor{\binits{H.} \bsnm{{Baumgardt}}},
\batitle{{The distribution of stars around the Milky Way's central black hole.
  I. Deep star counts}}.
\bjtitle{\aap}
\bvolume{609},
\bfpage{26}
(\byear{2018}).
doi:\doiurl{10.1051/0004-6361/201730451}
\end{barticle}
\endbibitem

\bibitem[\protect\citeauthoryear{{Gebhardt} et~al.}{2002}]{GRH02}
\begin{barticle}
\bauthor{\binits{K.} \bsnm{{Gebhardt}}},
\bauthor{\binits{R.M.} \bsnm{{Rich}}},
\bauthor{\binits{L.C.} \bsnm{{Ho}}},
\batitle{{A $20000 M_\odot$ Black Hole in the Stellar Cluster G1}}.
\bjtitle{ApJ Lett.}
\bvolume{578},
\bfpage{41}--\blpage{45}
(\byear{2002})
\end{barticle}
\endbibitem

\bibitem[\protect\citeauthoryear{{Genzel} et~al.}{2010}]{2010RvMP...82.3121G}
\begin{barticle}
\bauthor{\binits{R.} \bsnm{{Genzel}}},
\bauthor{\binits{F.} \bsnm{{Eisenhauer}}},
\bauthor{\binits{S.} \bsnm{{Gillessen}}},
\batitle{{The Galactic Center massive black hole and nuclear star cluster}}.
\bjtitle{Reviews of Modern Physics}
\bvolume{82}(\bissue{4}),
\bfpage{3121}--\blpage{3195}
(\byear{2010}).
doi:\doiurl{10.1103/RevModPhys.82.3121}
\end{barticle}
\endbibitem

\bibitem[\protect\citeauthoryear{{Gerssen} et~al.}{2002}]{GerssenEtAl02}
\begin{barticle}
\bauthor{\binits{J.} \bsnm{{Gerssen}}},
\bauthor{\binits{R.P.} \bsnm{{van der Marel}}},
\bauthor{\binits{K.} \bsnm{{Gebhardt}}},
\bauthor{\binits{P.} \bsnm{{Guhathakurta}}},
\bauthor{\binits{R.C.} \bsnm{{Peterson}}},
\bauthor{\binits{C.} \bsnm{{Pryor}}},
\batitle{{Hubble Space Telescope Evidence for an Intermediate-Mass Black Hole
  in the Globular Cluster M15. II. Kinematic Analysis and Dynamical Modeling}}.
\bjtitle{AJ}
\bvolume{124},
\bfpage{3270}--\blpage{3288}
(\byear{2002})
\end{barticle}
\endbibitem

\bibitem[\protect\citeauthoryear{{Gieles}}{2009}]{Gieles2009}
\begin{barticle}
\bauthor{\binits{M.} \bsnm{{Gieles}}},
\batitle{{The early evolution of the star cluster mass function}}.
\bjtitle{MNRAS}
\bvolume{394}(\bissue{4}),
\bfpage{2113}--\blpage{2126}
(\byear{2009}).
doi:\doiurl{10.1111/j.1365-2966.2009.14473.x}
\end{barticle}
\endbibitem

\bibitem[\protect\citeauthoryear{{Giersz}}{1998}]{Giersz98}
\begin{barticle}
\bauthor{\binits{M.} \bsnm{{Giersz}}},
\batitle{Monte carlo simulations of star clusters - i. first results}.
\bjtitle{MNRAS}
\bvolume{298},
\bfpage{1239}--\blpage{1248}
(\byear{1998})
\end{barticle}
\endbibitem

\bibitem[\protect\citeauthoryear{{Giersz} and {Heggie}}{1994}]{GH94a}
\begin{barticle}
\bauthor{\binits{M.} \bsnm{{Giersz}}},
\bauthor{\binits{D.C.} \bsnm{{Heggie}}},
\batitle{Statistics of n-body simulations - part one - equal masses before core
  collapse}.
\bjtitle{MNRAS}
\bvolume{268},
\bfpage{257}
(\byear{1994})
\end{barticle}
\endbibitem

\bibitem[\protect\citeauthoryear{{Giersz} et~al.}{2015}]{GierszEtAl2015}
\begin{barticle}
\bauthor{\binits{M.} \bsnm{{Giersz}}},
\bauthor{\binits{N.} \bsnm{{Leigh}}},
\bauthor{\binits{A.} \bsnm{{Hypki}}},
\bauthor{\binits{N.} \bsnm{{L{\"u}tzgendorf}}},
\bauthor{\binits{A.} \bsnm{{Askar}}},
\batitle{{MOCCA code for star cluster simulations - IV. A new scenario for
  intermediate mass black hole formation in globular clusters}}.
\bjtitle{MNRAS}
\bvolume{454}(\bissue{3}),
\bfpage{3150}--\blpage{3165}
(\byear{2015}).
doi:\doiurl{10.1093/mnras/stv2162}
\end{barticle}
\endbibitem

\bibitem[\protect\citeauthoryear{{Gillessen} et~al.}{2006}]{GillessenEtAl06}
\begin{bchapter}
\bauthor{\binits{S.} \bsnm{{Gillessen}}},
\bauthor{\binits{G.} \bsnm{{Perrin}}},
\bauthor{\binits{W.} \bsnm{{Brandner}}},
\bauthor{\binits{C.} \bsnm{{Straubmeier}}},
\bauthor{\binits{F.} \bsnm{{Eisenhauer}}},
\bauthor{\binits{S.} \bsnm{{Rabien}}},
\bauthor{\binits{A.} \bsnm{{Eckart}}},
\bauthor{\binits{P.} \bsnm{{Lena}}},
\bauthor{\binits{R.} \bsnm{{Genzel}}},
\bauthor{\binits{T.} \bsnm{{Paumard}}},
\bauthor{\binits{S.} \bsnm{{Hippler}}},
\bctitle{{GRAVITY: the adaptive-optics-assisted two-object beam combiner
  instrument for the VLTI}},
in \bbtitle{Society of Photo-Optical Instrumentation Engineers (SPIE)
  Conference Series}.
\bsertitle{Society of Photo-Optical Instrumentation Engineers (SPIE) Conference
  Series},
vol. \bseriesno{6268},
\byear{2006}.
doi:\doiurl{10.1117/12.671431}
\end{bchapter}
\endbibitem

\bibitem[\protect\citeauthoryear{{Glampedakis}
  et~al.}{2002a}]{GlampedakisEtAl2002}
\begin{barticle}
\bauthor{\binits{K.} \bsnm{{Glampedakis}}},
\bauthor{\binits{S.A.} \bsnm{{Hughes}}},
\bauthor{\binits{D.} \bsnm{{Kennefick}}},
\batitle{{Approximating the inspiral of test bodies into Kerr black holes}}.
\bjtitle{Phys. Rev. D}
\bvolume{66}(\bissue{6}),
\bfpage{064005}
(\byear{2002}a).
doi:\doiurl{10.1103/PhysRevD.66.064005}
\end{barticle}
\endbibitem

\bibitem[\protect\citeauthoryear{{Glampedakis}
  et~al.}{2002b}]{GlampedakisEtAl02}
\begin{barticle}
\bauthor{\binits{K.} \bsnm{{Glampedakis}}},
\bauthor{\binits{S.A.} \bsnm{{Hughes}}},
\bauthor{\binits{D.} \bsnm{{Kennefick}}},
\batitle{{Approximating the inspiral of test bodies into Kerr black holes}}.
\bjtitle{Phys. Rev. D}
\bvolume{66}(\bissue{6}),
\bfpage{064005}
(\byear{2002}b)
\end{barticle}
\endbibitem

\bibitem[\protect\citeauthoryear{{Glampedakis} and
  {Kennefick}}{2002}]{GlampedakisEtAl2002b}
\begin{barticle}
\bauthor{\binits{K.} \bsnm{{Glampedakis}}},
\bauthor{\binits{D.} \bsnm{{Kennefick}}},
\batitle{{Zoom and whirl: Eccentric equatorial orbits around spinning black
  holes and their evolution under gravitational radiation reaction}}.
\bjtitle{Phys. Rev. D}
\bvolume{66}(\bissue{4}),
\bfpage{044002}
(\byear{2002}).
doi:\doiurl{10.1103/PhysRevD.66.044002}
\end{barticle}
\endbibitem

\bibitem[\protect\citeauthoryear{{Gonz{\'a}lez}
  et~al.}{2007}]{GonzalezEtAl2007}
\begin{barticle}
\bauthor{\binits{J.A.} \bsnm{{Gonz{\'a}lez}}},
\bauthor{\binits{U.} \bsnm{{Sperhake}}},
\bauthor{\binits{B.} \bsnm{{Br{\"u}gmann}}},
\bauthor{\binits{M.} \bsnm{{Hannam}}},
\bauthor{\binits{S.} \bsnm{{Husa}}},
\batitle{{Maximum Kick from Nonspinning Black-Hole Binary Inspiral}}.
\bjtitle{Physical Review Letters}
\bvolume{98}(\bissue{9}),
\bfpage{091101}
(\byear{2007}).
doi:\doiurl{10.1103/PhysRevLett.98.091101}
\end{barticle}
\endbibitem

\bibitem[\protect\citeauthoryear{{Gravity Collaboration}
  et~al.}{2020}]{2020A&A...636L...5G}
\begin{barticle}
\bauthor{\bsnm{{Gravity Collaboration}}},
\bauthor{\binits{R.} \bsnm{{Abuter}}},
\bauthor{\binits{A.} \bsnm{{Amorim}}},
\bauthor{\binits{M.} \bsnm{{Baub{\"o}ck}}},
\bauthor{\binits{J.P.} \bsnm{{Berger}}},
\bauthor{\binits{H.} \bsnm{{Bonnet}}},
\bauthor{\binits{W.} \bsnm{{Brand ner}}},
\bauthor{\binits{V.} \bsnm{{Cardoso}}},
\bauthor{\binits{Y.} \bsnm{{Cl{\'e}net}}},
\bauthor{\binits{P.T.} \bsnm{{de Zeeuw}}},
\bauthor{\binits{J.} \bsnm{{Dexter}}},
\bauthor{\binits{A.} \bsnm{{Eckart}}},
\bauthor{\binits{F.} \bsnm{{Eisenhauer}}},
\bauthor{\binits{N.M.} \bsnm{{F{\"o}rster Schreiber}}},
\bauthor{\binits{P.} \bsnm{{Garcia}}},
\bauthor{\binits{F.} \bsnm{{Gao}}},
\bauthor{\binits{E.} \bsnm{{Gendron}}},
\bauthor{\binits{R.} \bsnm{{Genzel}}},
\bauthor{\binits{S.} \bsnm{{Gillessen}}},
\bauthor{\binits{M.} \bsnm{{Habibi}}},
\bauthor{\binits{X.} \bsnm{{Haubois}}},
\bauthor{\binits{T.} \bsnm{{Henning}}},
\bauthor{\binits{S.} \bsnm{{Hippler}}},
\bauthor{\binits{M.} \bsnm{{Horrobin}}},
\bauthor{\binits{A.} \bsnm{{Jim{\'e}nez-Rosales}}},
\bauthor{\binits{L.} \bsnm{{Jochum}}},
\bauthor{\binits{L.} \bsnm{{Jocou}}},
\bauthor{\binits{A.} \bsnm{{Kaufer}}},
\bauthor{\binits{P.} \bsnm{{Kervella}}},
\bauthor{\binits{S.} \bsnm{{Lacour}}},
\bauthor{\binits{V.} \bsnm{{Lapeyr{\`e}re}}},
\bauthor{\binits{J.-B.} \bsnm{{Le Bouquin}}},
\bauthor{\binits{P.} \bsnm{{L{\'e}na}}},
\bauthor{\binits{M.} \bsnm{{Nowak}}},
\bauthor{\binits{T.} \bsnm{{Ott}}},
\bauthor{\binits{T.} \bsnm{{Paumard}}},
\bauthor{\binits{K.} \bsnm{{Perraut}}},
\bauthor{\binits{G.} \bsnm{{Perrin}}},
\bauthor{\binits{O.} \bsnm{{Pfuhl}}},
\bauthor{\binits{G.} \bsnm{{Rodr{\'\i}guez-Coira}}},
\bauthor{\binits{J.} \bsnm{{Shangguan}}},
\bauthor{\binits{S.} \bsnm{{Scheithauer}}},
\bauthor{\binits{J.} \bsnm{{Stadler}}},
\bauthor{\binits{O.} \bsnm{{Straub}}},
\bauthor{\binits{C.} \bsnm{{Straubmeier}}},
\bauthor{\binits{E.} \bsnm{{Sturm}}},
\bauthor{\binits{L.J.} \bsnm{{Tacconi}}},
\bauthor{\binits{F.} \bsnm{{Vincent}}},
\bauthor{\binits{S.} \bsnm{{von Fellenberg}}},
\bauthor{\binits{I.} \bsnm{{Waisberg}}},
\bauthor{\binits{F.} \bsnm{{Widmann}}},
\bauthor{\binits{E.} \bsnm{{Wieprecht}}},
\bauthor{\binits{E.} \bsnm{{Wiezorrek}}},
\bauthor{\binits{J.} \bsnm{{Woillez}}},
\bauthor{\binits{S.} \bsnm{{Yazici}}},
\bauthor{\binits{G.} \bsnm{{Zins}}},
\batitle{{Detection of the Schwarzschild precession in the orbit of the star S2
  near the Galactic centre massive black hole}}.
\bjtitle{\aap}
\bvolume{636},
\bfpage{5}
(\byear{2020}).
doi:\doiurl{10.1051/0004-6361/202037813}
\end{barticle}
\endbibitem

\bibitem[\protect\citeauthoryear{{Guillochon} and
  {Ramirez-Ruiz}}{2013}]{GuillochonRamirez-Ruiz2013}
\begin{barticle}
\bauthor{\binits{J.} \bsnm{{Guillochon}}},
\bauthor{\binits{E.} \bsnm{{Ramirez-Ruiz}}},
\batitle{{Hydrodynamical Simulations to Determine the Feeding Rate of Black
  Holes by the Tidal Disruption of Stars: The Importance of the Impact
  Parameter and Stellar Structure}}.
\bjtitle{ApJ}
\bvolume{767},
\bfpage{25}
(\byear{2013}).
doi:\doiurl{10.1088/0004-637X/767/1/25}
\end{barticle}
\endbibitem

\bibitem[\protect\citeauthoryear{{G{\"u}ltekin}
  et~al.}{2004}]{GuelketinEtAl2004}
\begin{barticle}
\bauthor{\binits{K.} \bsnm{{G{\"u}ltekin}}},
\bauthor{\binits{M.C.} \bsnm{{Miller}}},
\bauthor{\binits{D.P.} \bsnm{{Hamilton}}},
\batitle{{Growth of Intermediate-Mass Black Holes in Globular Clusters}}.
\bjtitle{ApJ}
\bvolume{616},
\bfpage{221}--\blpage{230}
(\byear{2004}).
doi:\doiurl{10.1086/424809}
\end{barticle}
\endbibitem

\bibitem[\protect\citeauthoryear{{Haehnelt}}{1994}]{Haehnelt1994}
\begin{barticle}
\bauthor{\binits{M.G.} \bsnm{{Haehnelt}}},
\batitle{{Low-Frequency Gravitational Waves from Supermassive Black-Holes}}.
\bjtitle{MNRAS}
\bvolume{269},
\bfpage{199}
(\byear{1994}).
doi:\doiurl{10.1093/mnras/269.1.199}
\end{barticle}
\endbibitem

\bibitem[\protect\citeauthoryear{Hannuksela et~al.}{2019}]{Hannuksela:2018izj}
\begin{barticle}
\bauthor{\binits{O.A.} \bsnm{Hannuksela}},
\bauthor{\binits{K.W.K.} \bsnm{Wong}},
\bauthor{\binits{R.} \bsnm{Brito}},
\bauthor{\binits{E.} \bsnm{Berti}},
\bauthor{\binits{T.G.F.} \bsnm{Li}},
\batitle{{Probing the existence of ultralight bosons with a single
  gravitational-wave measurement}}.
\bjtitle{Nature Astron.}
\bvolume{3}(\bissue{5}),
\bfpage{447}--\blpage{451}
(\byear{2019}).
doi:\doiurl{10.1038/s41550-019-0712-4}
\end{barticle}
\endbibitem

\bibitem[\protect\citeauthoryear{{Hansen}}{1972}]{Hansen1972}
\begin{barticle}
\bauthor{\binits{R.O.} \bsnm{{Hansen}}},
\batitle{{Post-Newtonian Gravitational Radiation from Point Masses in a
  Hyperbolic Kepler Orbit}}.
\bjtitle{Phys. Rev. D}
\bvolume{5},
\bfpage{1021}--\blpage{1023}
(\byear{1972}).
doi:\doiurl{10.1103/PhysRevD.5.1021}
\end{barticle}
\endbibitem

\bibitem[\protect\citeauthoryear{{Haster} et~al.}{2016}]{HasterEtAl2016}
\begin{barticle}
\bauthor{\binits{C.-J.} \bsnm{{Haster}}},
\bauthor{\binits{F.} \bsnm{{Antonini}}},
\bauthor{\binits{V.} \bsnm{{Kalogera}}},
\bauthor{\binits{I.} \bsnm{{Mandel}}},
\batitle{{N-Body Dynamics of Intermediate Mass-ratio Inspirals in Star
  Clusters}}.
\bjtitle{ApJ}
\bvolume{832},
\bfpage{192}
(\byear{2016}).
doi:\doiurl{10.3847/0004-637X/832/2/192}
\end{barticle}
\endbibitem

\bibitem[\protect\citeauthoryear{{Helstrom}}{1968}]{Helstrom68}
\begin{bbook}
\bauthor{\binits{C.W.} \bsnm{{Helstrom}}},
\bbtitle{{Statistical Theory of Signal Detection}}
\byear{1968}
\end{bbook}
\endbibitem

\bibitem[\protect\citeauthoryear{{H{\'{e}}non}}{1973}]{Henon73}
\begin{bchapter}
\bauthor{\binits{M.} \bsnm{{H{\'{e}}non}}},
\bctitle{Collisional dynamics of spherical stellar systems},
in \bbtitle{Dynamical structure and evolution of stellar systems, Lectures of
  the 3rd Advanced Course of the Swiss Society for Astronomy and Astrophysics
  (SSAA)},
ed. by \beditor{\binits{L.} \bsnm{{Martinet}}},
\beditor{\binits{M.} \bsnm{{Mayor}}},
\byear{1973},
pp. \bfpage{183}--\blpage{260}
\end{bchapter}
\endbibitem

\bibitem[\protect\citeauthoryear{{H{\'e}non}}{1975}]{Henon75}
\begin{bchapter}
\bauthor{\binits{M.} \bsnm{{H{\'e}non}}},
\bctitle{Two Recent Developments Concerning the Monte Carlo Method},
in \bbtitle{IAU Symp. 69: Dynamics of Stellar Systems},
ed. by \beditor{\binits{A.} \bsnm{{Hayli}}},
\byear{1975},
p. \bfpage{133}
\end{bchapter}
\endbibitem

\bibitem[\protect\citeauthoryear{{Hild} et~al.}{2011}]{HildEtAl2011}
\begin{barticle}
\bauthor{\binits{S.} \bsnm{{Hild}}},
\bauthor{\binits{M.} \bsnm{{Abernathy}}},
\bauthor{\binits{F.} \bsnm{{Acernese}}},
\bauthor{\binits{P.} \bsnm{{Amaro-Seoane}}},
\bauthor{\bsnm{{et al}}},
\batitle{{Sensitivity studies for third-generation gravitational wave
  observatories}}.
\bjtitle{Classical and Quantum Gravity}
\bvolume{28}(\bissue{9}),
\bfpage{094013}
(\byear{2011}).
doi:\doiurl{10.1088/0264-9381/28/9/094013}
\end{barticle}
\endbibitem

\bibitem[\protect\citeauthoryear{{Hills}}{1988}]{Hills88}
\begin{barticle}
\bauthor{\binits{J.G.} \bsnm{{Hills}}},
\batitle{Hyper-velocity and tidal stars from binaries disrupted by a massive
  galactic black hole}.
\bjtitle{Nat}
\bvolume{331},
\bfpage{687}--\blpage{689}
(\byear{1988})
\end{barticle}
\endbibitem

\bibitem[\protect\citeauthoryear{{Hils} and {Bender}}{1995}]{HilsBender1995}
\begin{barticle}
\bauthor{\binits{D.} \bsnm{{Hils}}},
\bauthor{\binits{P.L.} \bsnm{{Bender}}},
\batitle{{Gradual approach to coalescence for compact stars orbiting massive
  black holes}}.
\bjtitle{ApJ Lett.}
\bvolume{445},
\bfpage{7}--\blpage{10}
(\byear{1995}).
doi:\doiurl{10.1086/187876}
\end{barticle}
\endbibitem

\bibitem[\protect\citeauthoryear{{Hong} and {Lee}}{2015}]{HongLee2015}
\begin{barticle}
\bauthor{\binits{J.} \bsnm{{Hong}}},
\bauthor{\binits{H.M.} \bsnm{{Lee}}},
\batitle{{Black hole binaries in galactic nuclei and gravitational wave
  sources}}.
\bjtitle{MNRAS}
\bvolume{448},
\bfpage{754}--\blpage{770}
(\byear{2015}).
doi:\doiurl{10.1093/mnras/stv035}
\end{barticle}
\endbibitem

\bibitem[\protect\citeauthoryear{{Hopman} and
  {Alexander}}{2006}]{2006ApJ...645.1152H}
\begin{barticle}
\bauthor{\binits{C.} \bsnm{{Hopman}}},
\bauthor{\binits{T.} \bsnm{{Alexander}}},
\batitle{{Resonant Relaxation near a Massive Black Hole: The Stellar
  Distribution and Gravitational Wave Sources}}.
\bjtitle{\apj}
\bvolume{645}(\bissue{2}),
\bfpage{1152}--\blpage{1163}
(\byear{2006}).
doi:\doiurl{10.1086/504400}
\end{barticle}
\endbibitem

\bibitem[\protect\citeauthoryear{{Hopman} et~al.}{2007}]{2007MNRAS.378..129H}
\begin{barticle}
\bauthor{\binits{C.} \bsnm{{Hopman}}},
\bauthor{\binits{M.} \bsnm{{Freitag}}},
\bauthor{\binits{S.L.} \bsnm{{Larson}}},
\batitle{{Gravitational wave bursts from the Galactic massive black hole}}.
\bjtitle{\mnras}
\bvolume{378}(\bissue{1}),
\bfpage{129}--\blpage{136}
(\byear{2007}).
doi:\doiurl{10.1111/j.1365-2966.2007.11758.x}
\end{barticle}
\endbibitem

\bibitem[\protect\citeauthoryear{{Hughes}}{2001}]{Hughes2001}
\begin{barticle}
\bauthor{\binits{S.A.} \bsnm{{Hughes}}},
\batitle{{Evolution of circular, nonequatorial orbits of Kerr black holes due
  to gravitational-wave emission. II. Inspiral trajectories and gravitational
  waveforms}}.
\bjtitle{Ph. Rev. D}
\bvolume{64}(\bissue{6}),
\bfpage{064004}
(\byear{2001}).
doi:\doiurl{10.1103/PhysRevD.64.064004}
\end{barticle}
\endbibitem

\bibitem[\protect\citeauthoryear{{Kepler}}{1619}]{Kepler1619}
\begin{botherref}
\oauthor{\binits{J.} \bsnm{{Kepler}}},
Harmonices mundi libri v.
Publisher Linz
(1619)
\end{botherref}
\endbibitem

\bibitem[\protect\citeauthoryear{{Kocsis} et~al.}{2006}]{KocsisEtAl2006}
\begin{barticle}
\bauthor{\binits{B.} \bsnm{{Kocsis}}},
\bauthor{\binits{M.E.} \bsnm{{G{\'a}sp{\'a}r}}},
\bauthor{\binits{S.} \bsnm{{M{\'a}rka}}},
\batitle{{Detection Rate Estimates of Gravity Waves Emitted during Parabolic
  Encounters of Stellar Black Holes in Globular Clusters}}.
\bjtitle{ApJ}
\bvolume{648},
\bfpage{411}--\blpage{429}
(\byear{2006}).
doi:\doiurl{10.1086/505641}
\end{barticle}
\endbibitem

\bibitem[\protect\citeauthoryear{{Konstantinidis}
  et~al.}{2013}]{KonstantinidisEtAl2013}
\begin{barticle}
\bauthor{\binits{S.} \bsnm{{Konstantinidis}}},
\bauthor{\binits{P.} \bsnm{{Amaro-Seoane}}},
\bauthor{\binits{K.D.} \bsnm{{Kokkotas}}},
\batitle{{Investigating the retention of intermediate-mass black holes in star
  clusters using N-body simulations}}.
\bjtitle{A\&A}
\bvolume{557},
\bfpage{135}
(\byear{2013}).
doi:\doiurl{10.1051/0004-6361/201219620}
\end{barticle}
\endbibitem

\bibitem[\protect\citeauthoryear{{Kormendy} and {Ho}}{2013}]{KormendyHo2013}
\begin{barticle}
\bauthor{\binits{J.} \bsnm{{Kormendy}}},
\bauthor{\binits{L.C.} \bsnm{{Ho}}},
\batitle{{Coevolution (Or Not) of Supermassive Black Holes and Host Galaxies}}.
\bjtitle{ARA\&A}
\bvolume{51},
\bfpage{511}--\blpage{653}
(\byear{2013}).
doi:\doiurl{10.1146/annurev-astro-082708-101811}
\end{barticle}
\endbibitem

\bibitem[\protect\citeauthoryear{{Kormendy}}{2004}]{2004cbhg.symp....1K}
\begin{bchapter}
\bauthor{\binits{J.} \bsnm{{Kormendy}}},
\bctitle{{The Stellar-Dynamical Search for Supermassive Black Holes in Galactic
  Nuclei}},
in \bbtitle{Coevolution of Black Holes and Galaxies},
ed. by \beditor{\binits{L.C.} \bsnm{{Ho}}},
\byear{2004},
p. \bfpage{1}
\end{bchapter}
\endbibitem

\bibitem[\protect\citeauthoryear{{Kormendy} and
  {Ho}}{2013}]{2013ARA&A..51..511K}
\begin{barticle}
\bauthor{\binits{J.} \bsnm{{Kormendy}}},
\bauthor{\binits{L.C.} \bsnm{{Ho}}},
\batitle{{Coevolution (Or Not) of Supermassive Black Holes and Host Galaxies}}.
\bjtitle{\araa}
\bvolume{51}(\bissue{1}),
\bfpage{511}--\blpage{653}
(\byear{2013}).
doi:\doiurl{10.1146/annurev-astro-082708-101811}
\end{barticle}
\endbibitem

\bibitem[\protect\citeauthoryear{{Kr{\'o}lak} et~al.}{1995}]{KrolakEtAl1995}
\begin{barticle}
\bauthor{\binits{A.} \bsnm{{Kr{\'o}lak}}},
\bauthor{\binits{K.D.} \bsnm{{Kokkotas}}},
\bauthor{\binits{G.} \bsnm{{Sch{\"a}fer}}},
\batitle{{Estimation of the post-Newtonian parameters in the gravitational-wave
  emission of a coalescing binary}}.
\bjtitle{Phys. Rev. D}
\bvolume{52},
\bfpage{2089}--\blpage{2111}
(\byear{1995}).
doi:\doiurl{10.1103/PhysRevD.52.2089}
\end{barticle}
\endbibitem

\bibitem[\protect\citeauthoryear{{Kroupa} et~al.}{2013}]{KroupaEtAl2013}
\begin{botherref}
\oauthor{\binits{P.} \bsnm{{Kroupa}}},
\oauthor{\binits{C.} \bsnm{{Weidner}}},
\oauthor{\binits{J.} \bsnm{{Pflamm-Altenburg}}},
\oauthor{\binits{I.} \bsnm{{Thies}}},
\oauthor{\binits{J.} \bsnm{{Dabringhausen}}},
\oauthor{\binits{M.} \bsnm{{Marks}}},
\oauthor{\binits{T.} \bsnm{{Maschberger}}},
{The Stellar and Sub-Stellar Initial Mass Function of Simple and Composite
  Populations},
ed. by T.D. {Oswalt}, G. {Gilmore}
2013,
p. 115.
doi:\doiurl{10.1007/978-94-007-5612-0\_4}
\end{botherref}
\endbibitem

\bibitem[\protect\citeauthoryear{{Kroupa}}{2001}]{2001MNRAS.322..231K}
\begin{barticle}
\bauthor{\binits{P.} \bsnm{{Kroupa}}},
\batitle{{On the variation of the initial mass function}}.
\bjtitle{\mnras}
\bvolume{322}(\bissue{2}),
\bfpage{231}--\blpage{246}
(\byear{2001}).
doi:\doiurl{10.1046/j.1365-8711.2001.04022.x}
\end{barticle}
\endbibitem

\bibitem[\protect\citeauthoryear{{Kupi} et~al.}{2006}]{KupiEtAl06}
\begin{botherref}
\oauthor{\binits{G.} \bsnm{{Kupi}}},
\oauthor{\binits{P.} \bsnm{{Amaro-Seoane}}},
\oauthor{\binits{R.} \bsnm{{Spurzem}}},
{Dynamics of compact object clusters: a post-Newtonian study}.
MNRAS,
77
(2006).
doi:\doiurl{10.1111/j.1745-3933.2006.00205.x}
\end{botherref}
\endbibitem

\bibitem[\protect\citeauthoryear{{Larson}}{1970}]{Larson70a}
\begin{barticle}
\bauthor{\binits{R.B.} \bsnm{{Larson}}},
\batitle{A method for computing the evolution of star clusters}.
\bjtitle{MNRAS}
\bvolume{147},
\bfpage{323}
(\byear{1970})
\end{barticle}
\endbibitem

\bibitem[\protect\citeauthoryear{{Lee} et~al.}{2010}]{LeeEtAl2010}
\begin{barticle}
\bauthor{\binits{W.H.} \bsnm{{Lee}}},
\bauthor{\binits{E.} \bsnm{{Ramirez-Ruiz}}},
\bauthor{\binits{G.} \bsnm{{van de Ven}}},
\batitle{{Short Gamma-ray Bursts from Dynamically Assembled Compact Binaries in
  Globular Clusters: Pathways, Rates, Hydrodynamics, and Cosmological
  Setting}}.
\bjtitle{ApJ}
\bvolume{720},
\bfpage{953}--\blpage{975}
(\byear{2010}).
doi:\doiurl{10.1088/0004-637X/720/1/953}
\end{barticle}
\endbibitem

\bibitem[\protect\citeauthoryear{{Leigh} et~al.}{2014}]{LeighEtAl2014}
\begin{barticle}
\bauthor{\binits{N.W.C.} \bsnm{{Leigh}}},
\bauthor{\binits{N.} \bsnm{{L{\"u}tzgendorf}}},
\bauthor{\binits{A.M.} \bsnm{{Geller}}},
\bauthor{\binits{T.J.} \bsnm{{Maccarone}}},
\bauthor{\binits{C.} \bsnm{{Heinke}}},
\bauthor{\binits{A.} \bsnm{{Sesana}}},
\batitle{{On the coexistence of stellar-mass and intermediate-mass black holes
  in globular clusters}}.
\bjtitle{MNRAS}
\bvolume{444},
\bfpage{29}--\blpage{42}
(\byear{2014}).
doi:\doiurl{10.1093/mnras/stu1437}
\end{barticle}
\endbibitem

\bibitem[\protect\citeauthoryear{{Lightman} and {Shapiro}}{1977}]{LS77}
\begin{barticle}
\bauthor{\binits{A.P.} \bsnm{{Lightman}}},
\bauthor{\binits{S.L.} \bsnm{{Shapiro}}},
\batitle{The distribution and consumption rate of stars around a massive,
  collapsed object}.
\bjtitle{ApJ}
\bvolume{211},
\bfpage{244}--\blpage{262}
(\byear{1977})
\end{barticle}
\endbibitem

\bibitem[\protect\citeauthoryear{{LIGO Scientific Collaboration} and {Virgo
  Collaboration}}{2020a}]{AbbottEtAlIMBH2020}
\begin{barticle}
\bauthor{\bsnm{{LIGO Scientific Collaboration}}},
\bauthor{\bsnm{{Virgo Collaboration}}},
\batitle{{GW190521: A Binary Black Hole Merger with a Total Mass of 150
  M$_{\odot}$}}.
\bjtitle{Ph. Rv. L}
\bvolume{125}(\bissue{10}),
\bfpage{101102}
(\byear{2020}a).
doi:\doiurl{10.1103/PhysRevLett.125.101102}
\end{barticle}
\endbibitem

\bibitem[\protect\citeauthoryear{{LIGO Scientific Collaboration} and {Virgo
  Collaboration}}{2020b}]{AbbottEtAlGWTC-2_2020}
\begin{botherref}
\oauthor{\bsnm{{LIGO Scientific Collaboration}}},
\oauthor{\bsnm{{Virgo Collaboration}}},
{GWTC-2: Compact Binary Coalescences Observed by LIGO and Virgo During the
  First Half of the Third Observing Run}.
arXiv e-prints,
2010--14527
(2020b)
\end{botherref}
\endbibitem

\bibitem[\protect\citeauthoryear{{LIGO Scientific Collaboration} and {Virgo
  Collaboration}}{2020c}]{AbbottEtAlIMBHb2020}
\begin{barticle}
\bauthor{\bsnm{{LIGO Scientific Collaboration}}},
\bauthor{\bsnm{{Virgo Collaboration}}},
\batitle{{Properties and Astrophysical Implications of the 150 M$_{\odot}$
  Binary Black Hole Merger GW190521}}.
\bjtitle{ApJ Lett.}
\bvolume{900}(\bissue{1}),
\bfpage{13}
(\byear{2020}c).
doi:\doiurl{10.3847/2041-8213/aba493}
\end{barticle}
\endbibitem

\bibitem[\protect\citeauthoryear{{Lin} and {Tremaine}}{1980}]{LinTremaine1980}
\begin{barticle}
\bauthor{\binits{D.N.C.} \bsnm{{Lin}}},
\bauthor{\binits{S.} \bsnm{{Tremaine}}},
\batitle{{A reinvestigation of the standard model for the dynamics of a massive
  black hole in a globular cluster}}.
\bjtitle{ApJ}
\bvolume{242},
\bfpage{789}--\blpage{798}
(\byear{1980}).
doi:\doiurl{10.1086/158513}
\end{barticle}
\endbibitem

\bibitem[\protect\citeauthoryear{{Lindblom} et~al.}{2008}]{LindblomEtAl2008}
\begin{barticle}
\bauthor{\binits{L.} \bsnm{{Lindblom}}},
\bauthor{\binits{B.J.} \bsnm{{Owen}}},
\bauthor{\binits{D.A.} \bsnm{{Brown}}},
\batitle{{Model waveform accuracy standards for gravitational wave data
  analysis}}.
\bjtitle{Phys. Rev. D}
\bvolume{78}(\bissue{12}),
\bfpage{124020}
(\byear{2008}).
doi:\doiurl{10.1103/PhysRevD.78.124020}
\end{barticle}
\endbibitem

\bibitem[\protect\citeauthoryear{Lousto and Zlochower}{2011}]{Lousto2011}
\begin{barticle}
\bauthor{\binits{C.O.} \bsnm{Lousto}},
\bauthor{\binits{Y.} \bsnm{Zlochower}},
\batitle{{Orbital Evolution of Extreme-Mass-Ratio Black-Hole Binaries with
  Numerical Relativity}}.
\bjtitle{Phys. Rev. Lett.}
\bvolume{106},
\bfpage{041101}
(\byear{2011}).
doi:\doiurl{10.1103/PhysRevLett.106.041101}
\end{barticle}
\endbibitem

\bibitem[\protect\citeauthoryear{{L{\"u}tzgendorf}
  et~al.}{2013}]{LuetzgendorfEtAl2013}
\begin{barticle}
\bauthor{\binits{N.} \bsnm{{L{\"u}tzgendorf}}},
\bauthor{\binits{M.} \bsnm{{Kissler-Patig}}},
\bauthor{\binits{N.} \bsnm{{Neumayer}}},
\bauthor{\binits{H.} \bsnm{{Baumgardt}}},
\bauthor{\binits{E.} \bsnm{{Noyola}}},
\bauthor{\binits{P.T.} \bsnm{{de Zeeuw}}},
\bauthor{\binits{K.} \bsnm{{Gebhardt}}},
\bauthor{\binits{B.} \bsnm{{Jalali}}},
\bauthor{\binits{A.} \bsnm{{Feldmeier}}},
\batitle{{M$_{•}$ - {$\sigma$}relation for intermediate-mass black holes in
  globular clusters}}.
\bjtitle{A\&A}
\bvolume{555},
\bfpage{26}
(\byear{2013}).
doi:\doiurl{10.1051/0004-6361/201321183}
\end{barticle}
\endbibitem

\bibitem[\protect\citeauthoryear{{Lynden-Bell} and
  {Kalnajs}}{1972}]{Lynden-BellKalnajs1972}
\begin{barticle}
\bauthor{\binits{D.} \bsnm{{Lynden-Bell}}},
\bauthor{\binits{A.J.} \bsnm{{Kalnajs}}},
\batitle{{On the generating mechanism of spiral structure}}.
\bjtitle{MNRAS}
\bvolume{157},
\bfpage{1}
(\byear{1972}).
doi:\doiurl{10.1093/mnras/157.1.1}
\end{barticle}
\endbibitem

\bibitem[\protect\citeauthoryear{{MacLeod} et~al.}{2016}]{MacLeodEtAl2016}
\begin{barticle}
\bauthor{\binits{M.} \bsnm{{MacLeod}}},
\bauthor{\binits{M.} \bsnm{{Trenti}}},
\bauthor{\binits{E.} \bsnm{{Ramirez-Ruiz}}},
\batitle{{The Close Stellar Companions to Intermediate-mass Black Holes}}.
\bjtitle{ApJ}
\bvolume{819},
\bfpage{70}
(\byear{2016}).
doi:\doiurl{10.3847/0004-637X/819/1/70}
\end{barticle}
\endbibitem

\bibitem[\protect\citeauthoryear{Maggiore}{2018}]{Maggiore022018}
\begin{bbook}
\bauthor{\binits{M.} \bsnm{Maggiore}},
\bbtitle{Gravitational Waves: Volume 2: Astrophysics and Cosmology}.
\bsertitle{Gravitational Waves}
(\bpublisher{Oxford University Press}, \blocation{???}, \byear{2018}).
\bisbn{9780198570899}.
\burl{https://books.google.de/books?id=3ZNODwAAQBAJ}
\end{bbook}
\endbibitem

\bibitem[\protect\citeauthoryear{{Magorrian} and {Tremaine}}{1999}]{MT99}
\begin{barticle}
\bauthor{\binits{J.} \bsnm{{Magorrian}}},
\bauthor{\binits{S.} \bsnm{{Tremaine}}},
\batitle{Rates of tidal disruption of stars by massive central black holes}.
\bjtitle{MNRAS}
\bvolume{309},
\bfpage{447}--\blpage{460}
(\byear{1999})
\end{barticle}
\endbibitem

\bibitem[\protect\citeauthoryear{{Maguire} et~al.}{2020}]{2020SSRv..216...39M}
\begin{barticle}
\bauthor{\binits{K.} \bsnm{{Maguire}}},
\bauthor{\binits{M.} \bsnm{{Eracleous}}},
\bauthor{\binits{P.G.} \bsnm{{Jonker}}},
\bauthor{\binits{M.} \bsnm{{MacLeod}}},
\bauthor{\binits{S.} \bsnm{{Rosswog}}},
\batitle{{Tidal Disruptions of White Dwarfs: Theoretical Models and
  Observational Prospects}}.
\bjtitle{\ssr}
\bvolume{216}(\bissue{3}),
\bfpage{39}
(\byear{2020}).
doi:\doiurl{10.1007/s11214-020-00661-2}
\end{barticle}
\endbibitem

\bibitem[\protect\citeauthoryear{{Mandel} et~al.}{2008}]{MandelEtAl2008}
\begin{barticle}
\bauthor{\binits{I.} \bsnm{{Mandel}}},
\bauthor{\binits{D.A.} \bsnm{{Brown}}},
\bauthor{\binits{J.R.} \bsnm{{Gair}}},
\bauthor{\binits{M.C.} \bsnm{{Miller}}},
\batitle{{Rates and Characteristics of Intermediate Mass Ratio Inspirals
  Detectable by Advanced LIGO}}.
\bjtitle{ApJ}
\bvolume{681},
\bfpage{1431}--\blpage{1447}
(\byear{2008}).
doi:\doiurl{10.1086/588246}
\end{barticle}
\endbibitem

\bibitem[\protect\citeauthoryear{{Menou} et~al.}{2008}]{2008NewAR..51..884M}
\begin{barticle}
\bauthor{\binits{K.} \bsnm{{Menou}}},
\bauthor{\binits{Z.} \bsnm{{Haiman}}},
\bauthor{\binits{B.} \bsnm{{Kocsis}}},
\batitle{{Cosmological physics with black holes (and possibly white dwarfs)}}.
\bjtitle{\nar}
\bvolume{51}(\bissue{10-12}),
\bfpage{884}--\blpage{890}
(\byear{2008}).
doi:\doiurl{10.1016/j.newar.2008.03.020}
\end{barticle}
\endbibitem

\bibitem[\protect\citeauthoryear{{Meynet} et~al.}{1994}]{MMSSC94}
\begin{barticle}
\bauthor{\binits{G.} \bsnm{{Meynet}}},
\bauthor{\binits{A.} \bsnm{{Maeder}}},
\bauthor{\binits{G.} \bsnm{{Schaller}}},
\bauthor{\binits{D.} \bsnm{{Schaerer}}},
\bauthor{\binits{C.} \bsnm{{Charbonnel}}},
\batitle{Grids of massive stars with high mass loss rates. v. from 12 to 120
  m\_$\{$sun$\}$\_ at z=0.001, 0.004, 0.008, 0.020 and 0.040}.
\bjtitle{A\&AS}
\bvolume{103},
\bfpage{97}--\blpage{105}
(\byear{1994})
\end{barticle}
\endbibitem

\bibitem[\protect\citeauthoryear{{Mezcua}}{2017}]{Mezcua2017}
\begin{barticle}
\bauthor{\binits{M.} \bsnm{{Mezcua}}},
\batitle{{Observational evidence for intermediate-mass black holes}}.
\bjtitle{International Journal of Modern Physics D}
\bvolume{26},
\bfpage{1730021}
(\byear{2017}).
doi:\doiurl{10.1142/S021827181730021X}
\end{barticle}
\endbibitem

\bibitem[\protect\citeauthoryear{{Miller} et~al.}{2005}]{MillerEtAl05}
\begin{barticle}
\bauthor{\binits{M.C.} \bsnm{{Miller}}},
\bauthor{\binits{M.} \bsnm{{Freitag}}},
\bauthor{\binits{D.P.} \bsnm{{Hamilton}}},
\bauthor{\binits{V.M.} \bsnm{{Lauburg}}},
\batitle{{Binary Encounters with Supermassive Black Holes: Zero-Eccentricity
  LISA Events}}.
\bjtitle{ApJ Lett.}
\bvolume{631},
\bfpage{117}--\blpage{120}
(\byear{2005}).
doi:\doiurl{10.1086/497335}
\end{barticle}
\endbibitem

\bibitem[\protect\citeauthoryear{{Misner}
  et~al.}{1973}]{MisnerThorneWheeler1973}
\begin{bbook}
\bauthor{\binits{C.W.} \bsnm{{Misner}}},
\bauthor{\binits{K.S.} \bsnm{{Thorne}}},
\bauthor{\binits{J.A.} \bsnm{{Wheeler}}},
\bbtitle{{Gravitation}}
\byear{1973}
\end{bbook}
\endbibitem

\bibitem[\protect\citeauthoryear{Mroue et~al.}{2013}]{BBHCatalog}
\begin{barticle}
\bauthor{\binits{A.H.} \bsnm{Mroue}}, \betal,
\batitle{{Catalog of 174 Binary Black Hole Simulations for Gravitational Wave
  Astronomy}}.
\bjtitle{Phys. Rev. Lett.}
\bvolume{111}(\bissue{24}),
\bfpage{241104}
(\byear{2013}).
doi:\doiurl{10.1103/PhysRevLett.111.241104}
\end{barticle}
\endbibitem

\bibitem[\protect\citeauthoryear{{O'Leary} et~al.}{2009}]{OlearyEtAl09}
\begin{barticle}
\bauthor{\binits{R.M.} \bsnm{{O'Leary}}},
\bauthor{\binits{B.} \bsnm{{Kocsis}}},
\bauthor{\binits{A.} \bsnm{{Loeb}}},
\batitle{{Gravitational waves from scattering of stellar-mass black holes in
  galactic nuclei}}.
\bjtitle{MNRAS}
\bvolume{395},
\bfpage{2127}--\blpage{2146}
(\byear{2009}).
doi:\doiurl{10.1111/j.1365-2966.2009.14653.x}
\end{barticle}
\endbibitem

\bibitem[\protect\citeauthoryear{{Panamarev}
  et~al.}{2019}]{2019MNRAS.484.3279P}
\begin{barticle}
\bauthor{\binits{T.} \bsnm{{Panamarev}}},
\bauthor{\binits{A.} \bsnm{{Just}}},
\bauthor{\binits{R.} \bsnm{{Spurzem}}},
\bauthor{\binits{P.} \bsnm{{Berczik}}},
\bauthor{\binits{L.} \bsnm{{Wang}}},
\bauthor{\binits{M.} \bsnm{{Arca Sedda}}},
\batitle{{Direct N-body simulation of the Galactic centre}}.
\bjtitle{\mnras}
\bvolume{484}(\bissue{3}),
\bfpage{3279}--\blpage{3290}
(\byear{2019}).
doi:\doiurl{10.1093/mnras/stz208}
\end{barticle}
\endbibitem

\bibitem[\protect\citeauthoryear{{Peebles}}{1972}]{Peebles1972}
\begin{barticle}
\bauthor{\binits{P.J.E.} \bsnm{{Peebles}}},
\batitle{{Star Distribution Near a Collapsed Object}}.
\bjtitle{ApJ}
\bvolume{178},
\bfpage{371}--\blpage{376}
(\byear{1972}).
doi:\doiurl{10.1086/151797}
\end{barticle}
\endbibitem

\bibitem[\protect\citeauthoryear{{Pei{\ss}ker}
  et~al.}{2020}]{2020ApJ...899...50P}
\begin{barticle}
\bauthor{\binits{F.} \bsnm{{Pei{\ss}ker}}},
\bauthor{\binits{A.} \bsnm{{Eckart}}},
\bauthor{\binits{M.} \bsnm{{Zaja{\v{c}}ek}}},
\bauthor{\binits{B.} \bsnm{{Ali}}},
\bauthor{\binits{M.} \bsnm{{Parsa}}},
\batitle{{S62 and S4711: Indications of a Population of Faint Fast-moving Stars
  inside the S2 Orbit{\textemdash}S4711 on a 7.6 yr Orbit around Sgr A*}}.
\bjtitle{\apj}
\bvolume{899}(\bissue{1}),
\bfpage{50}
(\byear{2020}).
doi:\doiurl{10.3847/1538-4357/ab9c1c}
\end{barticle}
\endbibitem

\bibitem[\protect\citeauthoryear{{Peters}}{1964}]{Peters64}
\begin{barticle}
\bauthor{\binits{P.C.} \bsnm{{Peters}}},
\batitle{{Gravitational Radiation and the Motion of Two Point Masses}}.
\bjtitle{Physical Review}
\bvolume{136},
\bfpage{1224}--\blpage{1232}
(\byear{1964})
\end{barticle}
\endbibitem

\bibitem[\protect\citeauthoryear{{Peters} and {Mathews}}{1963}]{PM63}
\begin{barticle}
\bauthor{\binits{P.C.} \bsnm{{Peters}}},
\bauthor{\binits{J.} \bsnm{{Mathews}}},
\batitle{{Gravitational Radiation from Point Masses in a Keplerian Orbit}}.
\bjtitle{Physical Review}
\bvolume{131},
\bfpage{435}--\blpage{440}
(\byear{1963})
\end{barticle}
\endbibitem

\bibitem[\protect\citeauthoryear{{Pierro} et~al.}{2001}]{PPSLR01}
\begin{barticle}
\bauthor{\binits{V.} \bsnm{{Pierro}}},
\bauthor{\binits{I.M.} \bsnm{{Pinto}}},
\bauthor{\binits{A.D.} \bsnm{{Spallicci}}},
\bauthor{\binits{E.} \bsnm{{Laserra}}},
\bauthor{\binits{F.} \bsnm{{Recano}}},
\batitle{{Fast and accurate computational tools for gravitational waveforms
  from binary stars with any orbital eccentricity}}.
\bjtitle{MNRAS}
\bvolume{325},
\bfpage{358}--\blpage{372}
(\byear{2001})
\end{barticle}
\endbibitem

\bibitem[\protect\citeauthoryear{Poisson et~al.}{2011a}]{PoissonEtal2011}
\begin{barticle}
\bauthor{\binits{E.} \bsnm{Poisson}},
\bauthor{\binits{A.} \bsnm{Pound}},
\bauthor{\binits{I.} \bsnm{Vega}},
\batitle{{The Motion of point particles in curved spacetime}}.
\bjtitle{Living Rev. Rel.}
\bvolume{14},
\bfpage{7}
(\byear{2011}a).
doi:\doiurl{10.12942/lrr-2011-7}
\end{barticle}
\endbibitem

\bibitem[\protect\citeauthoryear{Poisson et~al.}{2011b}]{Poisson2011}
\begin{barticle}
\bauthor{\binits{E.} \bsnm{Poisson}},
\bauthor{\binits{A.} \bsnm{Pound}},
\bauthor{\binits{I.} \bsnm{Vega}},
\batitle{{The Motion of point particles in curved spacetime}}.
\bjtitle{Living Rev. Rel.}
\bvolume{14},
\bfpage{7}
(\byear{2011}b).
doi:\doiurl{10.12942/lrr-2011-7}
\end{barticle}
\endbibitem

\bibitem[\protect\citeauthoryear{{Portegies Zwart} et~al.}{1998}]{PZHMMcM98}
\begin{barticle}
\bauthor{\binits{S.F.} \bsnm{{Portegies Zwart}}},
\bauthor{\binits{P.} \bsnm{{Hut}}},
\bauthor{\binits{J.} \bsnm{{Makino}}},
\bauthor{\binits{S.L.W.} \bsnm{{McMillan}}},
\batitle{On the dissolution of evolving star clusters}.
\bjtitle{A\&A}
\bvolume{337},
\bfpage{363}--\blpage{371}
(\byear{1998})
\end{barticle}
\endbibitem

\bibitem[\protect\citeauthoryear{{Press}}{1977}]{Press1977}
\begin{barticle}
\bauthor{\binits{W.H.} \bsnm{{Press}}},
\batitle{{Gravitational radiation from sources which extend into their own wave
  zone}}.
\bjtitle{Ph.Rv.D}
\bvolume{15}(\bissue{4}),
\bfpage{965}--\blpage{968}
(\byear{1977}).
doi:\doiurl{10.1103/PhysRevD.15.965}
\end{barticle}
\endbibitem

\bibitem[\protect\citeauthoryear{{Preto} and
  {Amaro-Seoane}}{2010}]{PretoAmaroSeoane10}
\begin{barticle}
\bauthor{\binits{M.} \bsnm{{Preto}}},
\bauthor{\binits{P.} \bsnm{{Amaro-Seoane}}},
\batitle{{On Strong Mass Segregation Around a Massive Black Hole: Implications
  for Lower-Frequency Gravitational-Wave Astrophysics}}.
\bjtitle{ApJ Lett.}
\bvolume{708},
\bfpage{42}--\blpage{46}
(\byear{2010}).
doi:\doiurl{10.1088/2041-8205/708/1/L42}
\end{barticle}
\endbibitem

\bibitem[\protect\citeauthoryear{{Quinlan} and
  {Shapiro}}{1989}]{QuinlanShapiro1989}
\begin{barticle}
\bauthor{\binits{G.D.} \bsnm{{Quinlan}}},
\bauthor{\binits{S.L.} \bsnm{{Shapiro}}},
\batitle{{Dynamical evolution of dense clusters of compact stars}}.
\bjtitle{ApJ}
\bvolume{343},
\bfpage{725}--\blpage{749}
(\byear{1989}).
doi:\doiurl{10.1086/167745}
\end{barticle}
\endbibitem

\bibitem[\protect\citeauthoryear{{Rauch} and {Tremaine}}{1996a}]{RT96}
\begin{barticle}
\bauthor{\binits{K.P.} \bsnm{{Rauch}}},
\bauthor{\binits{S.} \bsnm{{Tremaine}}},
\batitle{Resonant relaxation in stellar systems}.
\bjtitle{New Astronomy}
\bvolume{1},
\bfpage{149}--\blpage{170}
(\byear{1996}a)
\end{barticle}
\endbibitem

\bibitem[\protect\citeauthoryear{{Rauch} and
  {Tremaine}}{1996b}]{1996NewA....1..149R}
\begin{barticle}
\bauthor{\binits{K.P.} \bsnm{{Rauch}}},
\bauthor{\binits{S.} \bsnm{{Tremaine}}},
\batitle{{Resonant relaxation in stellar systems}}.
\bjtitle{New A.}
\bvolume{1}(\bissue{2}),
\bfpage{149}--\blpage{170}
(\byear{1996}b).
doi:\doiurl{10.1016/S1384-1076(96)00012-7}
\end{barticle}
\endbibitem

\bibitem[\protect\citeauthoryear{{Rees}}{1988}]{Rees88}
\begin{barticle}
\bauthor{\binits{M.J.} \bsnm{{Rees}}},
\batitle{{Tidal disruption of stars by black holes of 10 to the 6th-10 to the
  8th solar masses in nearby galaxies}}.
\bjtitle{Nat}
\bvolume{333},
\bfpage{523}--\blpage{528}
(\byear{1988}).
doi:\doiurl{10.1038/333523a0}
\end{barticle}
\endbibitem

\bibitem[\protect\citeauthoryear{{Rossi} et~al.}{2020}]{2020arXiv200512528R}
\begin{botherref}
\oauthor{\binits{E.M.} \bsnm{{Rossi}}},
\oauthor{\binits{N.C.} \bsnm{{Stone}}},
\oauthor{\binits{J.A.P.} \bsnm{{Law-Smith}}},
\oauthor{\binits{M.} \bsnm{{MacLeod}}},
\oauthor{\binits{G.} \bsnm{{Lodato}}},
\oauthor{\binits{J.L.} \bsnm{{Dai}}},
\oauthor{\binits{I.} \bsnm{{Mand el}}},
{The Process of Stellar Tidal Disruption by Supermassive Black Holes. The first
  pericenter passage}.
arXiv e-prints,
2005--12528
(2020)
\end{botherref}
\endbibitem

\bibitem[\protect\citeauthoryear{{Saslaw}}{1985}]{Saslaw85}
\begin{bbook}
\bauthor{\binits{W.C.} \bsnm{{Saslaw}}},
\bbtitle{Gravitational physics of stellar and galactic systems}
(\bpublisher{Cambridge, Cambridge University Press}, \blocation{???},
  \byear{1985})
\end{bbook}
\endbibitem

\bibitem[\protect\citeauthoryear{{Sathyaprakash}
  et~al.}{2012}]{2012CQGra..29l4013S}
\begin{barticle}
\bauthor{\binits{B.} \bsnm{{Sathyaprakash}}},
\bauthor{\binits{M.} \bsnm{{Abernathy}}},
\bauthor{\binits{F.} \bsnm{{Acernese}}},
\bauthor{\binits{P.} \bsnm{{Ajith}}},
\bauthor{\binits{B.} \bsnm{{Allen}}},
\bauthor{\binits{P.} \bsnm{{Amaro-Seoane}}},
\bauthor{\binits{N.} \bsnm{{Andersson}}},
\bauthor{\binits{S.} \bsnm{{Aoudia}}},
\bauthor{\binits{K.} \bsnm{{Arun}}},
\bauthor{\binits{P.} \bsnm{{Astone}}},
\bauthor{\bparticle{et} \bsnm{al.}},
\batitle{{Scientific objectives of Einstein Telescope}}.
\bjtitle{Classical and Quantum Gravity}
\bvolume{29}(\bissue{12}),
\bfpage{124013}
(\byear{2012}).
doi:\doiurl{10.1088/0264-9381/29/12/124013}
\end{barticle}
\endbibitem

\bibitem[\protect\citeauthoryear{{Schaller} et~al.}{1992}]{SSMM92}
\begin{barticle}
\bauthor{\binits{G.} \bsnm{{Schaller}}},
\bauthor{\binits{D.} \bsnm{{Schaerer}}},
\bauthor{\binits{G.} \bsnm{{Meynet}}},
\bauthor{\binits{A.} \bsnm{{Maeder}}},
\batitle{New grids of stellar models from 0.8 to 120 solar masses at z = 0.020
  and z = 0.001}.
\bjtitle{A\&AS}
\bvolume{96},
\bfpage{269}--\blpage{331}
(\byear{1992})
\end{barticle}
\endbibitem

\bibitem[\protect\citeauthoryear{{Schmidt}}{2002}]{Schmidt2002}
\begin{barticle}
\bauthor{\binits{W.} \bsnm{{Schmidt}}},
\batitle{{Celestial mechanics in Kerr spacetime}}.
\bjtitle{Classical and Quantum Gravity}
\bvolume{19},
\bfpage{2743}--\blpage{2764}
(\byear{2002}).
doi:\doiurl{10.1088/0264-9381/19/10/314}
\end{barticle}
\endbibitem

\bibitem[\protect\citeauthoryear{{Sch{\"o}del}
  et~al.}{2018}]{2018A&A...609A..27S}
\begin{barticle}
\bauthor{\binits{R.} \bsnm{{Sch{\"o}del}}},
\bauthor{\binits{E.} \bsnm{{Gallego-Cano}}},
\bauthor{\binits{H.} \bsnm{{Dong}}},
\bauthor{\binits{F.} \bsnm{{Nogueras-Lara}}},
\bauthor{\binits{A.T.} \bsnm{{Gallego-Calvente}}},
\bauthor{\binits{P.} \bsnm{{Amaro-Seoane}}},
\bauthor{\binits{H.} \bsnm{{Baumgardt}}},
\batitle{{The distribution of stars around the Milky Way's central black hole.
  II. Diffuse light from sub-giants and dwarfs}}.
\bjtitle{\aap}
\bvolume{609},
\bfpage{27}
(\byear{2018}).
doi:\doiurl{10.1051/0004-6361/201730452}
\end{barticle}
\endbibitem

\bibitem[\protect\citeauthoryear{{Schutz}}{1989}]{Schutz1989}
\begin{barticle}
\bauthor{\binits{B.F.} \bsnm{{Schutz}}},
\batitle{{REVIEW ARTICLE: Gravitational wave sources and their detectability}}.
\bjtitle{Classical and Quantum Gravity}
\bvolume{6}(\bissue{12}),
\bfpage{1761}--\blpage{1780}
(\byear{1989}).
doi:\doiurl{10.1088/0264-9381/6/12/006}
\end{barticle}
\endbibitem

\bibitem[\protect\citeauthoryear{{Sesana} et~al.}{2008}]{2008MNRAS.391..718S}
\begin{barticle}
\bauthor{\binits{A.} \bsnm{{Sesana}}},
\bauthor{\binits{A.} \bsnm{{Vecchio}}},
\bauthor{\binits{M.} \bsnm{{Eracleous}}},
\bauthor{\binits{S.} \bsnm{{Sigurdsson}}},
\batitle{{Observing white dwarfs orbiting massive black holes in the
  gravitational wave and electro-magnetic window}}.
\bjtitle{\mnras}
\bvolume{391}(\bissue{2}),
\bfpage{718}--\blpage{726}
(\byear{2008}).
doi:\doiurl{10.1111/j.1365-2966.2008.13904.x}
\end{barticle}
\endbibitem

\bibitem[\protect\citeauthoryear{{Sesana} et~al.}{2014}]{SesanaEtAl2014}
\begin{barticle}
\bauthor{\binits{A.} \bsnm{{Sesana}}},
\bauthor{\binits{E.} \bsnm{{Barausse}}},
\bauthor{\binits{M.} \bsnm{{Dotti}}},
\bauthor{\binits{E.M.} \bsnm{{Rossi}}},
\batitle{{Linking the Spin Evolution of Massive Black Holes to Galaxy
  Kinematics}}.
\bjtitle{ApJ}
\bvolume{794}(\bissue{2}),
\bfpage{104}
(\byear{2014}).
doi:\doiurl{10.1088/0004-637X/794/2/104}
\end{barticle}
\endbibitem

\bibitem[\protect\citeauthoryear{{Sigurdsson} and
  {Rees}}{1997}]{1997MNRAS.284..318S}
\begin{barticle}
\bauthor{\binits{S.} \bsnm{{Sigurdsson}}},
\bauthor{\binits{M.J.} \bsnm{{Rees}}},
\batitle{{Capture of stellar mass compact objects by massive black holes in
  galactic cusps}}.
\bjtitle{\mnras}
\bvolume{284}(\bissue{2}),
\bfpage{318}--\blpage{326}
(\byear{1997}).
doi:\doiurl{10.1093/mnras/284.2.318}
\end{barticle}
\endbibitem

\bibitem[\protect\citeauthoryear{{Sopuerta} and
  {Yunes}}{2011}]{SopuertaYunes2011}
\begin{barticle}
\bauthor{\binits{C.F.} \bsnm{{Sopuerta}}},
\bauthor{\binits{N.} \bsnm{{Yunes}}},
\batitle{{New Kludge scheme for the construction of approximate waveforms for
  extreme-mass-ratio inspirals}}.
\bjtitle{Phys. Rev. D}
\bvolume{84}(\bissue{12}),
\bfpage{124060}
(\byear{2011}).
doi:\doiurl{10.1103/PhysRevD.84.124060}
\end{barticle}
\endbibitem

\bibitem[\protect\citeauthoryear{{Spitzer}}{1987}]{Spitzer87}
\begin{bbook}
\bauthor{\binits{L.} \bsnm{{Spitzer}}},
\bbtitle{{Dynamical evolution of globular clusters}}
(\bpublisher{Princeton, NJ, Princeton University Press, 1987, 191 p.},
  \blocation{???}, \byear{1987})
\end{bbook}
\endbibitem

\bibitem[\protect\citeauthoryear{{Spitzer} and {Hart}}{1971}]{SH71a}
\begin{barticle}
\bauthor{\binits{L.J.} \bsnm{{Spitzer}}},
\bauthor{\binits{M.H.} \bsnm{{Hart}}},
\batitle{Random gravitational encounters and the evolution of spherical
  systems. i. method}.
\bjtitle{ApJ}
\bvolume{164},
\bfpage{399}
(\byear{1971})
\end{barticle}
\endbibitem

\bibitem[\protect\citeauthoryear{{Spurzem}}{1999}]{Spurzem99}
\begin{barticle}
\bauthor{\binits{R.} \bsnm{{Spurzem}}},
\batitle{Direct $n$-body simulations}.
\bjtitle{Journal of Computational and Applied Mathematics}
\bvolume{109},
\bfpage{407}--\blpage{432}
(\byear{1999})
\end{barticle}
\endbibitem

\bibitem[\protect\citeauthoryear{{Spurzem} and {Aarseth}}{1996}]{SA96}
\begin{barticle}
\bauthor{\binits{R.} \bsnm{{Spurzem}}},
\bauthor{\binits{S.J.} \bsnm{{Aarseth}}},
\batitle{Direct collisional simulation of 100000 particles past core collapse}.
\bjtitle{MNRAS}
\bvolume{282},
\bfpage{19}
(\byear{1996})
\end{barticle}
\endbibitem

\bibitem[\protect\citeauthoryear{{Sridhar} and
  {Touma}}{2016}]{2016MNRAS.458.4143S}
\begin{barticle}
\bauthor{\binits{S.} \bsnm{{Sridhar}}},
\bauthor{\binits{J.R.} \bsnm{{Touma}}},
\batitle{{Stellar dynamics around a massive black hole - II. Resonant
  relaxation}}.
\bjtitle{\mnras}
\bvolume{458}(\bissue{4}),
\bfpage{4143}--\blpage{4161}
(\byear{2016}).
doi:\doiurl{10.1093/mnras/stw543}
\end{barticle}
\endbibitem

\bibitem[\protect\citeauthoryear{{Stone} et~al.}{2020}]{StoneEtAl2020}
\begin{barticle}
\bauthor{\binits{N.C.} \bsnm{{Stone}}},
\bauthor{\binits{E.} \bsnm{{Vasiliev}}},
\bauthor{\binits{M.} \bsnm{{Kesden}}},
\bauthor{\binits{E.M.} \bsnm{{Rossi}}},
\bauthor{\binits{H.B.} \bsnm{{Perets}}},
\bauthor{\binits{P.} \bsnm{{Amaro-Seoane}}},
\batitle{{Rates of Stellar Tidal Disruption}}.
\bjtitle{Space Science Reviews}
\bvolume{216}(\bissue{3}),
\bfpage{35}
(\byear{2020}).
doi:\doiurl{10.1007/s11214-020-00651-4}
\end{barticle}
\endbibitem

\bibitem[\protect\citeauthoryear{{Syer} and {Ulmer}}{1999}]{SU99}
\begin{barticle}
\bauthor{\binits{D.} \bsnm{{Syer}}},
\bauthor{\binits{A.} \bsnm{{Ulmer}}},
\batitle{Tidal disruption rates of stars in observed galaxies}.
\bjtitle{MNRAS}
\bvolume{306},
\bfpage{35}--\blpage{42}
(\byear{1999})
\end{barticle}
\endbibitem

\bibitem[\protect\citeauthoryear{{Takahashi} and {Portegies
  Zwart}}{1998}]{TPZ98}
\begin{barticle}
\bauthor{\binits{K.} \bsnm{{Takahashi}}},
\bauthor{\binits{S.F.} \bsnm{{Portegies Zwart}}},
\batitle{The disruption of globular star clusters in the galaxy: A comparative
  analysis between fokker-planck and n-body models}.
\bjtitle{ApJ Lett.}
\bvolume{503},
\bfpage{49}
(\byear{1998})
\end{barticle}
\endbibitem

\bibitem[\protect\citeauthoryear{{Teukolsky}}{2015}]{Teukolsky2015}
\begin{barticle}
\bauthor{\binits{S.A.} \bsnm{{Teukolsky}}},
\batitle{{The Kerr metric}}.
\bjtitle{Classical and Quantum Gravity}
\bvolume{32}(\bissue{12}),
\bfpage{124006}
(\byear{2015}).
doi:\doiurl{10.1088/0264-9381/32/12/124006}
\end{barticle}
\endbibitem

\bibitem[\protect\citeauthoryear{{Thorne}}{1987}]{Thorne87}
\begin{botherref}
\oauthor{\binits{K.S.} \bsnm{{Thorne}}},
{Gravitational radiation.},
ed. by S.W. Hawking, W. Israel
1987,
pp. 330--458
\end{botherref}
\endbibitem

\bibitem[\protect\citeauthoryear{{Torres-Orjuela}
  et~al.}{2020}]{Torres-OrjuelaEtAl2020}
\begin{barticle}
\bauthor{\binits{A.} \bsnm{{Torres-Orjuela}}},
\bauthor{\binits{X.} \bsnm{{Chen}}},
\bauthor{\binits{P.} \bsnm{{Amaro-Seoane}}},
\batitle{{Phase shift of gravitational waves induced by aberration}}.
\bjtitle{Ph. Rv. D}
\bvolume{101}(\bissue{8}),
\bfpage{083028}
(\byear{2020}).
doi:\doiurl{10.1103/PhysRevD.101.083028}
\end{barticle}
\endbibitem

\bibitem[\protect\citeauthoryear{{Torres-Orjuela}
  et~al.}{2019}]{Torres-OrjuelaEtAl2019}
\begin{barticle}
\bauthor{\binits{A.} \bsnm{{Torres-Orjuela}}},
\bauthor{\binits{X.} \bsnm{{Chen}}},
\bauthor{\binits{Z.} \bsnm{{Cao}}},
\bauthor{\binits{P.} \bsnm{{Amaro-Seoane}}},
\bauthor{\binits{P.} \bsnm{{Peng}}},
\batitle{{Detecting the beaming effect of gravitational waves}}.
\bjtitle{Ph. Rv. D}
\bvolume{100}(\bissue{6}),
\bfpage{063012}
(\byear{2019}).
doi:\doiurl{10.1103/PhysRevD.100.063012}
\end{barticle}
\endbibitem

\bibitem[\protect\citeauthoryear{{Tremaine} et~al.}{1994}]{Tremaine94}
\begin{barticle}
\bauthor{\binits{S.} \bsnm{{Tremaine}}},
\bauthor{\binits{D.O.} \bsnm{{Richstone}}},
\bauthor{\binits{Y.} \bsnm{{Byun}}},
\bauthor{\binits{A.} \bsnm{{Dressler}}},
\bauthor{\binits{S.M.} \bsnm{{Faber}}},
\bauthor{\binits{C.} \bsnm{{Grillmair}}},
\bauthor{\binits{J.} \bsnm{{Kormendy}}},
\bauthor{\binits{T.R.} \bsnm{{Lauer}}},
\batitle{{A family of models for spherical stellar systems}}.
\bjtitle{AJ}
\bvolume{107},
\bfpage{634}--\blpage{644}
(\byear{1994})
\end{barticle}
\endbibitem

\bibitem[\protect\citeauthoryear{{Visser}}{2007}]{Visser2007}
\begin{botherref}
\oauthor{\binits{M.} \bsnm{{Visser}}},
{The Kerr spacetime: A brief introduction}.
arXiv e-prints,
0706--0622
(2007)
\end{botherref}
\endbibitem

\bibitem[\protect\citeauthoryear{{Wang} and {Merritt}}{2004}]{WM04}
\begin{barticle}
\bauthor{\binits{J.} \bsnm{{Wang}}},
\bauthor{\binits{D.} \bsnm{{Merritt}}},
\batitle{{Revised Rates of Stellar Disruption in Galactic Nuclei}}.
\bjtitle{ApJ}
\bvolume{600},
\bfpage{149}--\blpage{161}
(\byear{2004})
\end{barticle}
\endbibitem

\bibitem[\protect\citeauthoryear{{Wegg} et~al.}{2017}]{WeggEtAl2017}
\begin{barticle}
\bauthor{\binits{C.} \bsnm{{Wegg}}},
\bauthor{\binits{O.} \bsnm{{Gerhard}}},
\bauthor{\binits{M.} \bsnm{{Portail}}},
\batitle{{The Initial Mass Function of the Inner Galaxy Measured from OGLE-III
  Microlensing Timescales}}.
\bjtitle{ApJ Lett.}
\bvolume{843},
\bfpage{5}
(\byear{2017}).
doi:\doiurl{10.3847/2041-8213/aa794e}
\end{barticle}
\endbibitem

\bibitem[\protect\citeauthoryear{{Wen}}{2003}]{Wen2003}
\begin{barticle}
\bauthor{\binits{L.} \bsnm{{Wen}}},
\batitle{{On the Eccentricity Distribution of Coalescing Black Hole Binaries
  Driven by the Kozai Mechanism in Globular Clusters}}.
\bjtitle{ApJ}
\bvolume{598},
\bfpage{419}--\blpage{430}
(\byear{2003}).
doi:\doiurl{10.1086/378794}
\end{barticle}
\endbibitem

\bibitem[\protect\citeauthoryear{{Will}}{2011}]{Will2011}
\begin{barticle}
\bauthor{\binits{C.M.} \bsnm{{Will}}},
\batitle{{Inaugural Article: On the unreasonable effectiveness of the
  post-Newtonian approximation in gravitational physics}}.
\bjtitle{Proceedings of the National Academy of Science}
\bvolume{108}(\bissue{15}),
\bfpage{5938}--\blpage{5945}
(\byear{2011}).
doi:\doiurl{10.1073/pnas.1103127108}
\end{barticle}
\endbibitem

\bibitem[\protect\citeauthoryear{{Will}}{2014}]{Will2014}
\begin{barticle}
\bauthor{\binits{C.M.} \bsnm{{Will}}},
\batitle{{Incorporating post-Newtonian effects in N-body dynamics}}.
\bjtitle{Ph. Rv. D}
\bvolume{89}(\bissue{4}),
\bfpage{044043}
(\byear{2014}).
doi:\doiurl{10.1103/PhysRevD.89.044043}
\end{barticle}
\endbibitem

\bibitem[\protect\citeauthoryear{{Zwick} et~al.}{2019}]{ZwickEtAl2019}
\begin{botherref}
\oauthor{\binits{L.} \bsnm{{Zwick}}},
\oauthor{\binits{P.R.} \bsnm{{Capelo}}},
\oauthor{\binits{E.} \bsnm{{Bortolas}}},
\oauthor{\binits{L.} \bsnm{{Mayer}}},
\oauthor{\binits{P.} \bsnm{{Amaro-Seoane}}},
{Improved gravitational radiation time-scales: significance for LISA and
  LIGO-Virgo sources}.
arXiv e-prints,
1911--06024
(2019)
\end{botherref}
\endbibitem

\end{thebibliography}
\end{document}